\documentclass[12pt,english,final,letterpaper,notitlepage]{article}
\usepackage[T1]{fontenc}
\usepackage{amsmath}
\usepackage{graphicx}
\usepackage{amssymb}
\usepackage{verbatim}
\usepackage{dcolumn}
\usepackage{siunitx}
\usepackage{amsfonts}
\usepackage{amsthm}
\usepackage{enumerate}
\usepackage{rotating}
\usepackage[gen]{eurosym}

\usepackage{soul}

\usepackage{caption}
\usepackage{subcaption}

\usepackage{bbm}

\usepackage{pdflscape}

\usepackage[title]{appendix}

\usepackage{color,colortbl}

\definecolor{Yell}{rgb}{1,1,0}

\usepackage{babel}
\usepackage{setspace} %\doublespacing

\usepackage{natbib}
\usepackage{hyperref}

\def\sym#1{\ifmmode^{#1}\else\(^{#1}\)\fi}

\makeatletter

\newtheorem{theorem}{Theorem}
\newtheorem{proposition}{Proposition}

\newtheorem{remark}[theorem]{Remark}

	\usepackage{hyperref}
	\hypersetup{
  colorlinks = true,
  citecolor  = blue,
  linkcolor  =blue,
  urlcolor  =blue,
  filecolor=blue
}

\usepackage[margin=1.0in]{geometry}
\title{A Run on Fossil Fuel? \\
Climate Change and Transition Risk\footnote{Barnett: Business Administration, 300 E Lemon St, Tempe, AZ 85287; michael.d.barnett@asu.edu. I am particularly grateful for the guidance and support of Lars Peter Hansen and Pietro Veronesi, as well as Michael Greenstone and Bryan Kelly. I want to thank Ufuk Akcigit, Ravi Bansal, Buz Brock, Stefano Giglio, Amir Jina, Ryan Kellogg, Paymon Khorrami, Ralph Koijen, Moritz Lenel, Stefan Nagel, Alan Sanstad, Rob Townsend, Willem Van Vliet, and Amir Yaron for their comments and suggestions. This paper also greatly benefitted from the comments and suggestions of seminar participants at BYU, USU, USC, UNC, Duke, UBC, Yale, Imperial College, Texas A\&M, the Federal Reserve Board of Governors, Indiana, ASU, the Federal Reserve Bank of Dallas, as well as the Capital Theory Working Group, Booth School of Business Finance Brownbag, Booth School of Business Finance Seminar, and especially the Economic Dynamics and Financial Markets Working Group at the University of Chicago. I am grateful for the financial support of the National Science Foundation, the Fama-Miller Center, the Energy Policy Institute at the University of Chicago (EPIC), the Stevanovich Center for Financial Mathematics, and the University of Chicago. This paper is based on my PhD Thesis titled ``A Run on Oil: Climate Policy, Stranded Assets, and Asset Prices.''
}
}

\author{Michael Barnett \\ \small{Arizona State University }
}

\date{}

\begin{document}
\onehalfspacing
% \doublespacing

\maketitle

\thispagestyle{empty}

\renewcommand{\thefootnote}{\arabic{footnote}}

\begin{abstract}

I study the dynamic, general equilibrium implications of climate-change-linked transition risk on macroeconomic outcomes and asset prices. Climate-change-linked expectations of fossil fuel restrictions can produce a ``run on fossil fuels'' with accelerated production and decreasing spot prices, or a ``reverse run'' with restrained production and increased spot prices. The response depends on the expected economic consequences of the anticipated transition shock, and existing climate policies. Fossil fuel firm prices decrease in each case. I use a novel empirical measure of innovations in climate-related transition risk likelihood to show that dynamic empirical responses are consistent with a ``run on fossil fuel.'' 

\end{abstract}

\thispagestyle{empty}

\newpage
\pagenumbering{arabic}
\renewcommand{\thepage} {\arabic{page}}
\renewcommand{\thefootnote}{\fnsymbol{footnote}}

\cleardoublepage
\setcounter{page}{0}

\cleardoublepage
\setcounter{page}{1}

\renewcommand{\thefootnote}{\arabic{footnote}}

\section{Introduction}

% \onehalfspacing
\doublespacing

The ramifications of global climate change have already begun, and the impacts on households, financial markets, and aggregate social welfare are expected to become increasingly more severe in the near- and long-term. Importantly, potential physical consequences from climate change are only part of the story. Government, policy, and corporate decision-makers around the world are taking actions via policy implementation and investment in clean technologies in response to climate change, leading to the possibility of a ``climate transition'' or ``green transition'' away from fossil fuels. The implications of these actions are further influenced by risks and unknowns about the implementation and timing of this transition. Thus, in order to understand the full consequences of climate change on the global economy and financial markets requires analysis not just of possible physical climate risks, but also the potentially significant impacts of a transition to a green economy.

In this paper, I analyze the macroeconomic and financial market implications related to the expected arrival of a climate-change-linked green economic transition. To my knowledge, this paper provides one of the first explorations of the interconnected macro-asset pricing implications of the endogenous and dynamic feedback effects between production-generated emissions, economic damages resulting from climate change induced by by these emissions, and the likelihood of a climate change-linked transition shock for different types of transition shocks limiting future fossil-fuel production. Using an empirically-calibrated dynamic general equilibrium model, I show the quantitative implications related to various tax-related and technology-related transition risk scenarios, and the implications of extant climate policies for these scenarios. Finally, I use different empirical estimation exercises related to fossil fuel production and price movements around events that shift the transition risk likelihood to identify the theoretical transition risk scenarios most consistent with observable outcomes.

I start by constructing and solving a dynamic general equilibrium, production-based asset pricing framework where oil and coal fossil fuel production and extraction decisions are made based on the demand for fossil fuels, remaining reserves, and climate change impacts. Emissions result from the endogenous choice of fossil fuels used in production. The consequence of emissions is to increase climate change, which leads to economic damages and an increased likelihood of a climate-linked transition shock that will restrict the use of fossil fuels at an unknown future time. The climate-change-dependent arrival rate of the (Poisson jump induced) transition shock captures the dynamic and endogenous feedback effect of climate change on the transition risk that is the central model mechanism. Modeling the transition shock in this reduced-form manner, I am able to mimic either an abrupt green technological change, shifting the energy input share of fossil fuels, or the type of climate policy actions currently and historically proposed that significantly limit restrict the use of fossil fuels by substantially increasing the cost of generating emissions. In both cases, fossil fuel firm prices are negatively impacted. Anticipation of a technology-driven, climate-linked transition shocks leads to what I call a ``run on fossil fuel,'' where fossil fuel firms accelerate production and fossil fuel spot and prices dynamically decrease, even as temperature and climate damages rise. On the other hand, the anticipation of a taxation-driven, climate-linked transition shock leads to a ``reverse run,'' where production is restricted and spot prices are elevated in an effort to postpone the transition shock.

I investigate various transition scenarios, counterfactuals, and model extensions to provide further intuition about the model mechanism: when there is no risk of a climate transition action; when the arrival rate of climate transition is constant over time; when the climate-linked transition risk is only partial internalized; and when the transition shock is implemented in multiple steps as well as in heterogeneous forms. The model extensions focus on how additional complexities impact the main results: non-unitary elasticity of intertemporal substitution (EIS); log utility; imperfect fossil fuel sector competition; allowing for fossil fuel exploration; and when the climate change externality is only partial internalized. These counterfactuals and model extensions highlight important qualitative and quantitative features of the main model mechanism, emphasizing central implications of the model that can be empirically validated. 

Finally, I discuss the existing empirical evidence about transition risk and provide my own estimates to test whether model results for particular transition scenarios can be empirically validated. I first estimate an event-study analysis for climate-related transition events that shift the likelihood of a future green transition taking place. I find that sectors with the highest climate-related transition risk exposure experience the largest increases (decreases) in cumulative abnormal returns for events that decrease (increase) the likelihood of a future climate transition. I then construct a ``climate-related transition event index,'' exploiting the forward looking information contained in asset pricing returns around realized climate-related-transition events, to estimate the dynamic impact of changes in the likelihood of a future climate-related transition. I estimate a structural VAR for the global oil market augmented by this climate-linked transition event index and calculate impulse response functions for a shock to the index. I find that an increased likelihood of climate-linked transition risk leads to increased oil production and a decreased oil spot price. These results are amplified, more persistent, and only statistically significant during the recent time period (2009-2019), consistent with the state-dependent, dynamic transition risk mechanism in the model.

\subsubsection*{\textit{Contribution}}

The contribution of my analysis is threefold. First, I provide a rich qualitative and quantitative characterization of the endogenous, dynamic feedback effects between aggregate energy production, climate change and climate change consequences, and asset pricing outcomes generated by the stochastic, climate change-linked arrival of a future green economy transition. Second, I develop an innovative specification for transition risk in the form of a stochastic process whose arrival depends on the endogenous evolution of climate change, and implement an original calibration strategy of the transition risk mechanism based on existing work using rigorous climate-economics-based simulation analysis to enhance the quantitative analysis provided. While my model setting is stylized, somewhat by necessity due to the unknown nature and form of future climate changed-linked transition risk, by calibrating the model to macroeconomic and asset pricing empirical moments and incorporating scientifically-supported estimates of climate change impacts and the transition risk mechanism, my analysis makes valuable strides towards providing well-founded quantitative results. Finally, my empirical analysis uses a novel climate transition risk index variable which exploits the interaction of the forward-looking asset price returns of a climate-related-transition-risk-exposure weighted portfolio and the timing captured by a hand-collected narrative index of climate-change-related transition events. This new empirical evidence, which focuses on dynamic macroeconomic and asset pricing implications linked to likelihood of green transition risk, provides evidence for the type of transition risk likely anticipated by individuals when compared with model solutions across different transition shocks scenarios.

\subsubsection*{\textit{Examples of Transition Risk Events}}

To emphasize the relevance of climate-change-linked transition risk, consider recent climate-change-related policy actions and technological innovations. Prominent examples consistent with an increased likelihood of future abandonment of fossil fuels include: the 2015 Paris Climate Accords, an international agreement established during the 2015 UN Framework Convention on Climate Change to keep global mean temperature (GMT) well below a $2^{\circ}$ C anomaly relative to the pre-industrial level; the development and commercialization of modern battery electric vehicles, beginning with the EV1 from General Motors and significantly popularized in recent times by Tesla and other automotive companies; and breakthroughs in clean energy production achieved by scientists at Lawrence Livermore National Laboratory using nuclear fusion. Each of these events have generally been viewed as meaningful steps forward in limiting future climate change by transitioning towards a green economy. However, substantial unknowns remain about when and if such advances will be the catalysts for a full-scale ``green'' transition. Moreover, examples of significant policy action and technological change events enhancing support of continued fossil fuel production have also occurred: the development of horizontal drilling methods using hydraulic fracturing techniques (fracking) to access previously untapped shale oil and natural gas deposits; and the election as US president of Donald Trump, who campaigned on repealing the Clean Power Plan and other policies restricting greenhouse gas emissions, pulling out of the Paris Agreement, and rebuilding the manufacturing and coal sectors in the US. 

These events show the risks and unintended consequences that could potentially result from the expectation of a future climate-linked transition shock. For example, Norway, a climate-conscious country with sizeable oil reserves, proposed moving forward their carbon neutrality goal from 2050 to 2030 in response to the Paris Agreement, while at nearly the same time proposed policy to increase oil drilling and development of their considerable arctic reserves\footnote{New York Times, June 17, 2017, ``Both Climate Leader and Oil Giant? A Norwegian Paradox.''}. The US, initially a significant force in establishing the Paris Agreement by implementing the Clean Power Plan and renewable portfolio standards to restrict future greenhouse gas emissions and require increased renewable energy production, went on to withdraw from or repeal various climate policies under the Trump administration. The fairly recent IPO of Saudi Aramco, the state-owned oil producing firm of Saudi Arabia, was motivated in part by a desire to diversify away from oil\footnote{Financial Times, December 14, 2016, ``The privatisation of Saudi Aramco.''}, while the final valuation was much lower than initially proposed partially due to the potential risk of the country's oil reserves becoming a stranded asset\footnote{Financial Times, August 13, 2017, ``Saudi Aramco's value at risk from climate change policies.''}. And examining the annual 10-K filings\footnote{US firms are required to include Section 1.A - ``Risk Factors'' in their annual 10-K filings with the SEC, and list the ``most significant factors'' that affect future profitability. See \cite{koijen2016financial} for details.} of the 10 largest US oil firms (in terms of reserves held) shows that major US oil producers are increasingly aware of climate-linked-transition risks, including risk about impacts, timing, and the form of potential mandates and demand shifts away from fossil fuels towards clean energy sources.\footnote{Between 2004 and 2010, each of the 10 largest firms (Anadarko, Chevron, ConocoPhillips, EOG Energy, ExxonMobil, Halliburton, Marathon, Occidental, Phillips 66, and Valero) began including sections about climate transition risk, with key words such as climate change policy, carbon-constrained economy, mandate, reduced demand, alternative energy/fuels, and the Paris Agreement becoming increasingly more prevalent.}

Such responses highlight the significance of the potential transition risk for firms and countries. Demonstrating this further, cost-benefit analysis based on fossil fuel reserve estimates and the global warming potential of these reserves by \cite{mcglade2015geographical} shows that it would require stranding 30\% of oil reserves and 80\% of coal reserves in order to avoid exceeding a $2^{\circ} C$ temperature anomaly relative to the pre-industrial level. And these magnitudes have only been amplified in more recent evaluations such as \cite{semieniuk2022stranded}. Stranding this magnitude of fossil fuel reserves through policy or significant technological innovation could have drastic global economic and financial market implications.

\section{Related Literature}

There are various areas of research in economics, finance, and climate change to which my paper contributes, which I briefly highlight now. The climate-economics literature has explored theoretically the social cost of carbon and optimal carbon taxation \citep{stern_stern_2007, golosov_optimal_2014, hambel2015optimal, Nordhaus:2017, cai_social_2015}, directed technological change \citep{acemoglu2016transition}, climate disasters \citep{pindyck_economic_2013}, climate model uncertainty \citep{lemoine2012tipping, brock2017wrestling, barnett2020pricing, Rudik:2020, BarnettBrockHansen:2021, barnett2023climate, barnett2024uncertainty}, and suboptimal climate policy \citep{hassler2021presidential, hong2023welfare} using elements of macroeconomic and IAM modeling frameworks. 

Extensive research in the asset pricing literature has provided valuable insight about policy and innovation risk. This includes research focused on the asset prices implications of political uncertainty \citep{pastor2012uncertainty, kelly2016price}, tax and healthcare policy risk \citep{sialm2006stochastic, koijen2016financial}, and the adoption of new technologies \citep{pastor2009technological}. Related work examining economic and policy uncertainty impacts on the real economy and financial outcomes has used narrative indices of policy shocks \citep{romer2010macroeconomic, ramey2011identifying, mertens2014reconciliation}, narrative indices interacted with quantity or price measures \citep{arezki2017news}, and text-based indices \citep{baker2016measuring, giglio2016systemic}. Work specific to climate change and asset pricing includes exploring long-run discount rates and climate change mitigation efforts \citep{giglio2021climate}, long-run climate change risk implications \citep{bansal2019climate}, the elasticity of climate damages \citep{dietz2018climate}, and the largely negative, and often dynamically amplified, implications of climate change for house prices \citep{baldauf2020does, bernstein2019disaster}, options \citep{kruttli2019pricing, ilhan2021carbon}, corporate and municipal bonds \citep{painter2020inconvenient, goldsmith2022sea, huynh2021climate}, institutional investments \citep{krueger2020importance}, and equity prices \citep{hong2016climate, engle2020hedging, bolton2021investors, bolton2021global}. 

\cite{hotelling1931economics} and \cite{dasgupta1974optimal} provide the foundations for the literature on natural resource extraction, while recent contributions have identified important model mechanisms required to explain various stylized facts about oil prices and production \citep{carlson2007equilibrium, kogan2009oil, anderson2018hotelling, casassus2009equilibrium, ready2018oil}. Extensive work has also been done estimating the impact of economic and financial shocks on global oil market outcomes \citep{hamilton1983oil, kilian2009not, kilian2014role, baumeister2019structural}. 

A notable literature closely related to my analysis has focused on the consequences of alternative climate actions in various forms. \cite{mcglade2015geographical}, the Grantham Research Institute, and recently \cite{semieniuk2022stranded}, have used least-cost analysis to characterize the potential magnitude of stranded assets risk and possible ``carbon bubble,'' or overvaluation of fossil fuel firms exposed to temperature ceiling policies. Theoretical work has modeled the consequences of alternative policy actions and stranded assets risk on investment and exploration choices \citep{van2020risk, fried2021macro, caldecott2021stranded}, while highlighting the need to further explore risk and macro-asset pricing implications \citep{van2020stranded, campiglio2022macrofinancial}.

The  Green Paradox is another important area of focus in this literature. The central idea behind the Green Paradox is that expected future climate policy that will make using fossil fuels more costly can lead to an unintended impacts on fossil fuel production today. The seminal work in this area comes from \cite{sinn2008public, sinn2009green}. Recent theoretical contributions in this area \citep{van2012there, jensen2015introduction, kotlikoff2016will, baldwin2020build}\footnote{See \cite{pittel2015climate, van2015global} for earlier reviews on this literature.} have explored various dimensions related to the Green Paradox, including the impact of competition, alternative production and investment specifications, and varying policy settings. Recent empirical work in this area has shown how state-level climate policy has impacted corporate real activities \citep{bartram2022real}, the negative effect of the growing amount of undeveloped reserves on North American oil firm values \citep{atanasova2019stranded},  and that proposed climate bills limiting future oil use have persistent effects that shift future oil consumption towards the present \citep{norman2024empirical}.

Two specific papers are worth noting in more detail. \cite{hsu2023pollution} analyzes environmental litigation risk associated with toxic emissions to highlight the implications for asset prices from environmental policy risk. They use a reduced-form model with a static and exogenous risk mechanism for their analysis, limiting their ability to capture the endogenous production responses and dynamic feedback mechanisms associated with climate-related transition risk. \cite{hambel2024pricing} is a recent paper that explores stranded assets and transition risk, building on the extant structure developed by the current paper including using a closely-related transition risk functional form and transition risk calibration strategy. While their framework accounts for negative emissions technology, inter-sectoral adjustment costs, and recurring risks of climate-related disasters and exogenous macroeconomics disasters, they assume a particular costly policy structure and deterministic pathways for reductions in emissions intensity, green energy costs, and improvement in backstop technology that rule out relevant general equilibrium and economic forces. 

Relative to the this existing literature, the contribution of my paper is to provide one of the first general equilibrium characterizations of the dynamic responses to expectations about potential climate-linked transition risk scenarios. This is achieved by simultaneously including two different fossil fuel sources, limited supply ``oil'' produced a non-transferable input and unlimited supply ``coal'' produced with a transferable input, in a production-based framework and by exploring numerous transition risk scenarios with variation in the assumed taxation, technology, externality internalization, and dynamic risk realization features related to transition risk. From this setting I am able to characterize novel economic forces related to the dynamic and endogenous feedback effects between emissions resulting from production choices, the resulting climate change and economic damages from these emissions, the likelihood of climate-linked transition risk associated with these outcomes, and expectations about social welfare implications from the potential realization of a transition shock. Because much of the existing work has focused on reduced-form or partial equilibrium model settings with exogenous policy and transition risk scenarios, without considering broader macroeconomic implications and asset pricing results, previous work has been unable to demonstrate the feedback effects present in my analysis\footnote{This includes the critical work done using large-scale scenario and simulation-based analysis by the International Panel on Climate Change (IPCC), International Institute for Applied Systems Analysis (IIASA), and Postdam Institute for Climate Impact Research (PIK). While there is significant value from such work, the modular approaches they use separate discounting, damage, climate, and socioeconomic outcomes and therefore cannot account for general equilibrium dynamic feedback effects as I do here.}.

Exploring distinctive transition risk scenarios in this setting that embeds the endogenous, state-dependent transition risk mechanism in a stochastic, production-based macroeconomic-asset pricing framework allows me to identify novel qualitive production and asset pricing outcomes in response to green transition risk, as well as provide quantitative insights about the potential economic and welfare consequences resulting from alternative climate-linked transition risk scenarios. Specifically, expectations of more economically beneficial transition shocks, such as technology-related shocks, lead to an optimal ``run'' on fossil fuel response marked by a dynamic acceleration of fossil fuel extraction and production, as well as corresponding attenuation in fossil fuel spot prices. However, if the expected transition shock is anticipated to be economically costly, such as a taxation-related shock, then the optimal response is a ``reverse run'' where fuel extraction and production are suppressed and spot prices ramp up in an attempt to postpone the transition shock.  While the existing literature, such as the work on the Green paradox, has previously found static fossil fuel production and price implications in response to expectations about future climate policy, the dynamic acceleration of emissions and attenuation of prices as a result of accounting for endogenous, general-equilibrium feedback effects is novel to my setting. Moreover, the comparisons across various transition risk scenarios considered in my framework allows me to identify and characterize settings where a ``reverse run'' can occur, providing additional new insight on how differences in expected future climate-related transition risk consequences alter the optimal macroeconomic and asset pricing outcomes in the model.

\section[A Dynamic Model of Climate-Linked Transition Risk]{A Dynamic Model of Climate-Linked Transition Risk}

I now put forward a theoretical model to study the implications of expected future climate transition. The model consists of the following components: preferences, consumption, production, climate change, and climate transition risk. After outlining the details for the solution in the baseline setting, I explore alternative specifications that incorporate additional complexities in order to provide additional insight into the mechanism of interest.

\subsubsection*{Preferences and Consumption}

In the model, I assume a representative agent whose decision problem is to maximize discounted lifetime utility subject to economic constraints. I assume the agent has recursive utility of the Duffie-Epstein-Zin-Weil type over current period consumption:
\begin{eqnarray*}\label{HJB}
h(C, J) & = & \begin{cases} \frac{\rho}{1-\theta^{-1}}\left( \frac{C^{1-\theta^{-1}}}{((1-\gamma)J)^{\frac{\gamma - \theta^{-1}}{1-\gamma}}} -(1-\gamma)J \right) & \text{if } \theta \ne 1 \\
\rho (1-\gamma) V \left( \log C - \frac{1}{1-\gamma} \log\left( (1-\gamma) V \right) \right) &  \text{if } \theta = 1 \end{cases}
\end{eqnarray*}
where $\rho$ is the subjective discount rate, $\theta$ is the elasticity of intertemporal substitution (EIS), and $\gamma$ is the risk aversion (RA). I assume recursive preferences for the Social Planner in order to provide more quantitatively realistic outcomes related to climate transition risk, given the success of recursive preferences in helping theoretical models match key asset pricing and macroeconomic moments of interest in other settings\footnote{The effects of the transition risk mechanism still hold for simpler and more tractable utility functions. To demonstrate this, I provide model results for the log utility setting in the Online Appendix.}.

Consumption in the model is assumed to be a Cobb-Douglas (CD) aggregate of three types of inputs.\footnote{This specification matches the long-run estimates of \cite{hassler2021directed}. While I omit directed technical change, which underlies these estimates, the setting gives a tractable approximation for my analysis.} The first input is a capital-based good, the second is an energy-based good, and the third is a labor-based good. A climate change damage function is assumed to scale the level of the CD aggregate of the three input types so that as climate change increases consumption decreases. Therefore, aggregate consumption $C_t$ is given by:
\begin{eqnarray*}
C_t = D(T_t) C_{K,t}^{\alpha} C_{E,t}^{\psi} C_{L,t}^{1 - \alpha - \psi}
\end{eqnarray*}

where $D(T_t)$ is the climate damage function representing the impact of climate change on aggregate consumption, $C_{K,t}$ is the capital-based input, $C_{E,t}$ is the energy-based input, and $C_{L,t}$ is the labor-based input. The demand shares for the capital-, energy-, and labor-based inputs are given by $\alpha$, $\psi$, and $1 - \alpha -\psi$, respectively. 

\subsubsection*{Production}

The capital-based input good is produced using an AK technology such that 
\begin{eqnarray*}
C_{K,t} = A_C K_{C,t} - i_{C,t} K_{C,t}
\end{eqnarray*}
where $A_C$ is the factor productivity of the capital sector, $K_{C,t}$ is the aggregate capital stock for $C_{K,t}$ production\footnote{As in \cite{pindyck_climate_2013} and \cite{BarnettBrockHansen:2021}, capital in this model should be interpreted broadly to include not just physical capital, but also human, intangible, and organizational capital.}, and $i_{C,t}$ is the investment-to-capital ratio. The capital stock for $C_{K,t}$ production is subject to adjustment costs, as in \cite{lucas1971investment} and \cite{hayashi_tobins_1982}. For tractability, I use $\hat{K}_{C,t} = \log K_{C,t}$ in my analysis, which evolves as:
\begin{eqnarray*}
d \hat{K}_{C,t} & = &  (\mu_{C} + i_{C,t} - \frac{\phi_C}{2} (i_{C,t})^2 - \frac{1}{2}\sigma_C^2 ) dt + \sigma_C dW_{C,t} 
\end{eqnarray*}

The labor-based input is assumed to be of a linear form such that
\begin{eqnarray*}
C_{L,t} = L_{1,t}
\end{eqnarray*}
where I denote $L_{1,t}$ as the fraction of total labor used for the final output production. I assume that labor is in limited, fixed supply, and normalize the amount of labor to be unitary. Labor is split between final output and coal production, and so the labor market clearing condition $1 = L_{1,t} + L_{2,t}$ must hold as there is no disutility from supplying labor.

%%%%%%%%%%%%%%%%%%%%%%%%%%%%%%%%%%%
The energy input is produced using a constant elasticity of substitution (CES) technology based on three different energy types
\begin{eqnarray*}
C_{E,t} = ( \nu_1 E_{1,t}^\omega + \nu_2 E_{2,t}^\omega + \nu_3 E_{3,t}^\omega )^{1/\omega} 
\end{eqnarray*}

where $\omega$ is the energy elasticity of substitution, $E_{1,t}$ is a fossil fuel energy source with limited reserves (oil), $E_{2,t}$ is a fossil fuel energy source with unlimited reserves (coal), $E_{3,t}$ is a green energy source, $\nu_1$ is the demand share of oil, $\nu_2$ is the demand share of coal, and $\nu_3$ is the demand share of green energy. The green energy production technology is of the AK form with green capital as the only required input
\begin{eqnarray*}
G_t = A_{G} K_{G,t} - i_{G,t} K_{G,t}
\end{eqnarray*}
where $A_{G}$, $K_{G,t}$, and $i_{G,t}$ are the total factor productivity, capital stock, and the investment-to-capital ratio, respectively, for green energy production. The stock of green capital includes both land needed for large-scale green energy production, as well as physical, human, and intangible capital used in the production of green energy. As with $K_{C,t}$, the capital stock for $G_{t}$ production is subject to adjustment costs, and $\hat{K}_{G,t} = \log K_{G,t}$ evolves as:
\begin{eqnarray*}
d \hat{K}_{G,t} & = &  (\mu_{G} + i_{G,t} - \frac{\phi_G}{2} (i_{G,t})^2 - \frac{1}{2}\sigma_G^2 ) dt + \sigma_G dW_{G,t} 
\end{eqnarray*}

The oil energy input is produced produced using a linear technology based on the amount extracted from the existing stock of reserves, denoted $R_t$, such that
\begin{eqnarray*}
E_{1,t} = \mathcal{E}_t R_t
\end{eqnarray*}
where $\mathcal{E}_t$ is the fraction of reserves extracted at time t. The evolution of $\hat{R}_{t} = \log R_t$ is determined by extraction and Brownian shocks\footnote{I show in the Online Appendix that results are very similar when allowing for exploration of reserves.}:
\begin{eqnarray*}
d \hat{R}_{t} & = & (-\mathcal{E}_t -\frac{1}{2}\sigma_R^2) dt + \sigma_R dW_{R,t} 
\end{eqnarray*}
There are no explicit costs of extraction. However, extraction is limited by the implicit value of holding reserves for future production and the social cost of economic damages from climate change induced by the accumulation of carbon emissions in the atmosphere, which is internalized by the Social Planner.

The coal energy input is produced competitively with decreasing returns to scale technology. The production technology uses only labor as an input and is scaled by a productivity factor. The input production function is given as follows:
\begin{align}
E_{2,t} = A_{2,t} (L_{2,t})^{\alpha_2}
\end{align}
where $A_{2,t}$ is the productivity for coal production, which gives output in the same emissions units as oil, $L_{2,t}$ is the labor used for coal production, and $1>\alpha_2 > 0$ is the decreasing returns to scale parameter for labor used in coal production. The coal production technology implicitly assumes that coal is in infinite supply. This assumption, and others related to the fossil fuel inputs, I discuss in the following remarks.

\begin{remark}%[Supply of fossil fuel reserves]
The assumption of coal being in infinite supply is clearly not perfectly realistic. However, given the continuing discovery of new fossil fuel reserves and because coal reserves are ``large enough'' that they are unlikely to be exhausted before facing extreme climate impacts that would force us to abandon these energy sources\footnote{See \cite{hassler_energy_2012}, \cite{hassler_environmental_2016}, \cite{mcglade2015geographical}, and \cite{semieniuk2022stranded} for discussion and estimates related to reserves and discoveries.}, this assumption is not entirely unreasonable. Most importantly, incorporating into the model an ``oil'' fossil fuel input with limited reserves and a ``coal'' fossil fuel input with unlimited reserves allows me to provide an important comparison and decomposition of the central transition risk mechanism in my model and the ``Hotelling-rule'' mechanism so prevalent in models with fossil fuel energy inputs.
\end{remark}

\begin{remark}
Additional frictions or adjustment costs related to changes fossil fuel production, beyond the CES technology assumed, are potentially important for providing complete quantitative characterizations from this model. While I provide  details on the calibration for this setting based on external estimates and direct model calibration, parameter sensitivity analysis over $\omega$, which changes the substitutability of the clean and dirty inputs, shows that core model outcomes are relatively robust to various values of $\omega$. I discuss further the potential impact of additional frictions later in the paper. %, and will discuss further the potential impact of such frictions at that time.
\end{remark}

\begin{remark} \cite{caldecott2021stranded}, the Grantham Institute, and others, emphasize that potential stranded assets include existing ``dirty'' capital stocks used for fossil fuel production are most often non-redeployable. Explicitly including fossil fuel reserves allows my analysis to account for this effect to some extent, as reserves are non-redeployable and therefore ``written-off'' when determining fossil fuel firm valuations. The extent to which existing dirty capital matters beyond this means that the quantitative outcomes in my analysis potentially underestimate the implications of climate-linked transition and stranded assets risks.
\end{remark}

\subsubsection*{Climate Change} \label{climatesec}

Atmospheric temperature in excess of pre-industrial levels $T_t$ evolves as
\begin{eqnarray*}
dT_t & = & \beta_T \left(E_{1,t} + E_{2,t}\right)  dt + \sigma_T dW_{T,t} 
\end{eqnarray*}

which is a stochastic version of the carbon-climate response (CCR) or transient climate response to cumulative carbon emissions (TCRE) relationship estimated by \cite{matthews2009proportionality} and \cite{macdougall2015origin} that approximates complex, multi-dimensional models of climate dynamics. I use this approximation, typically viewed as an approximation best suited for longer time scales, in place of more complex climate dynamics for tractability and precisely because it captures the longer-term, near permanent effect of emissions on the atmosphere that my analysis is focused on.\footnote{\cite{dietz2021economists} highlight this specification as a way to make economic models consistent with climate science models for sound physical reasons. Furthermore, the estimated decay rate for atmospheric carbon is on the order of hundreds or even thousands of years. Thus the permanence of climate change impacts from carbon emissions is an important model feature. See \cite{Pierrehumbert:2014} for further details.}

Climate change is assumed to have negative aggregate welfare consequences in the model. These economic damages from climate change, or ``climate damages,'' multiplicatively scale final output and are represented by the convex function\footnote{This functional form for $D(T_t)$ is such that: $D(T_t) \in [0,1] \quad \forall T_t$, $D(0)=1$, $D(\infty)=0$, and $\frac{d D}{d T} < 0$.}
\begin{eqnarray*}
D(T_t) = \exp(-\eta T_t)
\end{eqnarray*}

These damages capture negative social externalities from climate change such as increases in extreme temperatures, rising sea levels, destructive weather events, and other costly outcomes scientists have associated with rising temperatures due to anthropogenic emissions.

\subsubsection*{Climate-Related Transition Risk} \label{climatepolsec}

The novel mechanism in my analysis is the climate-change-linked transition risk shock. I begin by focusing on two specific forms of the shock, with each shift modeled by a jump process to a single absorbing post-jump state. I highlight various other transition risk scenarios later on in my analysis. The first shock I consider is a technological innovation such that the input share for green energy goes to $\nu_3 = 1$ and the input shares for fossil fuel energies go to $\nu_1 = \nu_2 = 0$. The resulting energy production technology is given by
\begin{eqnarray*}
\textbf{[Technology Shock:] } \tilde{C}_{E,t} = \left( A_G K_{G,t} - i_{G,t} K_{G,t} \right) 
\end{eqnarray*}

The second is a carbon taxation shock where the fossil fuel energy inputs are taxed at a rate of $100\%$.\footnote{There is no need to assume how tax revenues will be distributed in this example as the planner will choose in equilibrium to simply not produce using fossil fuels and so there will be no tax revenues. In later examples, we will consider alternative tax shock scenarios where we will need to take an explicit stand on this issue. We leave further discussion of this issue for those examples.} For this to be well defined I must assume $\omega > 0$. The resulting energy production technology is given by
\begin{eqnarray*}
\textbf{[Taxation Shock:] } \tilde{C}_{E,t} = \nu_3^{1/\omega}\left( A_G K_{G,t} - i_{G,t} K_{G,t} \right)
\end{eqnarray*}

While the difference in production technologies appears relatively minor, the differences in outcomes could be important both quantitatively and qualitatively. Explicitly modeling both types of climate transition risk using a rigorous, dynamic general equilibrium framework allows me to characterize the transition risk in a way not previously done before.

The arrival rate of the transition shock is given by $\lambda_t$, whose evolution is given by
\begin{eqnarray*}
\Lambda(T_t) & = & \psi_0 ( \exp[ \frac{\psi_1}{2}(T_t-\underline{T})^{2}] -1 ) \mathbbm{1}\{T_t \ge \underline{T} \}   \\
d \lambda_t & = & \left(\mu_{\lambda} + \Lambda'(T_t) \beta_T \mathcal{E}_t \exp(\hat{R}_t) -\frac{1}{2}\Lambda''(T_t) \sigma_T^2 \right) dt + \left( \Lambda'(T_t) \sigma_T + \sigma_{\lambda} \right) \cdot dW 
\end{eqnarray*}

The arrival rate is assumed to depend on the endogenously evolving level of climate change. The interpretation for this specification of the arrival rate is that the probability of significant climate-change-related transition event increases as climate change and climate damages become more pronounced. The assumption that the likelihood of a climate-linked transition event is related to increasing climate change is critical to the model. Figure E1 in the Online Appendix provides qualitative empirical support for this relationship. The panels show a map of carbon prices for various countries, as well as the global average trend in price, and the time series of US Government Research, Development, and Deployment for different green sectors and technologies with the US temperature anomaly. The increasing trends and magnitudes of these outcomes show a positive correlation\footnote{The time series correlation for RD\&D with US temperature is $0.40$, and $0.72$ with global temperature.} consistent with the assumption that there is an increasing likelihood of significant climate action towards a low-carbon transition as climate change increases. 

While the functional form for my transition arrival rate is stylized, it is not without rationale. The choice of this functional form is in part motivated by the link between the implementation of a fossil-fuel-restricting transition and the severity of future climate damages\footnote{The functional form is consistent with the functional form used for the arrival rate of the uncertain damage function curvature in \cite{BarnettBrockHansen:2021}.}. In addition, recent work by \cite{moore2022determinants} has focused on modeling behavior related to socio-political-technical processes that determine climate policy and emissions trajectories. To calibrate my arrival rate function, I use the policy trajectories estimated by \citeauthor{moore2022determinants}, provided in Figure E2 in the Online Appendix, which show a non-linear, increasing likelihood of policy stringency as climate change and climate damages increase. 

Finally, this specification allows for an exogenous component increasing the likelihood of a significant transition from fossil fuels. Underlying this exogenous component could be ``surprise'' technological changes in production technology using green sources or other climate policy related forces increasing the likelihood of a climate transition. The framework captures the fact that climate transition risk is positively correlated with observed climate change and climate damages, though potentially influenced by other economic forces. 

\section{Equilibrium Solutions}

I now solve for the equilibrium solution to the planner's problem. The social planner maximizes social welfare through optimal choices of consumption, investment in the consumption and green capital stocks, as well as oil and coal production subject to the evolution of the stochastic processes for the state variables and resource constraints. Imposing market clearing conditions, the equilibrium solution can be characterized by optimal choices of investment in consumption capital $K_{C,t}$, investment in green capital $K_{G,t}$, extraction of fossil fuel reserves $R_t$, and the division of labor between final output and coal input production:
\begin{equation*}
\{i^*_{C,t}, i^*_{G,t}, \mathcal{E}^*_t, L^*_{1,t}\}
\end{equation*}
The solution is a recursive Markov equilibrium such that the optimal control and social welfare functions depend only on the current values of the vector of state variables 
% \begin{equation*}
% \{\log K_{C,t}, \log K_{G,t}, \log R_t, T_t, \lambda_t \}.  
% \end{equation*}
\begin{equation*}
\{\hat{K}_{C,t}, \hat{K}_{G,t}, \hat{R}_t, T_t, \lambda_t \}.  
\end{equation*}

The solution for the planner's value function is derived from a Hamilton-Jacobi-Bellman (HJB) equation representing the optimization problem in recursive form. The optimal choices of the planner are derived from the first order conditions for the HJB equation. The (shadow) prices supporting the optimal quantities in the decentralized economy with a decentralization mechanism for the climate externality can be derived from the planner's solution.

\subsection[Macroeconomic Outcomes]{Macroeconomic Outcomes}

From the social planner's problem described above, we can derive an HJB equation characterizing the planner's value function as follows:

\begin{proposition}

With climate transition risk determined by the temperature dependent arrival rate function $\lambda_t$, the value functions for the two regimes are given by:
\begin{align*}
V_{pre}(\hat{K}_{C,t}, \hat{K}_{G,t}, \hat{R}_t, T_t, \lambda_t) & =  \frac{\left( K_{C,t}^\alpha  \exp\left[v(\hat{K}_{G,t}, \hat{R}_t, T_t,\lambda_t) + c_{pre}\right] \right)^{1-\gamma}}{1-\gamma} \\
V_{post}(\hat{K}_{C,t}, \hat{K}_{G,t}, T_t) & = \frac{\left( K_{C,t}^\alpha K_{G,t}^{\psi} \exp\left[ - \eta T_t + c_{post}\right] \right)^{1-\gamma}}{1-\gamma}
\end{align*}
The FOCs for optimal investment, green energy labor, and extraction are given by
\begin{eqnarray*}
0 & = & \rho (A_C - i_{C,t})^{-1} - (1- \phi_C i_{C,t})   \\
0 & = & \rho \psi \nu_{1} \left(\frac{E_{1,t}}{C_{E,t}} \right)^{\omega} \mathcal{E}_t^{-1} - \left( v_{\hat{R}} -\beta_{f} R_t \left( v_{T} + \Lambda'(T) v_{\lambda} \right) \right) \\
0 & = &  \rho \psi \nu_2  \left(\frac{E_{2,t}}{C_{E,t}} \right)^\omega (1 - L_{1,t})^{-1} - \beta_f A_{2,t} (1- L_{1,t})^{\alpha_2-1} \left( v_{T} + \Lambda'(T) v_{\lambda} \right) + \frac{\rho \left(1- \alpha - \psi \right)}{\alpha_2} L_{1,t}^{-1}  \\
0 & = &  \rho \psi \nu_3 \left(\frac{E_{3,t}}{C_{E,t}} \right)^\omega (A_G - i_{G,t})^{-1} - \left(1- \phi_G i_{G,t}\right) v_{\hat{K}_G} 
\end{eqnarray*}

Note $v(\hat{K}_{G,t}, \hat{R}_t, T_t, \lambda_t)$ is the solution to the simplified planner's problem HJB equation. The notation $v_{x}$ denotes the partial derivative of $v$ with respect to the state variable $x$. Constants $(c_{pre},c_{post})$, given in the appendix, are functions of the model parameters. 

\end{proposition}

The main contributors determining oil extraction, beyond the model primitives, are the marginal value of atmospheric temperature $(v_T)$, the marginal value of oil reserves $(v_{\hat{R}})$, and the marginal value of the transition arrival rate $(v_{\lambda})$. Without transition risk, increased temperature leads to a higher marginal cost of climate change due to increased economic damages, and reduced reserves leads to a higher marginal benefit of holding reserves due to limited supply remaining for future production. For coal production, the contributors are also the marginal value of atmospheric temperature and the marginal value of the transition arrival rate, but now the planner considers the trade-offs of limited labor supply rather than concerns about reserve constraints. The novel effect altering these results in my model is the dynamic climate-linked feedback mechanism of the anticipated risk of a future green transition. The temperature-dependent transition shock arrival rate impacts the marginal cost of climate change for both types of emissions, for oil the marginal benefit of holding reserves, and for coal the marginal cost of forgoing labor in final output production. As temperature increases, the likelihood of a transition shock increases and drives down the marginal cost of climate change, and alters the marginal value of reserves and the marginal value of using labor for coal production. The quantitative relevance of this effect depends on the assumptions about the transition arrival rate, how the planner internalizes the economic consequences of this risk, and the type of transition shock that is expected to occur.

\subsection[Asset Prices]{Asset Prices}
\label{assetprices}

I now characterize the impact of an anticipated future climate-linked transition shock on asset prices. As asset prices incorporate expectations about future risks and uncertainty due to their forward looking nature, and because the most severe consequences of climate change and transition risk are expected to happen in the future, they should be particularly informative in this setting. First, asset prices from the model provide additional insight about the model mechanism beyond the macroeconomic effects. Second, tests of the model predictions related to asset prices should provide greater power to identify the effects from long-term, forward-looking risks and concerns about climate change and transition risk, even if the estimated impacts on macroeconomic outcomes due to climate-linked transition risk are relatively small or insignificant. I derive the asset pricing outcomes based on the solution to the macroeconomic side of the model and a decentralization of the planner's problem. Because fossil fuel firms are only valued before the climate transition shock in my main specification, I focus on asset prices in the pre-transition state in what follows.

\subsubsection{Decentralization}
In order to derive the prices for the planner's solution, I need to characterize the decentralized counterpart to the planner's problem where an optimal tax incentivizes the internalization of the climate externality, given that individuals do not account for their contribution to aggregate climate change and climate damages that result from the use of fossil fuel. I briefly discuss the decentralization mechanism that generates prices corresponding to the planner's solution in the following remark, leaving the derivation and details for the appendix:

\begin{remark}%[Optimal Carbon Tax]

A decentralized market with a tax on the production of the two types of fossil fuels, lump-sum rebated back to households gives the socially optimal outcomes, and the prices that support market clearing equilibrium. In my model, there is an expression that can be derived by equating the social planner's optimal oil production choice with the decentralized oil production choice, as well as an expression that can be derived by equating the social planner's optimal coal production choice with the decentralized coal production choice. 

What matters for these tax expressions are the marginal cost of emissions from oil production or coal production ($-\beta_{f} R_t \left( v_{T} + \Lambda'(T) v_{\lambda} \right)$ or $\left(-\beta_{f}\alpha_{2}A_{2,t}(1-L_{1,t})^{\alpha_{2}-1}(v_{T}+\Lambda'v_{\lambda})\right)$, which depend explicitly on marginal values of climate change ($v_{T}$) and transition risk ($\Lambda'(T) v_{\lambda}$), as well as the marginal value of forgoing oil reserves for future use ($v_R$) or the marginal value of forgoing labor for final output production to use for coal production ($\rho (1-\alpha-\psi) L_{1,t}^{-1}$). Without transition risk, the marginal cost of climate change would increase with temperature, the marginal benefit of reserves would decrease with temperature, and the marginal value of using labor for coal production would also decrease with temperature, reflecting increasing concerns for climate damages. Climate-linked transition risk will alter these marginal benefits, as will be shown in my numerical results. The optimal tax will reflect these changes.

\end{remark}

\subsubsection{Spot Price of Fossil Fuel}\label{spot}

Input prices are calculated directly from the first order conditions for the final output firm's profit maximization problem, applying the planner's optimal choices. The input prices of interest for my analysis are given by
\begin{eqnarray*}
P_{K,t} = \alpha \frac{C_{t}}{C_{K,t}}, \quad P_{1,t} = \psi \nu_{1}\frac{C_{t}}{E_{1,t}}(\frac{E_{1,t}}{C_{E,t}})^{\omega}, \quad P_{2,t} = \psi \nu_{2}\frac{C_{t}}{E_{2,t}}(\frac{E_{2,t}}{C_{E,t}})^{\omega}, \quad P_{3,t} = \psi \nu_{3}\frac{C_{t}}{E_{3,t}}(\frac{E_{3,t}}{C_{E,t}})^{\omega}
\end{eqnarray*}

These representations come from the first order conditions of the planner. The main input prices of interest for my analysis are the spot price of oil and coal, which in this form demonstrates a clear inverse relationship between fossil fuel production and the spot price of the fossil fuel. As a result, significant production impacts for each type of fossil fuel from transition risk directly and inversely influence the fossil fuel spot prices.

\subsubsection{Stochastic Discount Factor}\label{sdf}

The stochastic discount factor (SDF) for our recursive preference specification is given by the discounted marginal utility of consumption $\pi_{t}=\exp(\int h_{J}ds)h_{C}$. The SDF is essential to deriving asset prices because it incorporates the information necessary to properly discount firm profits over time and across states of nature. For this reason, the risk-free rate and the compensations required for holding certain risks, or the prices of risk, are derived from the SDF's drift and volatility, respectively. Applying Ito's lemma to $\pi_t$ provides the SDF evolution, $\frac{d \pi_t}{\pi_t} = h_{J}dt + \frac{\mathcal{D}h_{C}}{h_{C}}$, and the aforementioned prices as follows:

\begin{proposition}

The evolution of the stochastic discount is given by
\begin{align*}
\frac{d\pi_t}{\pi_t} = -r_{f,t} dt - \sigma_{\pi,\hat{K}_C} dW_C - \sigma_{\pi,\hat{K}_G} dW_G  - \sigma_{\pi,\hat{R}} dW_R - \sigma_{\pi,T} dW_T - \sigma_{\pi,\lambda} dW_{\lambda} - \Theta_{\pi} dN_t
\end{align*}

where $r_{f,t}$ is the risk-free rate, $\sigma_{\pi,\hat{K}_C}, \sigma_{\pi,\hat{K}_G}, \sigma_{\pi,\hat{R}}, \sigma_{\pi,T}, \sigma_{\pi,\lambda}$ are the compensations for the diffusive risks of consumption capital, green capital, fossil fuel reserves, temperature, and transition and stranded assets, respectively, and $ \Theta_{\pi}$ is the compensation for the jump risk related to transition and stranded assets risk. Note that $N_t$ is the Poisson process for the jump transition of ($\nu_1, \nu_2, \nu_3$). Expressions for these compensations are given by
\begin{eqnarray*}
\sigma_{\pi,\hat{K}_C} = \alpha \gamma \sigma_{C}, &
\sigma_{\pi,\hat{K}_G} = \left[ \partial(C_t)_{\hat{K}_G}/ C_t - (1-\gamma) v_{\hat{K}_G} \right] \sigma_{G}
\end{eqnarray*}

\vspace{-1.0cm}

\begin{eqnarray*}
\sigma_{\pi,\hat{R}} = \left[ \partial(C_t)_{\hat{R}}/ C_t - (1-\gamma) v_{\hat{R}}  \right] \sigma_{R}, & & \sigma_{\pi,T} = \left[ \partial(C_t)_{T}/ C_t - (1-\gamma)v_{T} \right] \sigma_{T},
\end{eqnarray*}

\vspace{-1.0cm}

\begin{eqnarray*}
\sigma_{\pi,\lambda} =  \left[ \partial(C_t)_{\lambda}/ C_t - (1-\gamma)v_{\lambda} \right] \left(\Lambda'(T) \sigma_{T}  + \sigma_{\lambda} \right), &&
\Theta_{\pi} = \left[ 1 - \frac{V_{post}}{V_{pre}}  \frac{C_{post}^{-1}}{C_{pre}^{-1}} \right]
\end{eqnarray*}

where $\partial (C_t)_{X}$ denotes the partial derivative of consumption with respect to state variable $X \in \{\hat{K}_C, \hat{K}_G, \hat{R}, T, \lambda\} $, and $V_{pre}, C_{pre}, V_{post}$ and $C_{post}$ denote pre-jump and post-jump value function and consumption values, respectively. 

\end{proposition}

I leave the cumbersome expression for the risk-free rate for the appendix. The expressions for the risk prices provide useful intuition, even though they require numerical solutions. Each diffusion risk price follows the standard asset pricing result of having a volatility and a risk aversion component. For each diffusion risk component, the production choices, and derivatives of the value function and production choices matter for the risk aversion component. As a result, expected future green transition risk matters through its impact on the planner's value function, its impacts on optimal fossil fuel production, its impact on climate change due to emissions choices, and directly through its volatility component. Transition risk contributes directly through the jump risk compensation, and depends on the magnitude of the value function and production changes between pre- and post-transition economies.

\subsubsection{Stock Prices}\label{stock_prices}

We can now derive firm prices. I assume firms are $100\%$ equity-financed and so profits correspond one-to-one with dividends, where dividends depend on optimal quantity choices and input prices derived from the HJB equation solution. From the asset pricing Euler equation, the price of the firm is the (stochastically) discounted value of all future dividends:
\begin{align*}
S_t &= E_t [ \int_t^{\infty} \mathcal{D}_s \frac{\pi_s}{\pi_t} ds] 
\end{align*}

From this, the stock prices for each firm of interest can be determined as follows:

\begin{proposition}[Stock Prices]

The stock price of firms $i \in \{K, 1, 2, 3\}$ 
\begin{eqnarray*}
    S^{(i)}(\hat{K}_C, \hat{K}_G, \hat{R}, T, \lambda) = s^{(i)}(\hat{K}_G, \hat{R}, T, \lambda) K_{C}^{\alpha}
\end{eqnarray*} 
solves a sector-dependent partial differential equation
\begin{align*}\label{eqn:AP_PDE}
\begin{split}
& 0 = - \left( \tilde{r}_{f,t} + \lambda_t(T_t) \Upsilon_{\pi,t} \right) s^{(i)}  + \tilde{\mathcal{D}}^{(i)}_t \cr
& + \frac{1}{dt} \left(\mathbb{E}[d \hat{K}_G] s^{(i)}_{\hat{K}_G} +\mathbb{E}[d \hat{R}] s^{(i)}_{\hat{R}} + \mathbb{E}[d T] s^{(i)}_T + \mathbb{E}[d \lambda] s^{(i)}_{\lambda} \right) \cr
&  - \sigma_{\pi,\hat{K}_G} ' \sigma_G  s^{(i)}_{\hat{K}_G}  - \sigma_{\pi,\hat{R}} ' \sigma_R  s^{(i)}_{\hat{R}} - \sigma_{\pi,T} ' \sigma_T s^{(i)}_T - \sigma_{\pi,\lambda} ' \left( \Lambda'(T) \sigma_T + \sigma_{\lambda} \right) s^{(i)}_{\lambda} \cr
& + \frac{|\sigma_G |^2}{2}  s^{(i)}_{\hat{K}_G, \hat{K}_G} + \frac{|\sigma_R |^2}{2}  s^{(i)}_{\hat{R}, \hat{R}} + \frac{|\sigma_T |^2}{2} s^{(i)}_{TT}  + \frac{|\Lambda'(T) \sigma_T + \sigma_{\lambda}|^2}{2}  s^{(i)}_{\lambda \lambda} 
\end{split} 
\end{align*}

where $\Upsilon_{\pi,t} = \{1-\frac{s^{(i)}_{post} \pi_{post,t}}{s^{(i)}_{pre} \pi_{pre,t}}\}$, $\tilde{r}_{f,t}$ is the risk-free rate net of the capital sector contributions and $\tilde{\mathcal{D}}^{(i)}_t$ are dividends net of consumption capital contribution (i.e., $\tilde{\mathcal{D}}^{(i)}_t = \mathcal{D}^{(i)}_t /K_C^{\alpha}$)\footnote{Expressions for dividends for each firm type are given in the appendix.}. % with:

The post-jump firm prices $S^{(i)}_{post}$ are given by
\begin{eqnarray*}
S^{(K)}_{post,t} = \frac{\alpha C_{post,t}}{\rho}, \quad S^{(3)}_{post,t} = \frac{\psi C_{post,t}}{\rho}, \quad S^{(2)}_{post,t} = S^{(1)}_{post,t} = 0 
\end{eqnarray*}

\end{proposition} 
%\vspace{0.5cm}

I solve the differential equation for a given stock price, which arises as a direct consequence of the Euler equation expression given above, using the respective expressions for each sector $i \in \{K, 1, 2, 3\}$ . All quantities and prices used as inputs for the PDE are from the planner's optimal choices and equilibrium outcomes.

The firm prices contain useful intuition even though full characterization requires computational solutions. The central takeaway from the pricing expressions is that these outcomes are dependent upon the macroeconomic outcomes related to consumption and energy. As a result, significant shifts in energy production decisions and related implications for household consumption that result from expectations about transition shocks in response to climate change generate meaningful variation in asset prices. A ``run'' response in one or both of the fossil fuel sectors would lead to accelerated losses in firm values for those sectors. The degree to which a ``run on fossil fuels'' leads to substitution to other types of energy production and impacts on overall output can either exacerbate or attenuate the overall financial market implications, which we analyze further when exploring the numerical outcomes later on.

\section{Numerical Solutions}

I now discuss the numerical results based on the theoretical results above. I first outline the model parameters and numerical method used to solve the model, and then delve into the computational solutions of the model. For the baseline computational results, I make the simplifying assumptions for tractability purposes that $\mu_{\lambda} = \sigma_{\lambda} = 0$.  This allows me to drop $\lambda_t$ as a state variable, leaving me to solve a three-dimensional PDE for the HJB equation. I examine the impacts of relaxing certain model assumptions in Section \ref{section:CounterfactualsAlternatives} in order to highlight how alternative model specifications impact the baseline model results.

\subsection{Model Parameters}

The parameters used in the model are chosen based on external estimates from the literature and calibrated values matching empirical macroeconomic and asset pricing moments with model moments from a simplified, no-climate, no-transition risk version of the baseline model with an analytical solution. As a result, the numerical results are intended to provide quantitatively reasonable values for the economic and financial outcomes of the model. Parameter values used for the solution are given in Table \ref{params}. A short summary of the parameter values chosen is given below, with further details left for the Online Appendix.

First are the parameters chosen based on external estimates. The consumption capital productivity parameter $A_C$, and the green capital productivity parameter $A_G$, match values used by \cite{pindyck_climate_2013} and \cite{barnett2020pricing}. The energy input demand share $\psi$ comes from \cite{hassler2021directed}, matching energy's share of income for a CD specification. The initial oil and coal energy demand shared $\nu_1, \nu_2$ are based on estimates from the \cite{BP:2020}, \cite{vaclav2017energy}, and \cite{energy2023statistical}. There is a wide range of values used in the literature for the elasticity of substitution $\omega$, see \cite{golosov_optimal_2014} and \cite{acemoglu2012environment} for examples. I choose a positive value\footnote{Assuming $\omega > 0$ allows for analysis of taxation transition risk scenarios that would be infeasible otherwise.} that is within this range to generate macroeconomic and asset pricing outcomes consistent with the data. 
The reserves volatility $\sigma_R$ matches the time series standard deviation of log changes for oil reserves from \cite{BP:2020}. The temperature volatility $\sigma_T$ matches the time series standard deviation of global temperature from the NASA-GISS database. The damage function parameter is consistent with \cite{golosov_optimal_2014} and \cite{Nordhaus:2017} for temperature anomalies ranging between one and five degrees Celsius. The climate sensitivity parameter matches the mean estimate from \cite{MacDougallSwartKnutti:2017}.

%\vspace{-0.85cm}

\begin{center}
\begin{table}[!ht]
%\caption{Parameters}
\begin{center}
\caption{Model Parameters} \label{params} 
%\vspace{-0.5cm}
\begin{tabular}{l c c }
\hline \hline
${\begin{array}{c} \text{Discount Rate} \end{array}}$  & $\rho$ & $0.01$  \\
${\begin{array}{c} \text{Risk Aversion} \end{array}}$  & $\gamma$ & $9.5$  \\
${\begin{array}{c} \text{EIS} \end{array}}$  & $\theta$ & $1$  \\
${\begin{array}{c} \text{Capital TFP} \end{array}}$  & $A_i, \, i \in \{C,G\}$ & $0.115$  \\
${\begin{array}{c} \text{Capital Adjustment Costs} \end{array}}$  & $\{\mu_i, \phi_i\}, \, i \in \{C,G\}$ & $\{-0.043, 6.67\}$  \\
${\begin{array}{c} \text{Capital Input Demand Share} \end{array}}$  & $\alpha$ & $0.35$  \\
${\begin{array}{c} \text{Coal TFP} \end{array}}$  & $A_2$ & $132.21$  \\
${\begin{array}{c} \text{Energy Input Demand Share} \end{array}}$  & $\psi$ & $0.05$  \\
${\begin{array}{c} \text{Oil Energy Demand Share} \end{array}}$  & $\nu_1$ & $ 0.55$  \\
${\begin{array}{c} \text{Coal Energy Demand Share} \end{array}}$  & $\nu_2$ & $ 0.25$  \\
${\begin{array}{c} \text{Green Energy Demand Share} \end{array}}$  & $\nu_3$ & $0.2$  \\
${\begin{array}{c} \text{Elasticity of Substitution} \end{array}}$  & $\omega$ & $0.5$  \\
%\hline
${\begin{array}{c} \text{Climate Sensitivity} \end{array}}$  & $\beta_T$ & $1.86 \times 10^{-3}$  \\
${\begin{array}{c} \text{Climate Damages Parameter} \end{array}}$  & $\eta$ & $0.02$  \\
%${\begin{array}{c} \text{Climate Internalization Parameter} \end{array}}$  & $\chi$ & $\{0.5, 1.0\}$  \\
${\begin{array}{c} \text{Transition Arrival Rate Parameters} \end{array}}$  & $\{ \psi_0, \psi_1, \psi_2, \underline{T} \}$ & $\{0.072, 0.79, 2.0, 1.25 \}$  \\
%\hline
${\begin{array}{c} \text{$K_C, K_G,R, T$ Volatilities} \end{array}}$  & $\{\sigma, \sigma_R, \sigma_T \}$ & $\{0.072, 0.034, 0.10 \}$   \\
\hline
${\begin{array}{c} \text{Initial Consumption Capital Value} \end{array}}$   & $K_{C,0}$ & $85/A_C$ \\
${\begin{array}{c} \text{Initial Green Capital Value} \end{array}}$   & $K_{G,0}$ & $2.5/A_G$ \\
%${\begin{array}{c} \text{Initial Fossil Fuel Reserves Value} \end{array}}$   & $R_0$ & $650$ \\
${\begin{array}{c} \text{Initial Fossil Fuel Reserves Value} \end{array}}$   & $R_0$ & $850$ \\

${\begin{array}{c} \text{Initial Mean Temperature Value} \end{array}}$   & $T_0$ & $1.2$ \\
\hline  \hline
%\multicolumn{2}{l}{\textsuperscript{a}\footnotesize{$K_0$ is derived from $K_0=Y_0/\alpha$}}
%\hline
%\hline
\end{tabular}
\end{center} %\vspace{0.2cm}
\begin{footnotesize}
%Table~\ref{table:params} presents ...
\end{footnotesize}%\vspace{1cm}
\end{table}
\end{center}

Next are the directly estimated parameters. I estimate values of $\psi_0, \psi_1$, $\psi_2$, and $\underline{T}$ to produce a transition risk probability pathway consistent with the ``Technical Challenges'' Policy Trajectory from \cite{moore2022determinants} for annual emissions constant at today's value. The remaining parameters are estimated to match empirical moments with moments from no-climate, no-transition risk versions of the model. Consumption and green capital adjustment costs and depreciation parameters $\phi_C, \mu_C$ and $\phi_G, \mu_G$ are set to generate growth rates and Tobin's q values that match World Bank and BEA data for aggregate world GDP. The subjective discount rate $\rho$ is set to generate global annual emissions similar to estimates by \cite{friedlingstein2023global} and \cite{Figueresetal:2018}. I assume a unitary EIS $\theta$, and set risk aversion $\gamma$ and the capital volatility $\sigma=\sigma_K=\sigma_G$ to generate a risk-free rate and Sharpe ratio that match values from Ken French's data library, assuming a 5-to-3 leverage ratio.\footnote{This matches the value used by \cite{boldrin2001habit}, \cite{papanikolaou2011investment}, and others.} 

%%%%%%%%%%%%%%%%%%%%%%%%%%%%%%%%%%%%%%%%%%%%%%%%%%%%%%%%%%%%%%%%%%%%%%%%%%%%%%%
%%%%%%%%%%%%%%%%%%%%%%%%%%%%%%%%%%%%%%%%%%%%%%%%%%%%%%%%%%%%%%%%%%%%%%%%%%%%%%%
%%%%%%%%%%%%%%%%%%%%%%%%%%%%%%%%%%%%%%%%%%%%%%%%%%%%%%%%%%%%%%%%%%%%%%%%%%%%%%%
%%%%%%%%%%%%%%%%%%%%%%%%%%%%%%%%%%%%%%%%%%%%%%%%%%%%%%%%%%%%%%%%%%%%%%%%%%%%%%%

Initial values used for the model solution simulations (annual global GDP, annual green energy production, global fossil fuel reserves, and global mean temperature) come from the World Bank, the Energy Information Administration (EIA), the Bureau of Economic Activity (BEA), and the IPCC AR6 and NASA-GISS, respectively. All initial values and parameter values are at an annual frequency.

%%%%%%%%%%%%%%%%%%%%%%%%%%%%%%%%%%%%%%%%%%%%%%%%%%%%%%%%%%%%%%%%%%%%%%%%%%%%%%%
%%%%%%%%%%%%%%%%%%%%%%%%%%%%%%%%%%%%%%%%%%%%%%%%%%%%%%%%%%%%%%%%%%%%%%%%%%%%%%%

\subsection{Numerical Algorithm}

I briefly discuss the numerical method used to solve the model for each of the different specified frameworks mentioned previously. The Hamilton-Jacobian-Bellman (HJB) equations that characterize the value functions for the various settings are nonlinear partial differential equations (PDEs). To solve these nonlinear PDEs, I use an iterative algorithm that implements the method of false transient with a finite difference scheme and conjugate gradient solver. The algorithm requires constructing a conditionally linear system whose solution is approximated using the conjugate gradient method. Any nonlinear components of the HJB equation are treated as given for each iteration so that the system is well represented by the linear system for which I apply the conjugate gradient method to solve. After finding the solution of the conditionally linear system, I update the value function and nonlinear components and repeat to solve the modified, conditionally linear system. This process is repeated until convergence is achieved. More details can be found in the Online Appendix.

%%%%%%%%%%%%%%%%%%%%%%%%%%%%%%%%%%%%%%
\subsection{Numerical Results}

\begin{figure}[!thp]
\begin{center}
\caption{Climate Transition Shock Arrival Rate Function} 
\includegraphics[width=0.5\textwidth]{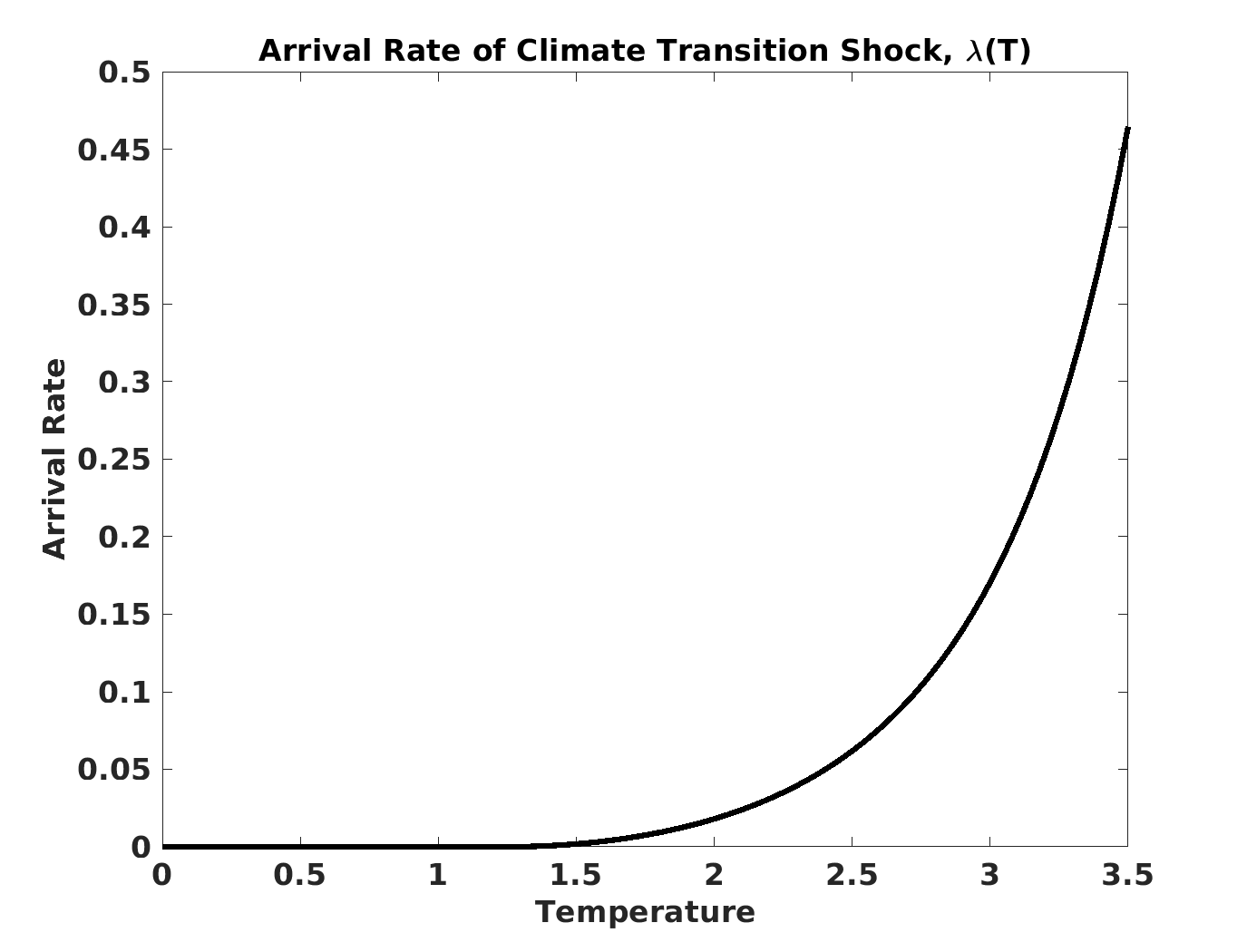}  \label{ArrivalRate} %\\ \includegraphics[width=0.495\textwidth]
\end{center}

\vspace{-0.1cm}

\begin{footnotesize}
Figure \ref{ArrivalRate} plots $\lambda(T_t)$, the arrival rate of the climate-dependent transition  as a function of temperature $T_t$. The function is parameterized as $\psi_0 = 0.072, \psi_1 = 0.79, \psi_2 = 2.0, \underline{T} = 1.25$ so that the transition risk probability pathway is consistent with the ``Technical Challenges'' Policy Trajectory from \cite{moore2022determinants}. %for annual emissions values constant at today's value.
\end{footnotesize}  
\end{figure}

For the numerical results, I first plot in Figure \ref{ArrivalRate} the arrival rate of the climate-dependent transition $\lambda(T_t)$ as a function of temperature $T_t$ to highlight the main model mechanism. The function is strictly increasing in $T$, and the likelihood goes from zero up until $1.25^{\circ} C$, gradually increasing through $1.5^{\circ} C$, and then takes off dramatically around $2^{\circ} C$. 

\begin{figure}[!t]

%\vspace{-1.0cm}
% \vspace{-0.5cm}

\caption{Macroeconomic and Asset Pricing Outcomes - ``Technology'' Shock} \label{fig:technology_model_sims}
\begin{center}
% {\scriptsize \textbf{Panel A: Hybrid Taxation/Technology Transition Scenario}}\\
        \begin{subfigure}[b]{0.328\textwidth}
            \centering
            \includegraphics[width=\textwidth]{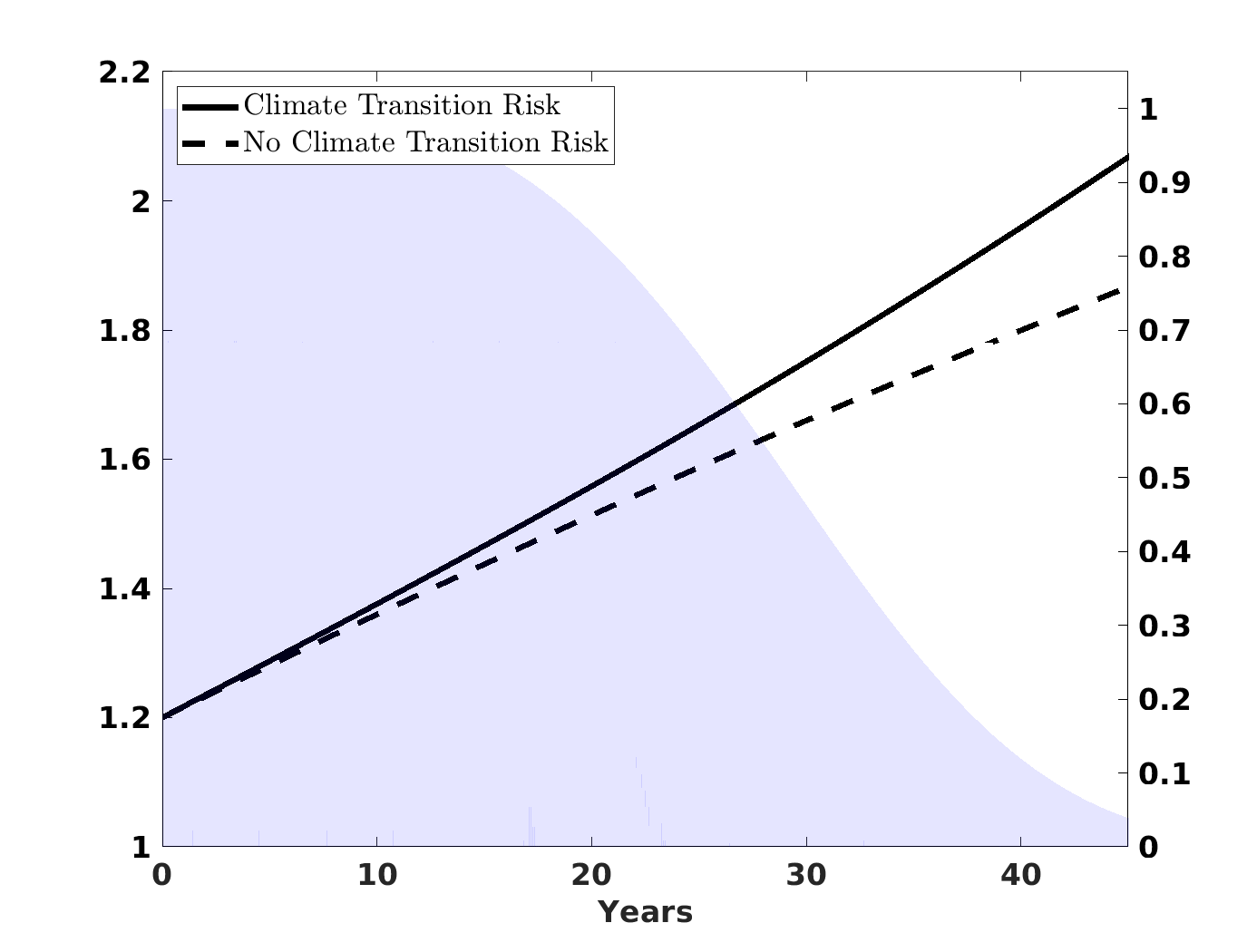}
            \caption[]{{\small Temperature: $Y_t$}}              
        \end{subfigure}
        \begin{subfigure}[b]{0.328\textwidth}  
            \centering 
            \includegraphics[width=\textwidth]{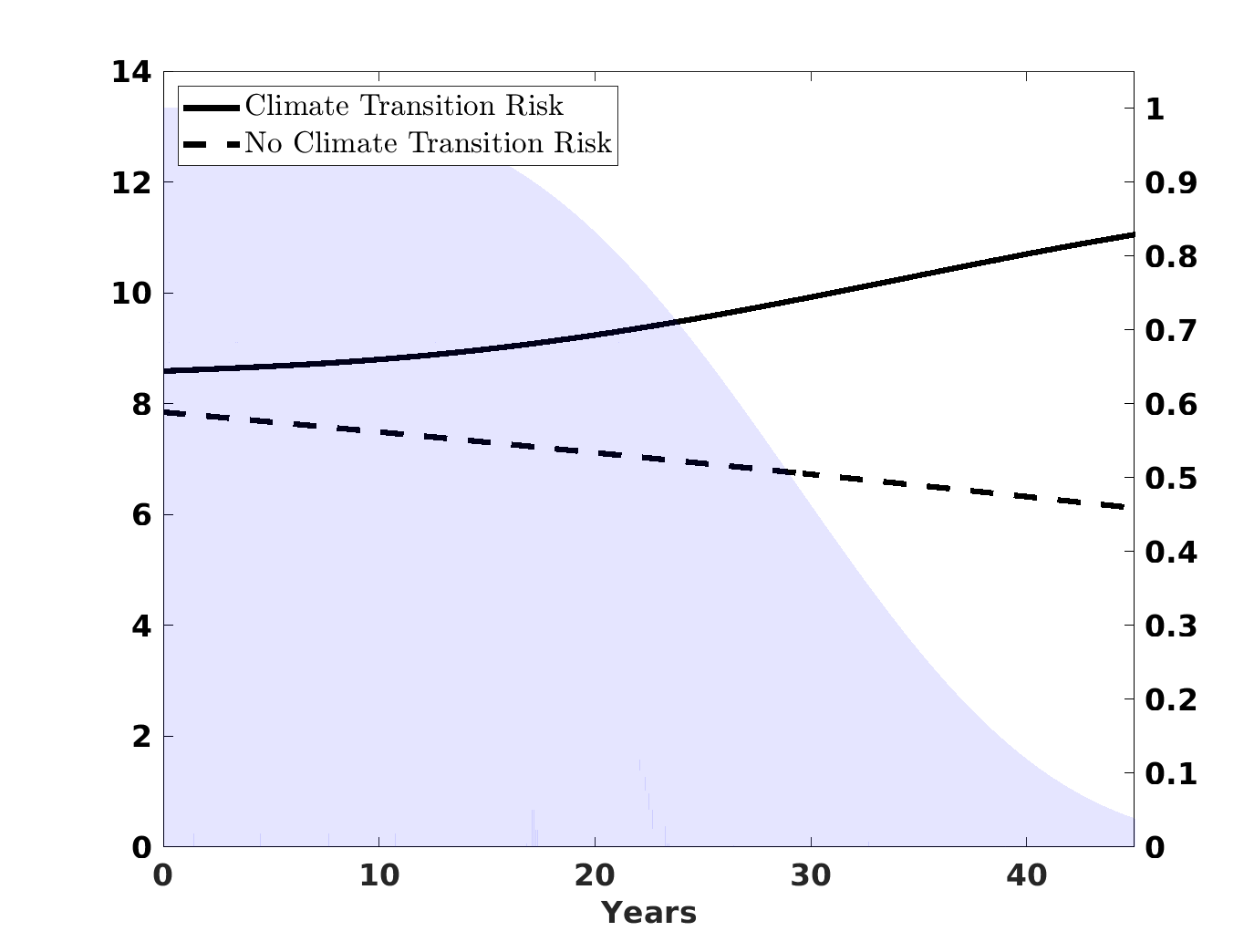}
           \caption[]{{\small Oil Production: $E_{1,t}$}}
        \end{subfigure}
        \begin{subfigure}[b]{0.328\textwidth}  
            \centering 
            \includegraphics[width=\textwidth]{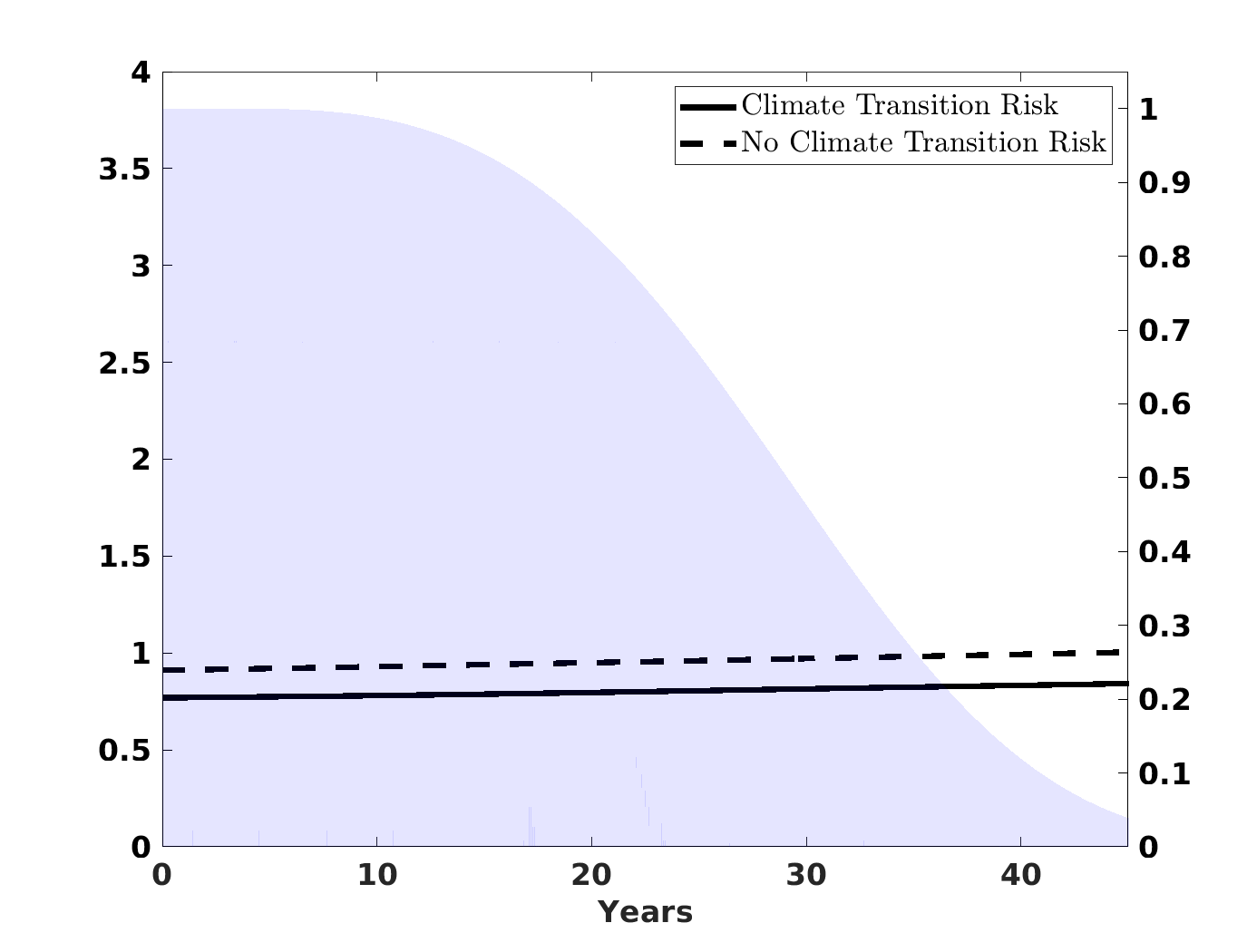}
           \caption[]{{\small Coal Production: $E_{2,t}$}}
        \end{subfigure}     
        
        \begin{subfigure}[b]{0.328\textwidth}
            \centering
            \includegraphics[width=\textwidth]{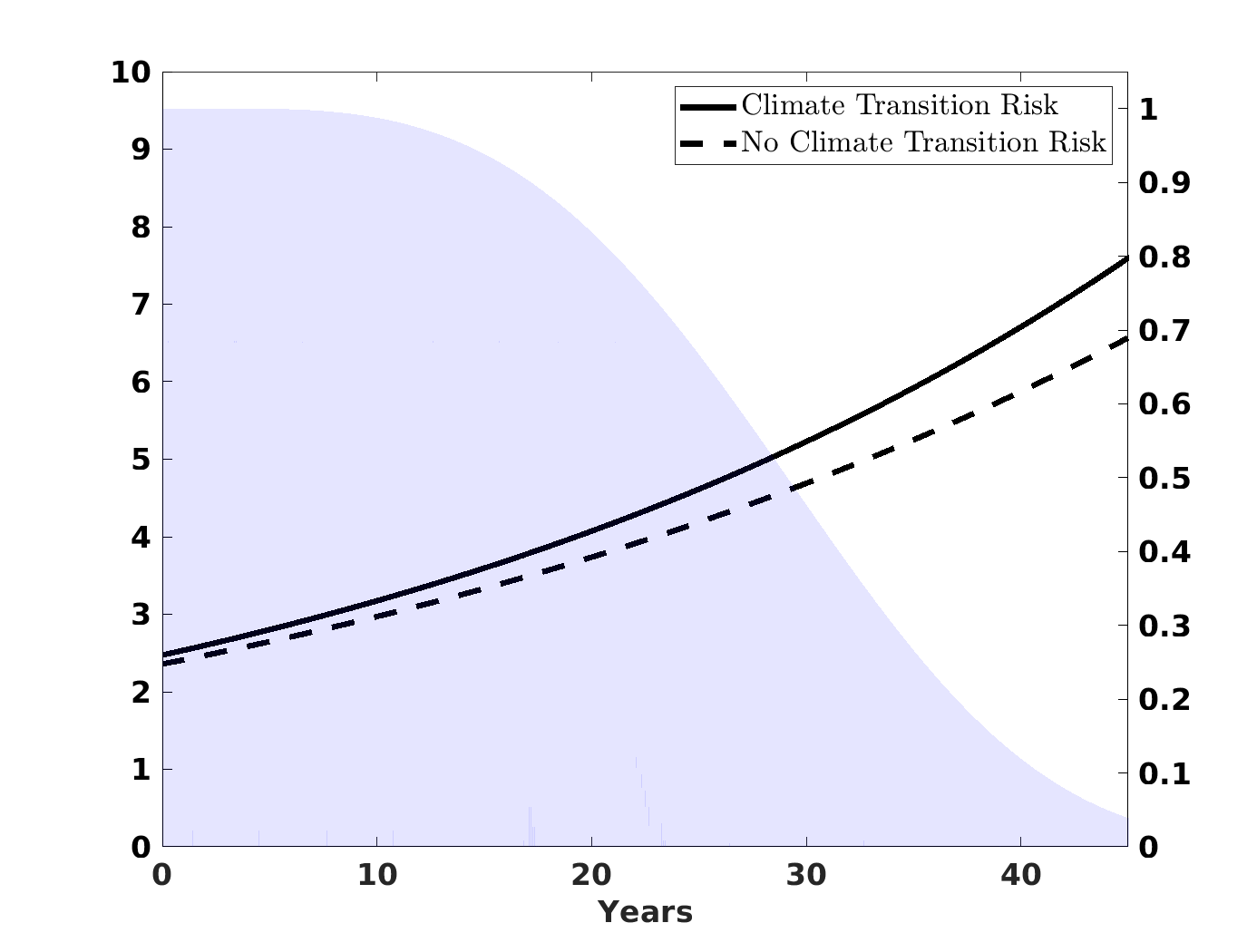}
            \caption[]{{\small Green Investment: $I_{G,t}$}}              
        \end{subfigure}
        \begin{subfigure}[b]{0.328\textwidth}  
            \centering 
            \includegraphics[width=\textwidth]{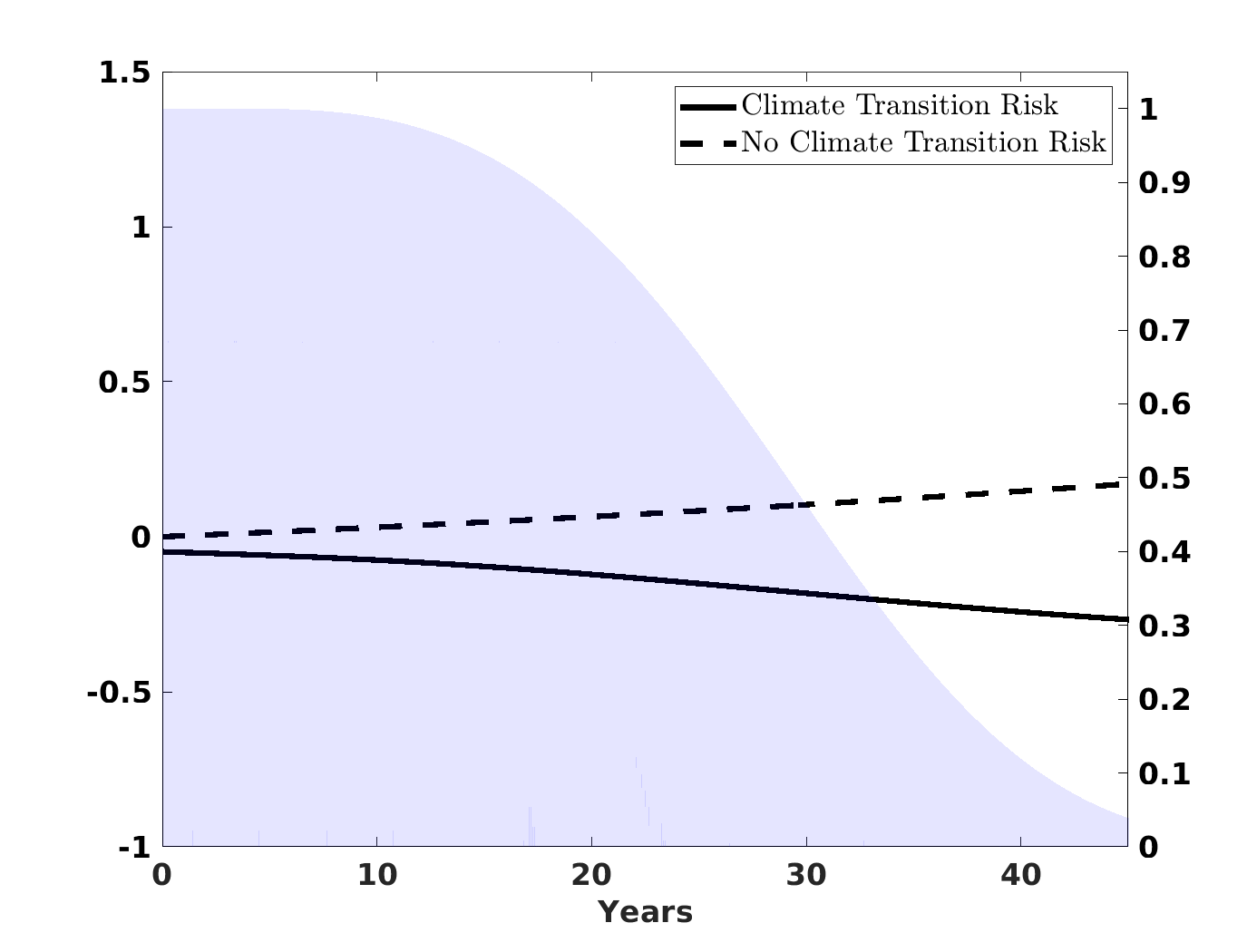}
           \caption[]{{\small Oil Spot Price: $P_{1,t}$}}
        \end{subfigure}
        \begin{subfigure}[b]{0.328\textwidth}  
            \centering 
            \includegraphics[width=\textwidth]{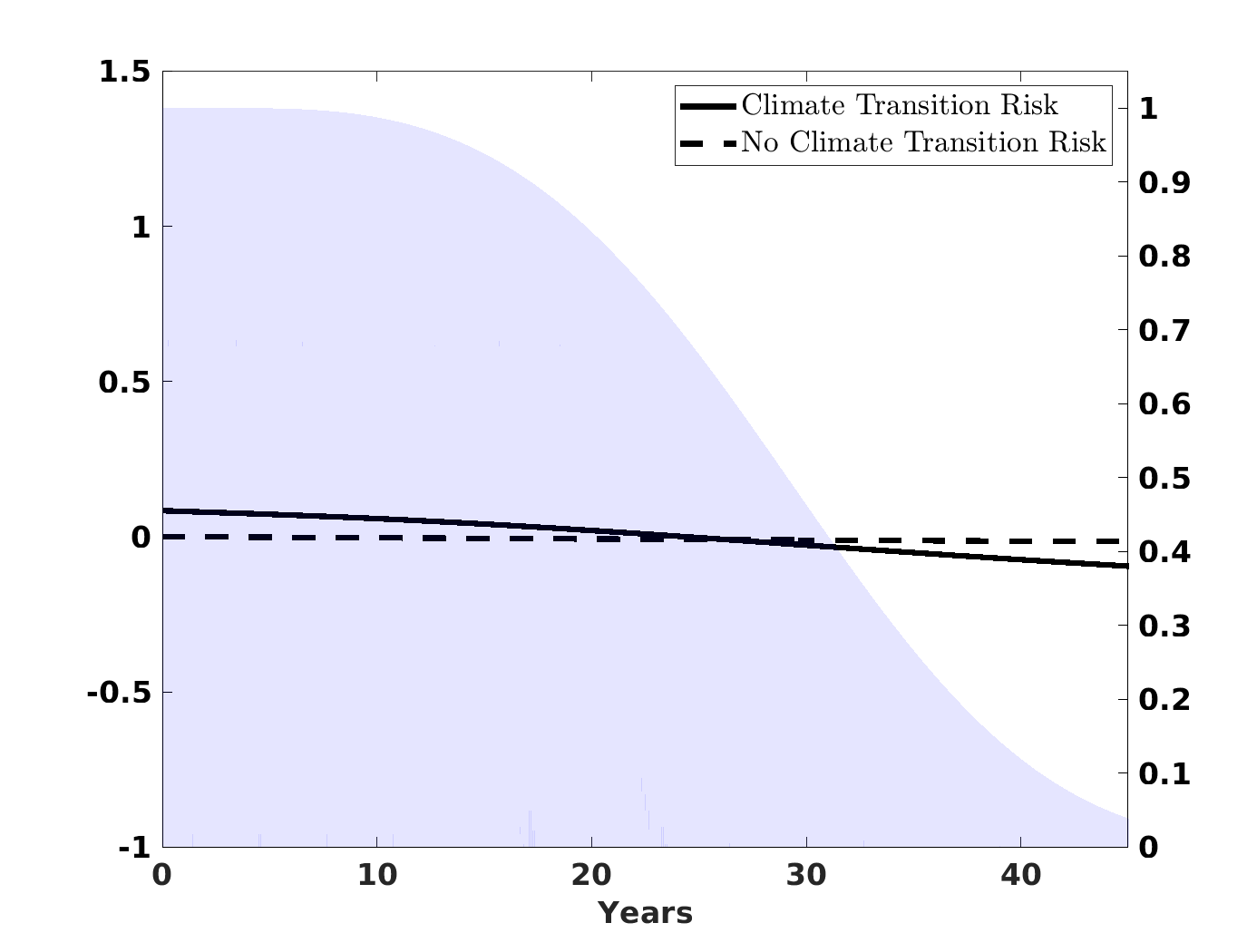}
           \caption[]{{\small Coal Spot Price: $P_{1,t}$}}
        \end{subfigure}

        \begin{subfigure}[b]{0.328\textwidth}
            \centering
            \includegraphics[width=\textwidth]{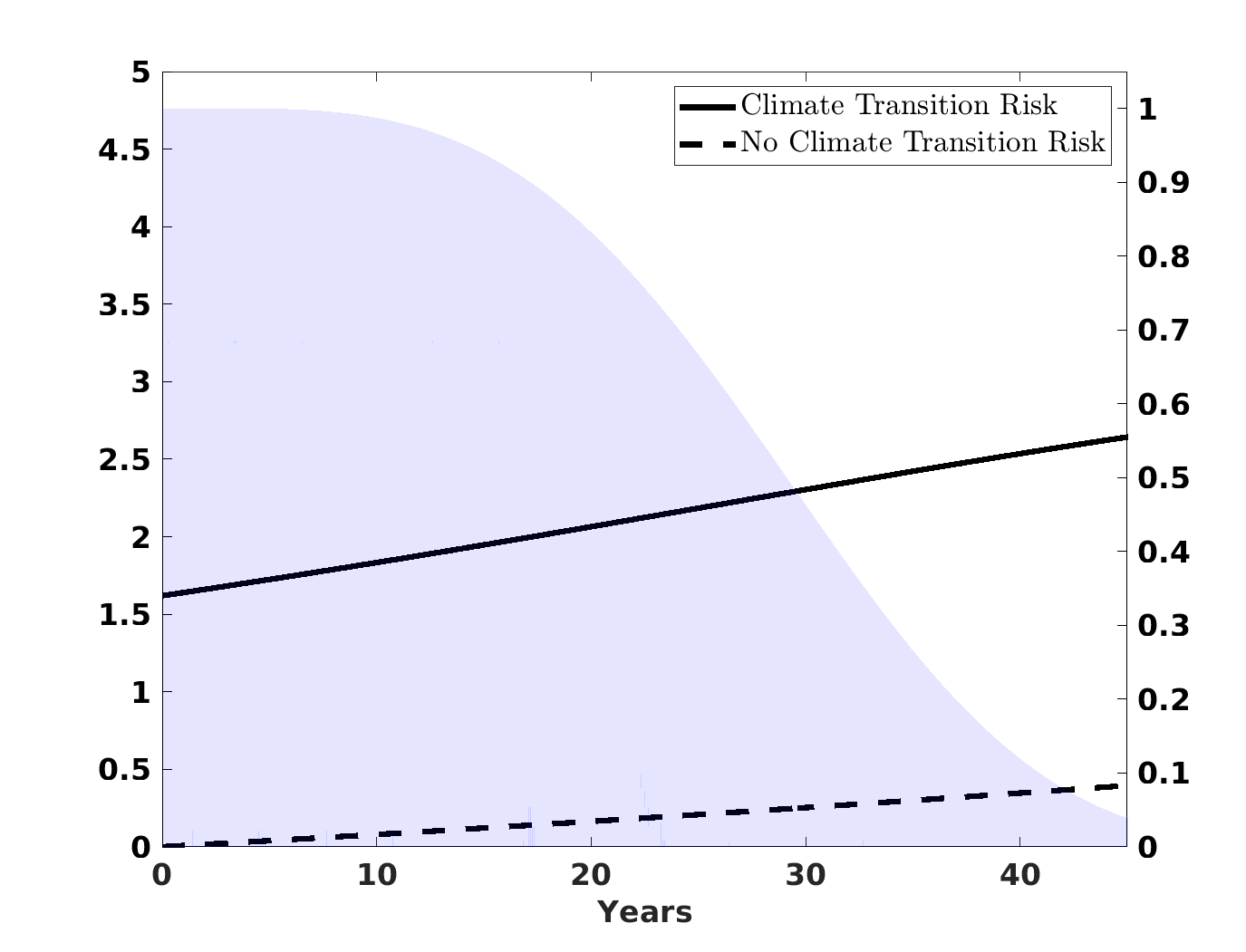}
            \caption[]{{\small Green Firm Price: $S^{(3)}_{t}$}}              
        \end{subfigure}
        \begin{subfigure}[b]{0.328\textwidth}  
            \centering 
            \includegraphics[width=\textwidth]{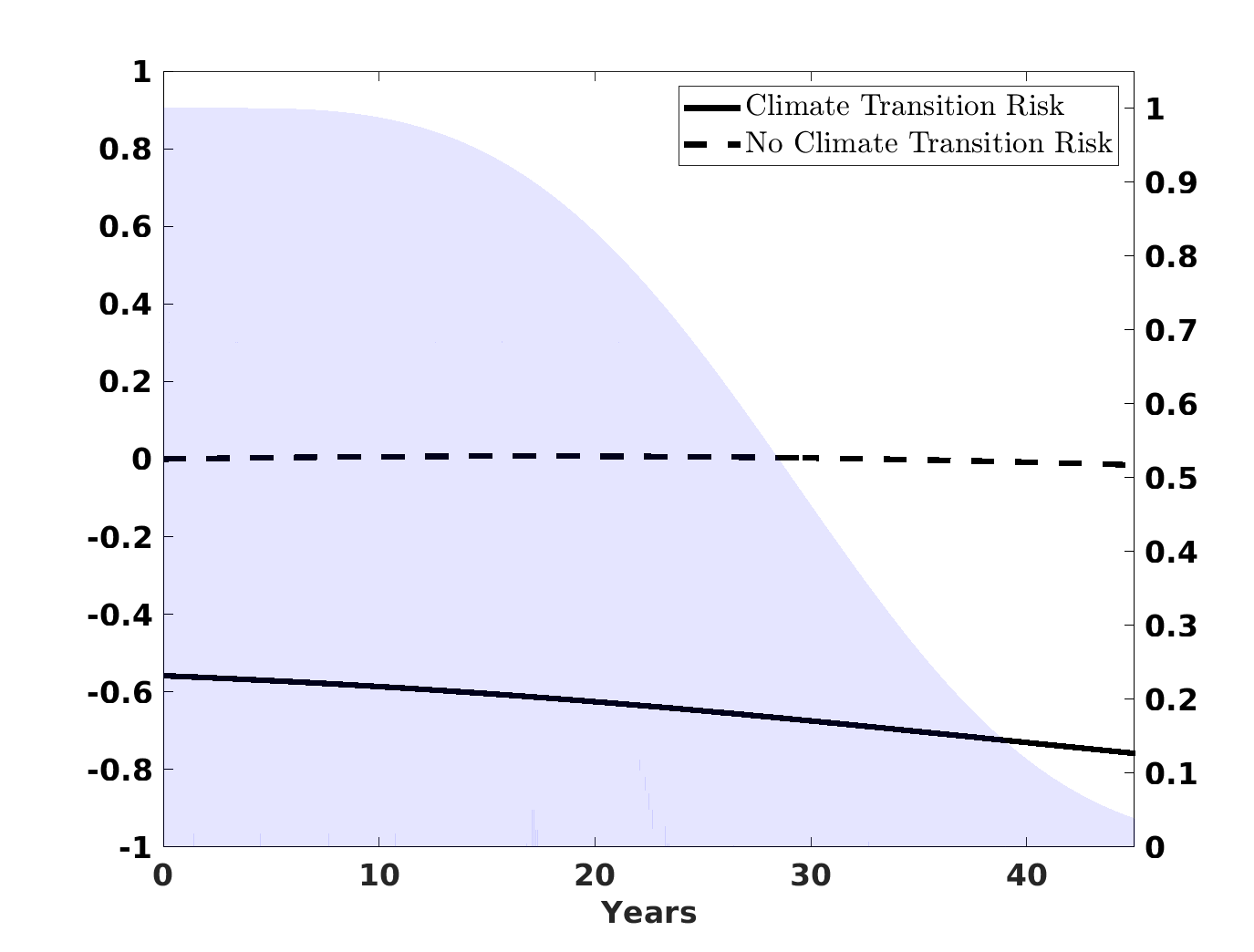}
           \caption[]{{\small Oil Firm Price: $S^{(1)}_{t}$}}
        \end{subfigure}
        \begin{subfigure}[b]{0.328\textwidth}  
            \centering 
            \includegraphics[width=\textwidth]{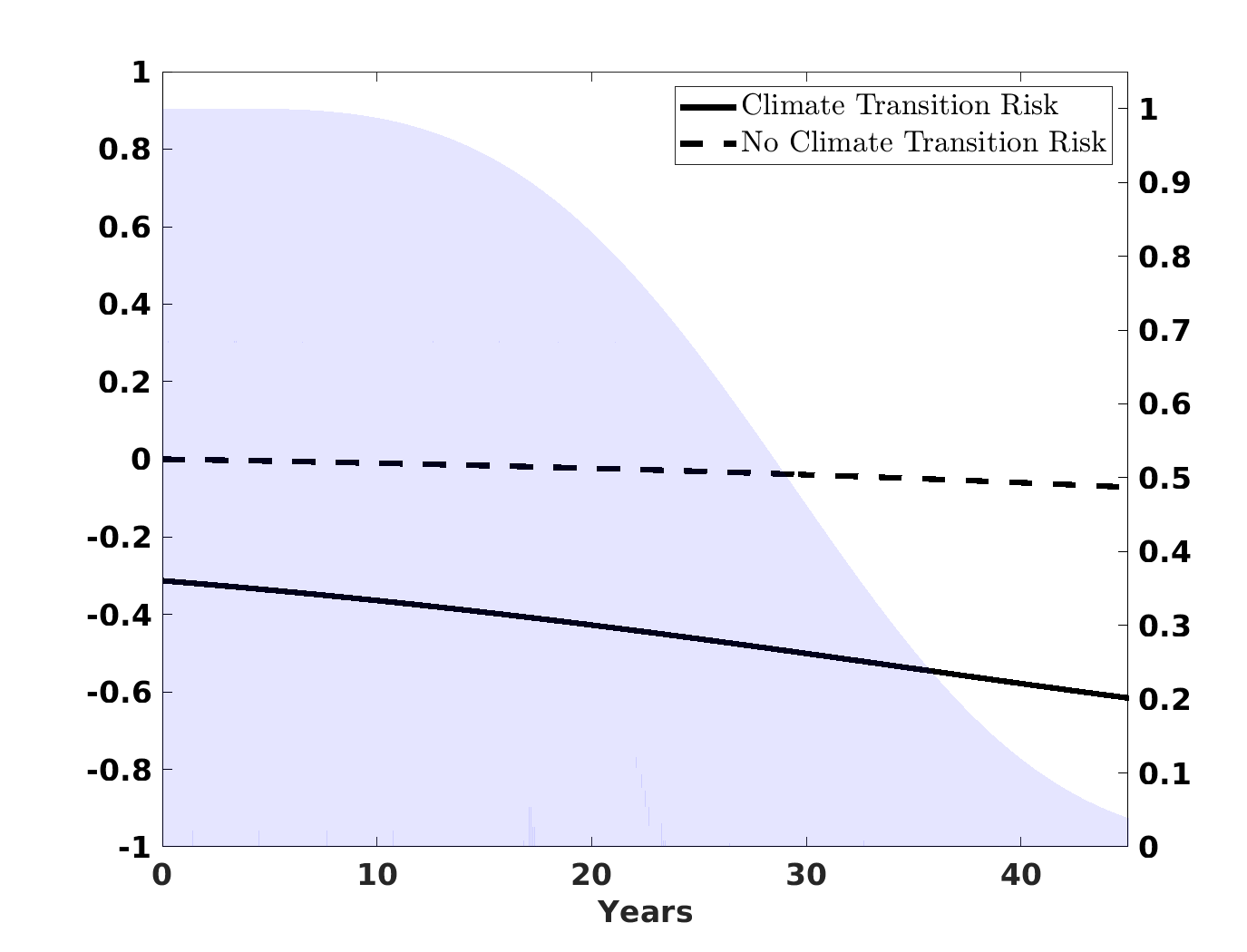}
           \caption[]{{\small Coal Firm Price: $S^{(2)}_{t}$}}
        \end{subfigure}        
        
        \vspace{-0.25cm}
\end{center}

\begin{footnotesize}
Figure \ref{fig:technology_model_sims} shows the simulated outcomes for the ``technology shock'' model based on the numerical solutions. Panels (a) through (c) show the temperature anomaly, oil production, and coal production. Panels (d) through (f) show the green investment choice, oil spot price, and coal spot price. Panels (g) and (i) show the green firm price, oil firm price, and coal firm price. Solid lines represent results for the Climate Transition Risk scenario where $\lambda_t = \lambda(T_t)$ and dashed lines represent results for the No Climate Transition Risk scenario where $\lambda_t = 0$. The blue shaded region shows the cumulative probability of no transition shock occurring.
\end{footnotesize} 

\end{figure}

In Figure \ref{fig:technology_model_sims}, I plot the (zero-shock) simulated time series using the model solutions for the key climate, macroeconomic, and asset pricing outcomes of the ``technology shock'' climate transition risk case. The solid lines labeled ``Climate Transition Risk'' give the results for the main ``technology shock'' climate transition risk scenario. I also plot two additional results: the counterfactual comparison setting without climate transition risk ($\lambda_t = 0$), labeled as ``No Climate Transition Risk'' and given by the dashed line; and the cumulative probability of the climate transition not yet occurring for each climate transition risk scenario, given by the blue shaded region to demonstrate the likelihood of the time-series realizations. The left y-axis corresponds to the simulated time series outcome, and the right y-axis corresponds to the cumulative probability of no transition shock having occurred yet.

For this ``Climate Transition Risk'' scenario, the likelihood of the climate transition shock not occurring, shown by the shaded blue region on each figure, is around $5\%$ by year 45. Panel (a) shows the temperature pathway which, as anticipated, increases in both cases. The temperature reaches approximately $2.05^{\circ} C$ in the ``Climate Transition Risk'' scenario, driving the increasing likelihood of a transition shock, whereas for the ``No Climate Transition Risk'' scenario the outcome is a more modest $1.85^{\circ} C$. Panels (b) and (c) show oil and coal production for each scenario. In the ``Climate Transition Risk'' scenario there is a clear ``run'' on oil, as the level of extraction is higher initially and dynamically increases as temperature increases, reserves decrease, and the likelihood of a climate transition shock increases. The lower initial extraction and gradual decrease in the ``No Climate Transition Risk'' scenario reflect the expected outcomes of a Hotelling-type model influenced by the optimal carbon tax underlying the social planner's solution. Importantly, this choice to ``run'' in the ``Climate Transition Risk'' scenario and maximize current gains from reserves before they become a stranded asset is socially optimal, even with potential climate and transition implications. The response is starkly different for coal production. In the ``Climate Transition Risk'' scenario, production is lower than in the ``No Climate Transition Risk'' and remains lower throughout the simulation. Thus, the run effect is concentrated in the oil sector and is not a feature of coal production in this setting. Panel (d) shows investment in the green capital stock (in levels) for each scenario. While investment in green capital is increasing over time in each case, the ``Climate Transition Risk'' scenario investment is elevated relative to the ``No Climate Transition Risk'' scenario. This response reflects increased expected importance of and dependence on the green sector in the future and increased availability of output to use for investment, both resulting from the climate transition risk shock mechanism.

%%%%%%%%%%%%%%%%%%%%%%%%%%%%%%%%%%%%%%%%%%%%%%%%%%
%%%%%%%%%%%%%%%%%%%%%%%%%%%%%%%%%%%%%%%%%%%%%%%%%%
%%%%%%%%%%%%%%%%%%%%%%%%%%%%%%%%%%%%%%%%%%%%%%%%%%

%%%%%%%%%%%%%%%%%%%%%%%%%%%%%%%%%%%%%%%%%%%%%%%%%%
%%%%%%%%%%%%%%%%%%%%%%%%%%%%%%%%%%%%%%%%%%%%%%%%%%
%%%%%%%%%%%%%%%%%%%%%%%%%%%%%%%%%%%%%%%%%%%%%%%%%%

 Panels (e) and (f) show the spot prices for oil and coal, respectively.  In the ``Climate Transition Risk'' setting the spot price of oil is lower relative to the ``No Climate Transition Risk'' setting, reflecting the increased supply of oil in the market. Furthermore, the run effect is seen in the spot price as the price has a downward sloping path over the time series. For the counterfactual setting, the spot price gradually and monotonically increases as extraction gradually decreases. For coal, the spot prices are quite similar in each scenario, though the spot price in decreasing at a faster rate for the ``Climate Transition Risk'' setting. Panels (h) and (i) show the firm values for the oil and coal firms, while Panel (g) gives the green firm price. For ease of interpretation, prices are normalized so that the counterfactual-setting firm value is zero at the initial period. The impact of climate transition risk is much more similar across the two types of fossil fuel firm valuations than for the production and spot price outcomes. The fossil fuel firm prices are both significantly diminished and decreasing over time in the ``Climate Transition Risk'' setting, whereas the counterfactual ``No Climate Transition Risk'' results show higher and gradually increasing firm values for the oil and coal firms. The similarity in fossil fuel firm price responses highlights the significance of the transition risk impact. The expectation of ``stranding'' these two types of fossil fuel inputs and production technologies amplifies the discounting of the future value of these firms in an accelerating way because of the increasing arrival probability of the transition shock occurring. In addition, the decreasing fossil fuel spot prices further exacerbates the diminishing firm valuations by reducing the value of the cash flows produced by these firms. For the green firm price, we see a substantial increase in firm value that grows over time, reflecting increased green capital investment and increased expected importance of and dependence on the green sector in the future in the ``Climate Transition Risk'' setting. While the magnitude of the increase in the value of the green firm is quite large, note that the price level is much smaller given the low demand share of green energy in the aggregate economy.

%%%%%%%%%%%%%%%%%%%%%%%%%%%%%%%%%%%%%%%%%%%%%%%%%%
%%%%%%%%%%%%%%%%%%%%%%%%%%%%%%%%%%%%%%%%%%%%%%%%%%
%%%%%%%%%%%%%%%%%%%%%%%%%%%%%%%%%%%%%%%%%%%%%%%%%%

% \begin{landscape}

\begin{figure}[!ht]

%\vspace{-1.0cm}
% \vspace{-0.5cm}

\caption{Macroeconomic and Asset Pricing Outcomes - ``Taxation'' Shock} \label{fig:taxation_model_sims}
\begin{center}
% {\scriptsize \textbf{Panel A: Hybrid Taxation/Technology Transition Scenario}}\\
        \begin{subfigure}[b]{0.328\textwidth}
            \centering
            \includegraphics[width=\textwidth]{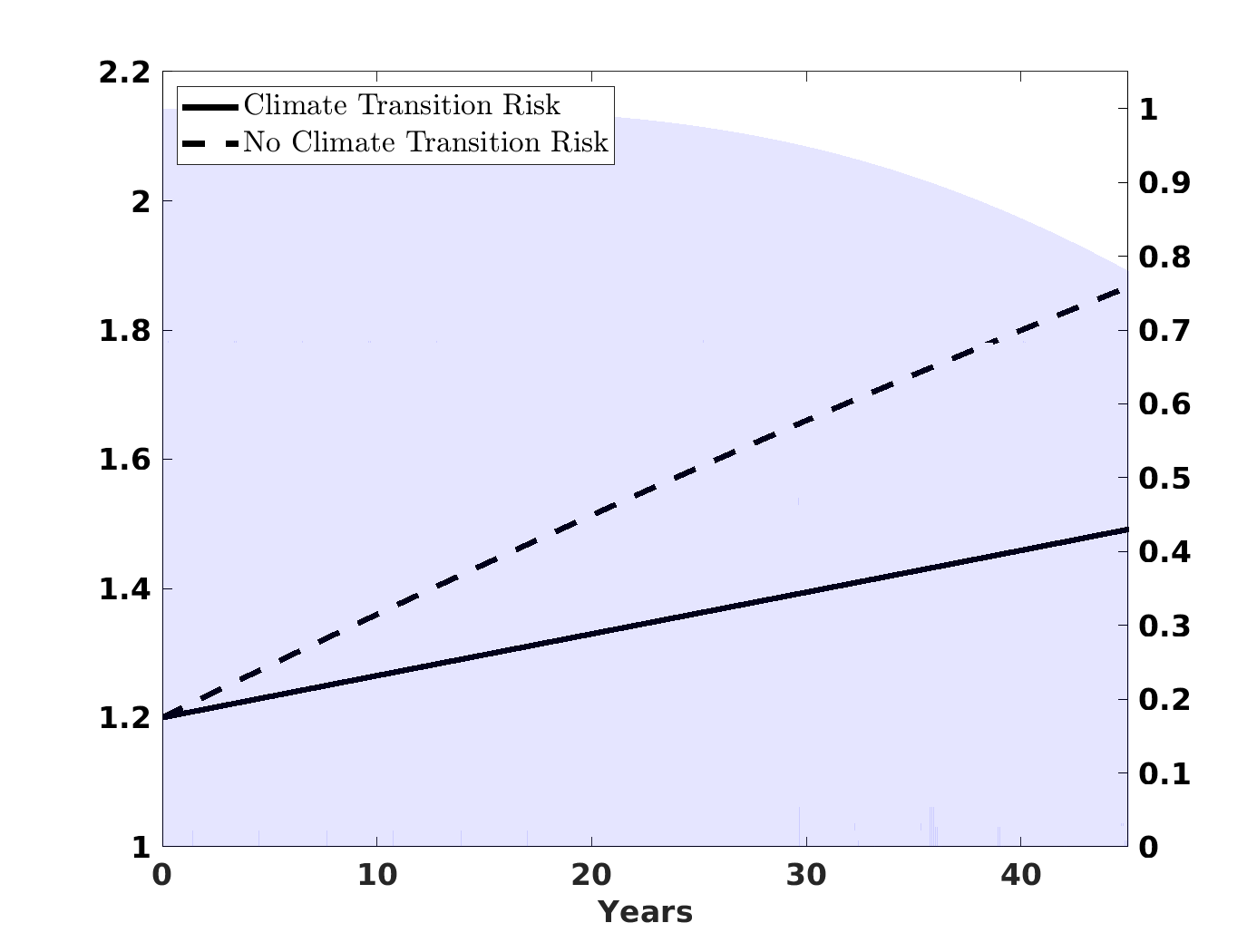}
            \caption[]{{\small Temperature: $Y_t$}}              
        \end{subfigure}
        \begin{subfigure}[b]{0.328\textwidth}  
            \centering 
            \includegraphics[width=\textwidth]{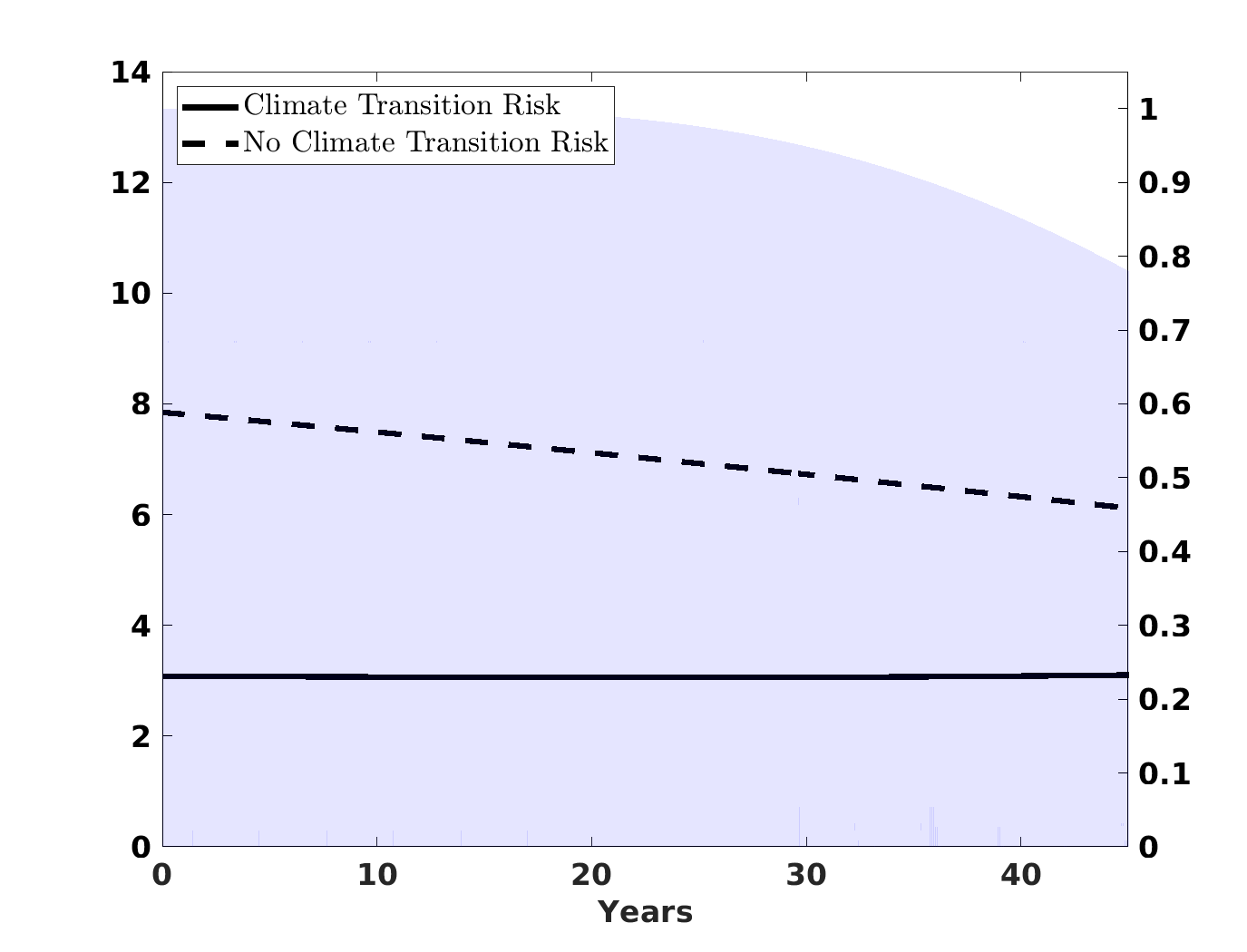}
           \caption[]{{\small Oil Production: $E_{1,t}$}}
        \end{subfigure}
        \begin{subfigure}[b]{0.328\textwidth}  
            \centering 
            \includegraphics[width=\textwidth]{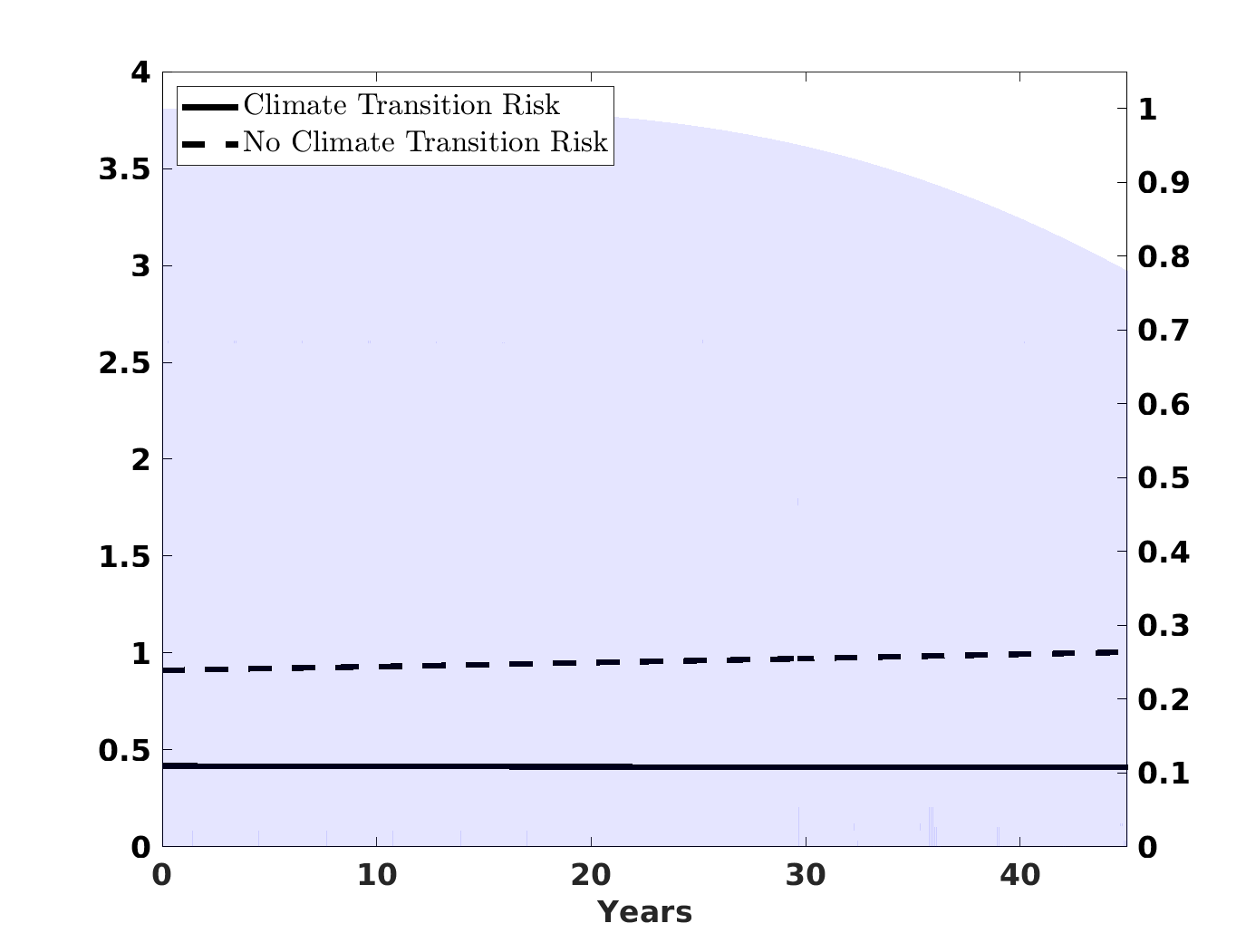}
           \caption[]{{\small Coal Production: $E_{2,t}$}}
        \end{subfigure}     
        
        \begin{subfigure}[b]{0.328\textwidth}
            \centering
            \includegraphics[width=\textwidth]{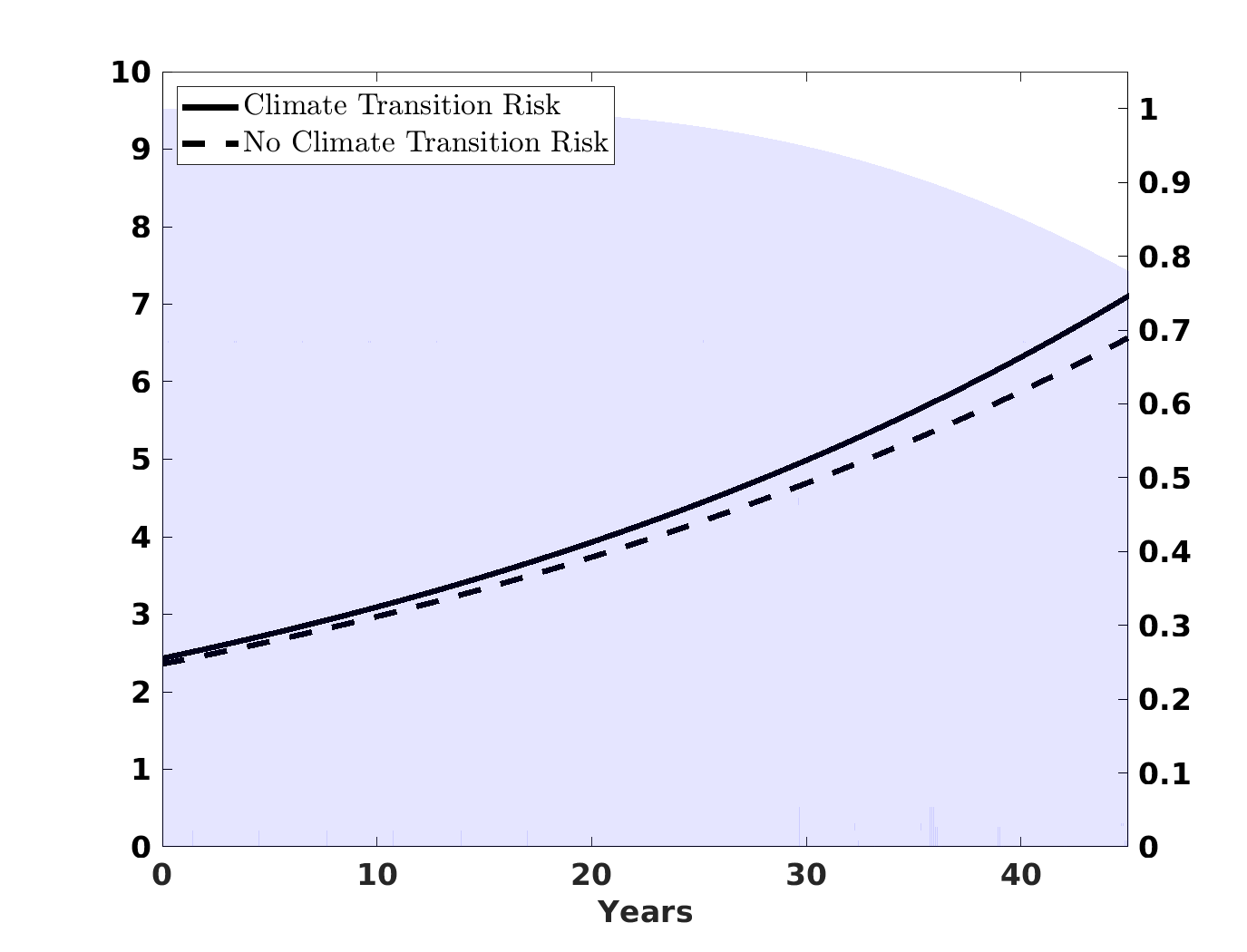}
            \caption[]{{\small Green Investment: $I_{G,t}$}}              
        \end{subfigure}
        \begin{subfigure}[b]{0.328\textwidth}  
            \centering 
            \includegraphics[width=\textwidth]{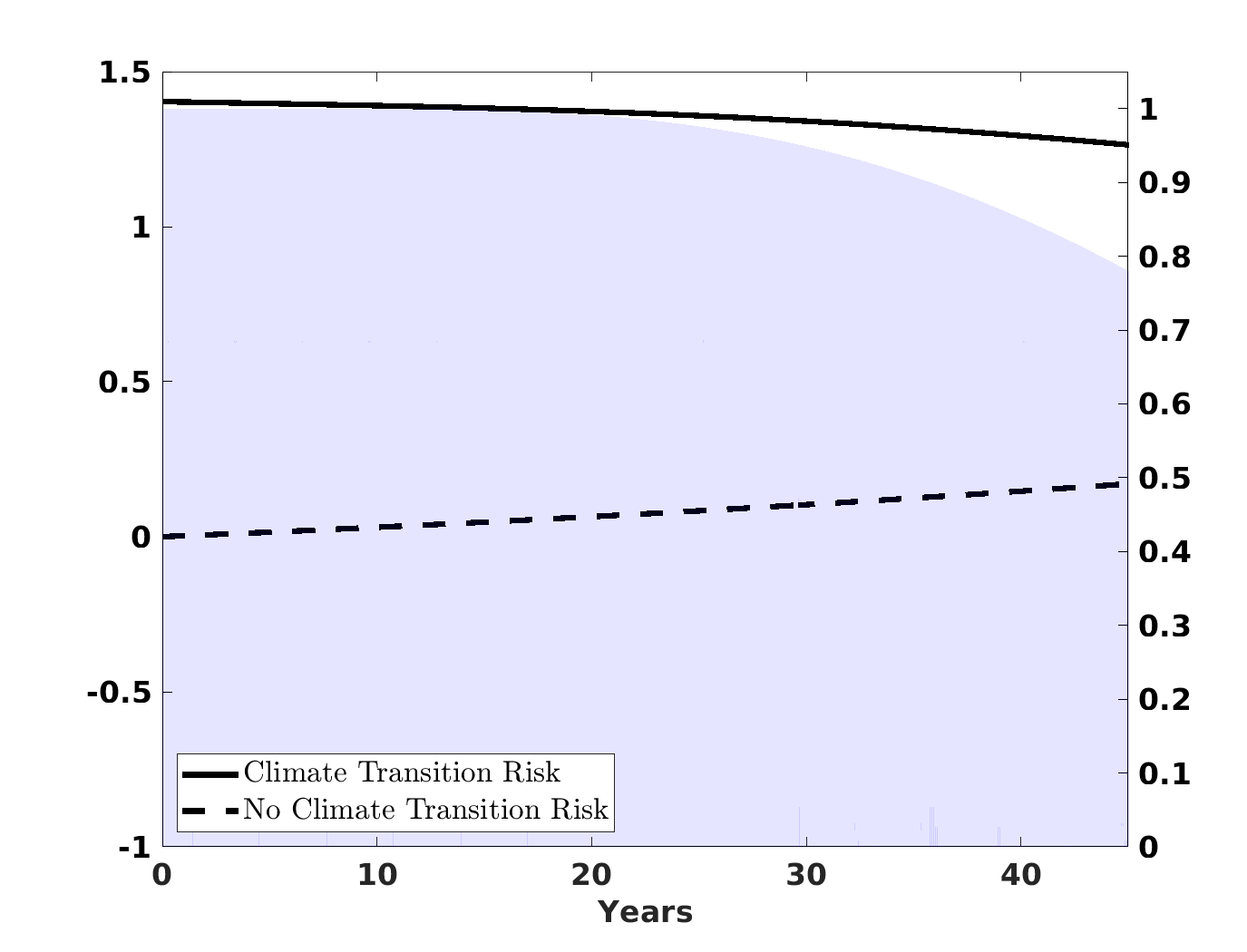}
           \caption[]{{\small Oil Spot Price: $P_{1,t}$}}
        \end{subfigure}
        \begin{subfigure}[b]{0.328\textwidth}  
            \centering 
            \includegraphics[width=\textwidth]{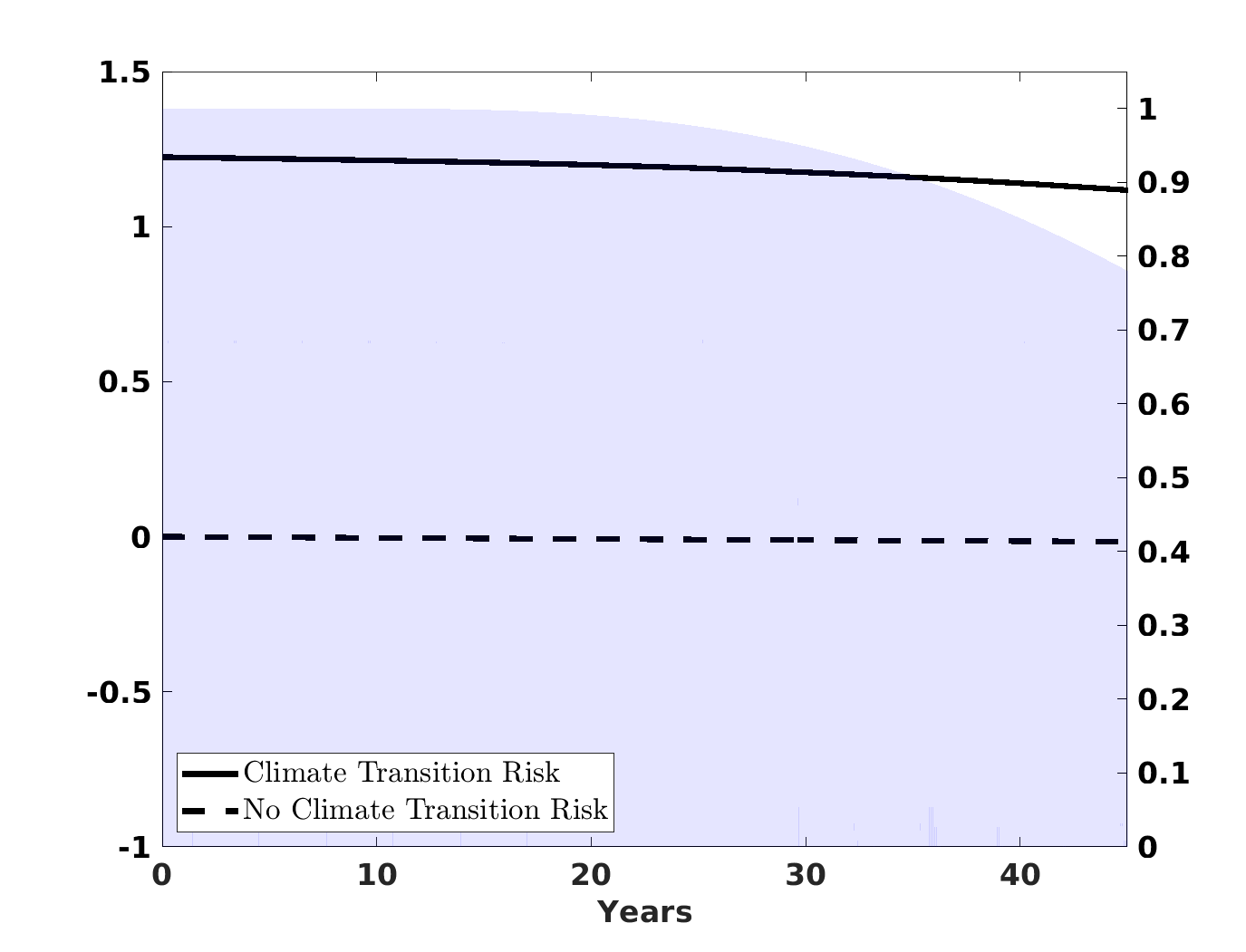}
           \caption[]{{\small Coal Spot Price: $P_{1,t}$}}
        \end{subfigure}

        \begin{subfigure}[b]{0.328\textwidth}
            \centering
            \includegraphics[width=\textwidth]{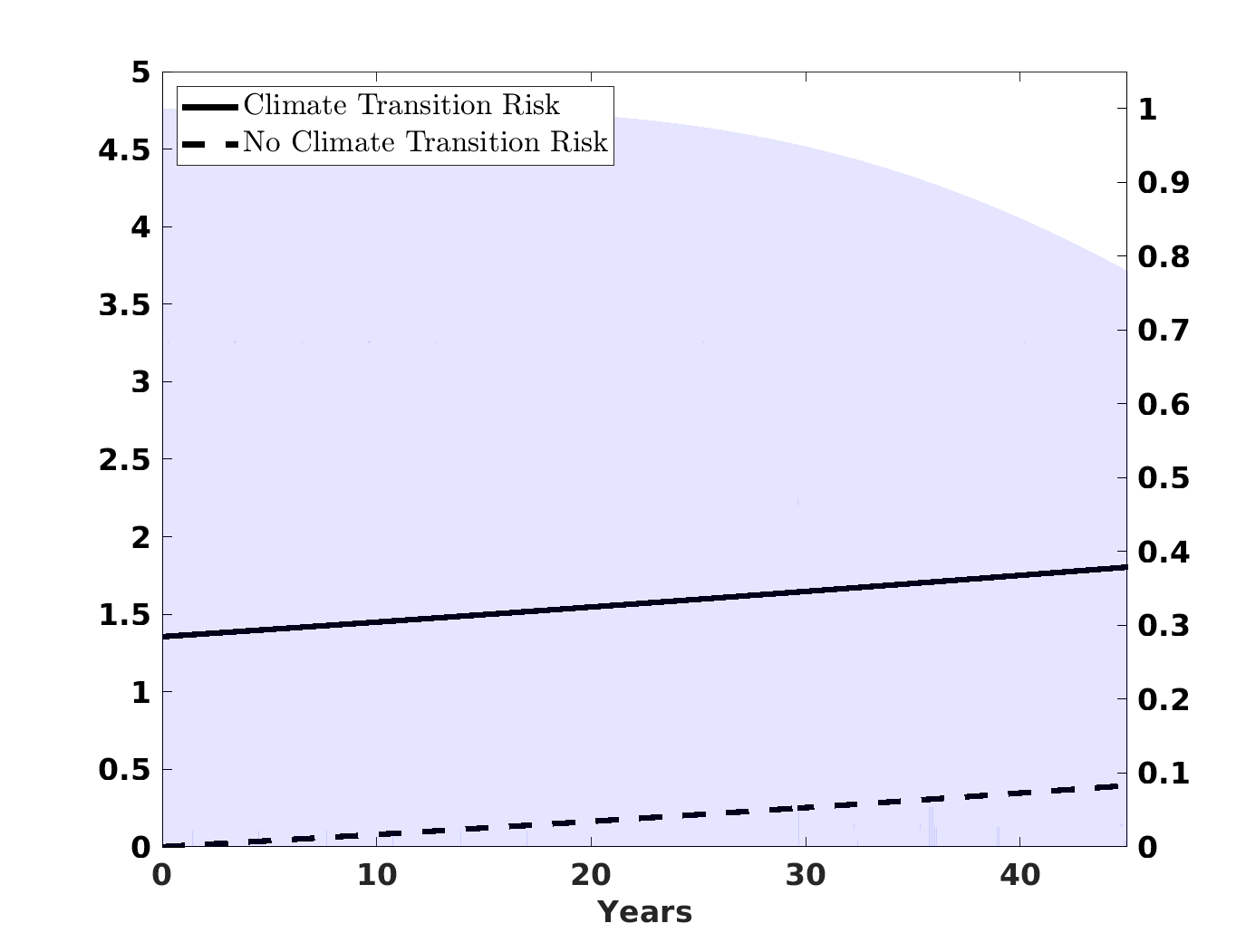}
            \caption[]{{\small Green Firm Price: $S^{(3)}_{t}$}}              
        \end{subfigure}
        \begin{subfigure}[b]{0.328\textwidth}  
            \centering 
            \includegraphics[width=\textwidth]{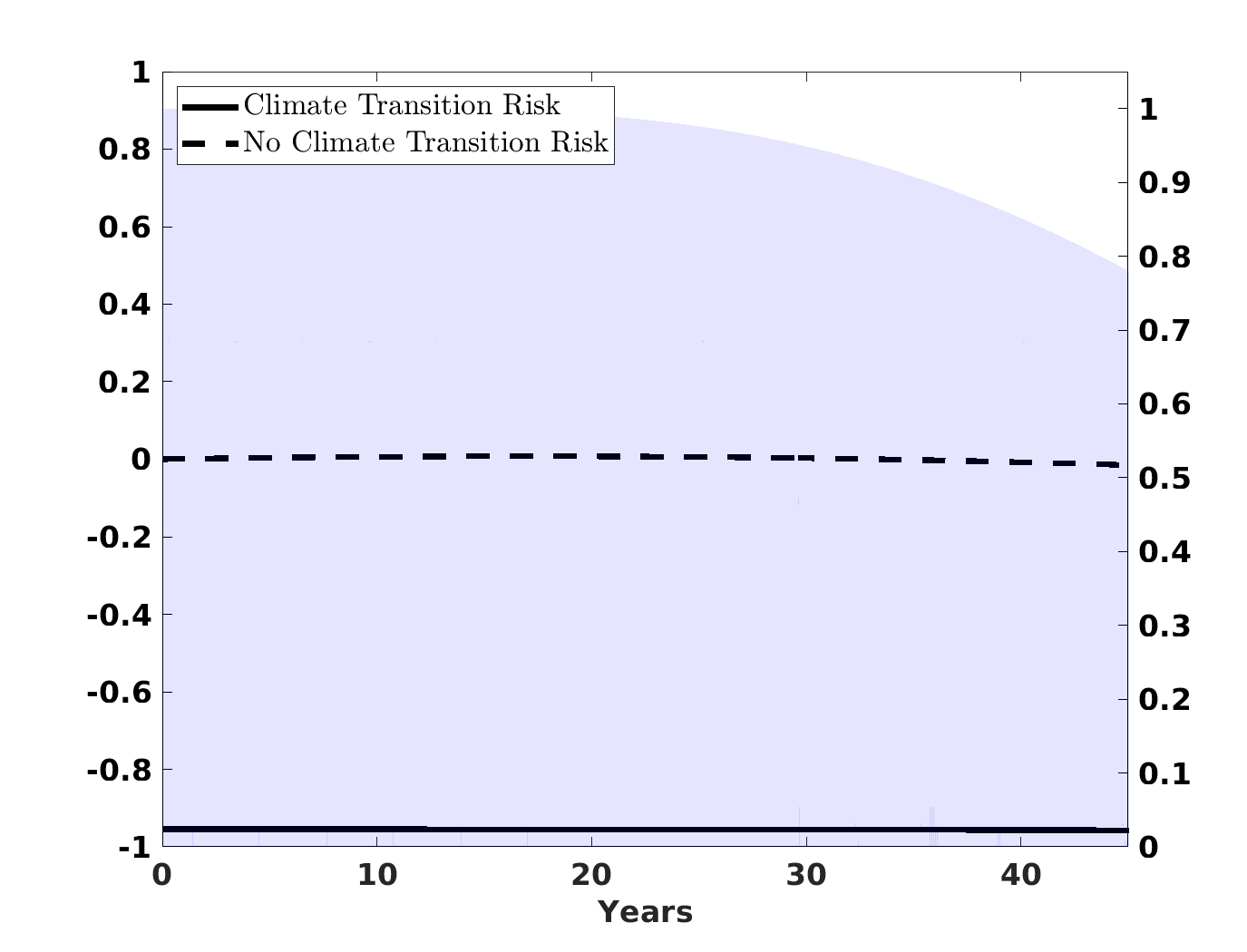}
           \caption[]{{\small Oil Firm Price: $S^{(1)}_{t}$}}
        \end{subfigure}
        \begin{subfigure}[b]{0.328\textwidth}  
            \centering 
            \includegraphics[width=\textwidth]{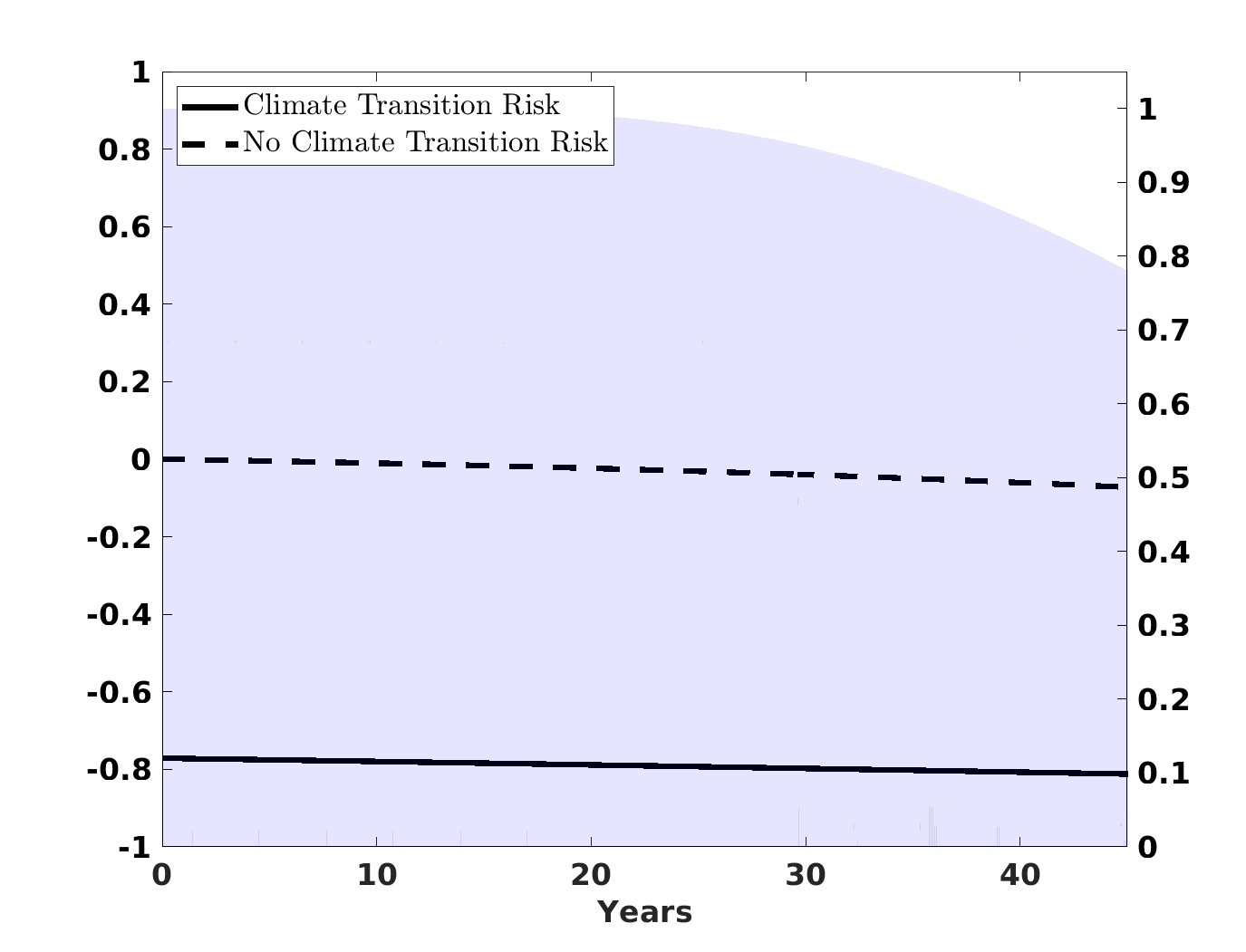}
           \caption[]{{\small Coal Firm Price: $S^{(2)}_{t}$}}
        \end{subfigure}      
        
        \vspace{-0.25cm}
\end{center}

\begin{footnotesize}
Figure \ref{fig:taxation_model_sims} shows the simulated outcomes for the ``taxation shock'' model based on the numerical solutions. Panels (a) through (c) show the temperature anomaly, oil production, and coal production. Panels (d) through (f) show the green investment choice, oil spot price, and coal spot price. Panels (g) and (i) show the green firm price, oil firm price, and coal firm price. Solid lines represent results for the Climate Transition Risk scenario where $\lambda_t = \lambda(T_t)$ and dashed lines represent results for the No Climate Transition Risk scenario where $\lambda_t = 0$. The blue shaded region shows the cumulative probability of no transition shock occurring.
\end{footnotesize} 

\end{figure}

%%%%%%%%%%%%%%%%%%%%%%%%%%%%%%%%%%%%%%%%%%%%%%%%%%
%%%%%%%%%%%%%%%%%%%%%%%%%%%%%%%%%%%%%%%%%%%%%%%%%%
%%%%%%%%%%%%%%%%%%%%%%%%%%%%%%%%%%%%%%%%%%%%%%%%%%
 
Next I consider the ``taxation shock'' scenario, with climate, macroeconomic, and asset pricing results shown in Figure \ref{fig:taxation_model_sims}. The ``taxation shock'' scenario outcomes differ strikingly from the ``technology shock'' setting. In particular, the ``Climate Transition Risk'' case now leads to much slower climate change, shown in Panel (a), reducing the amplification of transition risk. Note that the likelihood of a transition shock not occurring reaches only about $75\%$ through 45 years as the temperature anomaly increase is limited to about $1.5^{\circ}$ C. The level of oil and coal production are also significantly lower, shown in Panels (b) and (c), starting at a very low value of 3 GtC per year for oil and 0.5 GtC for coal, and remain close to constant for both. Green investment, shown in Plot (c), is amplified like in the ``Climate Risk Scenario'' as it was for the ``technology shock'' case given the continued anticipated importance of the green sector in the future, however the increase is not as large given the lower level of output without the run up in oil production. The spot and firm pricing implications, shown in Plots (e) through (i), are also notable when comparing the ``Climate Transition Risk'' scenario for the the ``taxation shock'' case to the ''technology shock'' case. The spot prices of the two fossil fuels are now significantly higher than in the ``No Climate Transition Risk'' due to reduced supply. However, the fossil fuel firm values in the ``Climate Transition Risk'' scenario are even lower than in the ''technology shock'' case. The green firm value is still significantly higher, though less so than in the ``technology shock'' setting. 

These results highlight that the type of transition shock is critical for determining the optimal response. The cost of the transition shock resulting from a taxation transition leads the planner to reduce fossil fuel production in an attempt to cut emissions, slow climate change, and delay the transition shock arrival. While this amplifies fossil fuel spot prices, the expectation of the transition shock still augments the discounting of future fossil firms' values. Moreover, the costly consequences of the shock coming in a taxation form, without an improvement in the value of the green input in the production technology, means that the fossil fuel firm values are depressed even more than in the technology shock case. In addition, because the production and spot price responses are flat, the accelerated impact on firm prices is no longer present in this setting. The green firm value still increases substantially because of the expected increase in dependence and importance of the green sector in the future, but the lack of innovation and run response means the increase is somewhat tempered.

   \begin{table}[!htb]
\begin{center}
    \caption{Post Transition Outcomes - Percent Change ($\% \Delta$)} \label{table:PostJump}
    \centering
\begin{tabular}{l | c c c }
\hline
\hline
Transition & $\Delta Y_{t}$ & $\Delta I_{G,t}$ & $\Delta S^{(G)}_t$ \\
 \hline
% Technological Shock  & $-4.4 \%$ & $-20.7 \%$ & $13.6 \%$ \\
Technological Shock  & $-4.1 \%$ & $-20.7 \%$ & $14.0 \%$ \\
% Taxation Shock & $-14.4 \%$ & $-17.7 \%$ & $ 30.4 \%$  \\
Taxation Shock & $-14.2 \%$ & $-17.7 \%$ & $ 28.5 \%$  \\
\hline
\hline
\end{tabular}
\end{center}
% \vspace{0.15cm}
\begin{footnotesize}
%\begin{flushleft}
Table \ref{table:PostJump} shows the percent change in consumption ($C_t$), green investment ($I_{G,t}$), and the green firm price ($S^{(G)}_t$) after the realization of the different transition shocks based on the numerical solutions. The top row gives the results for the ``technology shock'' transition scenario, and the bottom row gives the results for the ``taxation shock'' transition scenario. The transition shock is assumed to occur after 50 years in each case.
%\end{flushleft}
\end{footnotesize}  
\end{table}  

To give further intuition about the stark difference in model outcomes across transition risk scenarios, I show the change in output, green investment, and the green firm price after a transition shock occurs for both transition shock cases at year 50 in Table \ref{table:PostJump}. Most significantly, while there is a decrease in output in each case, the magnitude is over three times as large for the taxation shock ($-14.4\%$) compared to the technological shock ($-4.4\%$). There is a notable decrease in green investment of similar magnitude in each case, $-17.7\%$ for the taxation shock versus  $-20.7\%$ for the technological shock, reflecting the reduction in output. Finally, the green firm price increases substantially in each case, demonstrating the increased importance of the green sector as the sole energy provider in the post-climate transition state. While the green firm price increase is over twice as large in the taxation shock case ($30.4\%$) relative to the technological shock case ($13.6\%$), the pre-transition increase in the green firm value is much larger in the technological shock case. These outcomes further highlight the key feature of climate-linked transition risk and the potential magnitude of its macroeconomic and financial impacts. In particular, when the cost of the transition risk is expected to be lower, the incentive to amplify the ``run'' on fossil fuels to maximize the use of fossil fuels is larger. When the costs of the climate-linked transition risk are perceived as larger, the planner places more value on delaying the costs of climate change and the perceived negative future economic shock coming from transition risk.

\section{Counterfactuals and Alternative Specifications} \label{section:CounterfactualsAlternatives}

To enrich the theoretical analysis, I explore a number of counterfactual and alternative model settings which provide further insight and intuition into the impact of climate transition risk with respect to the financial and economic outcomes in the model. I briefly summarize the main results and intuition from these alternative specifications below, and leave most of the model solution details and figures of results for the Online Appendix.

\subsection{Alternative Transition Scenarios} %% Start here

For the first counterfactual, I consider the case where the arrival rate of the transition is assumed \emph{constant}, i.e., $\lambda(T_t) = \bar{\lambda} \ge 0$. This setting removes the climate-linked feedback effect that generates the dynamic implications associated with the anticipation of a climate transition event. The transition risk effect here leads to a constant level shift up in extraction and constant adjustment to asset pricing outcomes, in distinct contrast to the baseline model.

Next, I consider a \emph{two-step transition}. In the technological innovation case, the fossil fuel input share $\nu_2$ first goes to $\tilde{\nu}_2 = 0.5 \times \nu_2$ and then to zero. In the carbon taxation case, the fossil fuel energy is taxed first at a rate of $\tau = 50\%$ and then at a rate of $\tau = 100\%$.\footnote{As in the baseline case, I assume $\omega > 0$. I also assume government spending is entirely wasteful. This arguably extreme assumption is not uncommon in macroeconomic models with taxation (e.g., see \cite{liu2020risks} for one recent example), keeps the analysis tractable, and is without loss of generality for the $100\%$ taxation rate.} This setting alters the impact of the climate-linked feedback effect by dampening the abruptness of the transition shock. The qualitative responses of prices and production match those of the baseline model settings. Quantitatively, the run is slightly amplified in the ``technology shock'' case, whereas the protective response in the ``taxation shock'' case is a bit less severe.

I also consider the ``hybrid'' case where a counterfactual \emph{carbon tax policy that does not adjust for transition risk} is (partially) implemented before a transition shock occurs. This counterfactual carbon tax differs from the optimal tax supporting the planner's solution in the baseline transition shock setting because it does account for transition risk. To approximate this transition scenario, I use the term $\widehat{External}_t$ in the FOC for emissions, where $\widehat{External}_t = (1-\chi) ( v_{\hat{R}}^{(tr)} - \beta_T v_T^{(tr)} \exp(\hat{R}) \mathcal{E}_t^{(tr)}) + \chi ( v_{\hat{R}}^{(cf)} - \beta_T v_T^{(cf)} \exp(\hat{R}) \mathcal{E}_t^{(cf)})$, $\chi \in [0,1]$ is an internalization parameter that sets the fraction of the transition risk impact that is not internalized into the optimal emissions choice, $v_X^{(tr)}$ is the marginal value of an increase in state variable $X$ from the social planner's solution from the baseline technology shock model, and $v_X^{(cf)}$ is the marginal value of an increase in state variable X from the social planner's solution to the counterfactual model without transition risk. The central impact of pre-transition policy is to attenuate the transition risk effect, by either dampening the run effect for the technology shock case or loosening up the emissions pull back in the taxation shock cast, muting potentially unintended transition risk consequences.

Finally, I consider the potentially realistic transition risk scenario where \emph{transition risk first impacts the dirtiest sector (coal), followed by a secondary transition risk impact on the relatively cleaner fossil fuel sector (oil)}. There are a number of transition shock combinations to consider in this setting, such as when there as a coal carbon tax followed by a tax shock to oil or a technology shock to oil (and coal). I consider each of the potential combinations, and highlight the results for the most relevant and illuminating example in what follows below.

%%%%%%%%%%%%%%%%%%%%%%%%%%%%%%%%%%%%%%%%%%%%%%%%%%
%%%%%%%%%%%%%%%%%%%%%%%%%%%%%%%%%%%%%%%%%%%%%%%%%%
%%%%%%%%%%%%%%%%%%%%%%%%%%%%%%%%%%%%%%%%%%%%%%%%%%

% \begin{landscape}

\begin{figure}[!t]

%\vspace{-1.0cm}
% \vspace{-0.5cm}

\caption{Macroeconomic and Asset Pricing Outcomes - Hybrid ``Technology'' Shock} \label{fig:hybrid_model_sims}
\begin{center}
% {\scriptsize \textbf{Panel A: Hybrid Taxation/Technology Transition Scenario}}\\
        \begin{subfigure}[b]{0.328\textwidth}
            \centering
            \includegraphics[width=\textwidth]{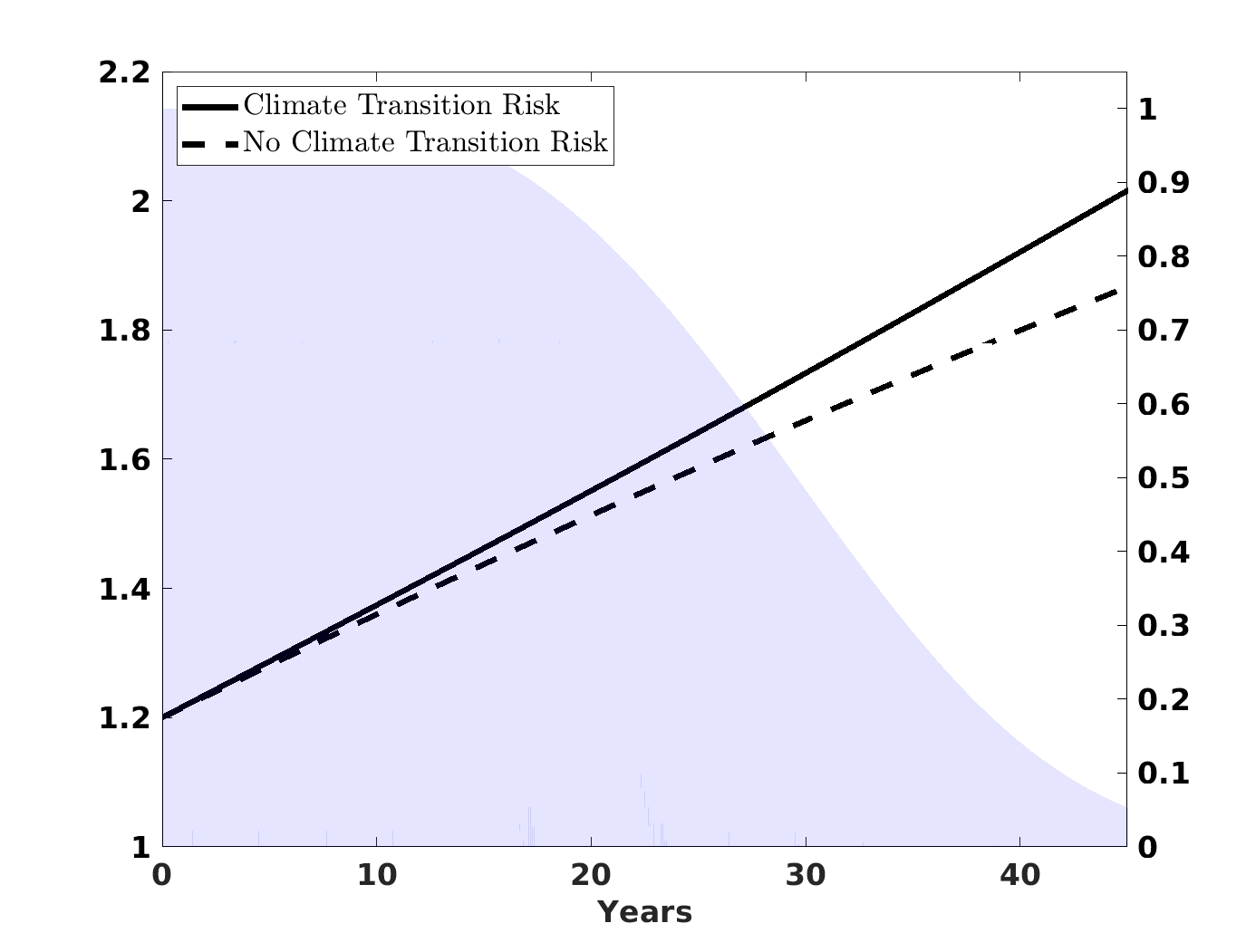}
            \caption[]{{\small Temperature: $Y_t$}}              
        \end{subfigure}
        \begin{subfigure}[b]{0.328\textwidth}  
            \centering 
            \includegraphics[width=\textwidth]{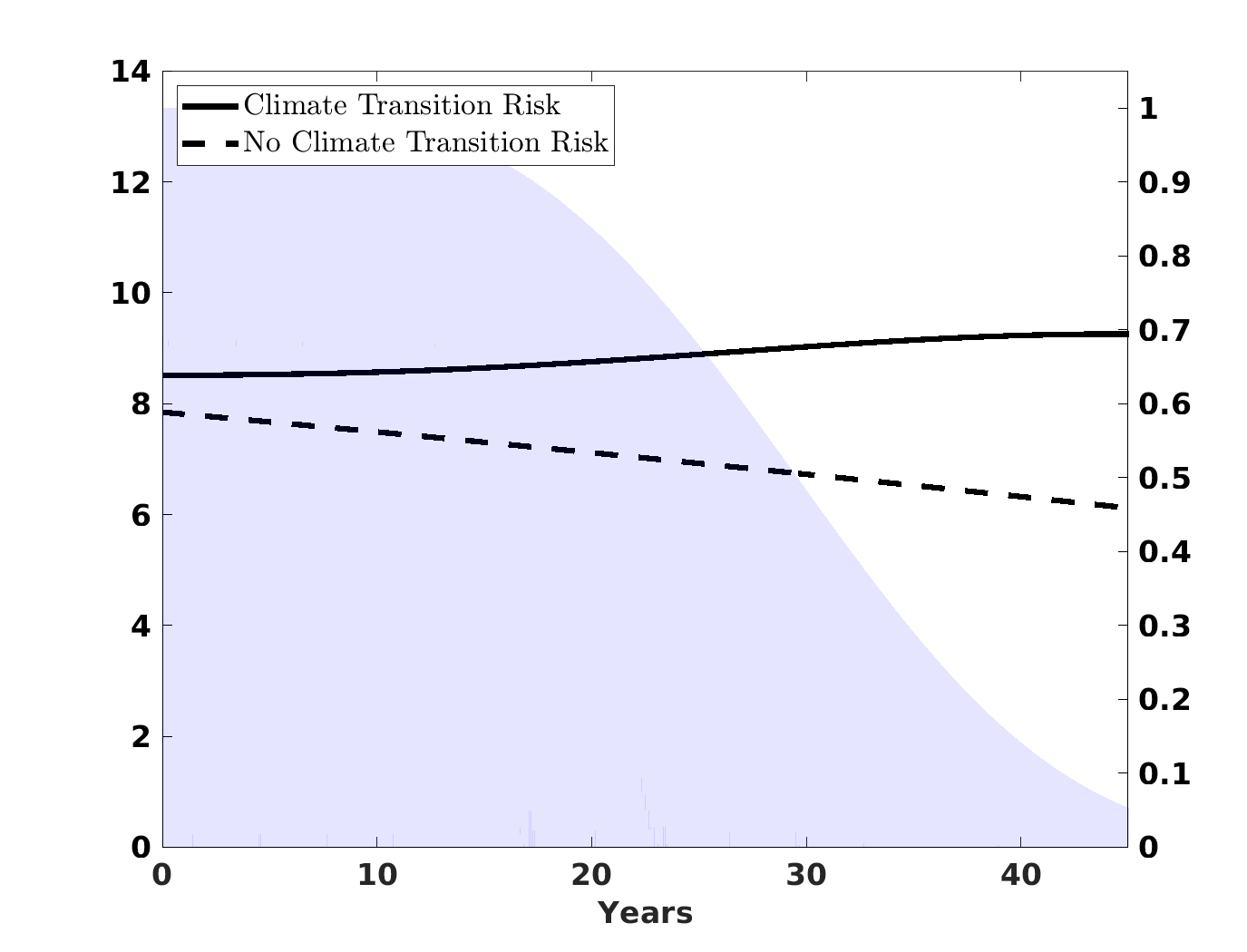}
           \caption[]{{\small Oil Production: $E_{1,t}$}}
        \end{subfigure}
        \begin{subfigure}[b]{0.328\textwidth}  
            \centering 
            \includegraphics[width=\textwidth]{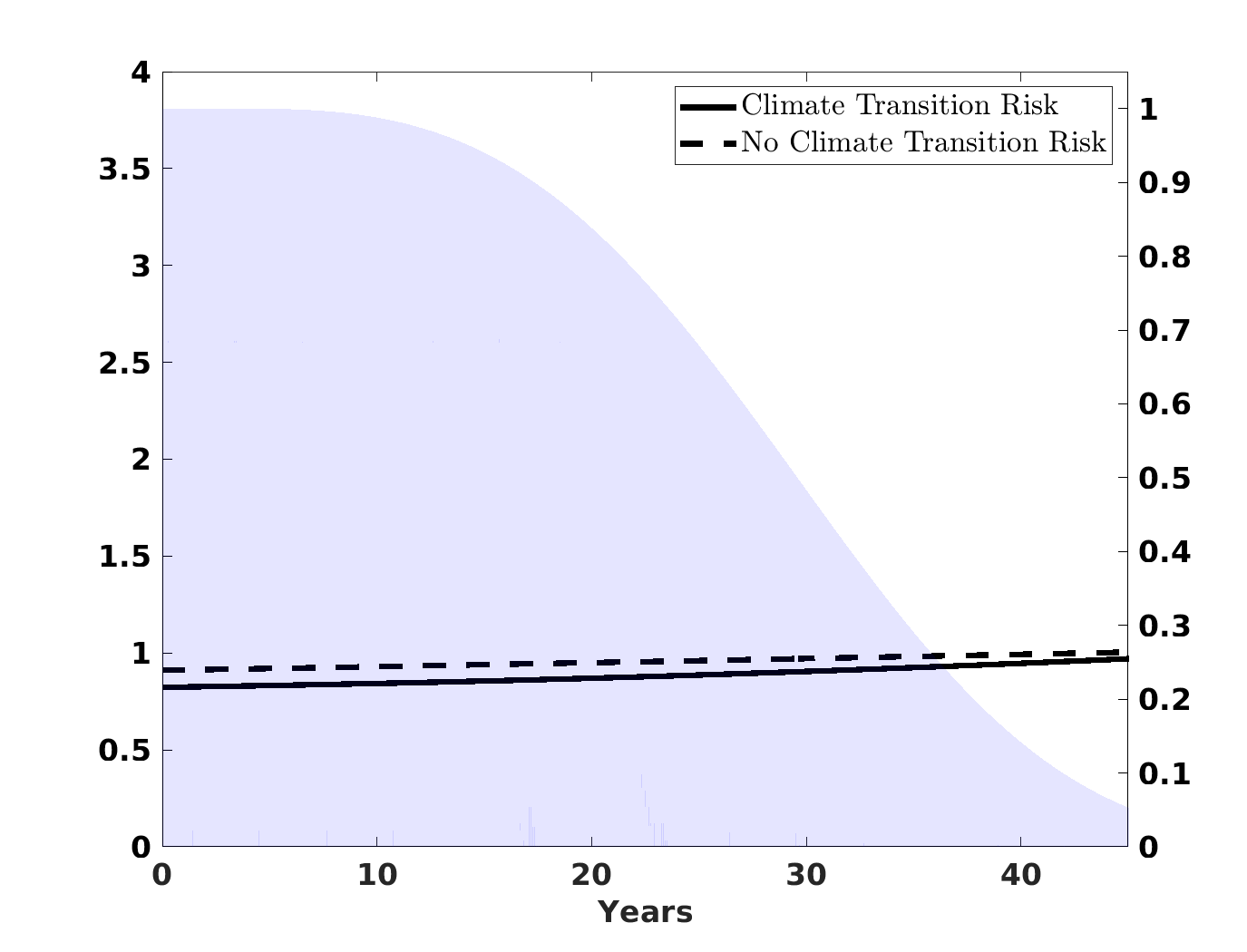}
           \caption[]{{\small Coal Production: $E_{2,t}$}}
        \end{subfigure}     
        
        \begin{subfigure}[b]{0.328\textwidth}
            \centering
            \includegraphics[width=\textwidth]{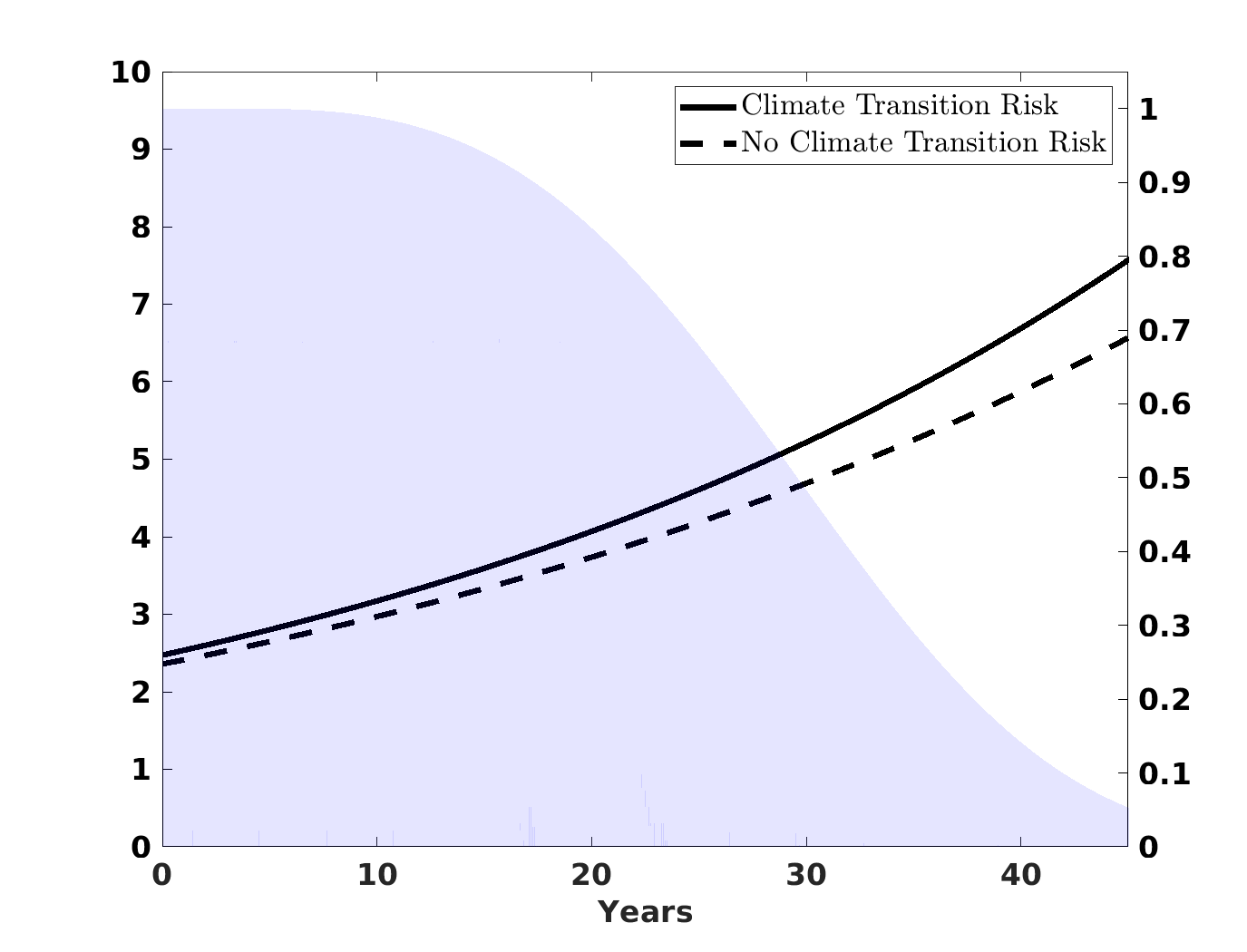}
            \caption[]{{\small Green Investment: $I_{G,t}$}}              
        \end{subfigure}
        \begin{subfigure}[b]{0.328\textwidth}  
            \centering 
            \includegraphics[width=\textwidth]{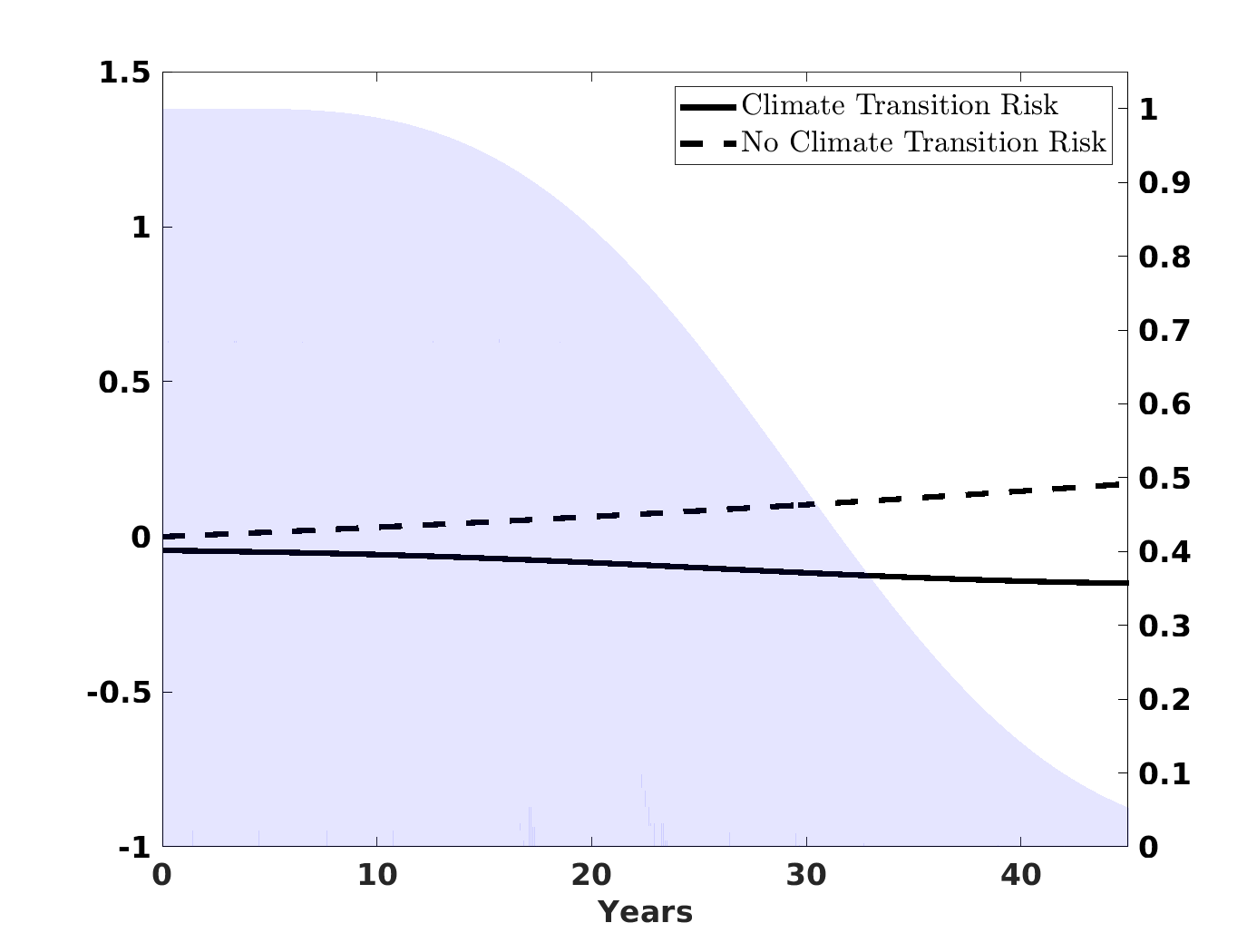}
           \caption[]{{\small Oil Spot Price: $P_{1,t}$}}
        \end{subfigure}
        \begin{subfigure}[b]{0.328\textwidth}  
            \centering 
            \includegraphics[width=\textwidth]{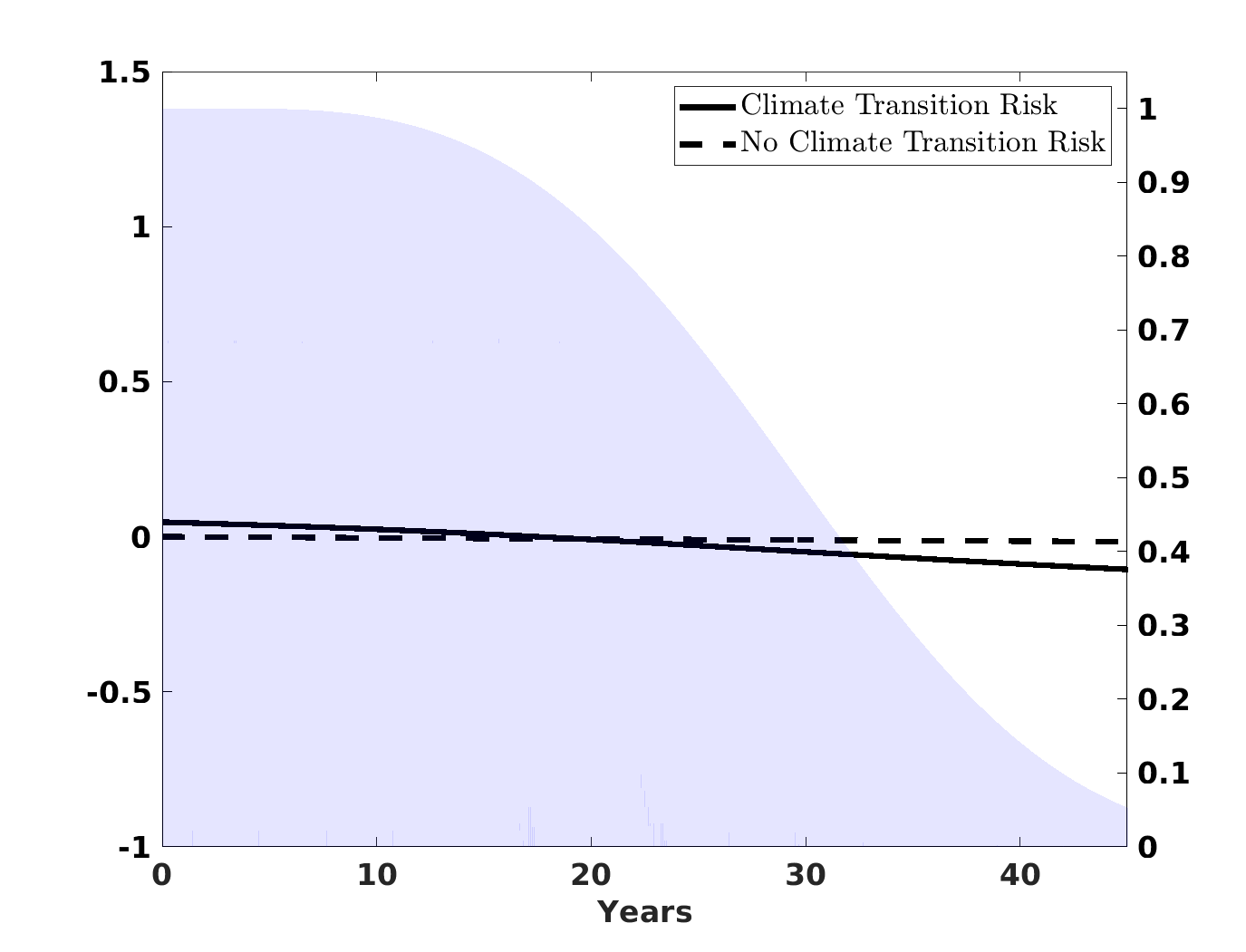}
           \caption[]{{\small Coal Spot Price: $P_{1,t}$}}
        \end{subfigure}

        \begin{subfigure}[b]{0.328\textwidth}
            \centering
            \includegraphics[width=\textwidth]{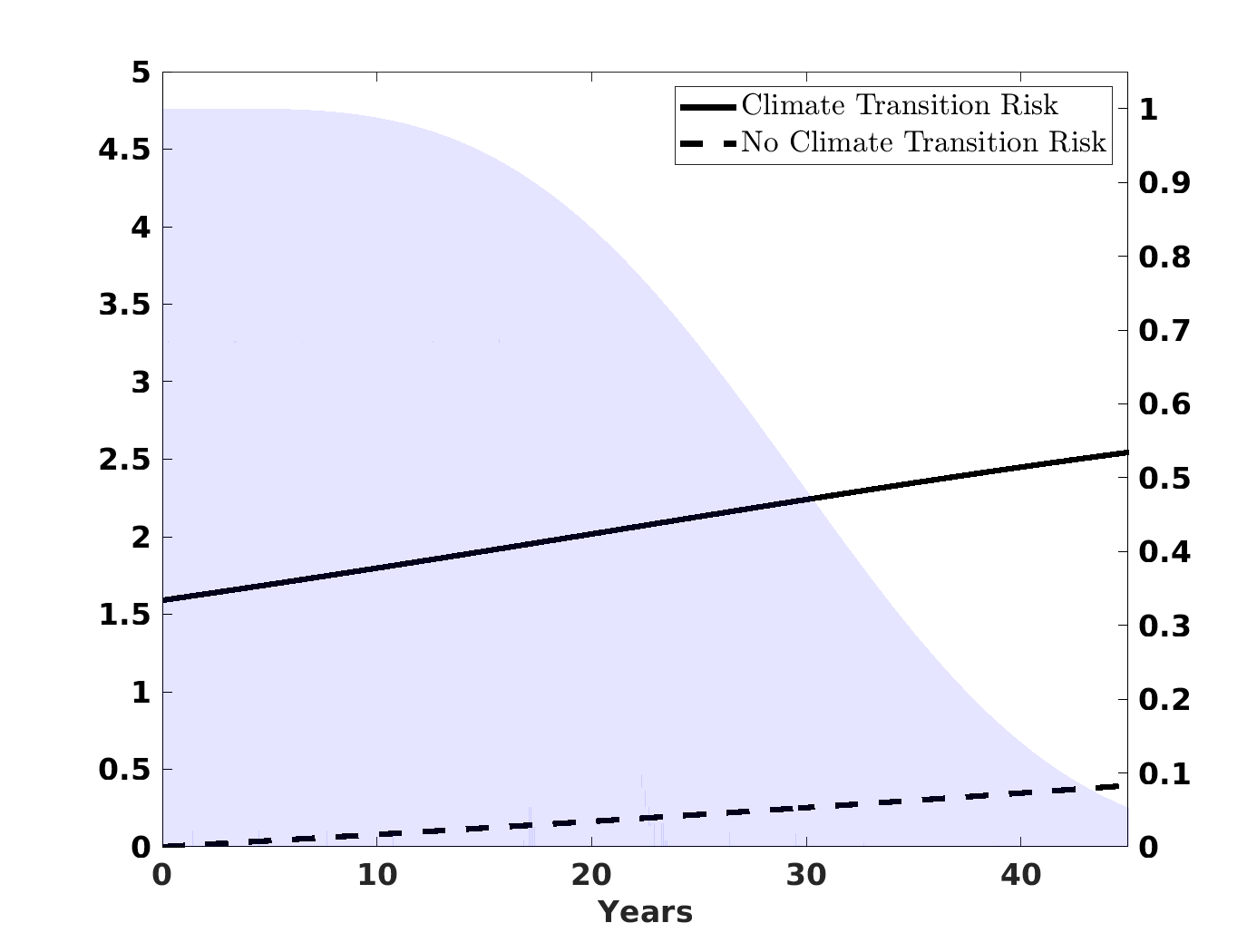}
            \caption[]{{\small Green Firm Price: $S^{(3)}_{t}$}}              
        \end{subfigure}
        \begin{subfigure}[b]{0.328\textwidth}  
            \centering 
            \includegraphics[width=\textwidth]{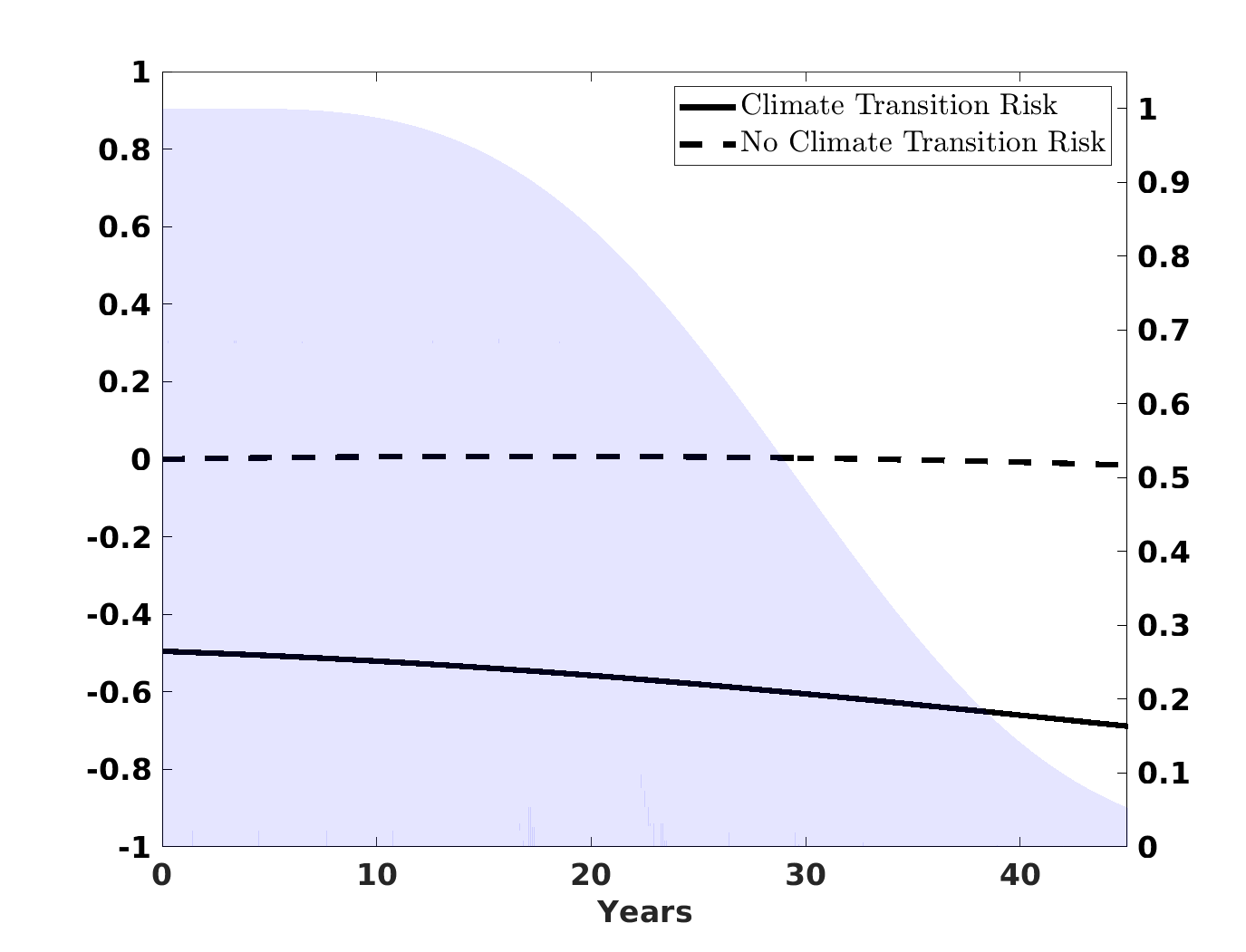}
           \caption[]{{\small Oil Firm Price: $S^{(1)}_{t}$}}
        \end{subfigure}
        \begin{subfigure}[b]{0.328\textwidth}  
            \centering 
            \includegraphics[width=\textwidth]{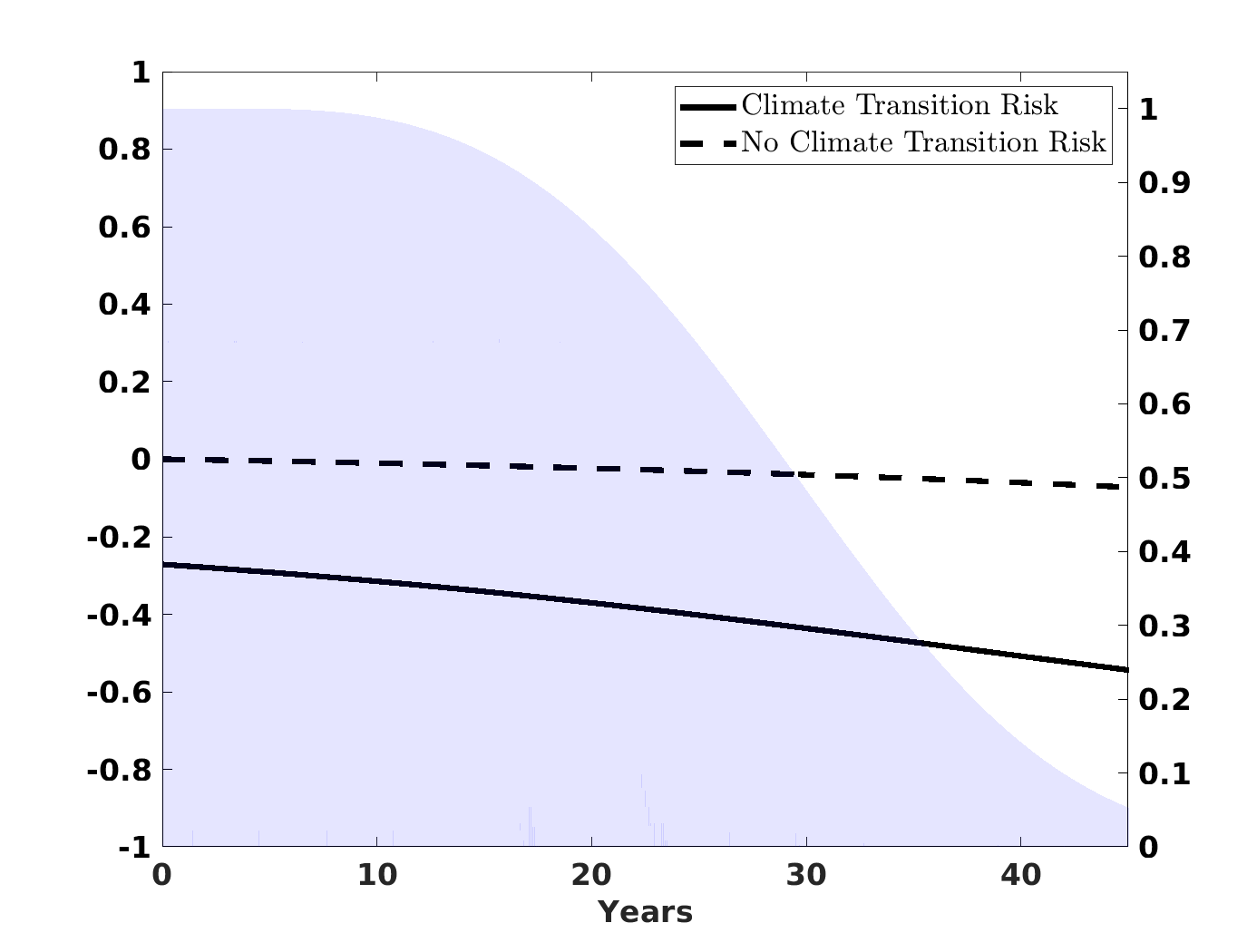}
           \caption[]{{\small Coal Firm Price: $S^{(2)}_{t}$}}
        \end{subfigure}          
        
        \vspace{-0.25cm}
\end{center}

\begin{footnotesize}
Figure \ref{fig:hybrid_model_sims} shows the simulated outcomes for the ``hybrid technology shock'' model based on the numerical solutions. Panels (a) through (c) show the temperature anomaly, oil production, and coal production. Panels (d) through (f) show the green investment choice, oil spot price, and coal spot price. Panels (g) and (i) show the green firm price, oil firm price, and coal firm price. Solid lines represent results for the Climate Transition Risk scenario where $\lambda_t = \lambda(T_t)$ and dashed lines represent results for the No Climate Transition Risk scenario where $\lambda_t = 0$. The blue shaded region shows the cumulative probability of no transition shock occurring.
\end{footnotesize} 

\end{figure}

% \end{landscape}

%%%%%%%%%%%%%%%%%%%%%%%%%%%%%%%%%%%%%%%%%%%%%%%%%%
%%%%%%%%%%%%%%%%%%%%%%%%%%%%%%%%%%%%%%%%%%%%%%%%%%
%%%%%%%%%%%%%%%%%%%%%%%%%%%%%%%%%%%%%%%%%%%%%%%%%%

% \begin{landscape}

\begin{figure}[!t]

%\vspace{-1.0cm}
% \vspace{-0.5cm}

\caption{Macroeconomic and Asset Pricing Outcomes - Coal ``Tax'' then ``Technology'' Shock} \label{fig:ctax_otech_model_sims}
\begin{center}
% {\scriptsize \textbf{Panel A: Hybrid Taxation/Technology Transition Scenario}}\\
        \begin{subfigure}[b]{0.328\textwidth}
            \centering
            \includegraphics[width=\textwidth]{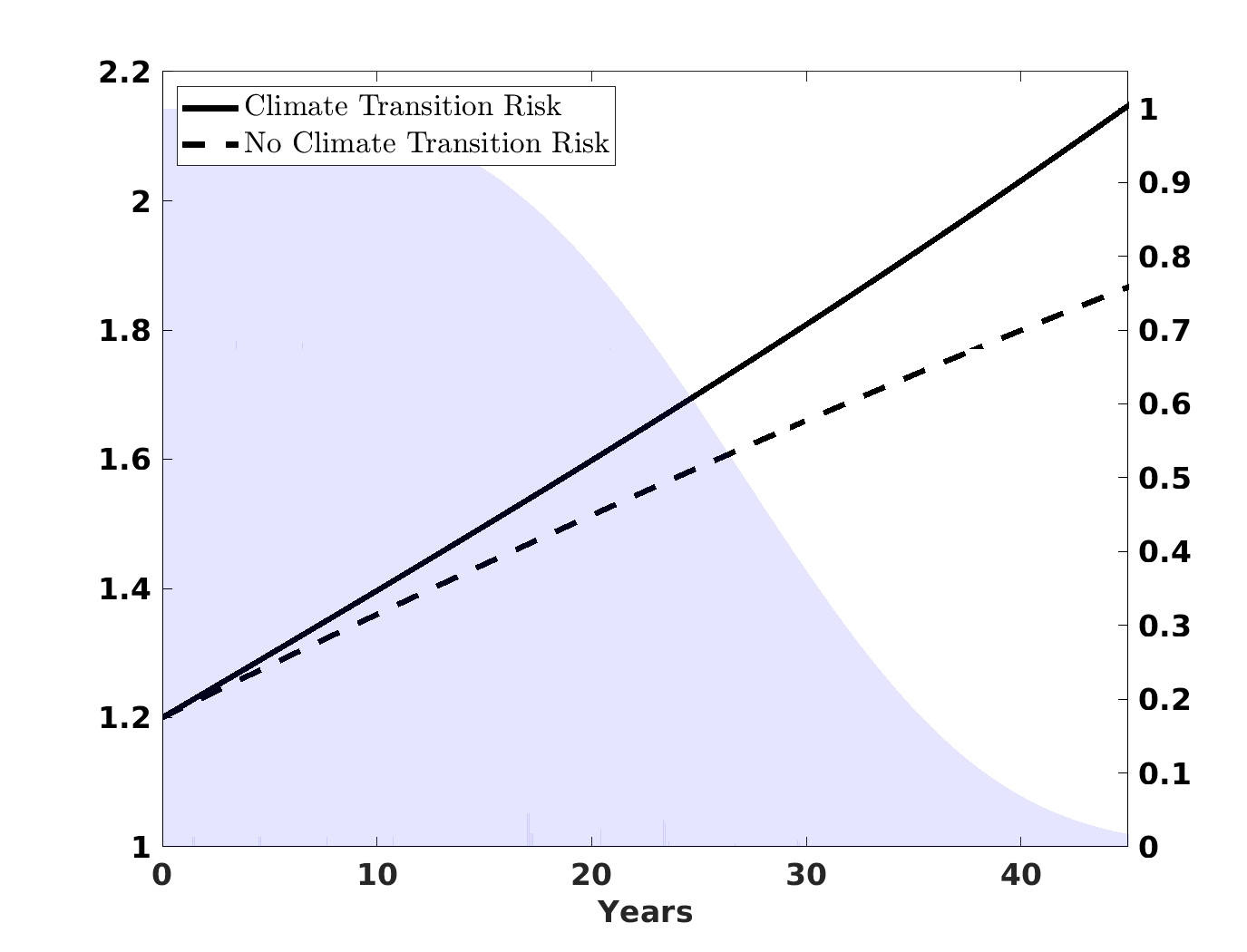}
            \caption[]{{\small Temperature: $Y_t$}}              
        \end{subfigure}
        \begin{subfigure}[b]{0.328\textwidth}  
            \centering 
            \includegraphics[width=\textwidth]{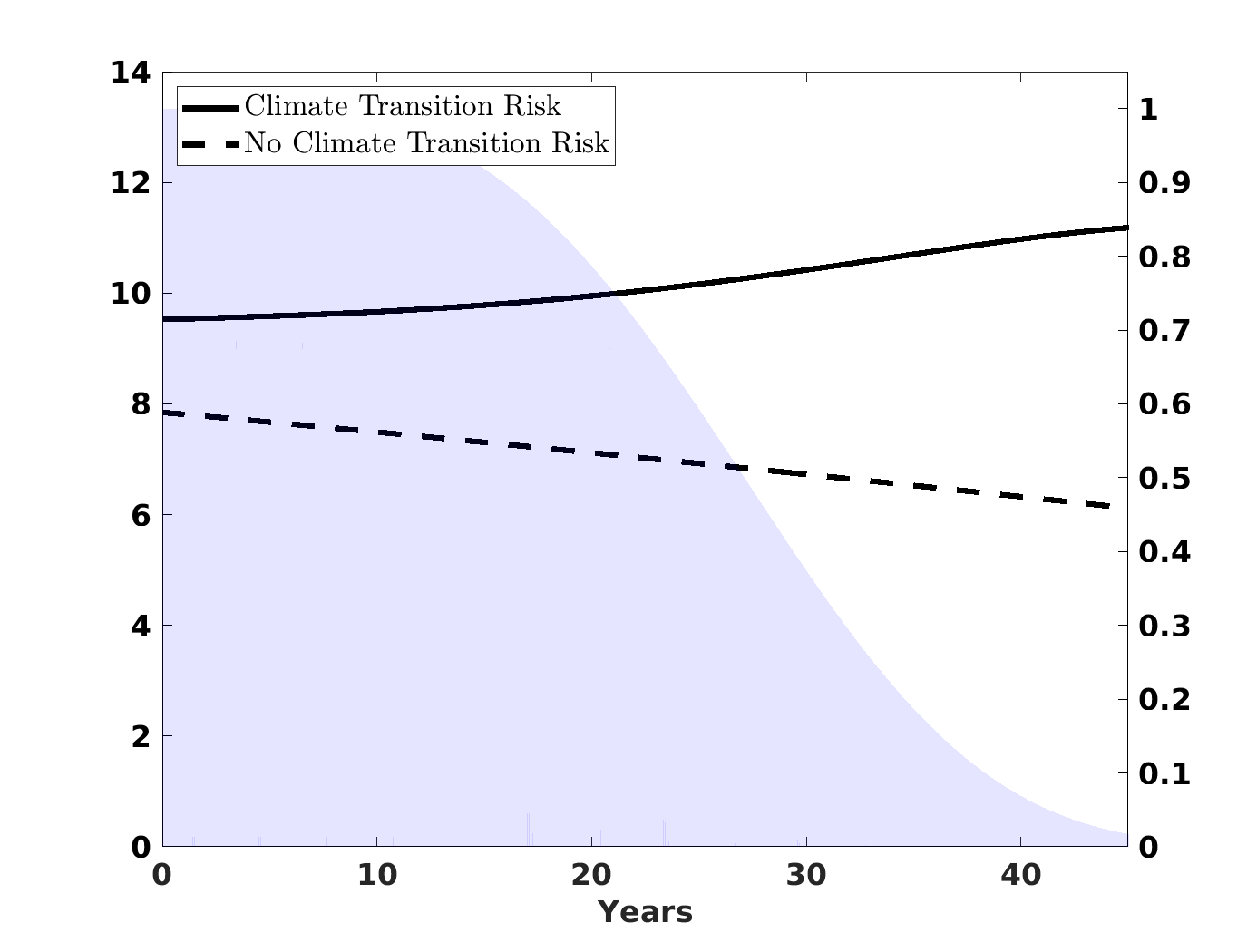}
           \caption[]{{\small Oil Production: $E_{1,t}$}}
        \end{subfigure}
        \begin{subfigure}[b]{0.328\textwidth}  
            \centering 
            \includegraphics[width=\textwidth]{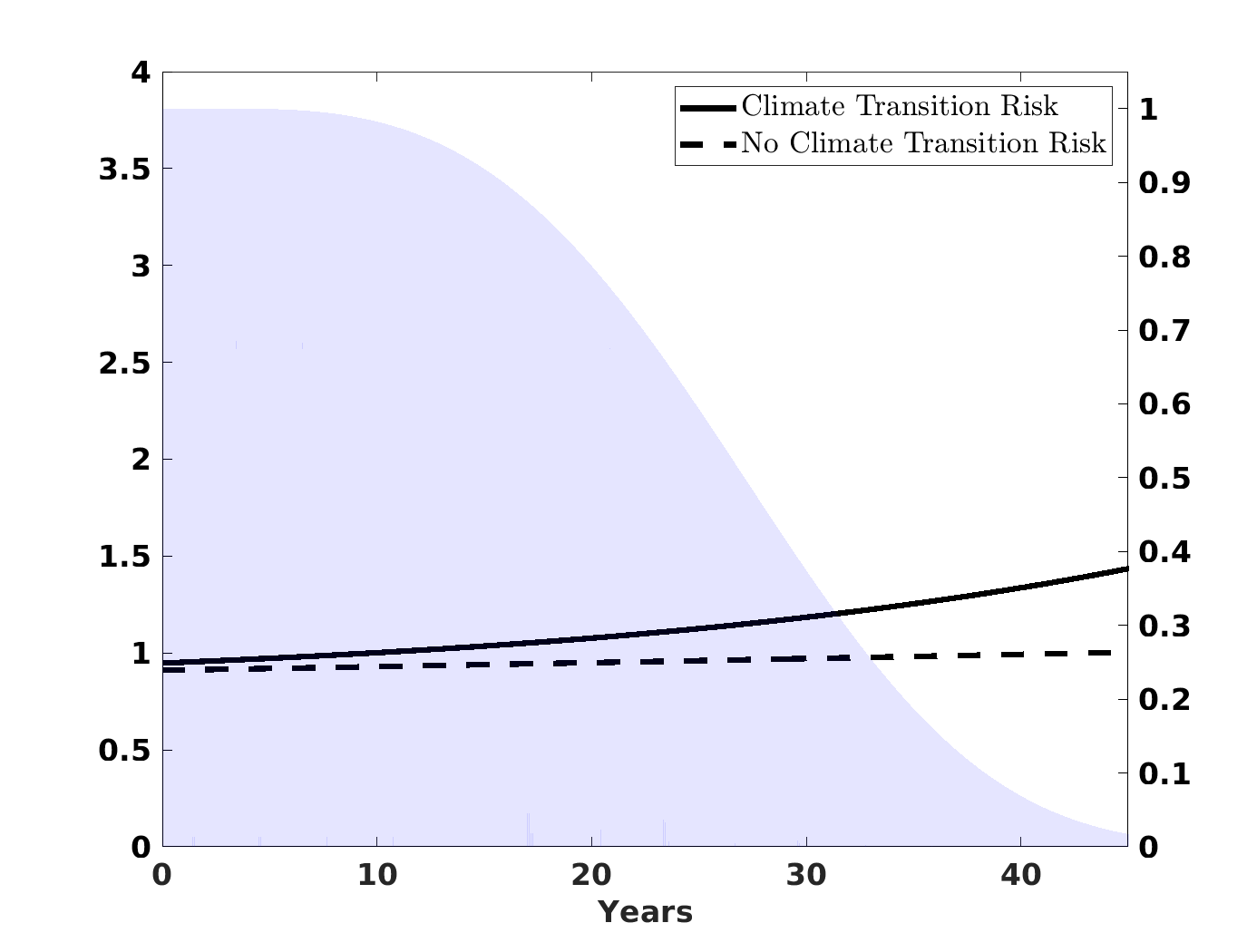}
           \caption[]{{\small Coal Production: $E_{2,t}$}}
        \end{subfigure}     
        
        \begin{subfigure}[b]{0.328\textwidth}
            \centering
            \includegraphics[width=\textwidth]{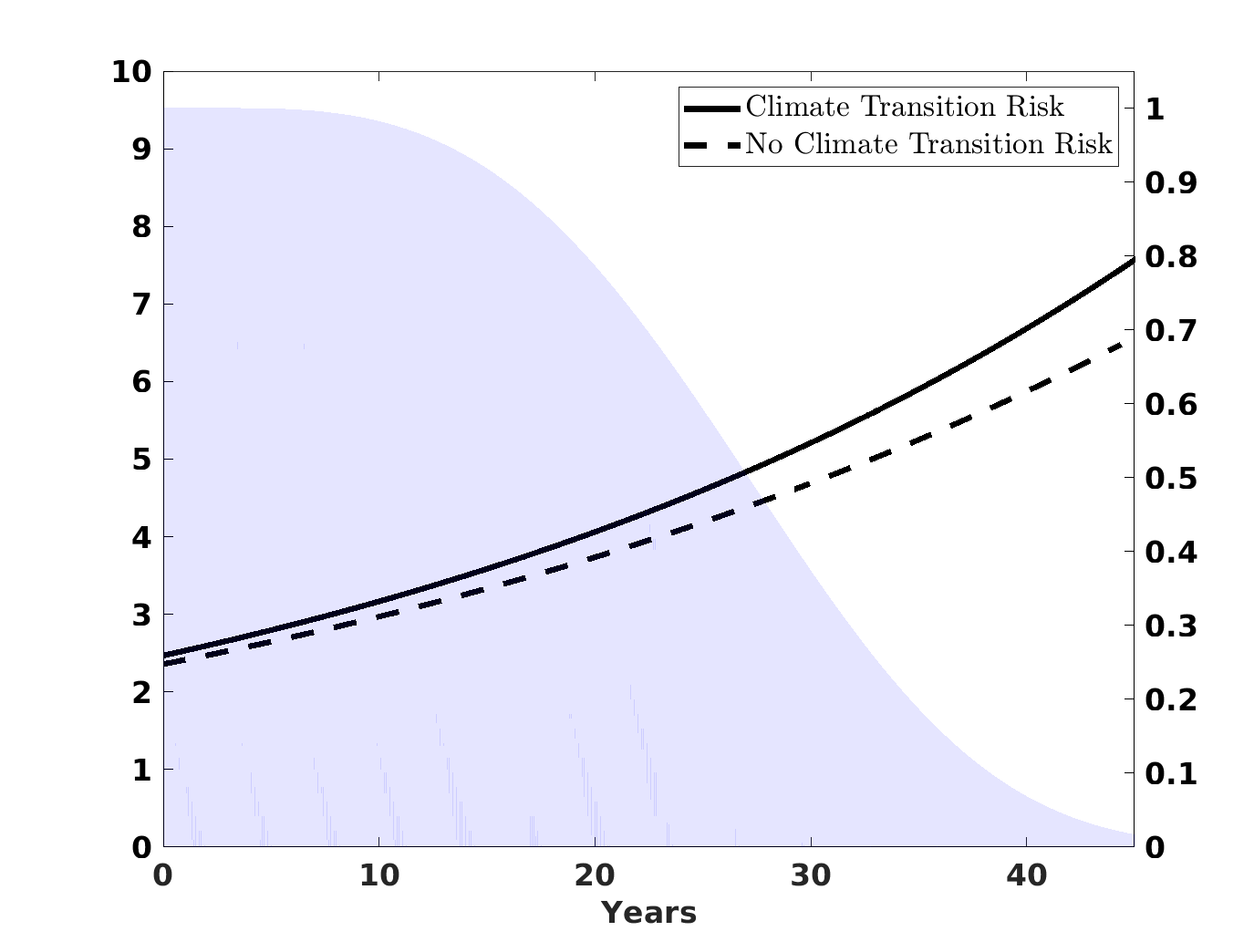}
            \caption[]{{\small Green Investment: $I_{G,t}$}}              
        \end{subfigure}
        \begin{subfigure}[b]{0.328\textwidth}  
            \centering 
            \includegraphics[width=\textwidth]{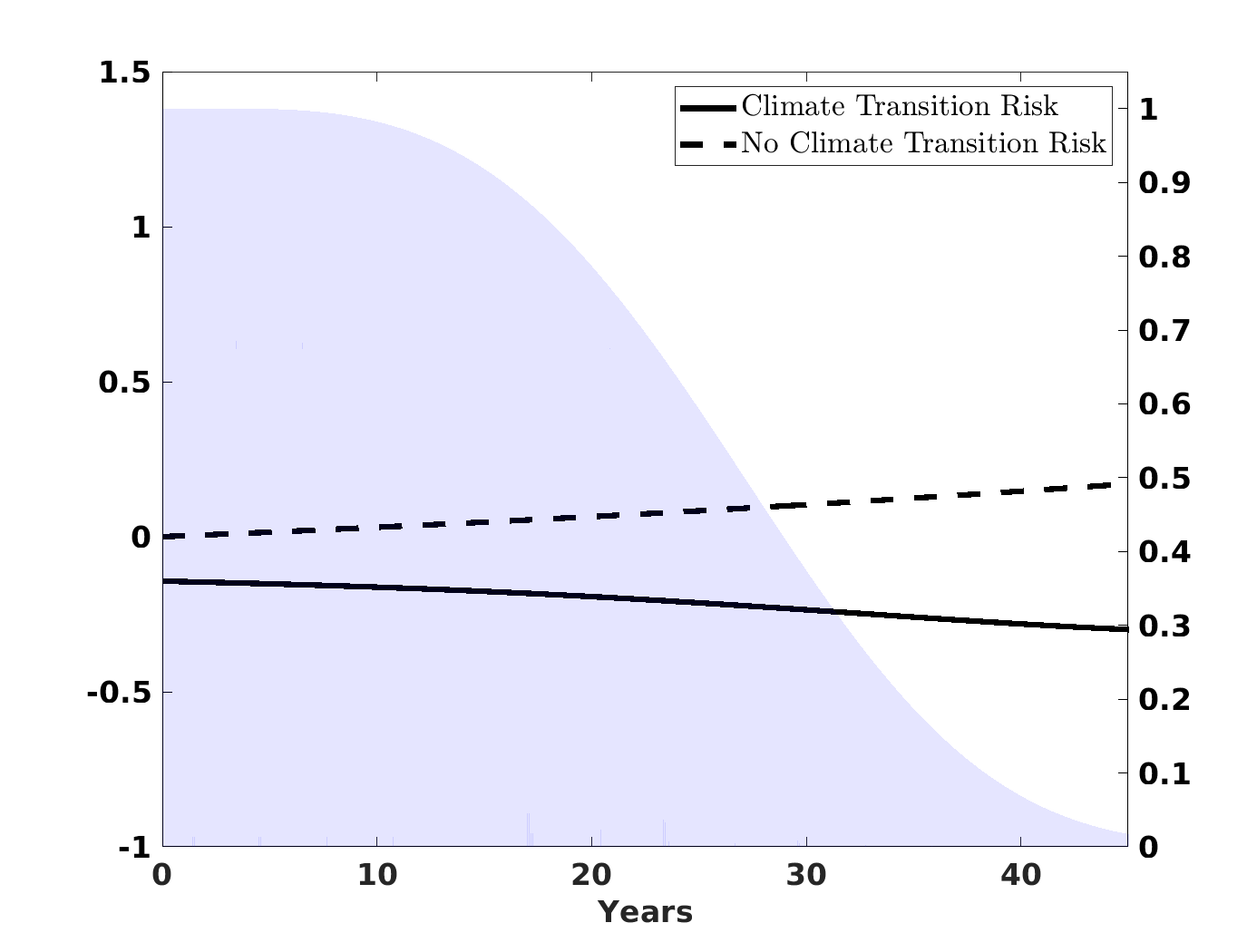}
           \caption[]{{\small Oil Spot Price: $P_{1,t}$}}
        \end{subfigure}
        \begin{subfigure}[b]{0.328\textwidth}  
            \centering 
            \includegraphics[width=\textwidth]{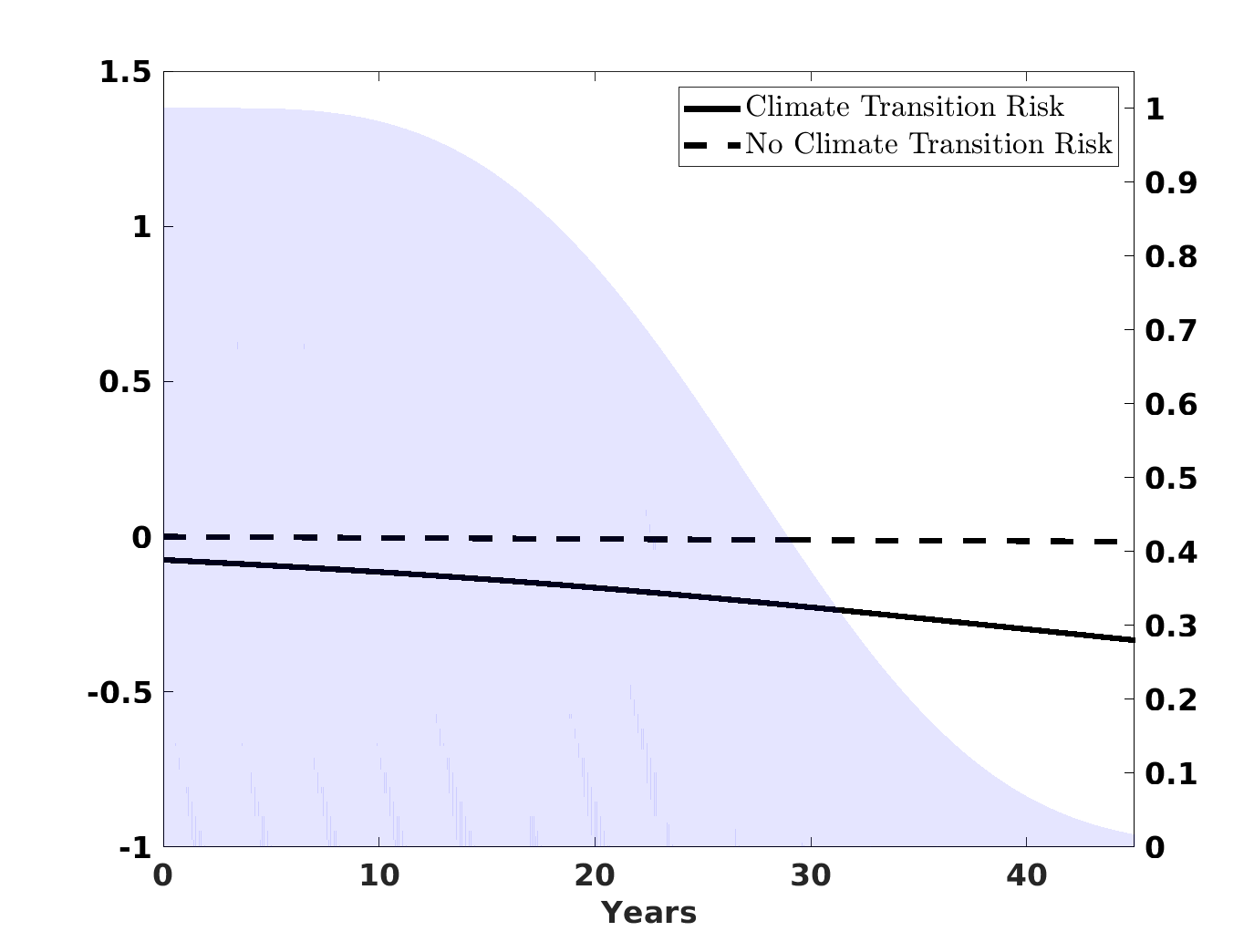}
           \caption[]{{\small Coal Spot Price: $P_{1,t}$}}
        \end{subfigure}

        \begin{subfigure}[b]{0.328\textwidth}
            \centering
            \includegraphics[width=\textwidth]{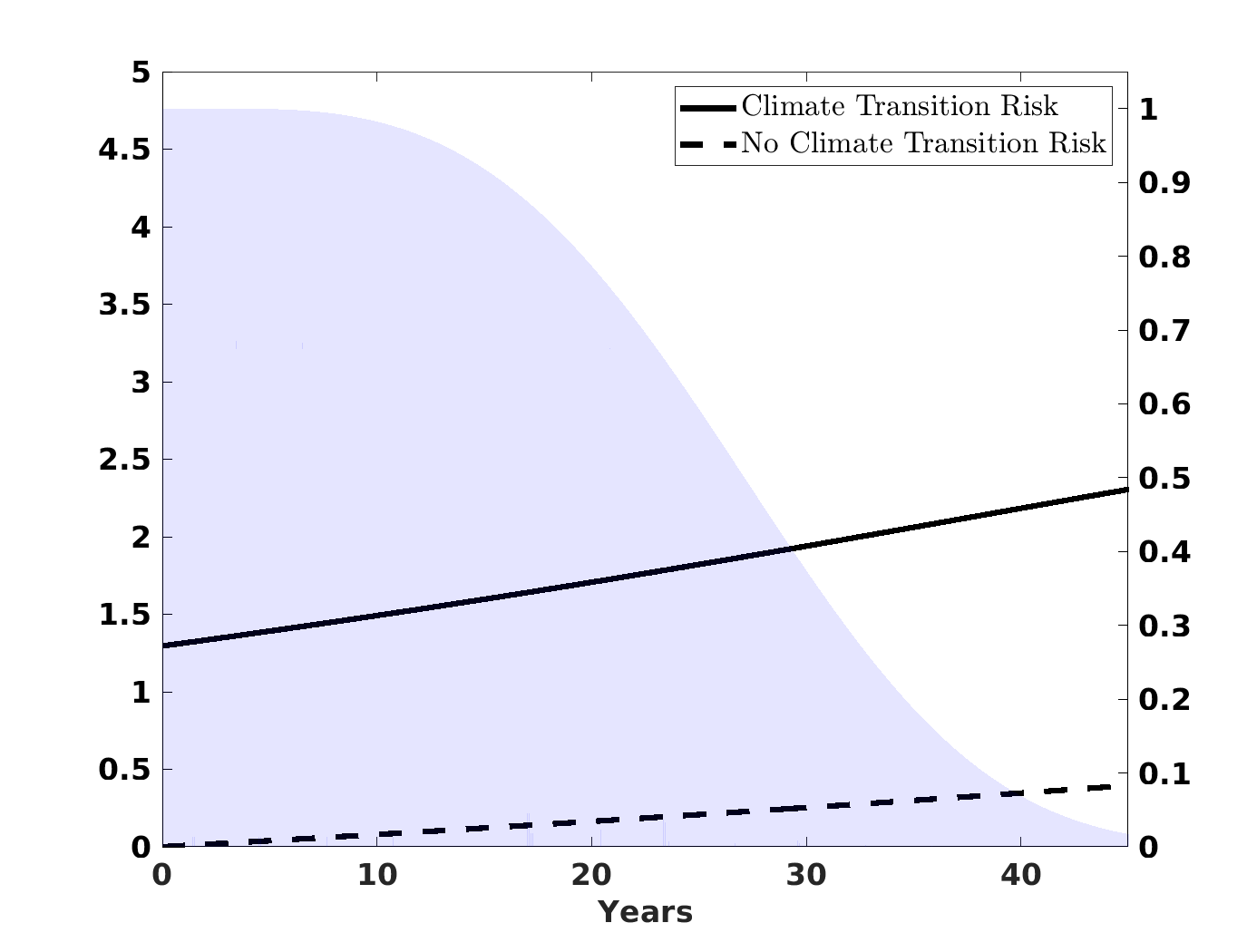}
            \caption[]{{\small Green Firm Price: $S^{(3)}_{t}$}}              
        \end{subfigure}
        \begin{subfigure}[b]{0.328\textwidth}  
            \centering 
            \includegraphics[width=\textwidth]{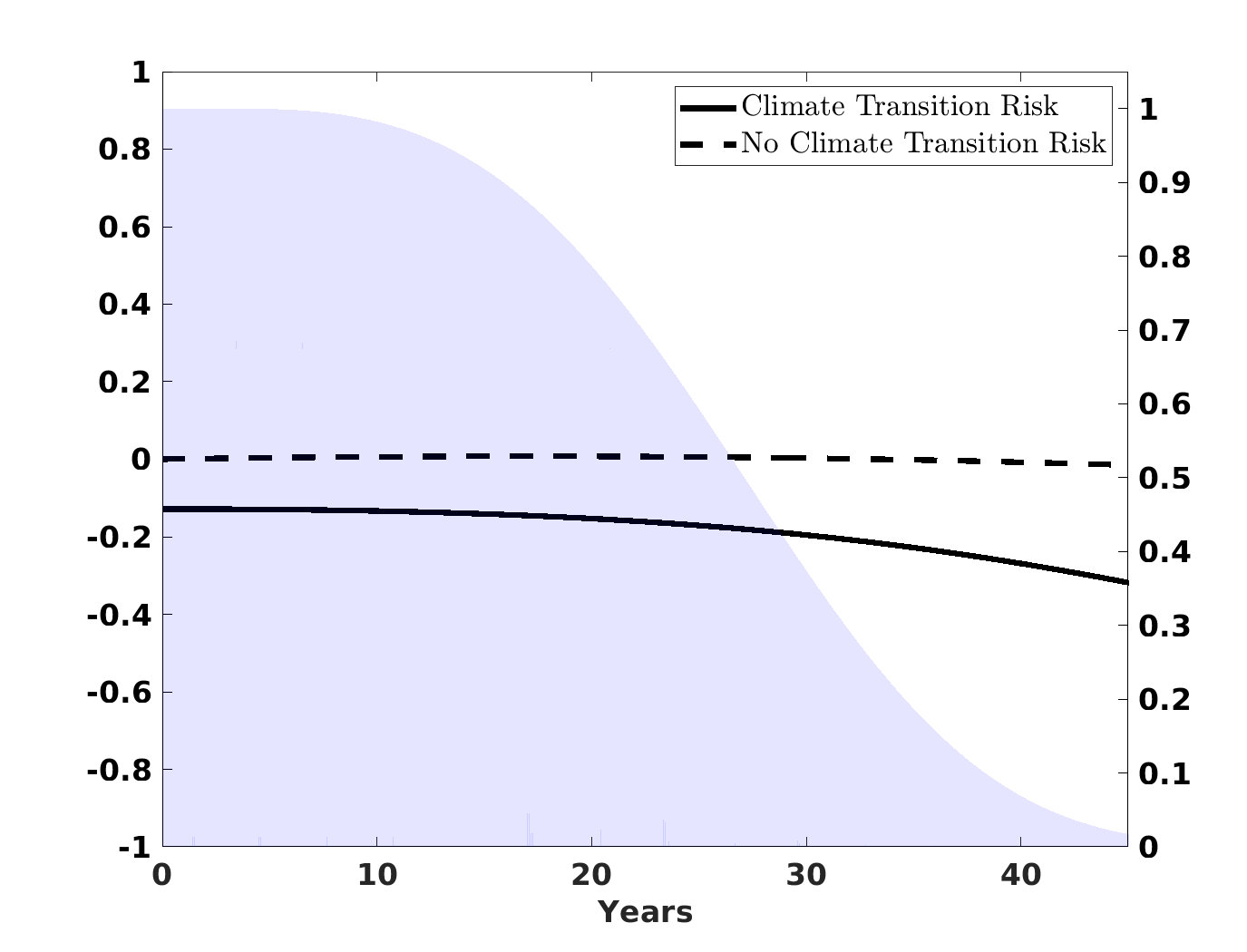}
           \caption[]{{\small Oil Firm Price: $S^{(1)}_{t}$}}
        \end{subfigure}
        \begin{subfigure}[b]{0.328\textwidth}  
            \centering 
            \includegraphics[width=\textwidth]{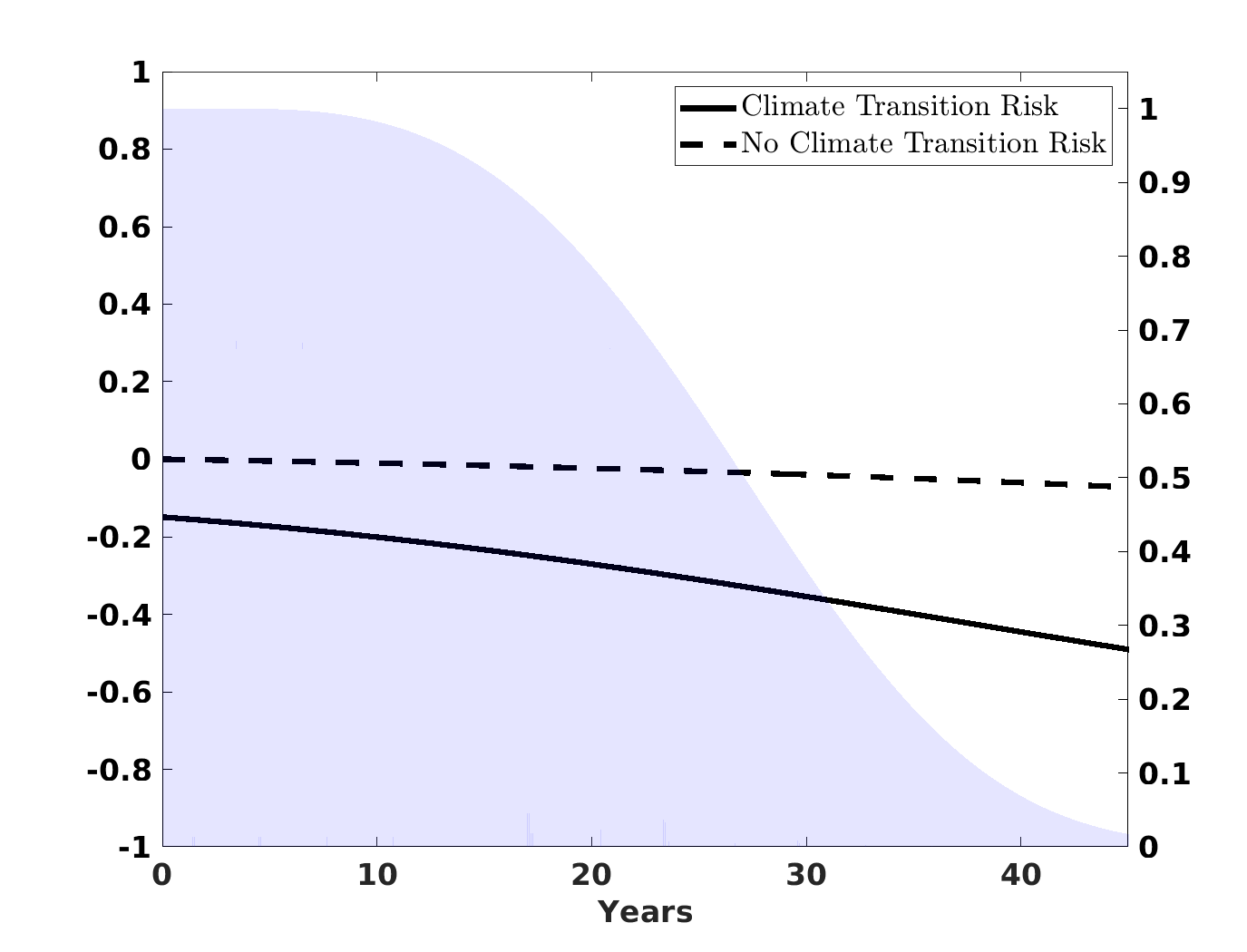}
           \caption[]{{\small Coal Firm Price: $S^{(2)}_{t}$}}
        \end{subfigure}  
        
        \vspace{-0.25cm}
\end{center}

\begin{footnotesize}
Figure \ref{fig:ctax_otech_model_sims} shows the simulated outcomes for the ``coal tax then technology shock''  model based on the numerical solutions. Panels (a) through (c) show the temperature anomaly, oil production, and coal production. Panels (d) through (f) show the green investment choice, oil spot price, and coal spot price. Panels (g) and (i) show the green firm price, oil firm price, and coal firm price. Solid lines represent results for the Climate Transition Risk scenario where $\lambda_t = \lambda(T_t)$ and dashed lines represent results for the No Climate Transition Risk scenario where $\lambda_t = 0$. The blue shaded region shows the cumulative probability of no transition shock occurring.
\end{footnotesize} 

\end{figure}

% \end{landscape}
%%%%%%%%%%%%%%%%%%%%%%%%%%%%%%%%%%%%%%%%%%%%%%%%%%
%%%%%%%%%%%%%%%%%%%%%%%%%%%%%%%%%%%%%%%%%%%%%%%%%%
%%%%%%%%%%%%%%%%%%%%%%%%%%%%%%%%%%%%%%%%%%%%%%%%%%

Figures \ref{fig:hybrid_model_sims}-\ref{fig:ctax_otech_model_sims} give the results for two specific alternative transition scenarios: first, where a fraction ($25\%$)\footnote{This value is consistent with estimates of the fraction of priced emissions by \cite{santikarn2021state}.} of the carbon tax before the transition shock ignores transition risk and the expected transition shock is in the form of a technology shock (Figure \ref{fig:hybrid_model_sims}); and second, a two stage-transition shock where the first stage is the implementation of a coal carbon tax and the second stage is a technology shock that eliminates the need for coal and oil in the energy input for final output production (Figure \ref{fig:ctax_otech_model_sims}). Each plot includes the ``Climate Transition Risk'' outcome, shown by the solid lines, the counterfactual ``No Climate Transition Risk'' outcome where there is no climate transition risk ($\lambda_t = 0$), shown by the dashed lines, and the cumulative probability of the climate transition not yet occurring for the alternative transition risk scenario, given by the blue shaded region.

For the ``hybrid technology shock,'' where a carbon tax policy that ignores transition risk is partially implemented before a technology transition shock occurs, the ``run'' effect is still present. However, the dynamic responses in production and prices are dynamically attenuated given the carbon tax cost prior to the technology realization.  For the ``coal tax then fossil fuel technology shock,'' where only coal is taxed after the first transition shock, followed by a second transition shock in the form of a technology shock, again the ``run'' effect is present. In this case, however, the dynamic responses in production are amplified and coal now showing a modest run response not present before. The spot price responses of oil and coal are now similar as they are both lower and dynamically decreasing. And while the value of both fossil fuel firms are lower, the impact is significantly less given the two-step transition. Both fossil fuel firm prices do demonstrate the dynamic attenuation driven by the run on fossil fuel response only present for oil in the baseline technology shock setting. In each of these two transition shock cases, the green investment and firm valuation are similar to what we saw in previous scenarios. These results further validate the robustness of the run on fossil fuel effect shown previously, and highlight key intuition about what is driving these responses. Specifically, when the cost of transition risk is perceived to be higher, as is the case with this hybrid technological transition risk scenario, greater value is placed on preventing this perceived negative future shock and emissions are restrained. If instead the cost is perceived to be low, such as in the coal tax then technology shock case, the incentive to maximize the use of fossil fuels by ``running'' is amplified.
%%%%%%%%%%%%%%%%%%%%%%%%%%%%%%%%%%%%%%%%%%%%%%%%%%%%%%%%%%%%%%%

\subsection{Alternative Utility Specifications}

The next set of alternative frameworks consider alternative utility function specifications: first is log utility, and second is recursive utility when the EIS is non-unitary. In each case, the marginal cost of climate change, marginal value of reserves, and marginal cost of forgoing labor for final output production are still present in the optimal choices of fossil fuel production. Thus, the same qualitative dynamic impacts of climate transition risk are present. However, the key differences are quantitative. For log utility, the value function structure is now additive, leading to more muted ``run'' effects without the multiplicative amplification related to the forward looking concerns about the resolution of uncertainty. For recursive utility when the EIS is non-unitary, the quantitative differences depend on the assumption about the value of the EIS. Consistent with the asset pricing results in other settings, an EIS greater than one leads to increased concern about the resolution of uncertainty and an amplification of the ``run'' effect on asset prices relative to unitary EIS. With an EIS less than unitary,  the ``run'' effect on asset prices is muted relative to the unitary EIS case. This result highlights the value of asset prices for characterizing climate change and transition risks, and provides insight for the expected macroeconomic outcomes, where a larger (smaller) EIS amplifies (diminishes) the transition risk mechanism.

\subsection{Alternative Fossil Fuel Sector Settings} %Stranded Assets and 

I consider three alternative fossil fuel sector frameworks. The first considers adding exploration in a stylized way, which leads to qualitatively similar outcomes, though quantitatively they are slightly larger as reserves are renewable, at a cost. Second, I solve a simplified approximation of imperfect competition by assuming a symmetric oligopoly in the oil sector. Firms in this setting account for pricing impacts that they have, approximating the production behavior of OPEC and numerous state-owned oil firms. The standard effect that market competition effect leads to higher extraction interacts with the incentive to run up oil production in expectation of transition risk. As a result, the run effect is still present, but fewer ``firms'' means better ``coordination'' to maintain lower production, limiting price and climate-linked transition risk impacts, and higher profits. Finally, I consider a technology shock for economies with just the coal sector or just the oil sector\footnote{In each case, I specify the relevant energy input demand share, $\nu_1$ or $\nu_2$, to be $0.8$.}. The oil only framework leads to results very similar to the main specification. The ``run'' effect is also present, to some degree, for the coal only framework, but substantially attenuated in magnitude compared to the baseline. Thus, the magnitude of the ``run'' also depends importantly on the resource transferability (reserves vs. labor), i.e., the magnitude of the stranded assets risk.
%%%%%%%%%%%%%%%%%%%%%%%%%%%%%%%%%%%%%%%%%%%%%%%%%%%%%%%%%%%%%%%

\subsection{Alternative Internalization Scenario}

In addition, I consider the case where the climate externality is only \emph{partially internalized} by the planner. An internalization parameter $\chi \in [0,1]$ determines the fraction of the climate impact that is internalized and therefore ``endogenous,'' $\chi \beta_T \mathcal{E}_t R_t$, and the fraction is not internalized and therefore ``exogenous'' to the decision maker, $(1-\chi) \beta_T \mathcal{E}_t R_t$. The solution to this model iterates on a planner-type solution until the equilibrium temperature drift $\mu_T$ matches the aggregation of the ``endogenous'' and ``exogenous'' components of the temperature drift, akin to a ``big K, little k'' type macroeconomic framework. As in \cite{barnett2023climate}, limited internalization of the climate externality reduces concerns about climate consequences. However, in this setting it also reduces concerns about influencing the likelihood of a transition shock. Thus, in each transition shock case emissions production is amplified, leading to either an augmented ``run'' response or an undoing of the ``reverse run'' response. 

\subsection{Carbon Bubble Implications?}

The Grantham Institute and others have highlighted the potential existence of a ``carbon bubble,'' or a potential overvaluation of fossil fuel firms resulting from the ignored risk that a substantial fraction of reserves and resources would need to be ``stranded'' in order to meet temperature ceiling policy targets such as those outlined in the Paris Climate Accords. Concerns about this overvaluation of fossil fuel firms resulting from ignoring possible stranded assets risk is precisely what has driven recent concerns of investors and regulators about Oil Majors.\footnote{See Wall Street Journal, October 21, 2019, ``Exxon's Climate-Change Accounting Goes on Trial''; New York Times, June 9, 2021, ``Exxon Mobil Defeated by Activist Investor Engine No. 1.''; Financial Times, June 15, 2020, ``BP to take up to \$17.5bn hit on assets after cutting energy price outlook''; Reuters, April 15, 2021, ``Shell plays down risk of stranded oil and gas reserves.''} The existing stranded assets and green paradox literature has identified a static effect that climate-linked transition risk reduces the expected usable fossil fuel reserves. My model highlights an additional dynamic general equilibrium feedback effect where the accelerating run up in fossil fuel production induced by transition risk further diminishes the value of fossil fuel reserves, resources, and technology. If investors and market participants do not account for potential transition and stranded assets risks in their valuations, the dynamic, non-linear implications present in my model highlight that a ``carbon bubble'' could be more severe than shown in analyses that omit these general equilibrium effects.

\section{Supporting Empirical Evidence}

The model studied in this paper provides a number of testable implications that can be empirically validated. Specifically, for an increase in the likelihood of a future climate transition that would restrict fossil fuel use, we should see:
% \vspace{0.25cm}
\begin{center}
        \begin{minipage}{.875\textwidth}
\begin{enumerate}
\item \emph{negative} returns for high climate transition risk exposure sectors
\item a dynamic and persistent \emph{decrease} in the spot price of fossil fuels 
\item \emph{amplified} impacts with increasing climate change and climate awareness 
\end{enumerate}
        \end{minipage}
\end{center}

The macroeconomic implications vary by transition shock type. For an anticipated technology shock leading to a climate transition, we should expect:
\begin{center}
        \begin{minipage}{.875\textwidth}
\begin{enumerate}
\item[4a.] a dynamic and persistent \emph{increase} in fossil fuel production
\end{enumerate}
        \end{minipage}
\end{center}

For an anticipated taxation shock leading to a climate transition, we should expect:
\begin{center}
        \begin{minipage}{.875\textwidth}
\begin{enumerate}
\item[4b.] a dynamic and persistent \emph{decrease} in fossil fuel production
\end{enumerate}
        \end{minipage}
\end{center}

\vspace{0.1cm}

In what follows, I first discuss the existing literature on climate-linked transition and stranded assets risks. I then summarize results from two empirical tests designed to show whether observable events related to changes in the likelihood of climate-linked transition risk lead to changes in production and prices consistent with predictions and mechanisms highlighted in the model. My focus is on inferring which, if any, form of transition risk has observable impacts by comparing whether the signs and statistical significance of the estimates are consistent with specific qualitative results of the model. The main estimates focus on the oil sector for two reasons: first, the model highlights that these predictions are most relevant for the oil sector; and second, we observe empirically that there is significant existing policy for coal and a view of natural gas as a ``cleaner'' alternative. I examine robustness of this assumption, and various others, in the Online Appendix.

\subsection{Existing Climate Transition Risk Empirical Evidence}

Within the climate finance and climate economics literature, researchers have only recently begun to identify and estimate climate-linked transition and stranded assets implications. \cite{atanasova2019stranded} find a negative effect on the value of North American oil firms due to the growing levels of undeveloped reserves they hold. \cite{delis2019being} show an increased cost of credit via syndicated corporate loans for fossil fuel firms, but only after the 2015 Paris Climate Accords. \cite{ilhan2021carbon} find a reduced cost of insurance against downside tail risks associated with climate policy uncertainty after an unexpected shift in expected future transition risk in the form of decreased implied volatility slopes for options on highly carbon-intense firms and sectors after the election of President Trump. \cite{bartram2022real} show regulatory arbitrage by California-based firms in response to a climate policy restricting fossil fuel emissions. \cite{seltzer2019climatereg} find more negative impacts from climate change risk for corporate bonds issued in states with stronger climate policy. \cite{norman2024empirical} estimate prevalent and persistent reductions in oil spot and futures prices, implying increases in current oil consumption, resulting from expectations about the future restriction of oil use, particularly around the Waxman-Markey bill in the US. \cite{hong2024great} find renewable portfolio standards (RPS), a popular green economy transition climate policy, negatively impacts corporate bond spreads and investment of affected utility firms.

Consistent with changing investor beliefs about climate transition risk leading to a bidding up of stock prices for low-emissions firms and a bidding down of stock prices for high-emissions firms,
\cite{bolton2021investors} find an increased premium associated with carbon emissions for US firms post-Paris Accords, though potentially resulting from the increasing sample size in their analysis, while \cite{bolton2021global} find an increased and significant carbon premium post-Paris Accords for global firms driven by considerations for long-term impacts of climate policy tightness. \cite{donadelli2021macro} analyze responses in the oil sector associated with climate transition risk, while \cite{jung2021crisk} assess the resilience of large global banks to climate transition risk, each exploiting the empirical measure I develop for the analysis in this paper to demonstrate dynamic and persistent responses in these sectors with respect to transition risk. \cite{ramelli2021investor} find that stock prices of carbon-intensive firms positively reacted to President Trump's election, though stock prices for climate-responsible firms also increased, while the expected ``boomerang'' back to strong climate policy with the election of President Biden saw carbon intensive firms' stock prices drop and climate responsible firms' stock prices soar. And while not explicitly associated with climate change risk, \cite{hsu2023pollution} analysis of environmental litigation risk associated with toxic pollution emissions shows that the cumulative abnormal returns (CARs) of high toxic pollution emissions firms were significantly positive following the election of Donald Trump as U.S. president, a decreasing environmental policy risk event.

While these results are qualitatively consistent with the predictions from my model, they do not fully address the dynamic general equilibrium effects related to transition and stranded assets risks central to my analysis. In addition, the existing literature often focuses either on asset pricing or production outcomes, not the interconnected risk implications simultaneously, or as in \cite{hsu2023pollution} rely on static and exogenous risk mechanisms that are unable to account for the dynamic feedback mechanisms of climate-linked transition risk as in my framework. Because of this gap in the existing empirical evidence on transition and stranded assets risk, I provide additional empirical analysis of the novel, dynamic macro-asset pricing implications of my model in what follows.

\subsection[Data Sources]{Data Sources}

My empirical analysis uses data from the following sources. Fossil fuel production and price data comes from the US Energy Information Administration (EIA) Monthly Review Database and FRED. Global atmospheric temperature data comes from the NASA Goddard Institute for Space Studies (NASA-GISS). Data on returns for the 49 sector portfolios and the market portfolio come from Ken French's website and the Compustat/CRSP merged database available from WRDS. Macroeconomic variables and indicators come from FRED, the BEA, and James Hamilton's and Lutz Kilian's websites.\footnote{For the oil production, oil price, and real economic activity data used in the VAR estimation, I use the updated replication data provided for \cite{baumeister2019structural} on James Hamilton's website \url{https://econweb.ucsd.edu/~jhamilto/software.htm} to ensure consistency. The results are identical when estimating the VAR using versions of the data series I construct myself.}

To captures changes in the probability of a future climate-linked transition occurring, which corresponds to changes in $\lambda_t$ in the model, I construct a proxy variable by first compiling a daily-frequency time series of major fossil fuel and alternative energy events, IPCC and UNFCCC meetings and related events, US presidential election results, other major global and national climate policies, and US energy policies from non-partisan government, academic, and non-profit informational websites (ProCon.org, IPCC and UNFCCC websites, and Wikipedia)\footnote{The specific sources used and websites for each are provided in the Online Appendix.}. After constructing the time-series event indicator variable of climate-related transition events, I interact it with the returns of a portfolio designed to track the asset pricing response for firms highly exposed to climate-related transition risk impacted by these events. This interaction variable incorporates the forward-looking information of relevant asset price returns with the event-based indicator to account for the magnitude, timing, and dynamics of the effects of climate-linked transition events that shift the likelihood of a future green transition on the oil price and oil production. I call this variable $ClimateTransition_t$, and provide additional details later in the empirical analysis discussion.

\subsection{Event Study Analysis}

My first empirical test builds on the event study framework of \cite{koijen2016financial} to estimate the impact of changes in the likelihood of a future climate transition event on US stock returns. Formally, I estimate the impact of an event that shifts climate transition risk expectations on cumulative abnormal returns around climate transition events by exploiting cross-sectional variation in climate transition risk exposure across different sectors. I use daily returns for the 49 sector portfolios provided on Ken French's website for the cross-section. I derive abnormal returns for sector portfolios as unexplained differences from the market model, and aggregate the residuals starting from 10 days before the event to 10 days after the event to get cumulative abnormal returns. I then compute a proxy measure of climate transition risk exposure for each sector based on their estimated exposure to oil price innovations, consistent with the theoretical model. Finally, I estimate the quantitative relationship between transition risk exposure and cumulative abnormal returns around shocks to the likelihood of a future climate-linked transition. Given the numerous climate-related transition events that occurred across time in my event list, and based on the model prediction that that asset pricing responses should increase over time as temperature increases and climate change concerns are higher, I estimate panel regressions to measure the aggregate or average effect of climate-linked transition events for three separate panels: all events in the full time sample of 1974-2019\footnote{The sample begins in 1973, but I omit the first year used as the initial lags in the VAR estimation as in \cite{baumeister2019structural}, and ends in 2019 to avoid incorporating implications of COVID-19.}, only events in the more recent time subsample starting in 2009\footnote{I start this subsample at 2009 since it is the first year post-Global Financial Crisis and includes major transition events such as significant ARRA funding for clean energy and, as shown by the WSJ Climate Change News Index of \cite{engle2020hedging}, the highly anticipated COP15 Copenhagen meeting.}, and only events in the earlier subsample before 2009.

 \begin{figure}[!t]
{    \centering
    \caption{Event Study Analysis of Climate-Linked Transition Events} \label{fig:PanelEventStudy}
    }
    \begin{subfigure}{\linewidth}
{    \centering
\begin{tabular}{l | c c c }
\hline
\hline
%& \multicolumn{3}{ c |}{+ Likelihood Shock} \\
%\hline
Time Period & $1974$-$2019$ Sample & $1974$-$2008$ Sample & $2009$-$2019$ Sample \\
 \hline
% $\hat{\delta}_1$ estimate & $\mathbf{-0.10}$ & $\mathbf{0.13}$ & $\mathbf{-0.41}$ \\
$\hat{\delta}_1$ estimate & $\mathbf{-0.08}$ & $\mathbf{0.12}$ & $\mathbf{-0.41}$ \\
% t-statistic & $-1.62$ & $2.25$ & $-2.85$  \\
t-statistic & $-1.50$ & $1.77$ & $-2.85$  \\
\hline
\hline
\end{tabular}
    \caption{Estimated Sensitivity for Transition Shocks}\label{table:PanelEventStudya}
    }
\end{subfigure}  

\medskip  

    \begin{center}
\begin{subfigure}{0.6\linewidth}
{    \centering
\includegraphics[width=\linewidth]{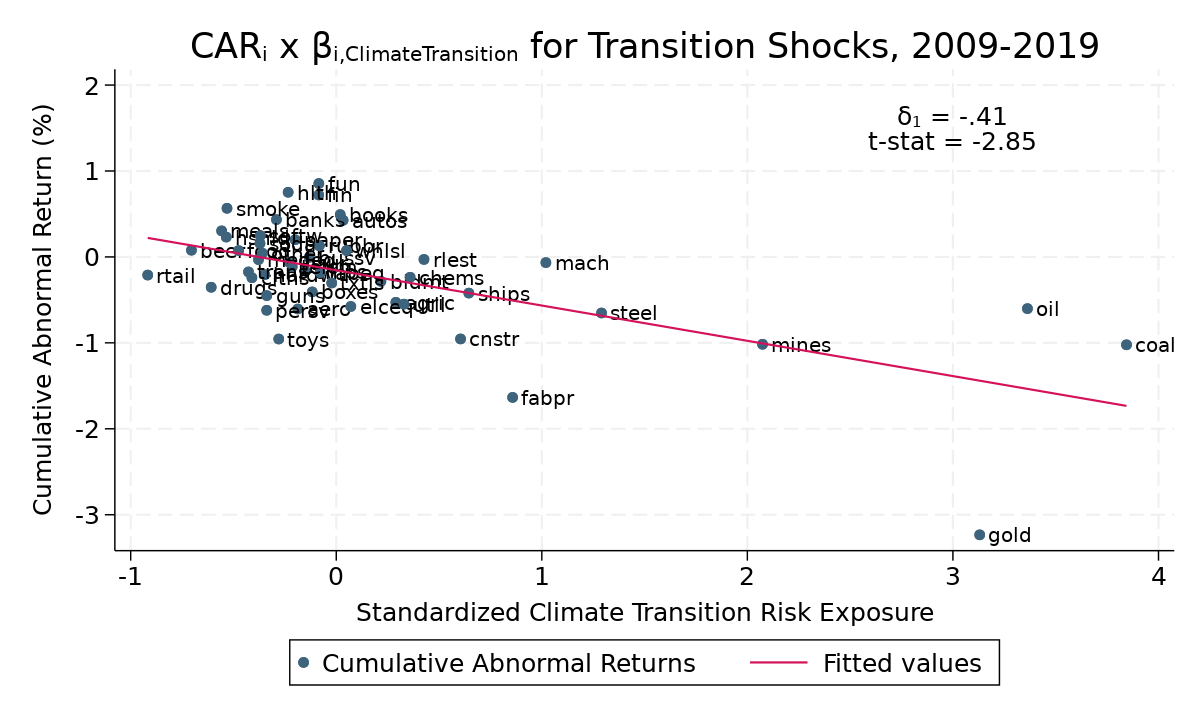}
\caption{$CAR_i$ Scatter Plot for Transition Shocks}\label{fig:PanelEventStudyb}
}
\end{subfigure}

\medskip

\begin{subfigure}{0.6\linewidth}
{    \centering
\includegraphics[width=\linewidth]{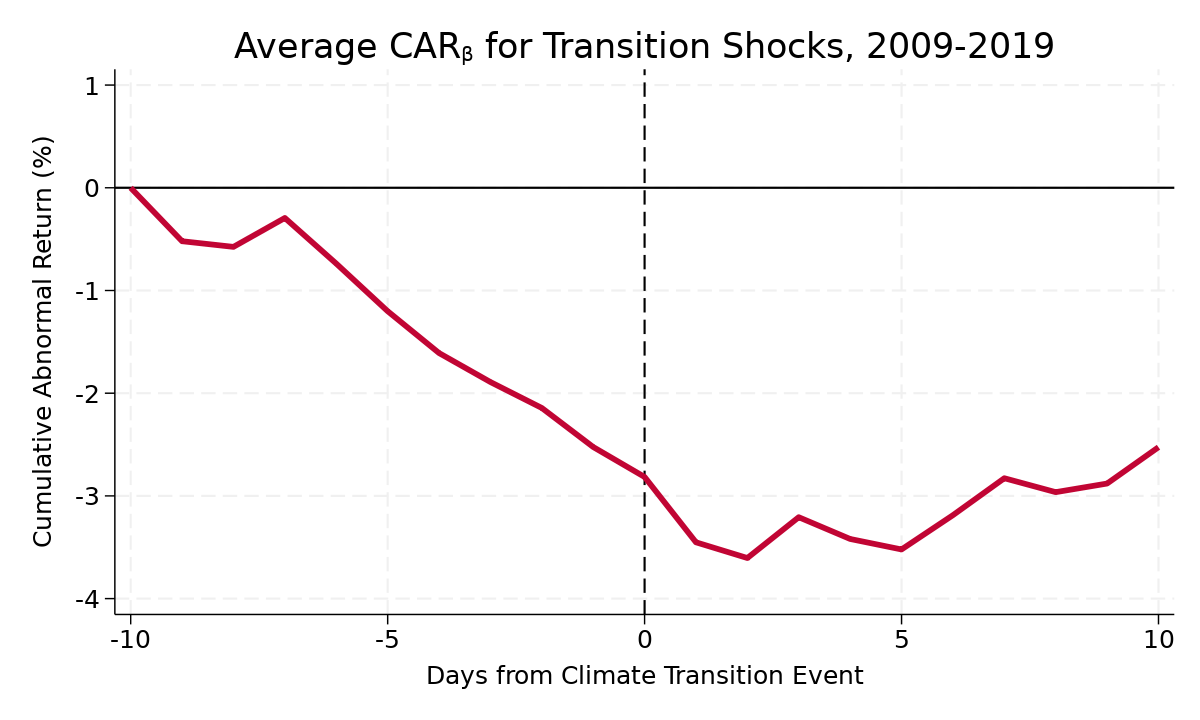}
\caption{$CAR_{\beta}$ Dynamics for Transition Shocks}\label{fig:PanelEventStudyc}
}
\end{subfigure}
    \end{center}

\vspace{0.15cm}

\begin{footnotesize}
%\begin{flushleft}
Figure \ref{fig:PanelEventStudy} highlights the relationship between cumulative abnormal returns and climate-related transition events. Panel (a) shows the coefficient estimate for the relationship between climate transition risk exposure and $CAR_{i,t-10 \rightarrow t+10}$ for all climate transition events. Panel (b) shows the scatter plot of the average $CAR_{i,t-10 \rightarrow t+10}$ by sector $i$ following a climate transition event. Panel (c) shows the average $CAR_{\beta,t-10 \rightarrow t+j}$ for $j = \{-10,..,10\}$ following a climate transition event. 
%\end{flushleft}
\end{footnotesize}  
    \end{figure}

Results for the aggregated panel estimates are given in Figure \ref{fig:PanelEventStudy}. Table \ref{table:PanelEventStudya} shows that for the full sample and recent, transition relevant sample, $\delta_1$ is negative. Also, the estimate is larger in magnitude and statistically significant in the recent, transition-relevant subsample of events. Finally, the early subsample estimate is positive and therefore differs in a statistically significant way from the estimate in the transition-relevant subsample. Figure \ref{fig:PanelEventStudyb} shows a scatter plot of the average climate transition risk exposure-cumulative abnormal return relationship across industries for the recent, transition-relevant sample. The scatter plot demonstrates the clear negative relationship shown by the negative $\delta_1$ estimate, and highlights that the ``dirtiest'' sectors with the most positive climate transition risk exposure (coal, oil, gold, mines, steel, machines, fabricated products) experienced the largest in magnitude return responses on average across all transition events. Figure \ref{fig:PanelEventStudyc} shows the average $CAR_{\beta,t-10 \rightarrow t+j}$ for $j = \{-10,...,10\}$ across all transition-related events of the cumulative abnormal returns for a transition-risk-exposure weighted average of sector portfolio returns. The clear result is that for an increase in the likelihood of a future climate-linked transition there is a significant and economically meaningful decrease in the returns of high transition-risk exposure firms that begins a few days into the event window, increases in magnitude for numerous days, and persists throughout the event window. Altogether, this empirical evidence confirms the asset pricing implications predicted by the theoretical model.

\subsection{Vector Autoregression Estimates}

While the event study results provide important evidence regarding climate-related transition risk implications, it is essential that I also analyze dynamic fossil fuel sector responses to climate-linked transition risk shocks to infer empirically which theoretical transition risk mechanism, if any, is anticipated by the aggregate economy. Therefore, I provide an additional set of empirical results by estimating a structural vector autoregression (VAR) for the global oil market. I estimate an augmented version of the global oil market VAR proposed by \cite{kilian2009not}, where the vector of endogenous state variable vector $y_t$ is defined by
\begin{eqnarray*}
y_{t} & = & [ClimateTransition_{t}, \Delta prod_t, \Delta rea_t, \Delta p_{t}^{oil} ] '
\end{eqnarray*}

 \begin{figure}[!t]
%    \centering
\caption{Impulse Response Functions for Transition Likelihood Shock}\label{fig:var_climpol_ret}
    
\begin{subfigure}{0.48\linewidth}
\includegraphics[width=\linewidth]{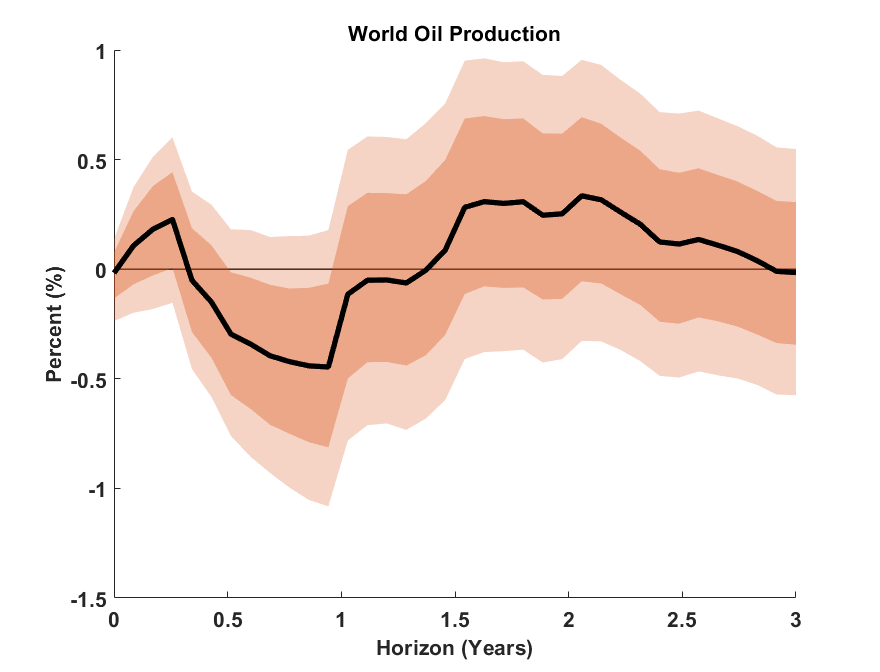}
\caption{Oil Production IRF, 1974-2008}
\end{subfigure}
    \hfill
\begin{subfigure}{0.48\linewidth}
\includegraphics[width=\linewidth]{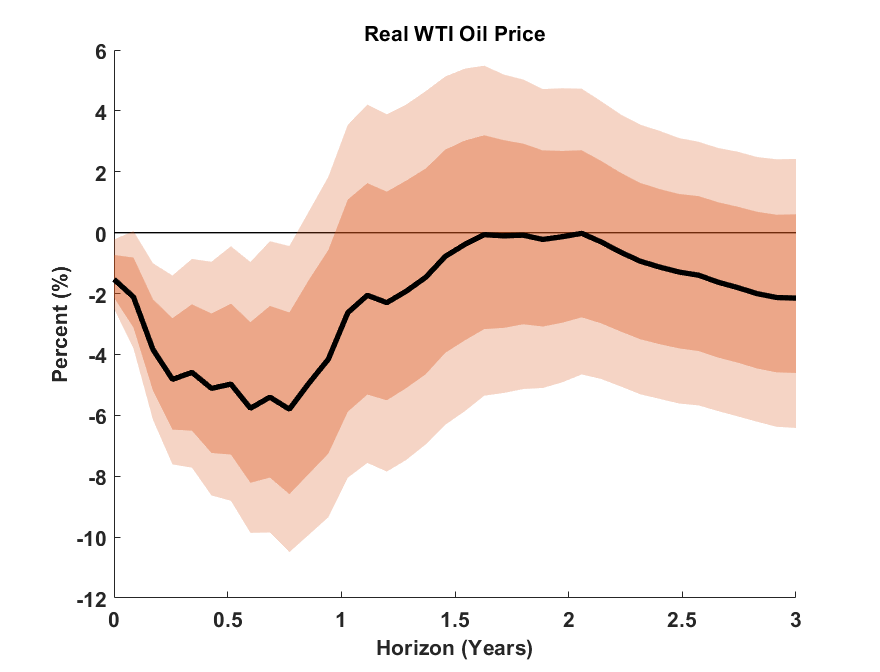}
\caption{Oil Price IRF, 1974-2008}
\end{subfigure}

\medskip

\begin{subfigure}{0.48\linewidth}
\includegraphics[width=\linewidth]{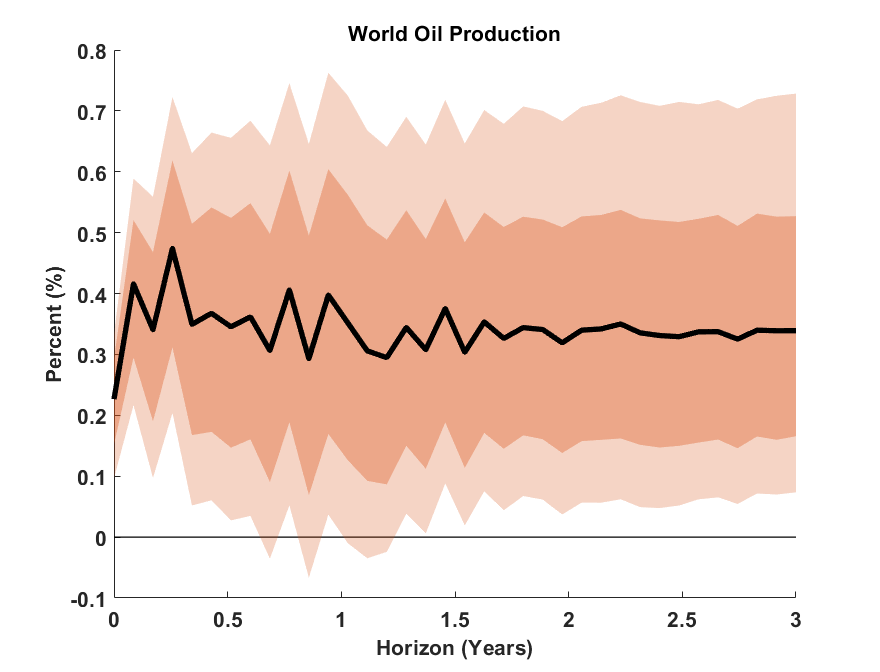}
\caption{Oil Production IRF, 2009-2019}
\end{subfigure}
    \hfill
\begin{subfigure}{0.48\linewidth}
\includegraphics[width=\linewidth]{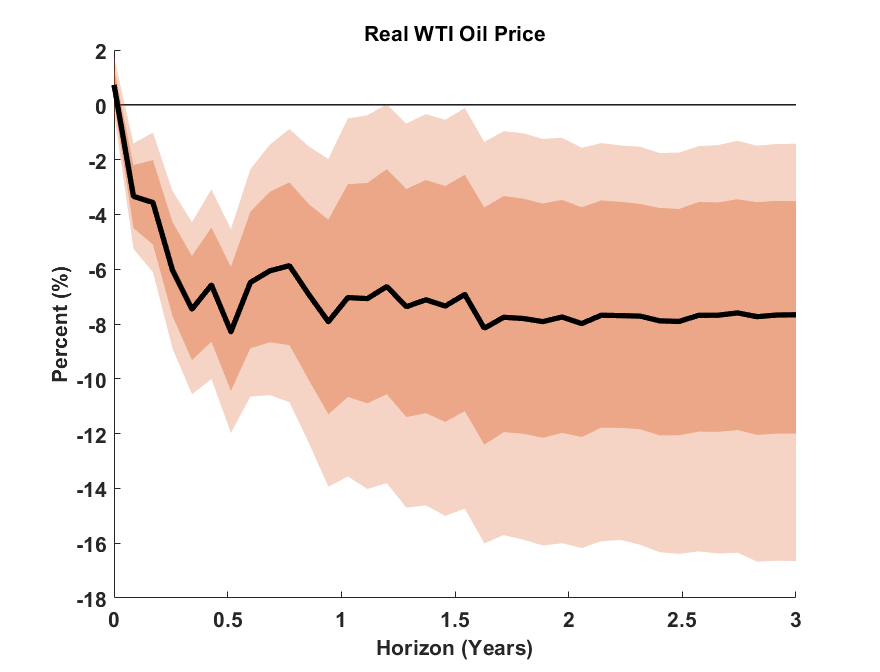}
\caption{Oil Price IRF, 2009-2019}
\end{subfigure}

\vspace{0.25cm}

\begin{footnotesize}

Figure \ref{fig:var_climpol_ret} shows the estimated impulse response functions for global oil production and the WTI spot price of oil for a shock to $ClimateTransition_t$. The black line is the estimated IRF, the dark red shaded region represents the one-standard deviation error band, and the light red shaded region represents the two-standard deviation error band. The top panels are for the early time subsample (1974-2008), and the bottom panels are for the transition-relevant time subsample (2009-2019). Bootstrapped error bands are calculated using 10,000 simulated samples. The VAR is estimated using 12 lags. %See text for the full VAR specification used and definition of variables.
\end{footnotesize}
\end{figure}

$ClimateTransition_{t}$ is the time-series of $CAR^{event}_{\beta,t-10 \rightarrow t+10}$ values for all climate-related-transition events, and zero otherwise, aggregated up to the monthly frequency\footnote{I discuss results for alternative $ClimateTransition_t$ definitions in the Online Appendix.} Following \cite{baumeister2019structural}, $\Delta prod_t$ is the percent change in global oil production available from the EIA, $\Delta rea_t$ is the log differences in a measure of real economic activity given by the log OECD industrial production index, and $\Delta p_{t}^{oil}$ is log differences in the real West Texas Intermediate (WTI) monthly closing price for crude oil\footnote{I include month dummies to control for seasonality, which has little impact on the results.}. The results are qualitatively the same and quantitatively similar if I use the logarithim of Kilian's updated measure of real economic activity based on an index of nominal shipping freight rates, as well as the real refiner's acquisition price for the oil price. I use a Cholesky decomposition of the estimated variance-covariance matrix for identification of the structural shocks, imposing a recursive interpretation of the impact of the shocks. 

Figure \ref{fig:var_climpol_ret} shows cumulative level impulse response functions of oil production and oil prices for a shock to the likelihood of a significant climate transition occurring based on the VAR estimates and recursive identification structure. The top panel is for the VAR estimated using the early time sample (1974-2008) and the bottom panel is for the VAR estimated using the more recent, transition-focused time sample (2009-2019). Each plot shows the individual impulse response functions with bootstrapped standard errors, where dark red shaded region is the one-standard deviation confidence interval, the light red shaded region is the two-standard deviation confidence interval, and the black line is point estimate for the IRF. The recent, transition-relevant subsample results show that a climate-linked transition shock leads to an increase in oil production and a decrease in the spot price of oil. Moreover, these impacts grow in magnitude and persist over time, and are statistically significant for almost the entirety of the 3-year horizon looked at. For the early sample IRFs, the outcomes are qualitatively consistent with the recent sample estimates along some dimensions, but are more transient, more muted, and statistically insignificant. These results confirms that climate-related transition risk has a statistically significant and economically meaningful impact on both prices and production for the recent subsample. This validates the importance and state-dependent effects of climate-linked transition risk predicted by the model, and suggests firms are likely anticipating a ``technology shock'' transition risk given the positive response of oil production and negative response of oil prices to the shock.\footnote{I provide additional robustness tests for validation and intuition in the Online Appendix.}

\section{Conclusions}

Using a dynamic, production-based asset pricing model I show how different climate-linked transition risk scenarios generate contrasting dynamic outcomes in macroeconomic production choices and asset prices. When the climate-linked transition shock is expected to have relatively low economics costs, such as for a technology-related shock, the optimal response is a characterized ``run on fossil fuels'' with dynamically increasing fossil fuel production and decreasing spot prices. This response occurs even as fossil fuel reserves diminish and the atmospheric temperature increases, because the expectation of the transition shocks leads to dynamically amplified discounting of the value of fossil fuel reserves and costs of future climate consequences that outweighs concerns about future economic consequences of the transition shock. If the climate-linked transition shock is instead expected to have relatively high economic costs, such as for a taxation-related shock, the optimal response is characterized by a ``reverse run'' with dynamically restrained fossil fuel production and amplified spot prices. In these settings, the concerns about future economic consequences of the expected transition shock outweighs the dynamic discounting effect of fossil fuel reserves and costs of future climate consequences, leading the planner to attempt to postpone or delay the transition shock. In both types of transition risk responses, there is a substantial level shift down and dynamic reduction in fossil fuel firm values. Empirical estimates from an event study analysis and an augmented VAR estimation for the global oil market suggests that individuals are anticipating a transition risk scenario that would produce a ``run on fossil fuel'' response. Potentially interesting extensions include theoretical and empirical exploration of the implications of climate-linked transition risk for settings that account for firm- or country-level heterogeneity, political economy frictions, the term structure of energy options and futures, or model uncertainty aversion. I leave such extensions for future work.

% \newpage
% \clearpage
%\singlespacing
%\footnotesize
\bibliographystyle{chicago}
\bibliography{MB_ROFF_JFE_Submit_Alt}

\newpage
\clearpage

\begin{appendices}

\onehalfspacing

\setcounter{page}{1}

\section*{Appendix}

This appendix is provided in support of the paper ``A Run on Fossil Fuel? Climate Change and Transition Risk.'' Included here are details on theoretical derivations, results for alternative model settings, numerical methods used in the paper, and additional details and robustness results for the empirical estimation.

\section{Theoretical Derivations} 

\subsection{Baseline Social Planner's Problem}

The social planner's problem for the pre-transition state can be written as an HJB equation of the following form:
\begin{eqnarray*}
0 & = & \rho(1-\gamma)V[ \alpha \log C_K + \psi \log C_E + (1-\alpha - \psi) \log C_L - \eta T - \frac{1}{1-\gamma} \log(1-\gamma)V] \\
 & & +(\mu_{C}+i_{C}-\frac{\phi_C}{2}i_{C}^{2} - \frac{1}{2}\sigma_{C}^{2})V_{\log K_C}+(\mu_{G}+i_{G}-\frac{\phi_G}{2}i_{G}^{2} - \frac{1}{2}\sigma_{G}^{2})V_{\log K_G} \\
 & & +\frac{1}{2}\sigma_{C}^{2} V_{\log K_C, \log K_C} +\frac{1}{2}\sigma_{G}^{2} V_{\log K_G, \log K_G} \\
 & &  +(-\mathcal{E} - \frac{1}{2}\sigma_{R}^{2})V_{\log R} + \beta_{T}\left( E_{1,t} + E_{2,t} \right) V_{T} \\
 & &  +\frac{1}{2}\sigma_{R}^{2} V_{\log R, \log R}+\frac{1}{2}\sigma_{T}^{2}V_{TT}+\lambda_{t}[V_{post}-V_{pre}]
\end{eqnarray*}

with the First Order Conditions (FOC) are given as in the main text. The planner's post-transition problem for the baseline shocks has two cases, each of which can be written as an HJB equation. The ``technology shock'' and ``taxation shock'' cases are given by
\begin{eqnarray*}
0 & = & \rho(1-\gamma)V[ \alpha \log C_K + \psi\log \hat{C}_E - \eta T + (1-\alpha-\psi) \log C_L - \frac{1}{1-\gamma} \log(1-\gamma)V] \\
 & & +(\mu_{C}+i_{C}-\frac{\phi_C}{2}i_{C}^{2} - \frac{1}{2}\sigma_{C}^{2})V_{\log K_C}+(\mu_{G}+i_{G}-\frac{\phi_G}{2}i_{G}^{2} - \frac{1}{2}\sigma_{G}^{2})V_{\log K_G} \\
 & & +\frac{1}{2}\sigma_{C}^{2} V_{\log K_C, \log K_C} +\frac{1}{2}\sigma_{G}^{2} V_{\log K_G, \log K_G} +\frac{1}{2}\sigma_{T}^{2}V_{TT}
\end{eqnarray*}

where $\hat{C}_{E,t}$ is defined as in the main text for each transition shock case. The First Order Conditions (FOC) for each post-transition-shock case are given by
\begin{eqnarray*}
\rho(1-\gamma)V  (A_C-i_{C})^{-1} \alpha & = & (1-\phi_C i_{C}) V_{\log K_C} \\
\rho(1-\gamma)V (A_G-i_{G})^{-1} \psi & = & (1-\phi_G i_{G}) V_{\log K_G} 
\end{eqnarray*}

I conjecture and verify that post-transition value functions are given by
\begin{eqnarray*}
V=\frac{\exp\left( (1-\gamma)(-\eta T + \alpha \log K_C + \psi \log K_G + c_{post,i}) \right)}{1-\gamma}
\end{eqnarray*}
where the value function coefficients $c_{post,i}$, $i \in \{tech, tax\}$ are given by
\begin{eqnarray*}
c_{post,tech} & = &  \log(A_C-i_{C})^{\alpha} (A_G - i_G)^{\psi}  + \frac{1-\gamma}{2 \rho}(\alpha^2 \sigma_{C}^{2} + \psi^2 \sigma_{G}^{2} + \eta^2 \sigma_{T}^{2})  \\
& & + \frac{\alpha}{\rho} (\mu_{C}+i_{C}-\frac{\phi}{2}i_{C}^{2} -\frac{\sigma_C^2}{2})  + \frac{\psi}{\rho} (\mu_{G}+i_{G}-\frac{\phi}{2}i_{G}^{2} -\frac{\sigma_G^2}{2}) \\
c_{post,tax} & = & \log(A_C-i_{C})^{\alpha} (A_G - i_G)^{\psi}(\nu_3)^{\psi/\omega}  + \frac{1-\gamma}{2 \rho}(\alpha^2 \sigma_{C}^{2} + \psi^2 \sigma_{G}^{2} + \eta^2 \sigma_{T}^{2})  \\
& & + \frac{\alpha}{\rho} (\mu_{C}+i_{C}-\frac{\phi}{2}i_{C}^{2} -\frac{\sigma_C^2}{2})  + \frac{\psi}{\rho} (\mu_{G}+i_{G}-\frac{\phi}{2}i_{G}^{2} -\frac{\sigma_G^2}{2})
\end{eqnarray*}

Returning to the pre-transition problem, I conjecture a value function solution in each case of the following form
\begin{eqnarray*}
V=  \frac{\exp\left((1-\gamma)\{c_{pre} + \alpha \log K_C + v(\log K_G, \log R, T) \} \right) }{1-\gamma}
\end{eqnarray*}

Plugging in terms, I arrive at a simplified pre-transition HJB equation of the form
\begin{eqnarray*}
0 & = & \rho[ \psi \log C_E + (1-\alpha - \psi) \log C_L  - \eta T - v] +\frac{1-\gamma}{2}\left[\sigma_{G}^{2} v_{\log K_G}^2 + \sigma_{R}^{2} v_{\log R}^2 +\sigma_{T}^{2}v_{T}^{2} \right] \\
 & & +(\mu_{G}+i_{G}-\frac{\phi_G}{2}i_{G}^{2} - \frac{1}{2}\sigma_{G}^{2})v_{\log K_G} +(-\mathcal{E} - \frac{1}{2}\sigma_{R}^{2})v_{\log R} + \beta_{T}\left( E_{1,t} + E_{2,t} \right) v_{T} \\
 & & +\frac{1}{2}\sigma_{G}^{2} v_{\log K_G, \log K_G} +\frac{1}{2}\sigma_{R}^{2} v_{\log R, \log R}+\frac{1}{2}\sigma_{T}^{2}v_{TT} \\
 & &+\frac{\lambda_{t}}{1-\gamma}[\exp((1-\gamma)\{ -\eta T + \psi \log K_G + c_{post,i} - c_{pre} - v \}) - 1] \\
c_{pre} & = & \alpha \log(A_C-i_{C}) + \frac{\alpha}{\rho} (\mu_{C}+i_{C}-\frac{\phi}{2}i_{C}^{2})+\frac{1-\gamma}{2 \rho}\alpha^2\sigma_{C}^{2} 
\end{eqnarray*}
%
%\newpage
%\clearpage

\subsection{Decentralized Economy and Asset Prices}

\subsubsection{Household}
The household optimization problem is given by
\begin{eqnarray*}
V_t & = & \max_{C_t} E[ \int_t^{\infty} \exp(-\rho t) h(C_t, V_t) dt ]
\end{eqnarray*}

subject to the evolution of household wealth
\begin{eqnarray*}
d\varpi_t/\varpi_t = [r_{f,t} (1-\vartheta_t) + \mu_{\varpi} \vartheta_t - C_t/\varpi_t ] dt + \sigma_{\varpi} \vartheta_t \varpi_t dW_t
\end{eqnarray*}

where $\rho$ is the subjective discount rate, $h(C_t, V_t)$ is the felicity function associated with recursive preferences, $\varpi_t$ is household wealth, $r_{f,t}$ is the risk-free rate, $\vartheta_t$ is the fraction of wealth invested in the market portfolio, $\mu_{\varpi}$ is the expected return on the market portfolio, $\sigma_{\varpi}$ is the volatility on the market portfolio, $C_t$ is the choice of household consumption, and $V_t$ is the Social Planner's value function.

In equilibrium, firms are 100\% equity financed and households will hold all the equity, and so the evolution of household wealth can be written as 
\begin{eqnarray*}
d\varpi_t/ \varpi_t = [ \mu_{\varpi} - C_t/\varpi_t ] dt + \sigma_{\varpi} dW_t
\end{eqnarray*}

where $\mu_{\varpi}$, $C_t/\varpi_t$, and $\sigma_{\varpi}$ are determined by optimal choices and endogenous firm price and state variable dynamics.

\subsubsection{Firms}

%\subsubsection{Consumption Sector Firm}
The consumption sector firm's profit maximization problem is given by
\begin{eqnarray*}
S^{(C)}_t =  \max_{C_{K,t}, C_{E,t}, L_{1,t}}  & &  E \int_0^{\infty} \pi_{t}\{ \exp(-\eta T_{t}) C_{K,t}^{\alpha} L_{1,t}^{1-\alpha-\psi} \left(\nu_1 E_{1,t}^\omega + \nu_2 E_{2,t}^\omega + \nu_3 E_{3,t}^\omega \right)^{\psi/\omega} \\
& &\qquad -P_{K,t}C_{K,t}-w_{t} L_{1,t}-P_{1,t}E_{1,t} -P_{2,t}E_{2,t} -P_{3,t}E_{3,t}\}ds
%   & \text{subject to } dK_t/K_t = \varphi_K(i_{K,t})  dt+\sigma_{K}dB_K
 \end{eqnarray*}

%\subsubsection{Capital Sector Firm}
The capital sector firm's profit maximization problem is given by
\begin{eqnarray*}
S^{(K)}_t & = & \max_{i_{K,t}} E \int_0^{\infty} \pi_t P_{K,t}(A_C - i_{C,t}) K_{C,t} ds %\\
% & s.t. & dK_t/K_t = \varphi_K(i_{K,t})  dt+\sigma_{K}dB_K
 \end{eqnarray*}

% \subsubsection{Brown Energy Sector Firm}
The oil fossil fuel sector firm's profit maximization problem is given by
\begin{eqnarray*}
S^{(1)}_t & = & \max_{\mathcal{E}_t} E \int_0^{\infty} \pi_t P_{1,t}  (1-\tau^{(1)}) \mathcal{E}_t \exp( \log R_t ) ds %\\
% & s.t. & dR_t/R_t=-n_t dt + \sigma_{R} dB_R \\
% &  & dT_t=\varphi_{T}n_t R_t dt+\sigma_{T} dB_T
\end{eqnarray*}

% \subsubsection{Brown Energy Sector Firm}
The coal fossil fuel sector firm's profit maximization problem is given by
\begin{eqnarray*}
S^{(2)}_t & = & \max_{L_{2,t}} E \int_0^{\infty} \pi_t \{ P _{2,t}  (1-\tau^{(2)}) A_2 L_{2,t}^{\alpha_2} - w_t L_{2,t} \} ds %\\
% & s.t. & dR_t/R_t=-n_t dt + \sigma_{R} dB_R \\
% &  & dT_t=\varphi_{T}n_t R_t dt+\sigma_{T} dB_T
\end{eqnarray*}

The green energy sector firm's profit maximization problem is given by
\begin{eqnarray*}
S^{(3)}_t & = & \max_{i_{G,t}} E \int_0^{\infty} \pi_t P_{3,t} (A_G - i_{G,t}) K_{G,t} ds %\\
% & s.t. & dR_t/R_t=-n_t dt + \sigma_{R} dB_R \\
% &  & dT_t=\varphi_{T}n_t R_t dt+\sigma_{T} dB_T
\end{eqnarray*}

The firms' profit maximization problems are subject to the evolution of the state variables, as well as the relevant market conditions (competitiveness and externality internalization) and market clearing conditions and constraints.

From the household's FOC, we derive the stochastic discount factor (SDF) as
\begin{eqnarray*}
\pi_{t} & = & \exp(\int_0^t h_{J} ds)h_{C}
\end{eqnarray*}

and from the FOC for the consumption firm, we derive the input prices $P_{K,t}$, $P_{L,t} = w_t$, $P_{1,t}$, $P_{2,t}$, and $P_{3,t}$ as given in the main text. By combining the input prices, the SDF, and the FOC from the firms, we can derive expressions for the optimal input choices:
\begin{eqnarray*}
0 & = & \rho (A_C - i_{C,t})^{-1} - (1- \phi_C i_{C,t})   \\
0 & = & \rho \psi \nu_{1} \left(\frac{E_{1,t}}{C_{E,t}} \right)^{\omega} \mathcal{E}_t^{-1}(1-\tau^{(1)}) - v_{\log R}  \\
0 & = &  \rho \psi \nu_2  \left(\frac{E_{2,t}}{C_{E,t}} \right)^\omega (1 - L_{1,t})^{-1}(1-\tau^{(2)}) - \frac{\rho \left(1- \alpha - \psi \right)}{\alpha_2} L_{1,t}^{-1}  \\
0 & = &  \rho \psi \nu_3 \left(\frac{E_{3,t}}{C_{E,t}} \right)^\omega (A_G - i_{G,t})^{-1} - \left(1- \phi_G i_{G,t}\right) v_{\log K_G} 
\end{eqnarray*}

Note that the optimal fossil fuel choices from the SP's problem are given by
\begin{eqnarray*}
0 & = & \rho \psi \nu_{1} \left(\frac{E_{1,t}}{C_{E,t}} \right)^{\omega} \mathcal{E}_t^{-1} - \left( v_{\log R} -\beta_{f} R_t \left( v_{T} + \Lambda'(T) v_{\lambda} \right) \right) \\
0 & = &  \rho \psi \nu_2  \left(\frac{E_{2,t}}{C_{E,t}} \right)^\omega (1 - L_{1,t})^{-1} - \beta_f A_2 (1- L_{1,t})^{\alpha_2-1} \left( v_{T} + \Lambda'(T) v_{\lambda} \right) + \frac{\rho \left(1- \alpha - \psi \right)}{\alpha_2} L_{1,t}^{-1}  
\end{eqnarray*}

Equating the SP and decentralized FOCs for the oil production choice gives us
\begin{eqnarray*}
(1-\tau_{opt}^{(1)}) & = &  \frac{v_{\log R}}{v_{\log R} -\beta_{f} R_t \left( v_{T} + \Lambda'(T) v_{\lambda} \right)}
\end{eqnarray*}

Equating the SP and decentralized FOCs for the coal production choice gives us
\begin{eqnarray*}
(1-\tau_{opt}^{(2)})& = & \frac{\rho (1-\alpha-\psi) L_{1}^{-1}}{\rho (1-\alpha-\psi) L_{1}^{-1}-\beta_{f}\alpha_{2}A_{2}L_{2,t}^{\alpha_{2}-1}(v_{T}+\Lambda'v_{\lambda})}
\end{eqnarray*}

Now, to derive firm prices, I apply the envelope theorem to the social planner's Lagrangian. This follows the approach used by \cite{papanikolaou2011investment}, among others. Note for the final output firm, the constant returns to scale and perfect competition assumptions mean that when we plug in the input prices, we see that the firm price is given by
\begin{eqnarray*}
S^{(C)}_t & = &  \max_{C_{K,t}, C_{E,t}, L_{1,t}}   E \int_0^{\infty} \pi_{t}\{\exp(-\eta T_{t}) C_{K,t}^{\alpha} L_{1,t}^{1-\alpha-\psi} \left(\nu_1 E_{1,t}^\omega + \nu_2 E_{2,t}^\omega + \nu_3 E_{3,t}^\omega \right)^{\psi/\omega} \\
& & -P_{K,t}C_{K,t}-w_{t} L_{1,t}-P_{1,t}E_{1,t} -P_{2,t}E_{2,t} -P_{3,t}E_{3,t}\}dt \\
& = & E\int_{t}^{\infty}\pi_{s}\{\exp(-\eta T_{t}) C_{K,t}^{\alpha} L_{1,t}^{1-\alpha-\psi} \left(\nu_1 E_{1,t}^\omega + \nu_2 E_{2,t}^\omega + \nu_3 E_{3,t}^\omega \right)^{\psi/\omega} \\
& & -\exp(-\eta T_{t}) C_{K,t}^{\alpha} L_{1,t}^{1-\alpha-\psi} \left(\nu_1 E_{1,t}^\omega + \nu_2 E_{2,t}^\omega + \nu_3 E_{3,t}^\omega \right)^{\psi/\omega}\}ds = 0
\end{eqnarray*}

Likewise, the consumption capital, green energy, and brown energy firm prices are
\begin{eqnarray*}
\pi_{t} S^{(K)}_t & = & E \int_t^{\infty} \pi_s P_{K,t}(A_C - i^*_{C,t}) K_{C,t} ds \\
\pi_{t} S^{(1)}_t & = & E \int_t^{\infty} \pi_s P_{1,t}  (1-\tau^{(1)}) \mathcal{E}^*_t R_t ds \\
\pi_{t} S^{(2)}_t & = & E \int_t^{\infty} \pi_s P_{2,t}  (1-\tau^{(2)}) (1-\alpha_2) A_2 L_{2,t}^{\alpha_2} ds \\
\pi_{t} S^{(3)}_t & = & E \int_t^{\infty} \pi_s P_{3,t} (A_G - i^*_{G,t}) K_{G,t} ds
\end{eqnarray*}

Note the Lagrangian for the social planner's problem is given by 
\begin{eqnarray*}
\mathcal{L}  & = &  E_{t}\int_{t}^{\infty}\{\exp( -\rho t) h(C_t,V_t) -\pi_{t}(C_t-\exp(-\eta T_{t}) C_{K}^{\alpha}C_{E}^{\psi} C_L^{1-\alpha-\psi} +P_{K}C_{K}+P_{E}C_{E} +P_L C_L) \\
& & -P_{K,t}\pi_{t}(C_{K,t} - (A_C - i_{C,t}) \exp(\log K_{C,t}) ) \\
& & -P_{1,t}\pi_{t}(E_{1,t} - \mathcal{E}_t\exp( \log R_t) (1-\tau^{(1)}) ) \\
& & -P_{2,t}\pi_{t}(E_{2,t} - A_2 L_{2,t}^{\alpha_2}(1-\tau^{(2)}) + w_t L_{2,t} ) \\
& & -P_{3,t}\pi_{t}(E_{3,t} - (A_G - i_{G,t}) \exp(\log K_{G,t}) ) \}dt 
\end{eqnarray*}

By application of the envelope theorem we know that
\begin{eqnarray*}
\frac{\partial\mathcal{L}}{\partial \log K_C} = \frac{\partial V}{\partial \log K_C}, \quad \frac{\partial\mathcal{L}}{\partial \log K_G} = \frac{\partial V}{\partial \log K_G}, \quad \frac{\partial\mathcal{L}}{\partial \log R} =  \frac{\partial V}{\partial \log R}%, \quad \frac{\partial\mathcal{L}}{\partial \log T} = \frac{\partial V}{\partial T}
\end{eqnarray*}

Calculating derivatives of the Lagrangian and comparing I find that 
\begin{eqnarray*}
S^{(K)}_t & = &  E \int_t^{\infty} \frac{\pi_s}{\pi_{t}} P_{K,t}(A_C - i^*_{C,t}) K_{C,t} ds = \frac{\partial V}{\partial\log K_C}\frac{C}{\rho(1-\gamma)V}  \\
S^{(1)}_t & = &  E \int_t^{\infty} \frac{\pi_s}{\pi_{t}} P_{1,t}  (1-\tau^{(1)}) \mathcal{E}^*_t R_t ds = \frac{\partial V}{\partial\log R}\frac{C}{\rho(1-\gamma)V} \\
S^{(3)}_t & = &  E \int_t^{\infty} \frac{\pi_s}{\pi_{t}} P_{3,t} (A_G - i^*_{G,t}) K_{G,t} ds = \frac{\partial V}{\partial\log K_G}\frac{C}{\rho(1-\gamma)V} 
\end{eqnarray*}

While there is no direct counterpart for $S^{(2)}_t$, I directly solve the firm price PDE derived and characterized in Section \ref{assetprices} of the main text. 

Finally, returning to the SDF $\pi_t$ the expression for risk free rate is given by 
\begin{eqnarray*}
r_{f} & = & \rho-\rho(1-\gamma)\left[ \psi \log C_{E} + (1-\alpha-\psi) \log C_L -\eta T -v\right] -\lambda_{t}(\frac{V_{post}}{V_{pre}}(\frac{C_{post}}{C_{pre}})^{-1}-1)\\
& & +\alpha(\mu_{C}+i_{C}-\frac{\phi_{C}}{2}(i_{C})^{2}-\frac{1}{2}\sigma_{C}^{2})+\frac{\alpha^{2}\sigma_{C}^{2}(1-2\gamma)}{2} \\
& & -[(1-\gamma)v_{\log K_{G}}-\frac{\partial(C)_{\log K_{G}}}{C}](\mu_{G}+i_{G}-\frac{\phi_{G}}{2}(i_{G})^{2}-\frac{1}{2}\sigma_{G}^{2}) \\
& & -[(1-\gamma)v_{\log R}-\frac{\partial(C)_{\log R}}{C}](-\mathcal{E}-\frac{1}{2}\sigma_{R}^{2}) -[(1-\gamma)v_{T}-\frac{\partial(C)_{T}}{C}](E_{1,t} + E_{2,t})\beta_{T} \\
& & -[(1-\gamma)v_{\lambda}-\frac{\partial(C)_{\lambda}}{C}](\mu_{\lambda} + \Lambda'(T)(E_{1,t} + E_{2,t})\beta_{T} - \frac{1}{2}\Lambda''(T)\sigma_T^2) \\
& & -\frac{\sigma_{G}^{2}}{2}\{(1-\gamma)v_{\log K_{G}\log K_{G}}-(\frac{\partial(C)_{\log K_{G}\log K_{G}}}{C}-\frac{\partial(C)_{\log K_{G}}^{2}}{C^{2}})+[(1-\gamma)v_{\log K_{G}}-\frac{\partial(C)_{\log K_{G}}}{C}]^{2}\} \\
& & -\frac{\sigma_{R}^{2}}{2}\{(1-\gamma)v_{\log R\log R}-(\frac{\partial(C)_{\log R\log R}}{C}-\frac{\partial(C)_{\log R}^{2}}{C^{2}})+[(1-\gamma)v_{\log R}-\frac{\partial(C)_{\log R}}{C}]^{2}\} \\
& & -\frac{\sigma_{T}^{2}}{2}\{(1-\gamma)v_{TT}-(\frac{\partial(C)_{TT}}{C}-\frac{\partial(C)_{T}^{2}}{C^{2}})+[(1-\gamma)v_{T}-\frac{\partial(C)_{T}}{C}]^{2}\} \\
& & -\frac{|\Lambda'(T)\sigma_T + \sigma_{\lambda}|^{2}}{2}\{(1-\gamma)v_{\lambda\lambda}-(\frac{\partial(C)_{\lambda\lambda}}{C}-\frac{\partial(C)_{\lambda}^{2}}{C^{2}})+[(1-\gamma)v_{\lambda}-\frac{\partial(C)_{\lambda}}{C}]^{2}\} 
\end{eqnarray*}

which comes from the no arbitrage condition and application of Ito's Lemma to derive
\begin{eqnarray*}
\frac{d\pi_t}{\pi_t} = -r_{f,t} dt - \sigma_{\pi,\log K_C} dW_C - \sigma_{\pi,\log K_G} dW_G - \sigma_{\pi,\log R} dW_R - \sigma_{\pi,T} dW_T - \sigma_{\pi,\lambda} dW_{\lambda} - \Theta_{\pi} dN_t
\end{eqnarray*}

The compensations for the diffusive risks of consumption capital $(\sigma_{\pi,\log K_C})$, green capital $(\sigma_{\pi,\log K_G})$, fossil fuel reserves $(\sigma_{\pi,\log R})$, temperature $(\sigma_{\pi,T})$, and transition and stranded assets $(\sigma_{\pi,\lambda})$, as well as the compensation for the jump risk related to transition and stranded assets risk $(\Theta_{\pi})$ are as given in the main text.

%\newpage
%\clearpage

\section{Alternative Specifications}

\subsection[Alternative Preferences]{Alternative Preferences}

\subsubsection[EIS \texorpdfstring{$\neq 1$}{TEXT}]{EIS \texorpdfstring{$\neq 1$}{TEXT}}
While the model with recursive preferences and non-unitary EIS becomes quite unwieldy, I highlight the potential impact of relaxing the unitary EIS assumption by characterizing the transition jump risk premium. When $\theta = 1$, we saw that this risk price was given by
\begin{eqnarray*}
\Theta_{\pi} & = & \{1-\frac{V_{post}}{V_{pre}} (\frac{C_{post}}{ C_{pre}})^{-1} \}
\end{eqnarray*}

Without the EIS restriction, and denoting the EIS as $\psi^{-1}$, preferences are given by
\begin{eqnarray*}
h(C,V) & = & \frac{\rho}{1-\psi^{-1}}(C^{1-\psi^{-1}}((1-\xi)V)^{\frac{\psi^{-1}-\xi}{1-\xi}}-(1-\xi)V)
 \end{eqnarray*}
The stochastic discount factor given by $\pi_{t}=\exp(\int_{0}^{t}h_{V})h_{C}$, but these derivatives are now
\begin{eqnarray*}
h_{V} & = & -\rho\frac{(\xi-\psi^{-1})}{1-\psi^{-1}}C^{1-\psi^{-1}}((1-\xi)V)^{\frac{\psi^{-1}-1}{1-\xi}}-\rho\frac{(1-\xi)}{1-\psi^{-1}} \\
h_{C} & = & \rho C^{-1}C^{1-\psi^{-1}}((1-\xi)V)^{\frac{\psi^{-1}-\xi}{1-\xi}}
 \end{eqnarray*}

Therefore, the climate policy jump risk price would therefore be given by 
\begin{eqnarray*}
\Theta_{\pi}' & = & \{1-(\frac{V_{post}}{V_{pre}})^{\frac{\psi^{-1}-\xi}{1-\xi}} (\frac{C_{post}}{C_{pre}})^{-\psi^{-1}}\}
\end{eqnarray*}
 
Note that for the model simulations results the climate-linked transition shock leads to reduced consumption and a more negative continuation value and so $\frac{C_{post}}{C_{pre}}<1$ and $\frac{V_{post}}{V_{pre}}>1$. As a result, holding all else constant, when $\psi^{-1}>1$, we see that 
\begin{eqnarray*}
(\frac{C_{post}}{C_{pre}})^{-\psi^{-1}} &<& (\frac{C_{post}}{C_{pre}})^{-1} \\
(\frac{V_{post}}{V_{pre}})^{\frac{\psi^{-1}-\xi}{1-\xi}}	&<&(\frac{V_{post}}{V_{pre}})
 \end{eqnarray*}
On the other hand, holding all else constant, when $\psi^{-1}<1$, we see that 
\begin{eqnarray*}
(\frac{C_{post}}{C_{pre}})^{-\psi^{-1}}	&>&(\frac{C_{post}}{C_{post}})^{-1} \\
(\frac{V_{post}}{V_{pre}})^{\frac{\psi^{-1}-\xi}{1-\xi}}	&>&(\frac{V_{post}}{V_{pre}})
 \end{eqnarray*}
 
 Therefore, the result of relaxing the EIS from being unitary is that when $\psi^{-1}>1$, holding all else constant, the transition jump risk premium is diminished, i.e., $|\Theta_{\pi}'|  <  |\Theta_{\pi}|$, whereas when $\psi^{-1}<1$, all else constant, the green transition jump risk premium is amplified, i.e., $|\Theta_{\pi}'|  >  |\Theta_{\pi}|$. This comparative static or partial equilibrium analysis highlights the role of the EIS. Consistent with the asset pricing literate, an EIS greater than one leads to increased concern about the resolution of uncertainty and amplifies the magnitude of the risk price of the climate-linked transition jump. However, an EIS less than one leads to decreased concern about the resolution of uncertainty and diminishes the magnitude of the risk price of the green transition jump. Such an analysis highlights the value of using asset prices in analyzing the impact of climate change and transition risk, and provides insight for the expected macroeconomic outcomes, where a larger EIS should amplify the run on oil we would expect and a smaller EIS should diminish this effect.

\subsubsection[Log Utility]{Log Utility}

I conjecture and verify that the value functions for the technology and taxation shock cases, pre- and post-transition shock are of the form
\begin{eqnarray*}
V_{pre} & = & \alpha \log K_C + v(\log K_G, \log R, T) + c_{pre} \\
V_{post} & = & -\eta T + \alpha \log K_C + \psi \log K_G + c_{post,i}
\end{eqnarray*}
where the value function coefficients $c_{pre}$ and $c_{post,i}$, $i \in \{tech, tax\}$ are given by
\begin{eqnarray*}
c_{pre} & = & \alpha \log(A_C-i_{C}) + \frac{\alpha}{\rho} (\mu_{C}+i_{C}-\frac{\phi}{2}i_{C}^{2}) \\
c_{post,tech} & = & \alpha \log(A_C-i_{C}) + \psi \log(A_G - i_G) \\
& & + \frac{\alpha}{\rho} (\mu_{C}+i_{C}-\frac{\phi}{2}i_{C}^{2})  + \frac{\psi}{\rho} (\mu_{G}+i_{G}-\frac{\phi}{2}i_{G}^{2}) \\
c_{post,tax} & = & \alpha \log(A_C-i_{C}) + \psi \log(A_G - i_G) + \frac{\psi}{\omega} \log \nu_1  \\
& & + \frac{\alpha}{\rho} (\mu_{C}+i_{C}-\frac{\phi}{2}i_{C}^{2}) + \frac{\psi}{\rho} (\mu_{G}+i_{G}-\frac{\phi}{2}i_{G}^{2})
\end{eqnarray*}

Plugging in terms, I arrive at a simplified pre-transition HJB equation of the form
\begin{eqnarray*}
0 & = & \rho \psi \log C_E + \rho (1-\alpha-\psi)\log C_L - (\rho + \lambda_t) \eta T + \lambda_{t} [ \psi \log K_G + c_{post,i} ] - (\rho + \lambda_t) v \\
 & & +(\mu_{G}+i_{G}-\frac{\phi_G}{2}i_{G}^{2} - \frac{1}{2}\sigma_{G}^{2})v_{\log K_G} +(-\mathcal{E} - \frac{1}{2}\sigma_{R}^{2})v_{\log R} + \beta_{T}\left( E_{1,t} + E_{2,t}\right)v_{T} \\
 & & +\frac{1}{2}\sigma_{G}^{2} v_{\log K_G, \log K_G} +\frac{1}{2}\sigma_{R}^{2} v_{\log R, \log R}+\frac{1}{2}\sigma_{T}^{2}v_{TT}
\end{eqnarray*}

There key difference is that the value function is now additively separable rather than  multiplicatively separable. Importantly, the main drivers previously highlighted, the marginal value of reserves, the marginal cost of forgoing labor for final output production, and the marginal cost of climate change, are still central components for the optimal choice of extraction. These results highlight that the recursive utility specification introduces an additional amplification effect related to the continuation value and forward-looking concerns about the resolution of uncertainty that the log utility setting does not have. The same dynamic run impacts of the risk of the climate-linked transition shock that strands fossil fuel production are still in effect, though they are likely more muted quantitatively.

\subsection{Constant Policy Arrival Rate}

With a constant policy arrival rate, the post-transition results match the baseline setting. For the pre-transition problem, I guess and verify the simplified value function solution
\begin{eqnarray*}
V_{pre} & = & \frac{\exp((1-\gamma)\left[ -\eta T + \alpha \log K + v(\log K_{G,t}, \log R_t) + c_{pre} \right] )}{1-\gamma} 
\end{eqnarray*}

The optimal FOC for oil and coal production simplify to
% \begin{eqnarray*}
% 0 & = & \frac{\rho (1 - \alpha)\nu_{2}}{v_{\log R} + \beta_{f} R_t \eta } \left(\frac{B_t}{C_{E,t}} \right)^{\omega}- \mathcal{E}_t %\\
% \end{eqnarray*}
\begin{eqnarray*}
0 & = & \rho \psi \nu_{1} \left(\frac{E_{1,t}}{C_{E,t}} \right)^{\omega} \mathcal{E}_t^{-1} - \left( v_{\log R} -\beta_{f} R_t \eta \right) \\
0 & = &  \rho \psi \nu_2  \left(\frac{E_{2,t}}{C_{E,t}} \right)^\omega (1 - L_{1,t})^{-1} - \beta_f A_2 (1- L_{1,t})^{\alpha_2-1} \eta + \frac{\rho \left(1- \alpha - \psi \right)}{\alpha_2} L_{1,t}^{-1} 
\end{eqnarray*}

These results highlight a clear distinction for the constant arrival rate. The climate effect is constant and $v_{\log R}$ is independent of temperature. As a result, there is only a level shift up in the optimal choice of emissions. This result matches previous analysis related to the Green Paradox, and highlights the endogenous feedback effect and dynamic transition risk mechanism highlighted in my main analysis that is not present in alternative settings, which use a constant policy arrival rate structure as in this special case.

\subsection{Reserves Exploration} \label{sec:}

Given the already complicated nonlinear optimal control outcomes without exploration, I turn to a simplification of the model to provide an intuitive analysis of the impact of allowing for exploration in the model. In particular, I consider the limiting case where $\omega = 0$, leading to Cobb-Douglas substitutability of energy inputs. In addition, I alter the emissions input production and reserves evolution such that 
\begin{eqnarray*}
E_{1,t} & = & (\mathcal{E}_t - i_{R,t})  \exp(r_t) \\
d \log R_t & = & (\mathcal{E}_t + \Gamma i_{R,t}^{\phi} - \frac{1}{2} \sigma_R^2) dt + \sigma_R dW_R
\end{eqnarray*} 
In this setting, exploration of new reserves is costly, captured by the adjustment cost parameters $\Gamma, \eta$, and requires sacrificing part of the fossil fuel extracted. The HJB equations in each jump state are very similar to the baseline, but the first order conditions now include an additional equation for exploration which can be written as
\begin{eqnarray*}
i_R & = & \left( \frac{\Gamma \phi  v_{\log R} }{v_{\log R} -\beta_{f} R_t \left( v_{T} + \Lambda'(T) v_{\lambda} \right)} \right)^{1/(1-\phi)}
\end{eqnarray*}

This choice of exploration feeds into the optimal choice of emissions, now given by 
\begin{eqnarray*}
0 & = & \frac{\rho \psi \nu_{1}}{v_{\log R} -\beta_{f} R_t \left( v_{T} + \Lambda'(T) v_{\lambda} \right)} \left(\frac{E_{1,t}}{C_{E,t}} \right)^{\omega} + i_{R,t}- \mathcal{E}_t %\\
\end{eqnarray*}

In this case, we still explicitly see that extraction will increase dynamically as in the baseline model as the expectation of future transition drives down the marginal value of reserves and the marginal cost of climate change. For reasonable parameter values ($\Gamma>0, 1 >\eta > 0$), the optimal choice for exploration is positive. Given the diminishing value of holding reserves, exploration will be a low value that decreases over time but remains positive. So, allowing for exploration in the model simply boosts the potential for the planner to run on fossil fuels such as u=oil because of the boost that comes from access to new potential reserves. Therefore, the macroeconomic and asset pricing results should be qualitatively consistent with or without exploration, with the incentive to run potentially even higher.

\subsection[Imperfect Oil Sector Competition]{Imperfect Oil Sector Competition}

Another important assumption to explore is that of a perfectly competitive oil sector. The existence of OPEC and numerous state-owned oil firms suggests that a model of imperfect competition or market power in the oil sector may more accurately approximate the real world. A simplified approximation of imperfect competition can be derived by assuming a symmetrical oligopolist in the oil sector where firms account for pricing impacts that they have. Here I also abstract from the coal sector for simplicity. In particular, I assume J firms use a common pool of reserves, they are homogeneous in production technology, and they internalize their impact on global reserves to ensure a symmetric solution. The evolution of oil reserves and temperature are adjusted as follows:
\begin{eqnarray*}
d \log R_t & = & (\sum_{j=1}^J -\mathcal{E}_{1,j,t} )dt - \frac{1}{2}\sigma_R^2 dt + \sigma_R  dW_R  \\
dT_t & = & \beta_T (\sum_{j=1}^J \mathcal{E}_{1,j,t} ) \exp(r_t)  dt + \sigma_T dW_T
\end{eqnarray*}

For this setting, I consider a constrained planner's problem  where oil firms optimize as price setters rather than as socially optimal price takers but the impact on climate of emissions production is internalized in this planner's setting. As with reserve exploration, I consider a simplified model with Cobb-Douglas substitutability of energy inputs to provide an intuitive analysis of the impact I'm interested in. To derive the optimal extraction choice used in this setting, consider the oil firm's problem as a price setter that internalizes climate impacts via an optimal carbon tax. The price of the oil input is given by
\begin{eqnarray*}
P_{1,t} & = & \psi \nu_{1} C_t (\frac{\sum_{j=1}^J \mathcal{E}_{1,j,t}}{C_{E,t}})^{\omega} \left(\sum_{j=1}^J \mathcal{E}_{1,j,t}\right)^{-1}
\end{eqnarray*}

and we use this to write out the profit-maximizing problem for the energy firm price
\begin{eqnarray*}
S^{(1)}_{i,t} & = & \max_{\mathcal{E}_{1,i,t}}E\int\exp(-\rho t)\rho(1-\gamma)V \psi \nu_1[\nu_{1}E_{3,t}^{\omega}+\nu_{1}(\sum_{j}^J\mathcal{E}_{1,j,t}R_t)^{\omega}]^{-1}(\sum_{j}\mathcal{E}_{1,j,t})^{\omega-1}(\mathcal{E}_{1,i,t})R_t^{\omega}dt \\
\text{ s.t. } &  &  d \log R_t = (\sum_{j=1}^J -\mathcal{E}_{1,j,t} )dt - \frac{1}{2}\sigma_R^2 dt + \sigma_R  dW_R\\
 &  & dT_t = \beta_T (\sum_{j=1}^J \mathcal{E}_{1,j,t} ) R_t  dt + \sigma_T dW_T
\end{eqnarray*}

Solving the first order conditions for this problem, and applying the SDF and Lagrangian multiplier definitions as before in the decentralized economy setting, we find that the optimal extraction choice in this setting is given by
\begin{eqnarray*}
0 & = & \frac{\rho \psi \nu_{1}}{v_{\log R} -\beta_{f} R_t \left( v_{T} + \Lambda'(T) v_{\lambda} \right)} \left(\frac{E_{1,t}}{C_{E,t}} \right)^{\omega} \left[J^{\omega-1} + J^{\omega-2}(\omega-1) - \nu_2 \left(\frac{E_{1,t}}{C_{E,t}} \right)^{\omega} J^{2(\omega-1)} \omega \right] - \mathcal{E}_t %\\
\end{eqnarray*}

The expression for the optimal choice of extraction is similar in functional form regardless of whether we are in a perfectly competitive or oligopolistic fossil fuel sector. Therefore, as was the case before, the green transition risk has the potential to generate a strong non-linearity in the value of holding reserves by firms and of temperature, thus driving the potential for a run on fossil fuel by the oil firms. An important distinction between the optimal extraction in this setting and the competitive setting is the scaling factor
\begin{eqnarray*}
\left[J^{\omega-1} + J^{\omega-2}(\omega-1) - \nu_2 \left(\frac{E_{1,t}}{C_{E,t}} \right)^{\omega} J^{2(\omega-1)} \omega \right]
\end{eqnarray*} 
that is a function of the number of firms in the oligopoly. Holding the value function constant, the optimal extraction by each firm decreases as the number of firms goes to infinity. However, there are two other critical effects to keep in mind. As aggregate demand scales extraction by the number of firms, in the limit production would actually go to $R_t$, the maximum amount of extraction possible, assuming the value function is held constant. The decrease in production per firm is less than the aggregation scaling, and so increased competition actually amplifies the run through this channel. The other impact of increasing the number of firms is the potential impact it could have on the value function and the marginal value of reserves and temperature, which this simple comparative static ignored. This effect, interacted with the impact of uncertain climate policy, can serve to either amplify or dampen the run on oil effect generated by climate policy.  A numerical solution is required to determine the full quantitative role that competition has in this model.

\subsection{Partial Internalization}

In the simple transition shock cases where the climate externality is only partially internalized by the planner, the post-transition results match the baseline setting. For the pre-transition problem, I guess and verify the simplified value function solution
\begin{eqnarray*}
V_{pre} & = & \frac{\exp((1-\gamma)\left[ -\eta T + \alpha \log K + v(\log K_{G,t}, \log R_t) + c_{pre} \right] )}{1-\gamma} 
\end{eqnarray*}

The optimal FOC for oil and coal production are slightly modified such that
\begin{eqnarray*}
0 & = & \rho \psi \nu_{1} \left(\frac{E_{1,t}}{C_{E,t}} \right)^{\omega} \mathcal{E}_t^{-1} - \left( v_{\log R} - \chi \beta_{f} R_t \eta \right) \\
0 & = &  \rho \psi \nu_2  \left(\frac{E_{2,t}}{C_{E,t}} \right)^\omega (1 - L_{1,t})^{-1} -  \chi \beta_f A_2 (1- L_{1,t})^{\alpha_2-1} \eta + \frac{\rho \left(1- \alpha - \psi \right)}{\alpha_2} L_{1,t}^{-1} 
\end{eqnarray*}

where $\chi \in [0,1]$ determines the fraction of the climate impact that is internalized and therefore ``endogenous,'' $\chi \beta_T \mathcal{E}_t R_t$, and the remaining fraction is not internalized and therefore ``exogenous'' to the decision maker, $(1-\chi) \beta_T \mathcal{E}_t R_t$. The critical difference is that the planner now only partially accounts for the impact of the impact of their production choices on climate change, which impacts to key mechanisms in our transition risk setting. First, concerns about how generating emissions influences climate change damages, which is the standard concern internalized by social planner's in socially optimal climate economic settings, are minimized. Second, the planner's concerns about their impact on the likelihood of a transition shock occurring are also diminished, though they still know the true rate of climate change taking place. As a result, both incentives to limit the ``run on fossil fuel'' response, concerns about climate damages and trying to postpone the transition shock, are attenuated. Therefore, the settings with relatively lower expected costs of a transition shock, such as a technology shock, see an amplification of emissions production and the ``run'' response. Settings with relatively higher expected costs of a transition shock, such as a taxation shock, also see an amplification of emissions production due to the limited perceived ability delay the transition shock and so the ``reverse run'' response in these settings is either diminished or potentially overturned to a ``run'' response.\footnote{I confirm this intuition in unreported computational results for select partial internalization scenarios.}

 \section{Numerical Solutions}

\subsection{Model Parameters} \label{Params}

In what follows, I detail the calibration strategy for the parameters used in the numerical solution of the model. I combine external parameter estimates and empirically measured values, both from the literature on climate finance and climate economics, as well as parameter values that allow a special case of my model to match empirical moments. This calibration strategy for the baseline model used in my analysis allows me to provide quantitative results that are justified by observable empirical outcomes and estimates for my analysis.

\subsubsection{External Estimates}

The production parameters are chosen to match estimates in the literature. Estimates from \cite{hassler2021directed} find that the Cobb-Douglas specification for final output production is consistent with estimates based on longer-term frequency (i.e., a decade or more) due to R\&D investments in directed change change to limit the fraction of income devoted to energy expenses. While shorter term estimates from \cite{hassler2021directed} are lower than the Cobb-Douglas value, assuming a value slightly larger allows us to explore the tax and technology transition risk scenarios in a tractable setting that would otherwise be very computationally costly. This specification is consistent with an implicit assumption that sufficient unmodeled directed technical change occurs in the background to allow for this level of substitution. The energy input demand share is set to $\psi = 0.05$, which equals the value used in \cite{hassler2021directed} and matches energy's share of income for a CD specification such as is used in our specification. The \cite{BP:2020} estimates of the empirical split for energy production between dirty and clean sources are approximately $80\%$ dirty to $20\%$ clean, and so I set pre-transition energy input demand share values at $\nu_1 + \nu_2 = 0.80$ and $\nu_3 = 0.20$. The specific values for $\nu_1$ and $\nu_2$ are based on data for global primary energy consumption by source from \cite{energy2023statistical} and \cite{vaclav2017energy} (with major processing by Our World in Data)\footnote{See \url{https://ourworldindata.org/energy-mix} for further information.}. The elasticity of substitution between energy inputs is set to $\omega = 0.5$. There is a wide range of values used in the literature for the elasticity of substitution $\omega$, e.g., \cite{golosov_optimal_2014} and \cite{acemoglu2012environment} explore values between $-0.058$ and $0.9$. Assuming a positive value allows for analysis of tax and technology transition risk scenarios that would be infeasible otherwise, and allows the model to produce outcomes that are consistent with the data for macroeconomic quantities and asset prices. The fossil fuel reserves volatility $\sigma_R = 0.034$ is chosen to match the empirically measured annual changes in the time series of oil reserves from \cite{BP:2020}, matching the value used in \cite{barnett2020pricing}.

The parameters relating to the climate part of the model are also all chosen based on external estimates. The temperature volatility $\sigma_T = 0.1$ is chosen to match the empirically measured monthly changes in the time series of global temperature from the NASA-GISS database and is consistent with a monthly counterpart of the value estimated by \cite{hambel2015optimal}. The damage function parameter $ \eta = 0.02$ is chosen to be consistent with the damage functions used by \cite{golosov_optimal_2014} and \cite{Nordhaus:2017} over the range of one to five degrees Celsius above the global mean temperature preindustrial value. The climate sensitivity parameter $\beta_T = 0.00186$ matches the mean value from estimates by \cite{MacDougallSwartKnutti:2017}, the mean value calculated by \cite{BarnettBrockHansen:2021} using pulse responses from \cite{Geoffroy:2013} and \cite{Joosetal:2013}, and is within the range of estimates provided by \cite{matthews2009proportionality}, \cite{matthews2012cumulative}, and \cite{macdougall2015origin}. The initial value for the atmospheric temperature anomaly relative to the pre-industrial value $T_0 = 1.2$ comes from the most recent IPCC reports, and matches the recent value from the NASA-GISS database.

There is little in terms of existing estimates or previously specified counterparts in the literature to use for setting the climate policy shock arrival rate. One exception to this is the work by \cite{moore2022determinants}, who explore the likelihood of climate policy and transition pathways across various socioeconomic and climate change scenarios. Therefore, I use the projected outcome in the ``Technical Challenges'' Policy Trajectory from \cite{moore2022determinants} to calibrate the values of $\psi_0, \psi_1$, $\psi_2$, and $\underline{T}$. To do this, I first set $\underline{T} = 1.25$, $\psi_2 = 2$, and assume annual emissions for current and future years to be constant at today's value of $10$ GtC. Conditional on these assumptions, I estimate the values of $\psi_0 = 0.0719$ and $\psi_1 = 0.7939$ that lead to a $50\%$ probability of a jump occurring within 60 years from the beginning of the model simulations and a $99\%$ probability of a jump occurring within 100 years from the beginning of the model simulations.

\begin{figure}[!phtb]
%    \centering
\caption{Climate-Linked Transition Risk Trends and Trajectories} \label{fig:transition_risk}
    
\begin{subfigure}{0.48\linewidth}
\includegraphics[width=\linewidth]{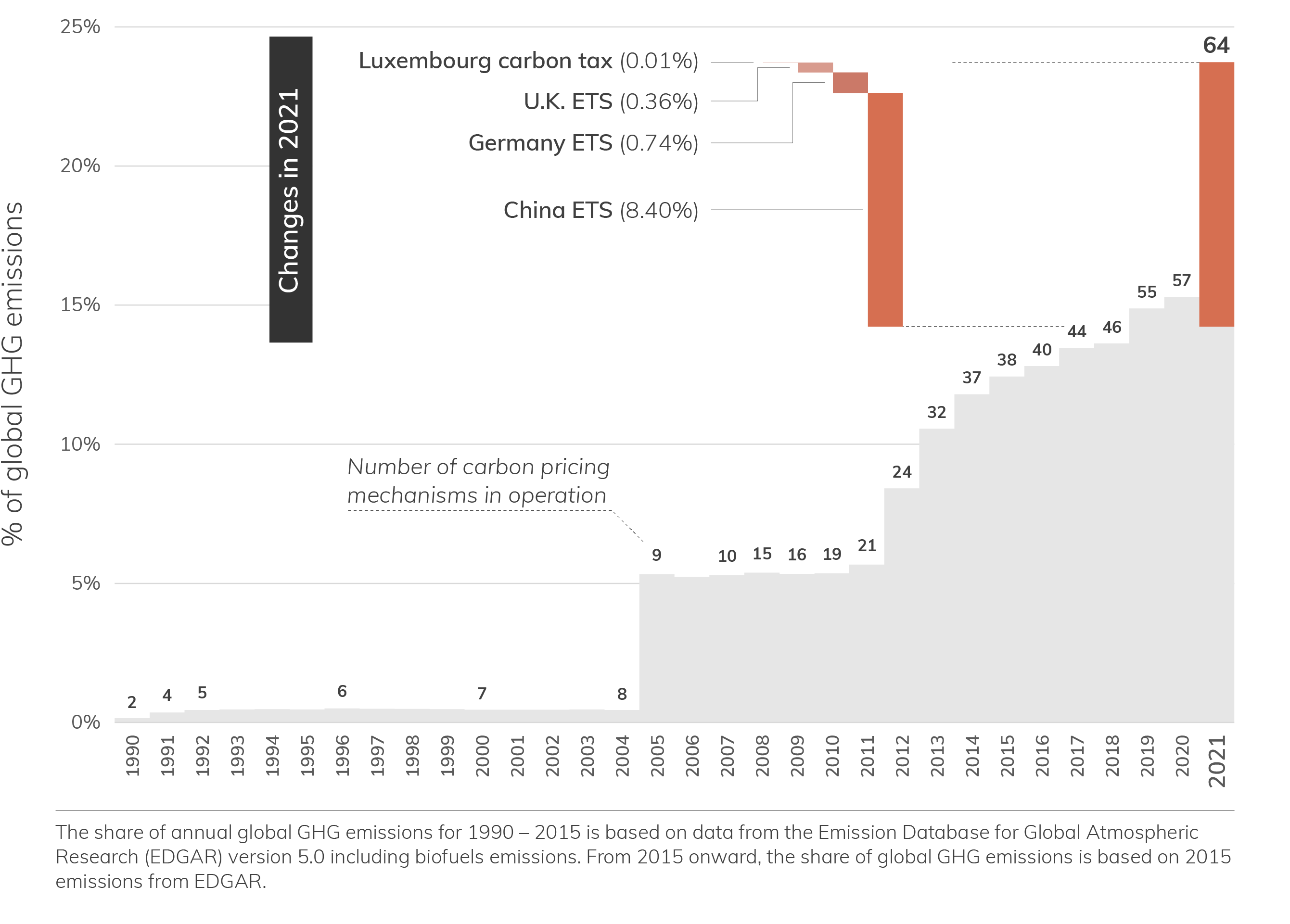}
\caption{Source: \cite{santikarn2021state}}\label{transition_risk1}  
\end{subfigure}
    \hfill
\begin{subfigure}{0.48\linewidth}
\includegraphics[width=\linewidth]{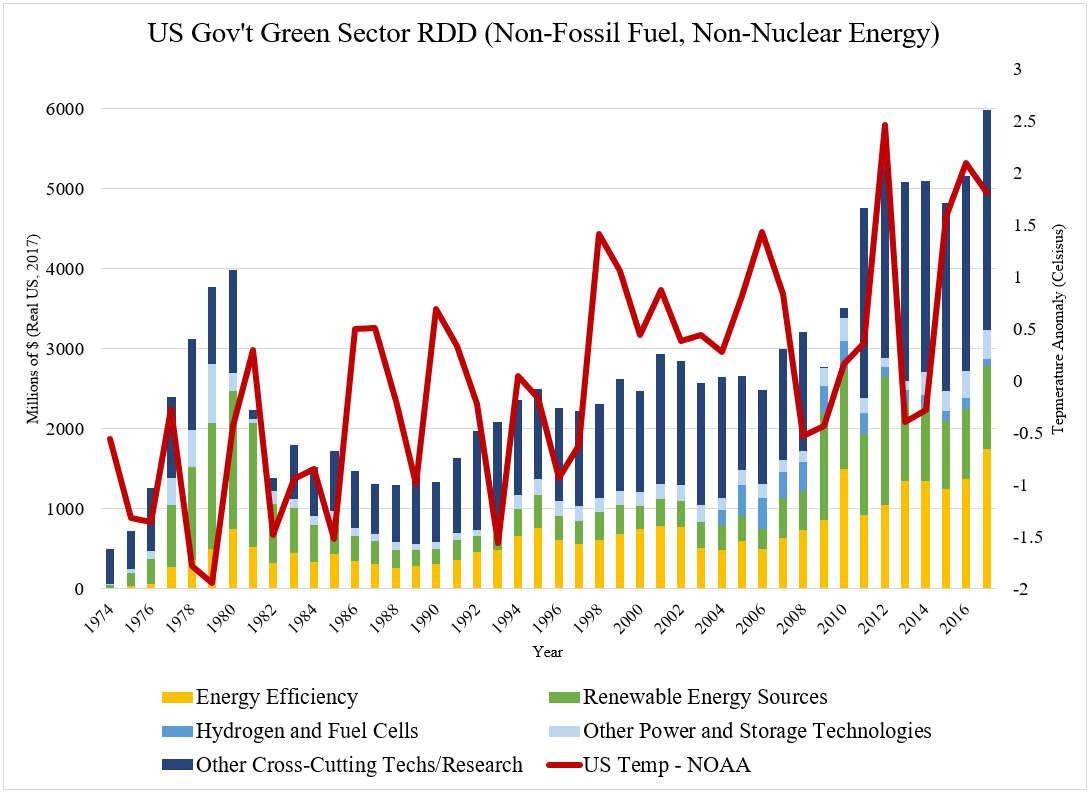}
\caption{Source: IEA and NOAA}\label{transition_risk2}  
\end{subfigure}

\medskip

\begin{center}

\begin{subfigure}{0.55\linewidth}
\includegraphics[width=\linewidth]{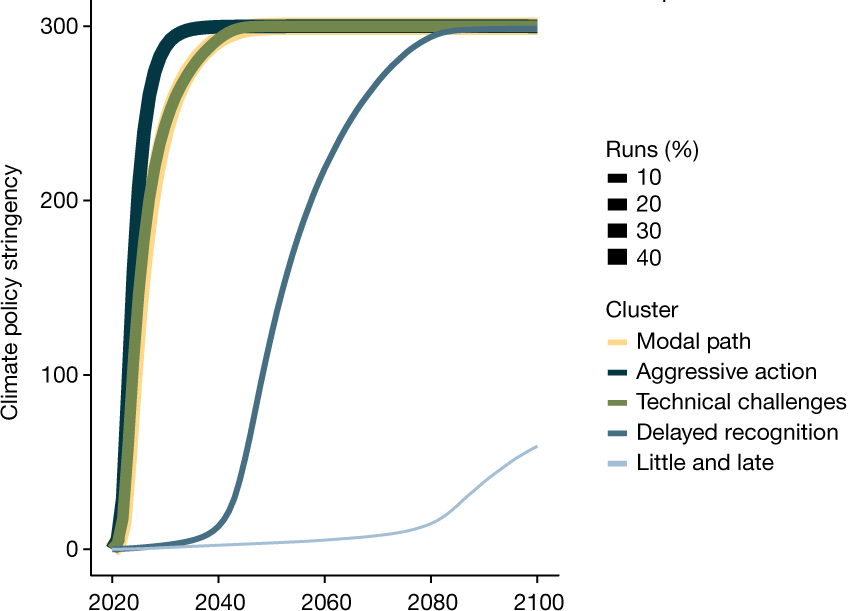}
\caption{Source: \cite{moore2022determinants}}\label{fig:transition_risk3}     
\end{subfigure}

\end{center}

\vspace{0.25cm}

\begin{footnotesize}
Figure \ref{fig:transition_risk} highlights trends in policy and technology changes over time related to climate-linked transition risk. The top left panel shows the increase time-series trend in the share of global emissions being priced estimated by \cite{santikarn2021state}. The top right panel shows the increase in Green RD\&D (right panel) estimated by the IEA and NOAA, highlighting the potentially increasing probability of significant green innovation. The bottom panel shows the left panel of Figure 3 from \cite{moore2022determinants}, which gives the future policy trajectories from 100,000 Monte Carlo runs of the coupled climate–social model, clustered into 5 clusters using k-means clustering. The line thickness corresponds to the size of the cluster.
\end{footnotesize}

\end{figure}

The partial internalization parameter scales either the carbon tax (for the hybrid technology shock case) or the marginal cost of climate change (for the partial internalization case) in the HJB equation and first-order conditions and is critical for determining the importance of climate, transition, and stranded assets risks. There is no precise direct evidence of the level of internalization of the climate externality and estimating such values is beyond the scope of this paper. However, anecdotal external evidence provides a reasonable estimate of the value of the parameter $\chi$ in my model. Figure \ref{fig:transition_risk} shows the number of carbon pricing policies in place and the fraction of emissions taxed globally as well as a map of where carbon prices exist or are planned. The increasing fraction of carbon emissions being taxed increased from $0\%$ in 1990 to nearly $25\%$ in 2021. In my analysis, the partial internalization parameter values are set to provide a stylized demonstration of how results change across different scenarios of climate concern. The case of $\chi = 1$ is the valued used for the social planner's problem setting and shows socially optimal responses when fully internalizing both the climate externality and climate policy risk. For the hybrid  technology shock case I use $\chi = 0.25$ to highlight how a carbon tax that only partially accounts for transition risk influences the optimal production choices and the prices resulting from these choices. %For the partial internalization setting, I consider an intermediate value of $\chi = 0.5$ to demonstrate the qualitative implications of this speification. 

% %%%%%%%%%%%%%%%%%%%%%%%%%%%%%%%%%%%%%%%%%%%
% %%%%%%%%%%%%%%%%%%%%%%%%%%%%%%%%%%%%%%%%%%%

\subsubsection{Direct Calibration}

In order to calibrate the model in a quantitatively meaningful way, the special case of my model that I consider is the no-climate, no-transition risk version with Cobb-Douglas substitution across energy inputs. This setting has an analytical solution:
\begin{proposition}
In the no-climate, no transition risk version of the model with Cobb-Douglas energy substitution, the Planner's value function and first order conditions are given by
\begin{eqnarray*}
V(\hat{K}_C, \hat{R}, \hat{K}_G) & = & \frac{\exp\left( (1-\gamma) (c_0 + \alpha \hat{K}_C +\psi \kappa_1 \hat{R} + \psi \kappa_3 \hat{K}_G) \right)}{1-\gamma} \\
c_0	& = & \log(A_C-i_{C})^{\alpha}(\mathcal{E})^{\psi \kappa_1} (A_2(1-L_1)^{\alpha_2})^{\psi \kappa_2} (A_G - i_G)^{\psi \kappa_3} L_{1}^{1-\alpha-\psi} \\
& & + \frac{\alpha}{\rho}[\mu_{C}+i_{C}-\frac{\phi_C}{2}(i_{C})^{2} - \frac{1}{2}\sigma_{C}^{2}]+\frac{(1-\gamma)\sigma_{C}^{2}\alpha^2}{2} \\
& & + \frac{\psi \kappa_3}{\rho}[\mu_{G}+i_{G}-\frac{\phi_G}{2}(i_{G})^{2} - \frac{1}{2}\sigma_{G}^{2}]+\frac{(1-\gamma)\sigma_{G}^{2}(\psi \kappa_3)^2}{2} \\
& & + \frac{\psi \kappa_1}{\rho}[-\mathcal{E} - \frac{1}{2}\sigma_R^2] + \frac{(1-\gamma)\sigma_{r}^{2} (\psi \kappa_1)^2}{2}  \\
i_{C} & = & \frac{(1+\phi_C A_C)-\sqrt{(1+\phi_C A_C)^{2}-4\phi_C(A_C-\rho)}}{2\phi_C} \\
i_{G} & = & \frac{(1+\phi_G A_G)-\sqrt{(1+\phi_G A_G)^{2}-4\phi_G(A_G-\rho \kappa_3)}}{2\phi_G} \\
L_1 & = & \frac{(1-\alpha-\psi)}{\rho\psi\kappa_{2}\alpha_{2}+(1-\alpha-\psi)} \\
\mathcal{E}_{t}	& = & \rho 
\end{eqnarray*}

\end{proposition}

From this solution, we see that the optimal emissions rate is uniquely determined by the subjective discount rate as in the canonical \cite{hotelling1931economics} solution. \cite{friedlingstein2023global}, for the Global Carbon Project, estimates annual global emissions near or above $10-11$ GtC, similar to values reported by \cite{Figueresetal:2018}. To be consistent with this empirical target, I set the initial value of fossil fuel reserves to be $R_0 = 850$ GtC, which is well within the the range of values for existing and potential recoverable reserves from the Energy Information Agency, \cite{BP:2020}, \cite{Rogner:1997}, and \cite{mcglade2015geographical}, and set the subjective discount rate at a commonly used value of $\rho = 0.01$.\footnote{I note that this value of $\rho$ is consistent with the values used in the macroeconomics, production-based asset pricing, and climate-finance/climate-economics literature, though it substantially exceeds the annual value of $0.1\%$ used by \cite{stern_stern_2007}, who advocates for much smaller value of the subjective discount rate when quantifying the social costs of climate change in his analysis.}

%%%%%%%%%%%%%%%%%%%%%%%%%%%%%%%%%%%%%%%%%%%%%%%%%%%%%%%%%%%%%%%%%%%%%%%%%%%%%%%
%%%%%%%%%%%%%%%%%%%%%%%%%%%%%%%%%%%%%%%%%%%%%%%%%%%%%%%%%%%%%%%%%%%%%%%%%%%%%%%
%%%%%%%%%%%%%%%%%%%%%%%%%%%%%%%%%%%%%%%%%%%%%%%%%%%%%%%%%%%%%%%%%%%%%%%%%%%%%%%
%%%%%%%%%%%%%%%%%%%%%%%%%%%%%%%%%%%%%%%%%%%%%%%%%%%%%%%%%%%%%%%%%%%%%%%%%%%%%%%
Second, given the value of $\rho$ and a choice of the productivity parameter $A_C$, we can estimate the capital adjustment cost parameters $\mu_C$ and $\phi_C$ conditional on two macroeconomic moments. For the productivity parameter $A_C$, I choose a value that matches the values used by \cite{pindyck_climate_2013} and \cite{barnett2020pricing}. For the two empirical moments, we choose an economic growth rate of $2\%$ and a Tobin's q of $2.5$, which are consistent with the empirical values for aggregate world GDP and capital from the World Bank and BEA databases. I then use the expressions for investment $i_C$, economic growth $dK_C/K_C$, and Tobin's q from the model to estimate the values $\mu_C = -0.043$ and $\phi_C = 6.667$, which allow the model to match the specified empirical moments. For simplicity, I then assume that the productivity and adjustment cost parameters for green capital match the values used for the consumption capital. The initial values for consumption and green capital are then backed out using the assumed productivity parameters, annual global GDP from the World Bank database, and annual green energy production from the IEA and EIA databases.

% \footnote{The empirical moments and model parameter values are consistent ...} 
%%%%%%%%%%%%%%%%%%%%%%%%%%%%%%%%%%%%%%%%%%%%%%%%%%%%%%%%%%%%%%%%%%%%%%%%%%%%%%%
%%%%%%%%%%%%%%%%%%%%%%%%%%%%%%%%%%%%%%%%%%%%%%%%%%%%%%%%%%%%%%%%%%%%%%%%%%%%%%%
%%%%%%%%%%%%%%%%%%%%%%%%%%%%%%%%%%%%%%%%%%%%%%%%%%%%%%%%%%%%%%%%%%%%%%%%%%%%%%%
%%%%%%%%%%%%%%%%%%%%%%%%%%%%%%%%%%%%%%%%%%%%%%%%%%%%%%%%%%%%%%%%%%%%%%%%%%%%%%%

I have chosen up-front to set the elasticity of intertemporal substitution to be $\theta=1$, a value that lies within the range of the various empirical estimates of the EIS found in the literature and that is consistent with studies such as \cite{vissing2002limited} who estimate the value to be near or equal to 1. As a result, the only remaining parameters to set are the risk aversion $\gamma$ and the volatility associated with the capital stocks $\sigma = \sigma_C = \sigma_G$. These parameters are set so that the model-derived values of the market price of risk and the risk-free rate for the special no-climate, no-transition risk version of the model with Cobb-Douglas aggregation over the energy sectors to match the empirical estimates of these asset-pricing moments. A second proposition shows that the model targets of interest are given as follows:
\begin{proposition}
In the no-climate, no transition risk version of the model with Cobb-Douglas energy substitution, the stochastic discount factor $\pi_t$ for the Planner's problem is given by the following dynamic evolution process
\begin{eqnarray*}
\frac{d\pi_{t}}{\pi_{t}} & = & -r_f dt - \sigma_{\pi} dW
\end{eqnarray*}
where the risk-free rate $r_f$ and the market price of risk $|\sigma_{\pi}|$ are given as
\begin{eqnarray*}
r_{f} & = & \rho + \alpha[\mu_{C}+i_{C}-\frac{\phi_C}{2}(i_{C})^{2}  - \frac{1}{2}\sigma_C^2]  + \frac{ \left(1 - 2 \gamma \right)\sigma_{C}^{2}\alpha^2}{2} \\
& & + \psi \kappa_3 [\mu_{G}+i_{G}-\frac{\phi_G}{2}(i_{G})^{2} - \frac{1}{2}\sigma_G^2]  + \frac{ \left(1 - 2 \gamma \right)\sigma_{G}^{2}(\psi \kappa_3)^2}{2}  \\
& & + \psi \kappa_1 [-\mathcal{E} - \frac{1}{2}\sigma_R^2] + \frac{ \left(1 - 2 \gamma \right)\sigma_{r}^{2}(\psi \kappa_1)^2}{2}\\
|\sigma_{\pi}| & = & \gamma \sqrt{(\alpha\sigma_{C})^2 + (\psi \kappa_1 \sigma_{R})^2 + (\psi \kappa_3 \sigma_{G})^2}
\end{eqnarray*}

\end{proposition}

For the empirical targets of the market price of risk and risk free rate, call them $MPR_{data}$ and $r_{f,data}$, I choose values at the annual frequency of $MPR_{data} = 0.4$ and $r_{f,data} = 0.01$, which are in line with the average values over the time series from Ken French's data library and other empirical estimates. Finally, I solve for values of $\gamma$ and $\sigma_C$ to satisfy $MPR_{model} = MPR_{data}$ and $ r_{f,model} = r_{f,data}$. As with the adjustments costs for the two types of capital, I assume the volatilities for the two types of capital are equivalent so that $\sigma_C = \sigma_G = \sigma$. The calibrated value of $\sigma$ from this calibration exercise exceeds the time series volatilities of GDP or capital stock implied by data from the World Bank, for example. However, as is well known the volatility of output, capital, and consumption are much smaller than the values needed to generate asset pricing outcomes in a production-based model setting that match the data. One potential justification for the higher volatility is the omission of leverage from the model. Firms in the model are financed entirely through equity, though empirical measurements show that firms are financed by $40\%$ debt and $60\%$ equity. Others in the production-based asset pricing literature, including \cite{boldrin2001habit} and \cite{papanikolaou2011investment}, scale up model risk premia ex post using a 5-to-3 leverage ratio to compare outcomes to empirical estimates. I apply this same scaling to estimate $\gamma = 9.5140$ and $\sigma = 0.0720$.
%%%%%%%%%%%%%%%%%%%%%%%%%%%%%%%%%%%%%%%%%%%
%%%%%%%%%%%%%%%%%%%%%%%%%%%%%%%%%%%%%%%%%%%

%%%%%%%%%%%%%%%%%%%%%%%%%%%%%%%%%%%%%%%%%%%%%%%%%%%%%%%%%%%%%%%%%%%%%%%%%%%%%%%
%%%%%%%%%%%%%%%%%%%%%%%%%%%%%%%%%%%%%%%%%%%%%%%%%%%%%%%%%%%%%%%%%%%%%%%%%%%%%%%

\subsection{Numerical Solution Method} \label{num_methods}

To solve the HJB equations, which are nonlinear partial differential equations, I use the method of false transient with a finite difference scheme and conjugate gradient solver\footnote{Joseph Huang, Paymon Khorrami, Fabrice Tourre, and the research professionals at the Macro Finance Research Program helped in developing the software for this solution method.}. The PDEs can be expressed in a conditionally linear form:
\begin{eqnarray*}
0 & = & V_t(x) + A(x;V,V_x,V_{xx})V(x) + B(x;V,V_x,V_{xx})V_x(x) \cr 
& & + \frac{1}{2}tr[C(x;V,V_x,V_{xx})V_{xx}(x)C(x;V,V_x,V_{xx})] + D(x;V,V_x,V_{xx})
\end{eqnarray*}

where $x$ is a state variable vector and $ V_x(x)=\frac{\partial V}{\partial x}(x), V_{xx}(x)=\frac{\partial^2 V}{\partial x \partial x'}(x) $ are used for notational simplicity. Note that the terms accounting for jumps are split into the $A(x;V,V_x,V_{xx})$ discounting coefficient and the $D(x;V,V_x,V_{xx})$ flow utility coefficient. The agent has an infinite horizon and so the problem is time stationary. Thus, $ V_t(x)=\frac{\partial V}{\partial t}(x) $ has been added as a ``false transient'' in order to construct the iterative solution algorithm. In particular, the solution comes by finding a $V(x)$ such that the above equality holds and $ V_t(x) = 0$.

The solution is found by first guessing a value function $V^0(x)$. Approximate derivatives $\widetilde{V^0_x}(x)$ and $\widetilde{V^0_{xx}}(x)$ are calculated from $V^0(x)$ using central finite differences (except at the boundaries where central differences require points outside the discretized state space and so appropriate forward or backward differences are used). These derivatives are used to calculate the coefficients $A, B, C$ and $D$, and depend on the value function and its derivatives because of the maximization from choosing optimal controls $\{ i_K, i_G, \mathcal{E}, L_1 \}$ in order to maximize utility. Applying a backward difference for $V^0_t(x)$, plugging in the calculated coefficients to the conditionally linear system, and rearranging gives
\begin{eqnarray*}
V^1(x) & = & V^0(x) + [ A(x,V^0,\widetilde{V^0_x},\widetilde{V^0_{xx}})V^0(x) + B(x,V^0,\widetilde{V^0_x},\widetilde{V^0_{xx}})\widetilde{V^0_x}(x) \cr 
& & + \frac{1}{2}tr[C(x,V^0,\widetilde{V^0_x},\widetilde{V^0_{xx}})\widetilde{V^0_{xx}}(x)C(x,V^0,\widetilde{V^0_x},\widetilde{V^0_{xx}})] + D(x,V^0,\widetilde{V^0_x},\widetilde{V^0_{xx}}) ] \Delta t
\end{eqnarray*}

We then solve numerically for $V^1(x)$ and repeat this process, with the solution at each iteration $k$ serving as the guess for the next iteration $k+1$, until $\max_x \frac{|V^{k+1}(x)-V^{k}(x)|}{\Delta t} < tol$ for a specified $tol>0$. The choice of $\Delta t$ is made by trading off increases in the speed of convergence, by increasing the size of $\Delta t$, and maintaining stability of the iterative algorithm, by decreasing $\Delta t$.

The equation for $V^1(x)$ can be expressed as a linear system $\Lambda \chi = \pi$. The solution at each iteration is found using a conjugate gradient solver to minimize the quadratic expression $\frac{1}{2} \chi' \Lambda \chi - \chi' \pi$. The $\chi$ that minimizes this expression is equivalent to the solution of the linear system if $\Lambda$ is positive definite as the first-order condition for the minimization problem requires $ \Lambda \chi - \pi = 0$. While $\Lambda$ is not necessarily symmetric, because it is invertible we can transform our system to $\Lambda' \Lambda \chi = \Lambda' \pi \iff \hat{\Lambda} \chi = \hat{\pi}$ which satisfies the necessary conditions and has the same solution as our original linear system.

\subsubsection{Implementation Details}

In order to implement the proposed algorithm, I need to specify hyperparameters regarding the state space size, discretization, false transient step size, error tolerance, and FOC relaxation parameter. Those values are given in Table \ref{hyperparams}. Choices are made trading-off between algorithm speed, stability, and accuracy, and are in line with other work using this algorithm including \cite{barnett2020pricing, BarnettBrockHansen:2021, barnett2023climate}.

\begin{center}
\begin{table}[!ht]
%\caption{Parameters}
\begin{center}
\caption{Numerical Algorithm Hyperparameters} \label{hyperparams} 
%\vspace{-0.5cm}
\begin{tabular}{l c c }
\hline \hline
${\begin{array}{c} \text{$\log K_G$ discretization} \end{array}}$  & $nk$ & $35$  \\
${\begin{array}{c} \text{$Y$ discretization} \end{array}}$  & $ny$ & $100$  \\
${\begin{array}{c} \text{$\log R$ discretization} \end{array}}$  & $nr$ & $200$  \\
${\begin{array}{c} \text{false transient step size} \end{array}}$  & $\Delta t$ & $0.25$  \\
${\begin{array}{c} \text{HJB error tolerance} \end{array}}$  & $tol_{HJB}$ & $1e-12$  \\
${\begin{array}{c} \text{FOC relaxation parameter} \end{array}}$  & $\epsilon_{FOC}$ & $0.05$  \\
% \hline
% ${\begin{array}{c} \text{Partial internalization error tolerance} \end{array}}$  & $tol_{PI}$ & $1e-8$  \\
% ${\begin{array}{c} \text{Partial internalization relaxation parameter} \end{array}}$  & $\epsilon_{PI}$ & $0.2$  \\
\hline
${\begin{array}{c} \text{Green Capital} \end{array}}$   & $[\log K_G^{(min)},\log K_G^{(max)}]$ & [2.0, 7.0] \\
${\begin{array}{c} \text{Temperature Anomaly} \end{array}}$   & $[Y^{(min)},Y^{(max)}]$ & [0.0, 3.5] \\
${\begin{array}{c} \text{Oil Reserves} \end{array}}$   & $[\log R^{(min)},\log R^{(max)}]$ & [-4.0, 7.0] \\
\hline  \hline
\end{tabular}
\end{center} %\vspace{0.2cm}
\begin{footnotesize}
%Table~\ref{table:params} presents ...
\end{footnotesize}%\vspace{1cm}
\end{table}
\end{center}

% ... hyperparameters table
% Quick notes on coding constraints imposed:
% \begin{itemize}
%     \item iter > 20000 -> shift dt = 0.0001 (oscillating near tol)
%     \item   $\frac{\partial V}{\partial Y} \le 1e-16$    
%     \item   $\frac{\partial V}{\partial R} \ge 1e-16$ 
%     \item  $0 \le L \le 1$
%     \item $\mathcal{E} \ge 1e-16$
%     \item relaxation parameter updating FOCs
% \end{itemize}

 In addition, I implement various numerical constraints to maintain stability and ensure convergence for the algorithm. In particular, the false transient step size $\Delta t$ is reduced to $0.0001$ if the algorithm reaches more than 20000 iterations. This guarantees convergences when the algorithm is oscillating very close to the error tolerance value $tol_{HJB}$. I also impose the constraints $\frac{\partial V}{\partial Y} \le 1e-16$, $\frac{\partial V}{\partial R} \ge 1e-16$, $0 \le L \le 1$, and $\mathcal{E} \ge 1e-16$. The economic intuition and motivation for the first two constraints are related to the negative externality associated with climate change and the economic benefit from, and substitutability of, oil in production. Both constraints help the algorithm avoid numerical approximation errors for near zero derivative values in economically uninteresting or implausible parts of the state space that can oscillate between very small positive and negative values, which can cause the algorithm to break down. The second two constraints are consistent with assumptions in the model about nonnegative emissions and physical limits for the unit supply of labor. 
 
 Finally, given the nonlinear equations characterizing the FOCs, I use a ``relaxation technique'' to update the FOC in each iteration. Thus, for a given value function update, the previous optimal control is only updated by a fractional amount of the new optimal control value, given by the relaxation parameter $\epsilon_{FOC}$. This again ensures stability of the algorithm.

% \newpage
% \clearpage

\subsection{Additional Numerical Results}

% I now provide additional numerical result for various alternative transition shock scenarios: i) partial externality internalization for (a) a technology shock and (b) a taxation shock with parameter $\chi = 0.5$; ii) multiple-jump transition shocks for (a) a two-step tax transition and (b) a two-step technology transition; iii) single fossil fuel type economy for (a) a technology shock in an oil only economy and (b) a technology shock in a coal only economy; iv) coal then oil transition shocks for (a) coal tax shock then oil tax shock scenario and (b) a coal technology shock then oil technology shock. Each of these cases are provided as additional "robustness checks" under various economic and policy scenarios of interest, further highlighting the economic intuition related to the the central model mechanism.

I now provide additional numerical result for various alternative transition shock scenarios: 

\begin{enumerate}
    \item Multiple-Jump Transition Shocks
    \begin{enumerate}
        \item Two-step tax transition
        \item Two-step technology transition
    \end{enumerate}
    \item Single Fossil Fuel Type
    \begin{enumerate}
        \item Oil only economy, technology shock
        \item Coal only economy, technology shock
    \end{enumerate}
    \item Coal then Oil Transition Shocks
    \begin{enumerate}
        \item Coal tax shock then oil tax shock
        \item Coal technology shock then oil technology shock
    \end{enumerate}    
\end{enumerate}

Each of these transition shock scenarios are provided as additional "sensitivity analysis" of the transition risk mechanism under various economic and policy scenarios of interest, further highlighting the economic intuition related to the the central model mechanism.

% \newpage
% \clearpage

% \subsubsection{Multiple-Jump Transition Shocks}

% \begin{landscape}

\begin{figure}[!pht]

%\vspace{-1.0cm}
% \vspace{-0.5cm}

\caption{Macroeconomic and Asset Pricing Outcomes - Two-Step ``Taxation'' Shock} \label{fig:multi2_model_sims}
\begin{center}
% {\scriptsize \textbf{Panel A: Hybrid Taxation/Technology Transition Scenario}}\\
        \begin{subfigure}[b]{0.328\textwidth}
            \centering
            \includegraphics[width=\textwidth]{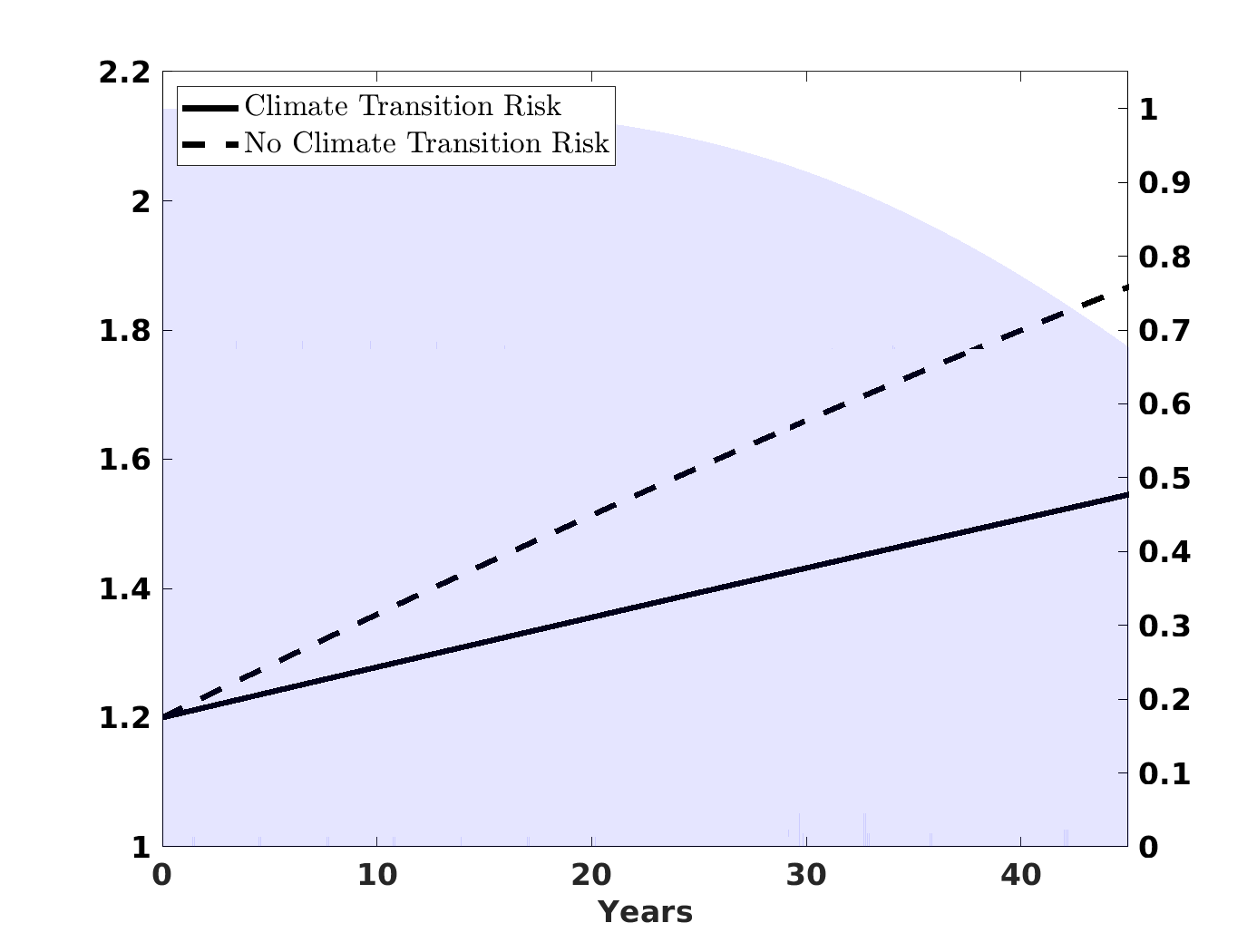}
            \caption[]{{\small Temperature: $Y_t$}}              
        \end{subfigure}
        \begin{subfigure}[b]{0.328\textwidth}  
            \centering 
            \includegraphics[width=\textwidth]{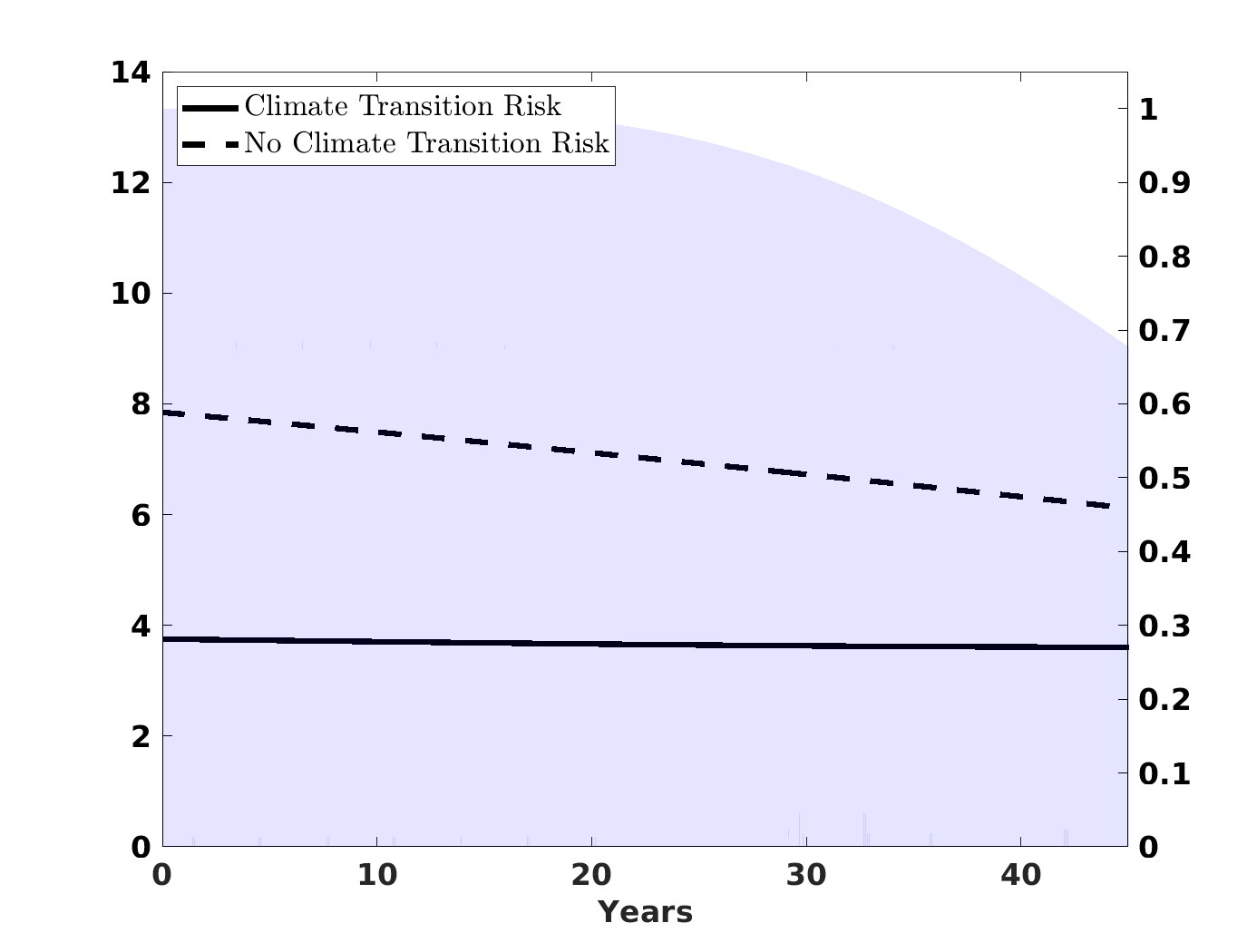}
           \caption[]{{\small Oil Production: $E_{1,t}$}}
        \end{subfigure}
        \begin{subfigure}[b]{0.328\textwidth}  
            \centering 
            \includegraphics[width=\textwidth]{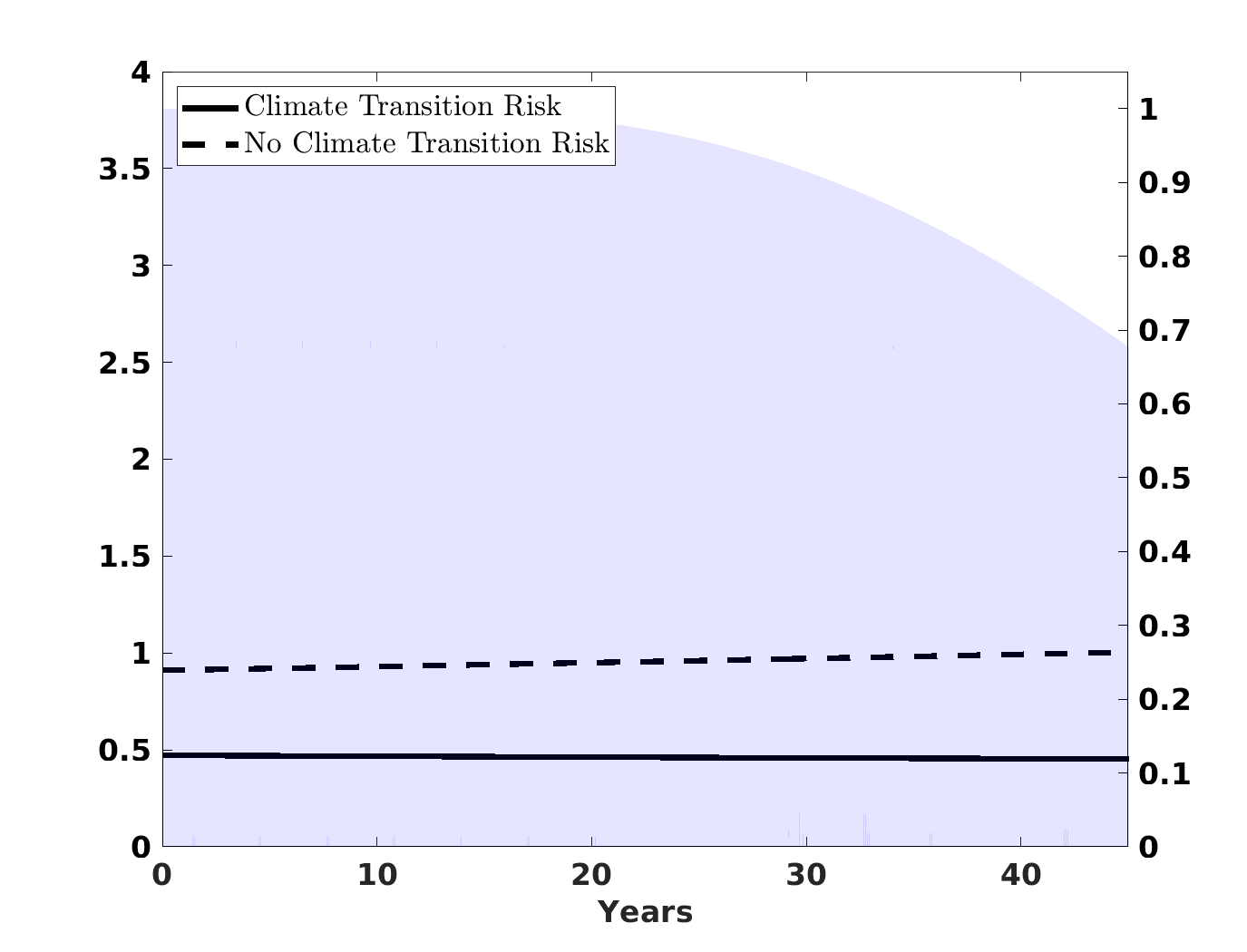}
           \caption[]{{\small Coal Production: $E_{2,t}$}}
        \end{subfigure}     
        
        \begin{subfigure}[b]{0.328\textwidth}
            \centering
            \includegraphics[width=\textwidth]{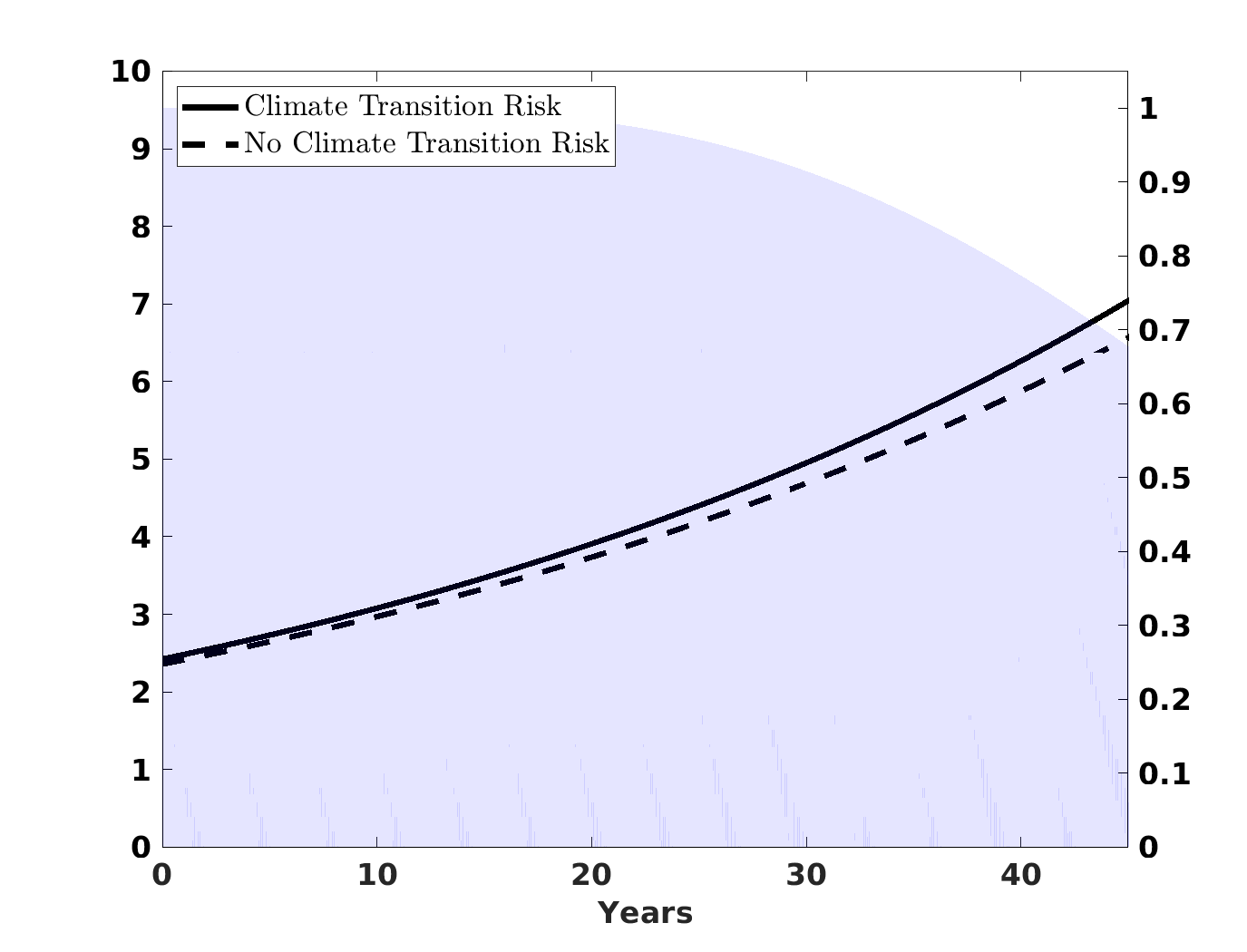}
            \caption[]{{\small Green Investment: $I_{G,t}$}}              
        \end{subfigure}
        \begin{subfigure}[b]{0.328\textwidth}  
            \centering 
            \includegraphics[width=\textwidth]{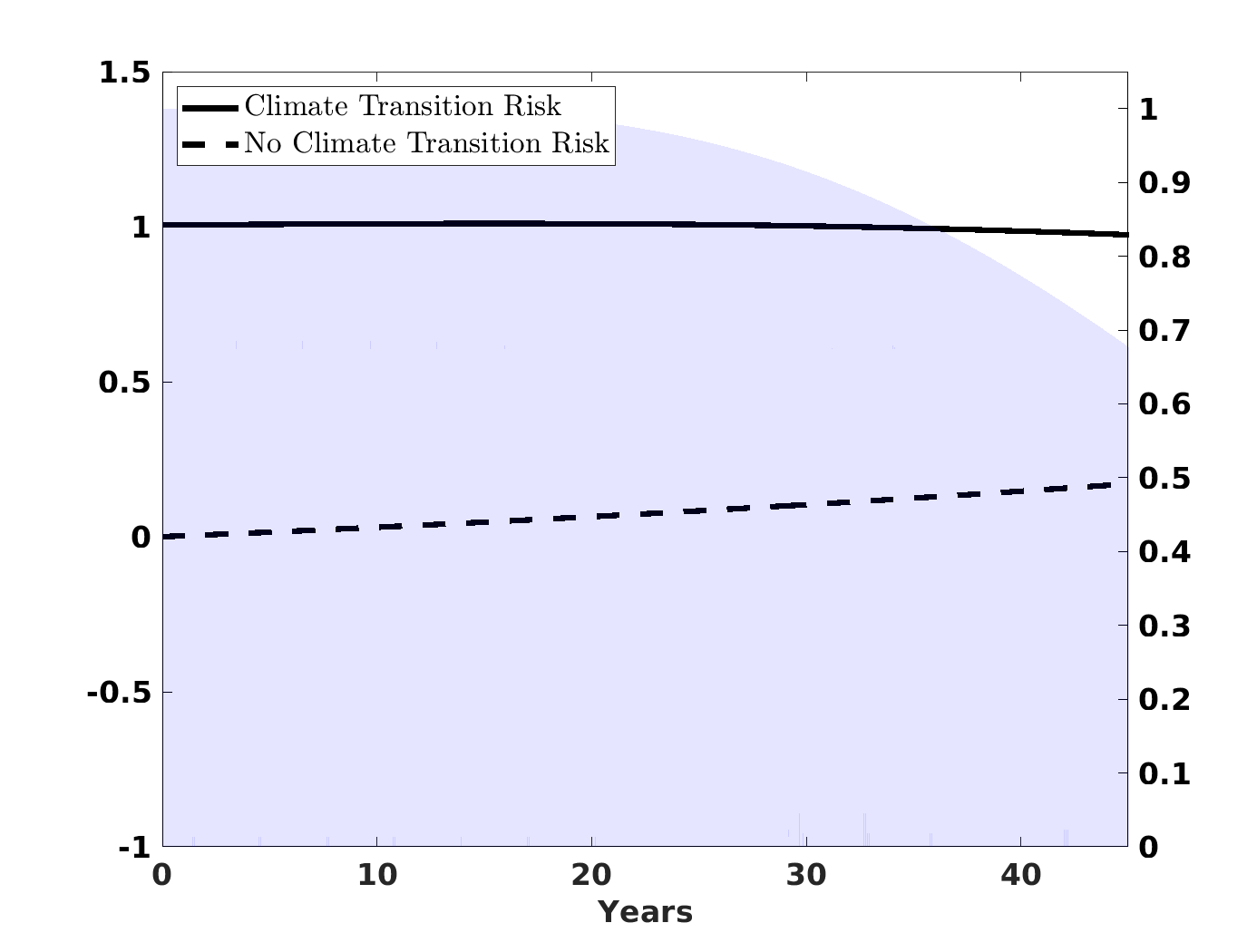}
           \caption[]{{\small Oil Spot Price: $P_{1,t}$}}
        \end{subfigure}
        \begin{subfigure}[b]{0.328\textwidth}  
            \centering 
            \includegraphics[width=\textwidth]{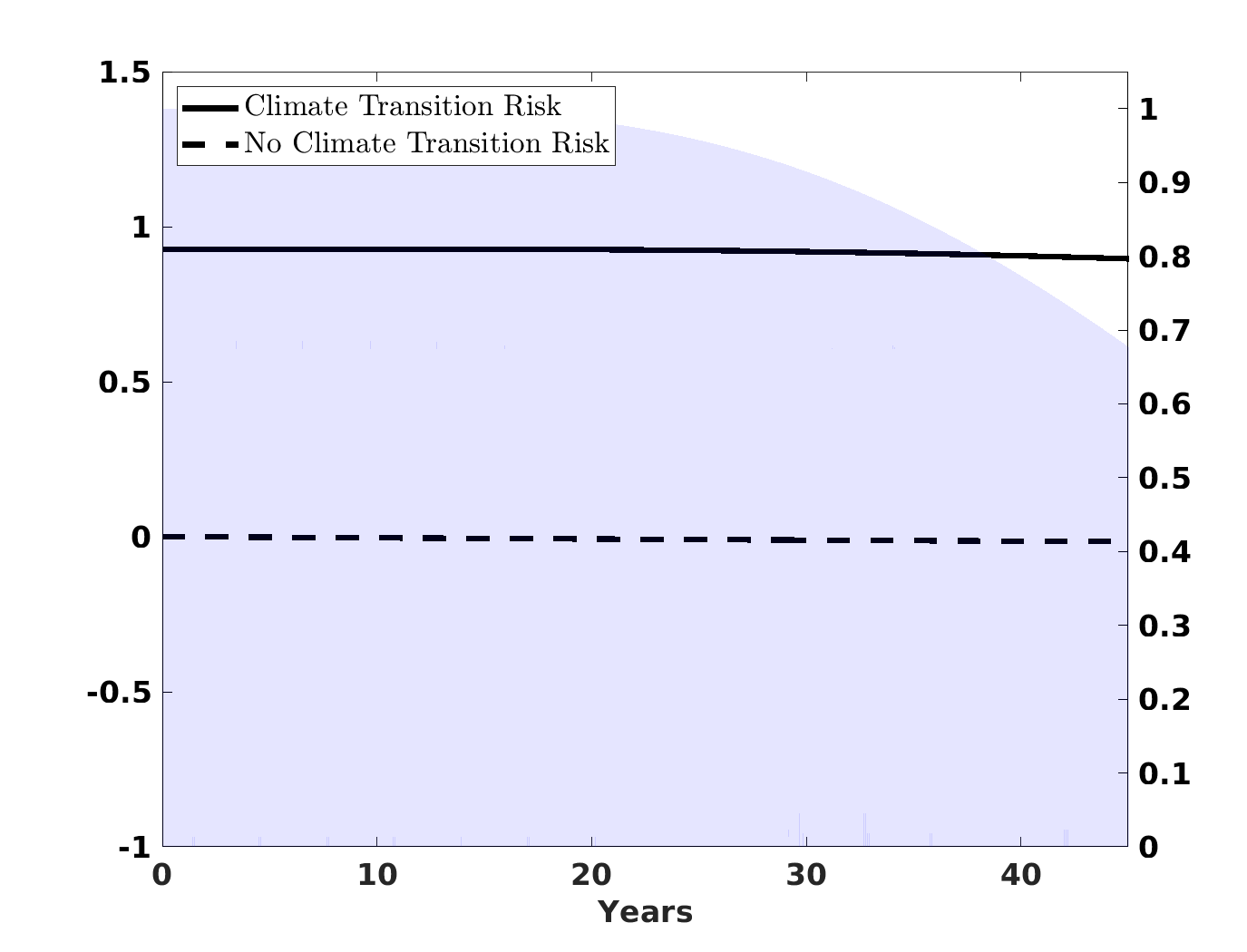}
           \caption[]{{\small Coal Spot Price: $P_{1,t}$}}
        \end{subfigure}

        \begin{subfigure}[b]{0.328\textwidth}
            \centering
            \includegraphics[width=\textwidth]{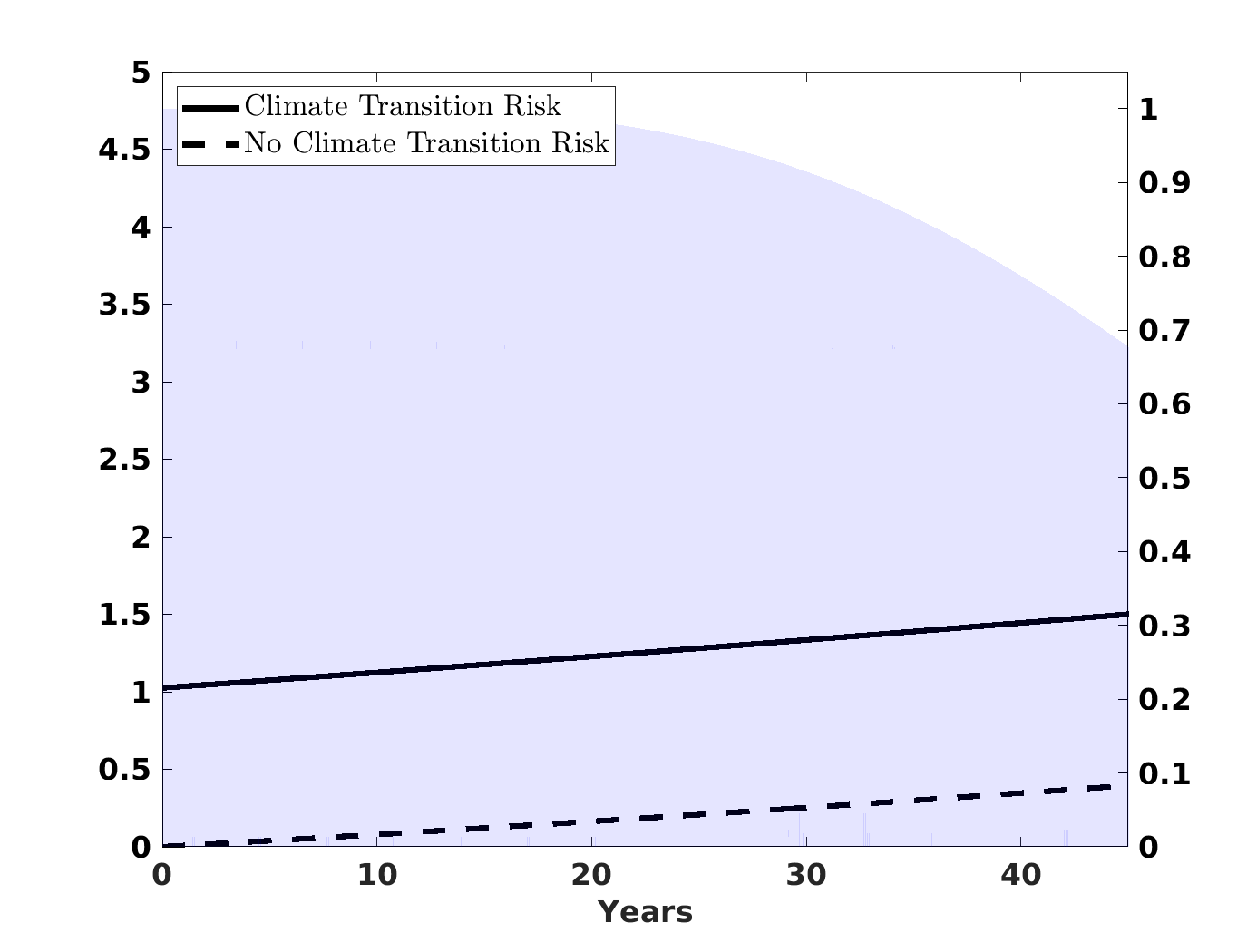}
            \caption[]{{\small Green Firm Price: $S^{(3)}_{t}$}}              
        \end{subfigure}
        \begin{subfigure}[b]{0.328\textwidth}  
            \centering 
            \includegraphics[width=\textwidth]{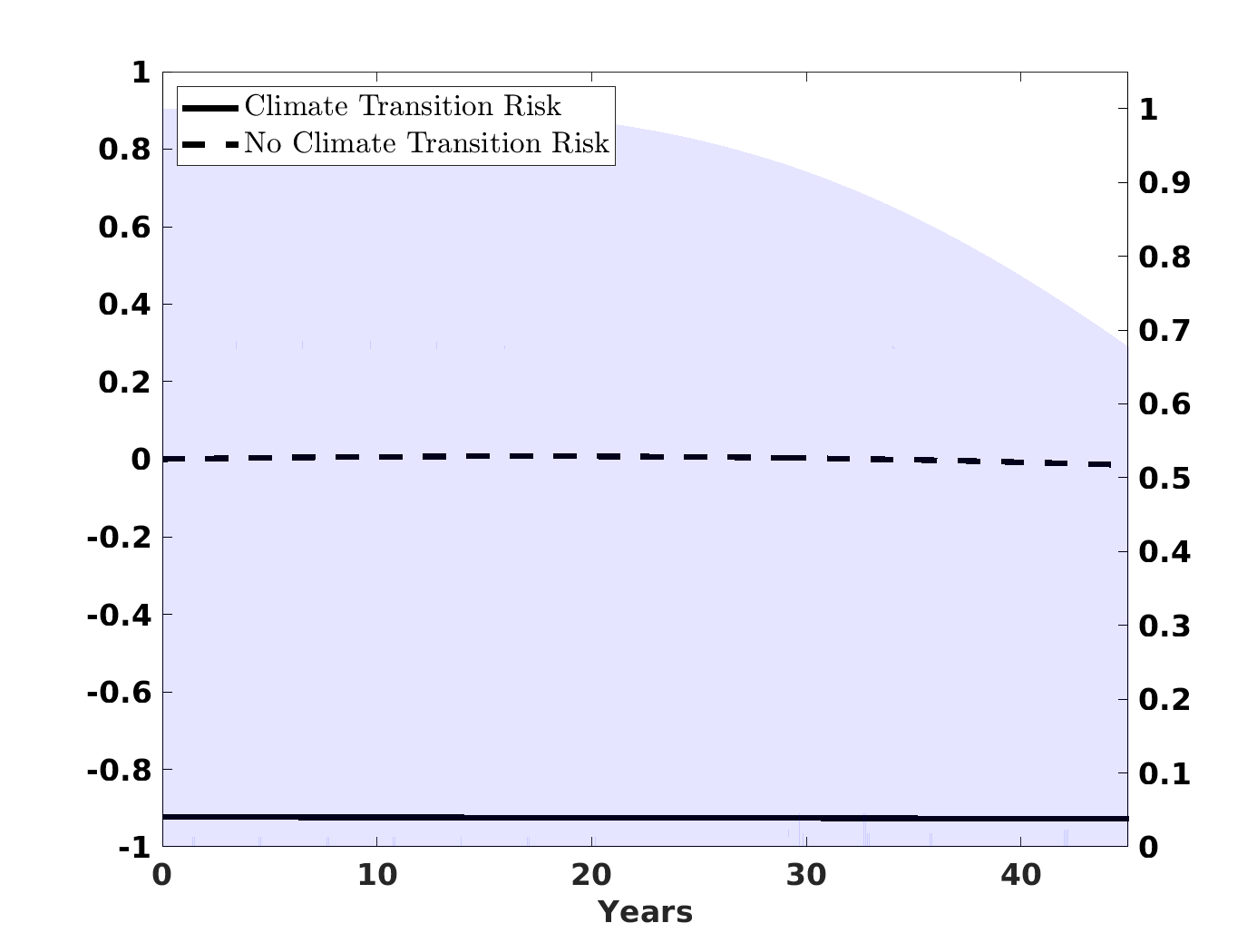}
           \caption[]{{\small Oil Firm Price: $S^{(1)}_{t}$}}
        \end{subfigure}
        \begin{subfigure}[b]{0.328\textwidth}  
            \centering 
            \includegraphics[width=\textwidth]{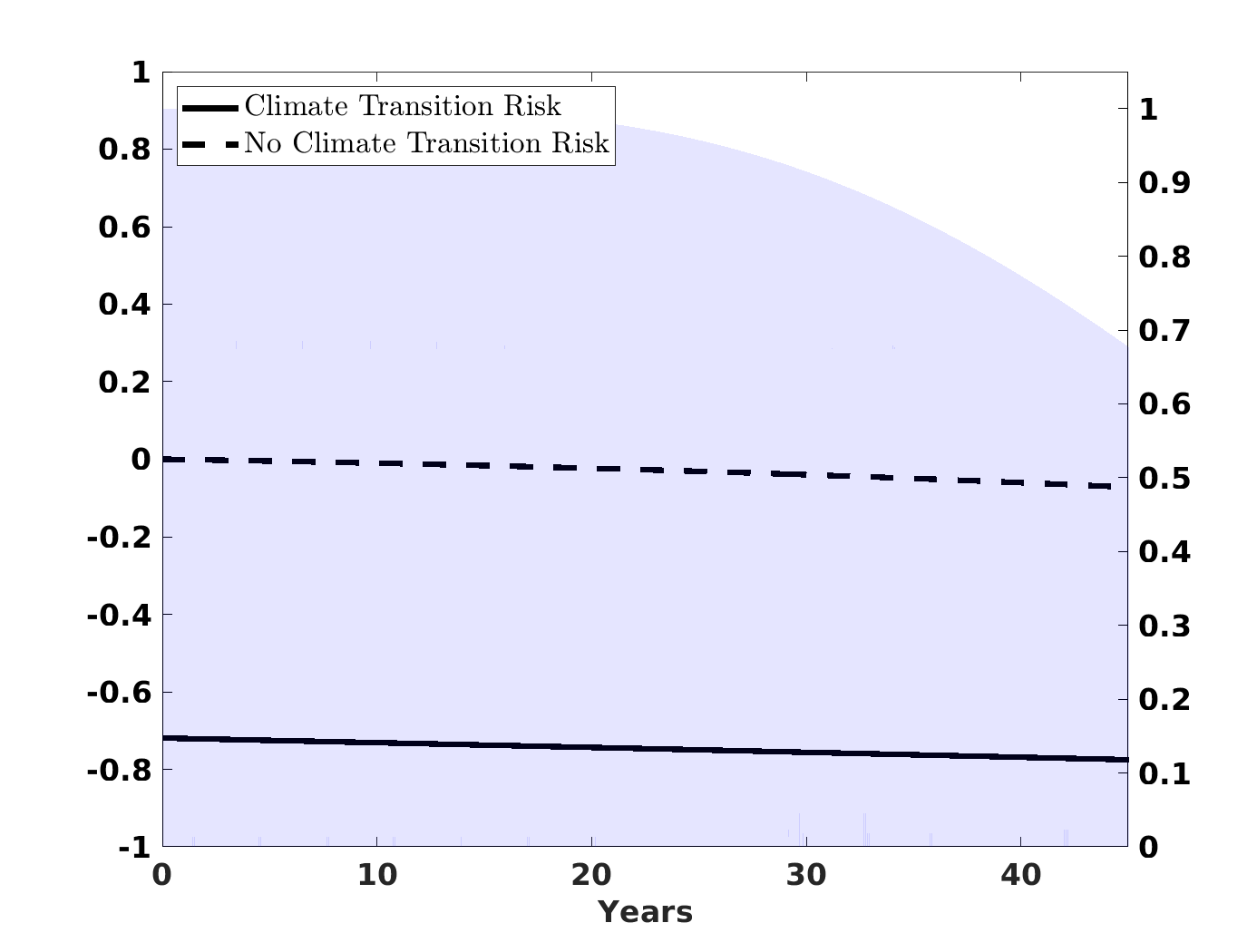}
           \caption[]{{\small Coal Firm Price: $S^{(2)}_{t}$}}
        \end{subfigure}  
        
        \vspace{-0.25cm}
\end{center}

\begin{footnotesize}
Figure \ref{fig:multi2_model_sims} shows the simulated outcomes for the two-step ``taxation shock'' model based on the numerical solutions. Panels (a) through (c) show the temperature anomaly, oil production, and coal production. Panels (d) through (f) show the green investment choice, oil spot price, and coal spot price. Panels (g) and (i) show the green firm price, oil firm price, and coal firm price. Solid lines represent results for the Climate Transition Risk scenario where $\lambda_t = \lambda(T_t)$ and dashed lines represent results for the No Climate Transition Risk scenario where $\lambda_t = 0$. The blue shaded region shows the cumulative probability of no transition shock occurring.
\end{footnotesize} 

\end{figure}

% \begin{landscape}

\begin{figure}[!pht]

%\vspace{-1.0cm}
% \vspace{-0.5cm}

\caption{Macroeconomic and Asset Pricing Outcomes - Two-Step ``Technology'' Shock} \label{fig:multi2_model_sims}
\begin{center}
% {\scriptsize \textbf{Panel A: Hybrid Taxation/Technology Transition Scenario}}\\
        \begin{subfigure}[b]{0.328\textwidth}
            \centering
            \includegraphics[width=\textwidth]{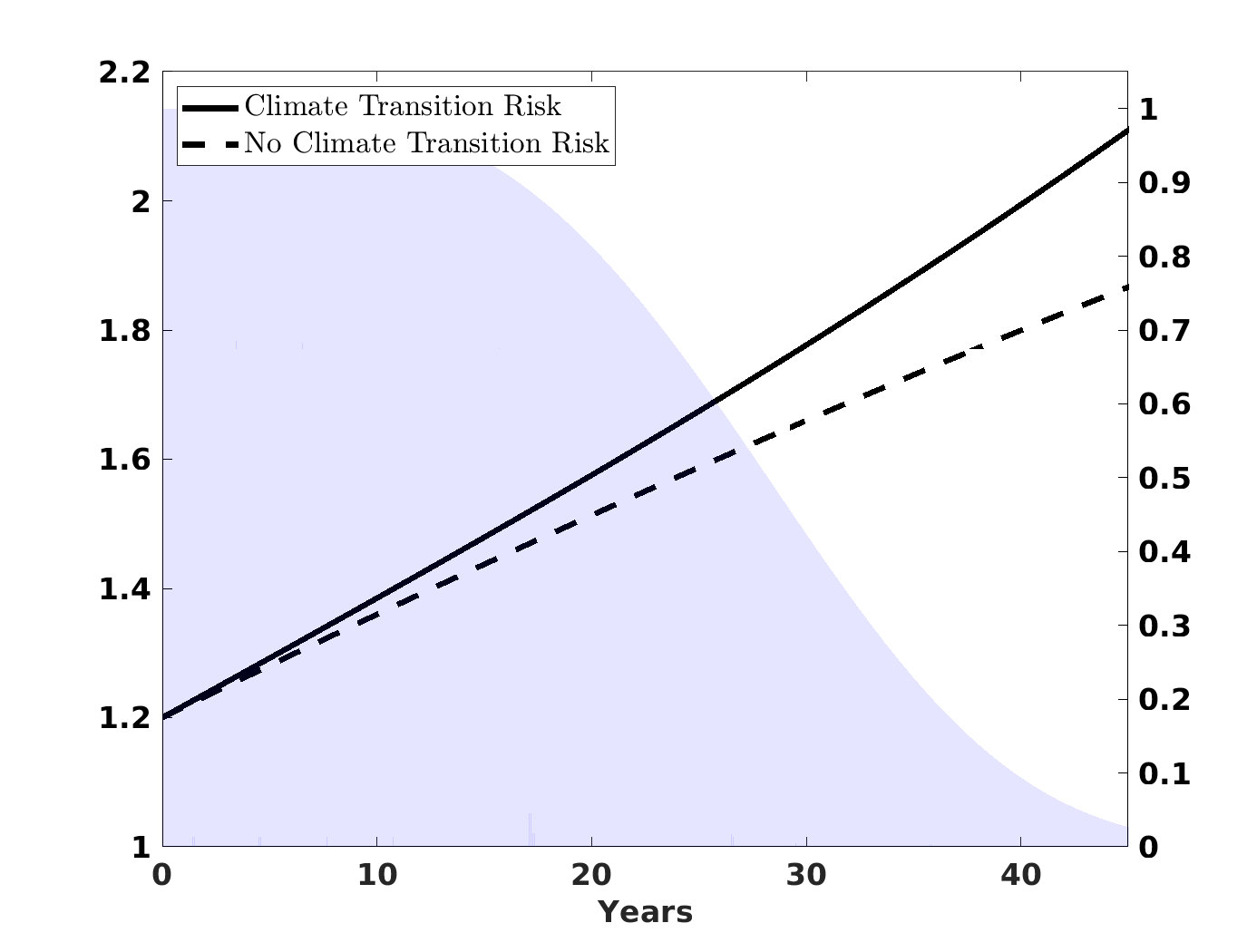}
            \caption[]{{\small Temperature: $Y_t$}}              
        \end{subfigure}
        \begin{subfigure}[b]{0.328\textwidth}  
            \centering 
            \includegraphics[width=\textwidth]{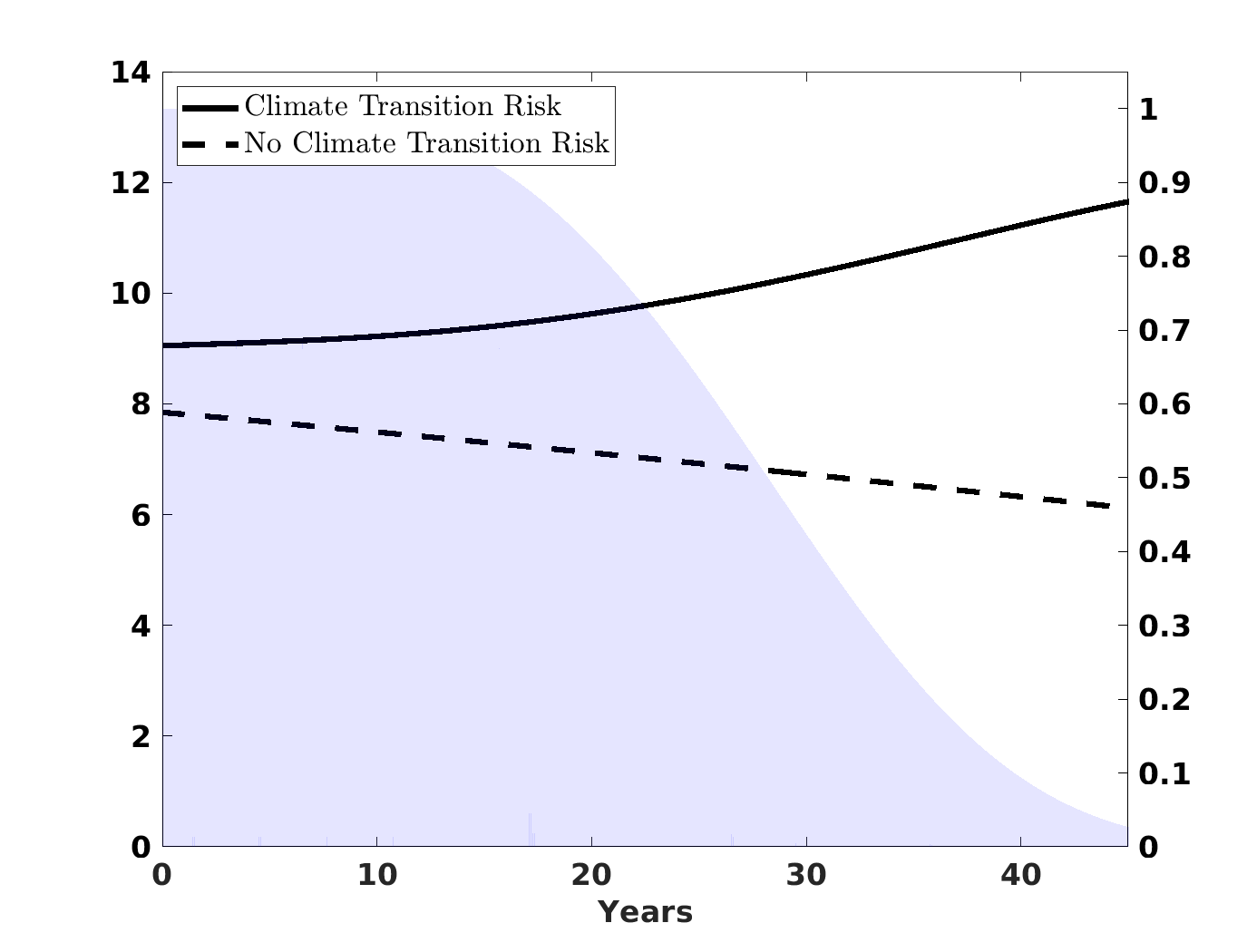}
           \caption[]{{\small Oil Production: $E_{1,t}$}}
        \end{subfigure}
        \begin{subfigure}[b]{0.328\textwidth}  
            \centering 
            \includegraphics[width=\textwidth]{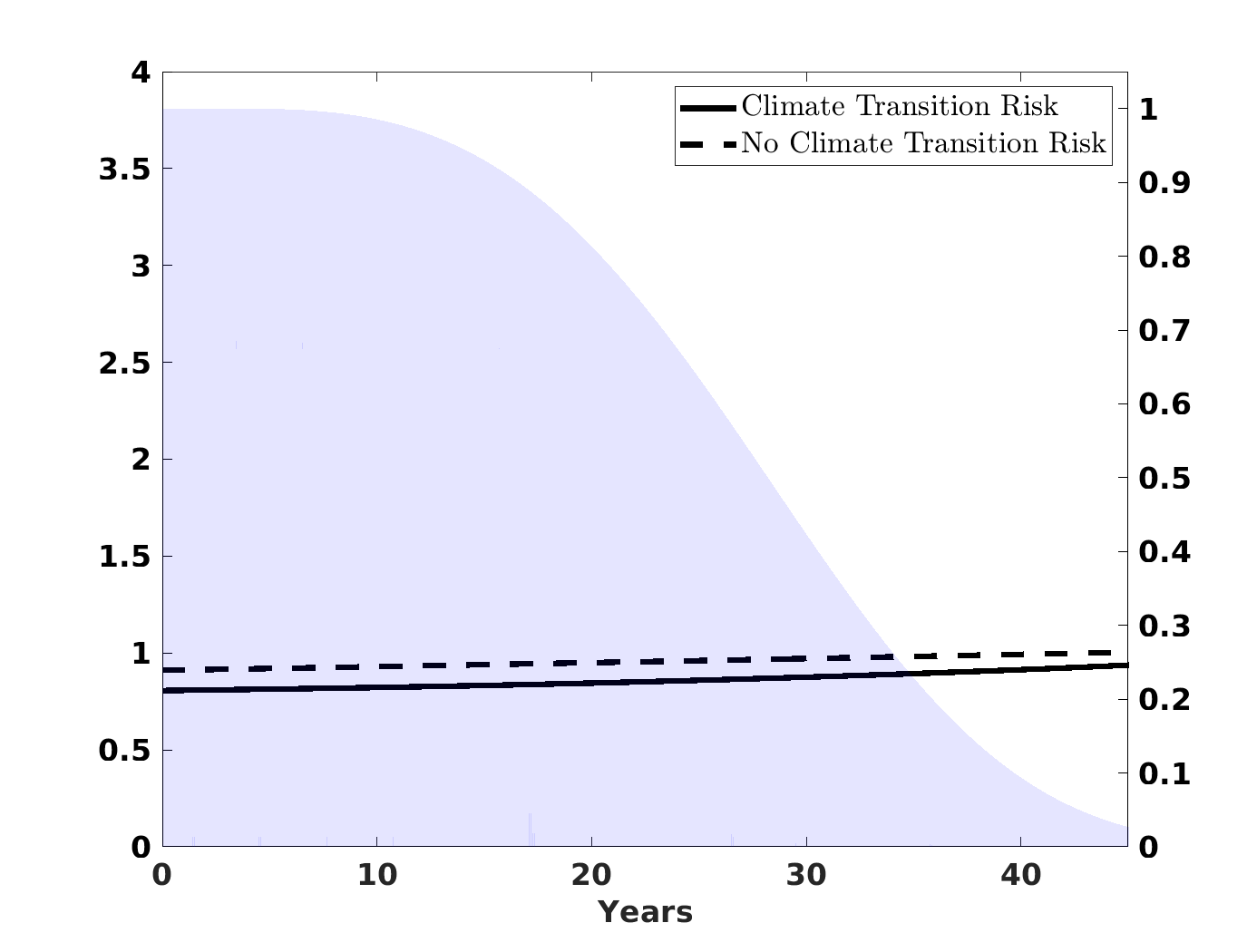}
           \caption[]{{\small Coal Production: $E_{2,t}$}}
        \end{subfigure}     
        
        \begin{subfigure}[b]{0.328\textwidth}
            \centering
            \includegraphics[width=\textwidth]{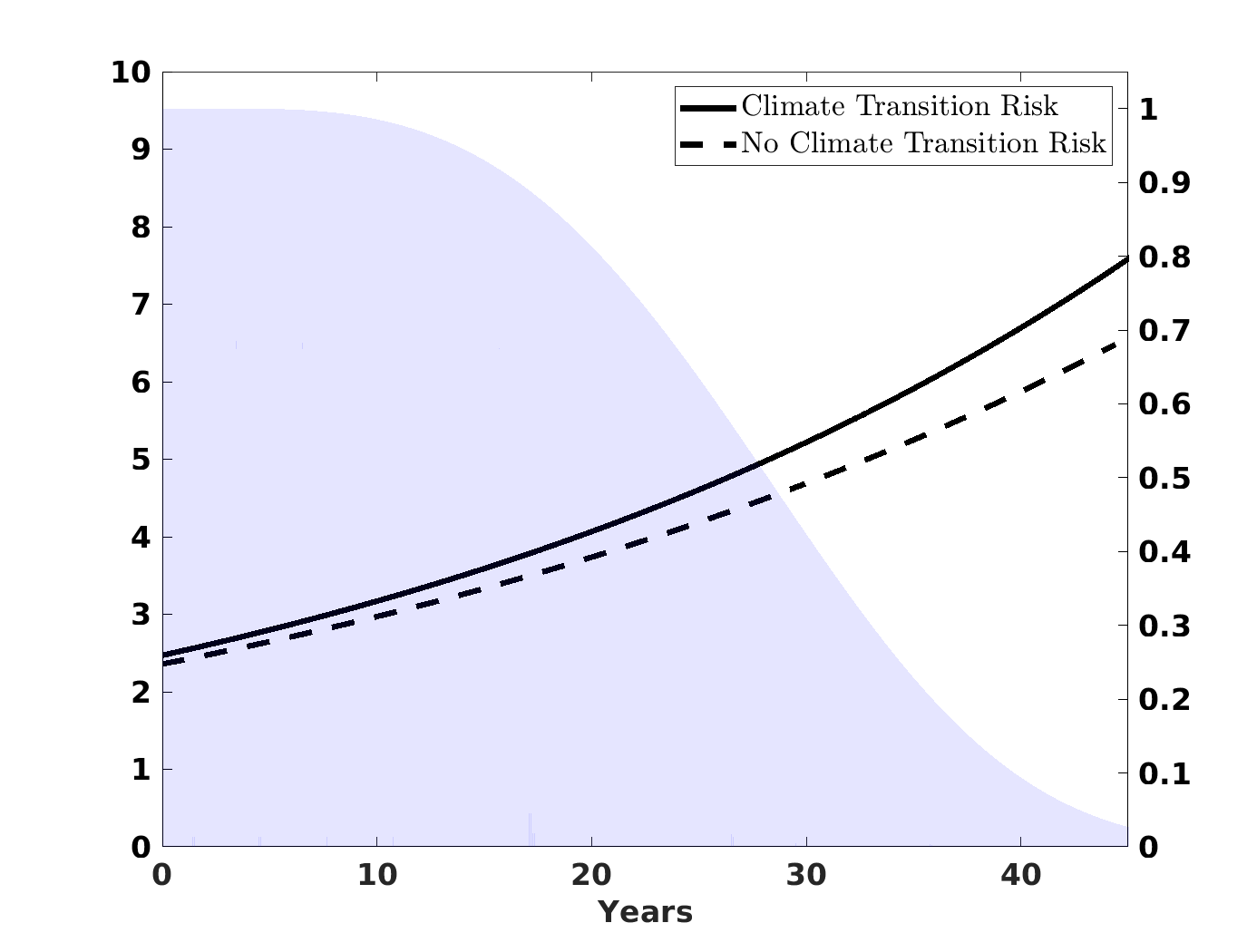}
            \caption[]{{\small Green Investment: $I_{G,t}$}}              
        \end{subfigure}
        \begin{subfigure}[b]{0.328\textwidth}  
            \centering 
            \includegraphics[width=\textwidth]{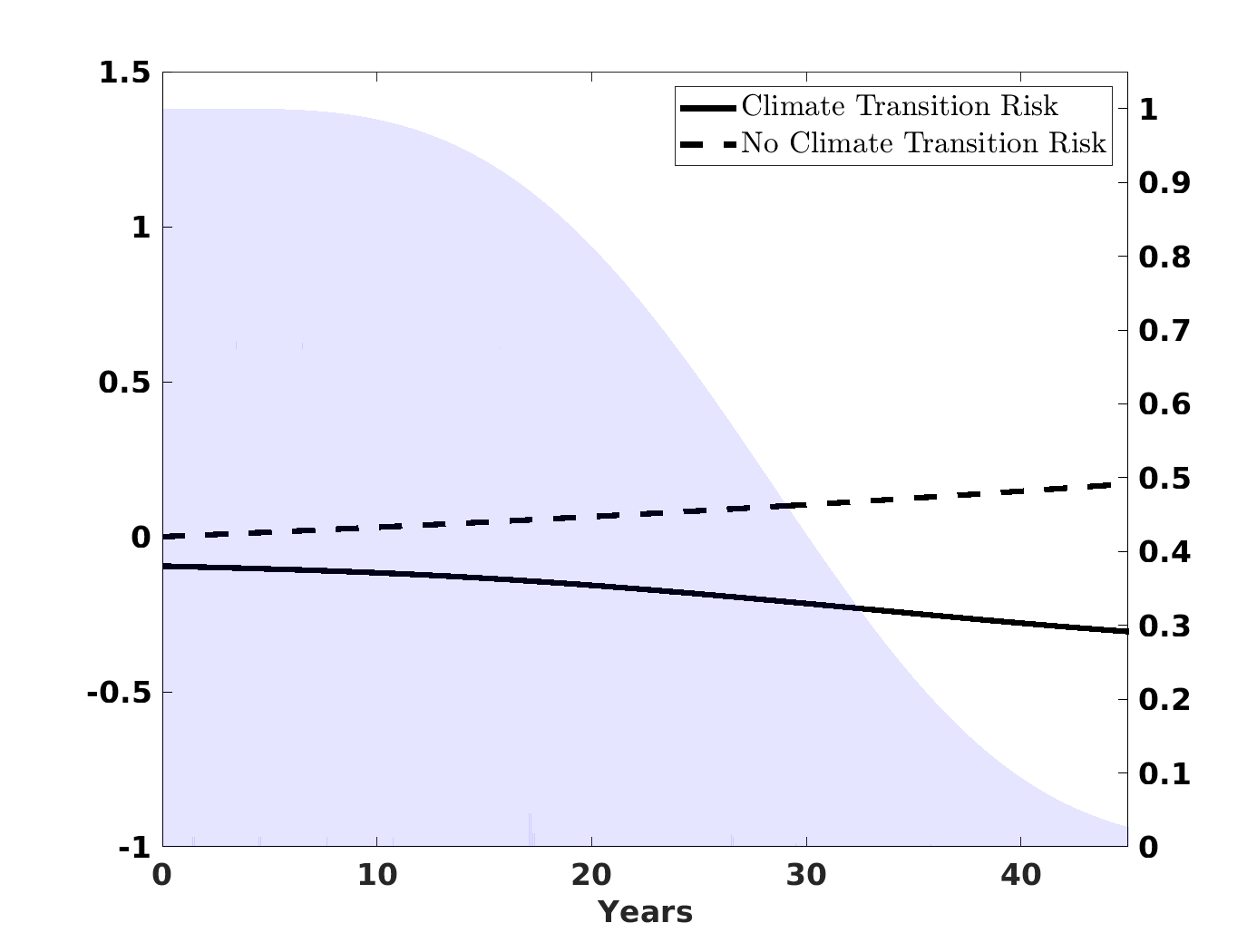}
           \caption[]{{\small Oil Spot Price: $P_{1,t}$}}
        \end{subfigure}
        \begin{subfigure}[b]{0.328\textwidth}  
            \centering 
            \includegraphics[width=\textwidth]{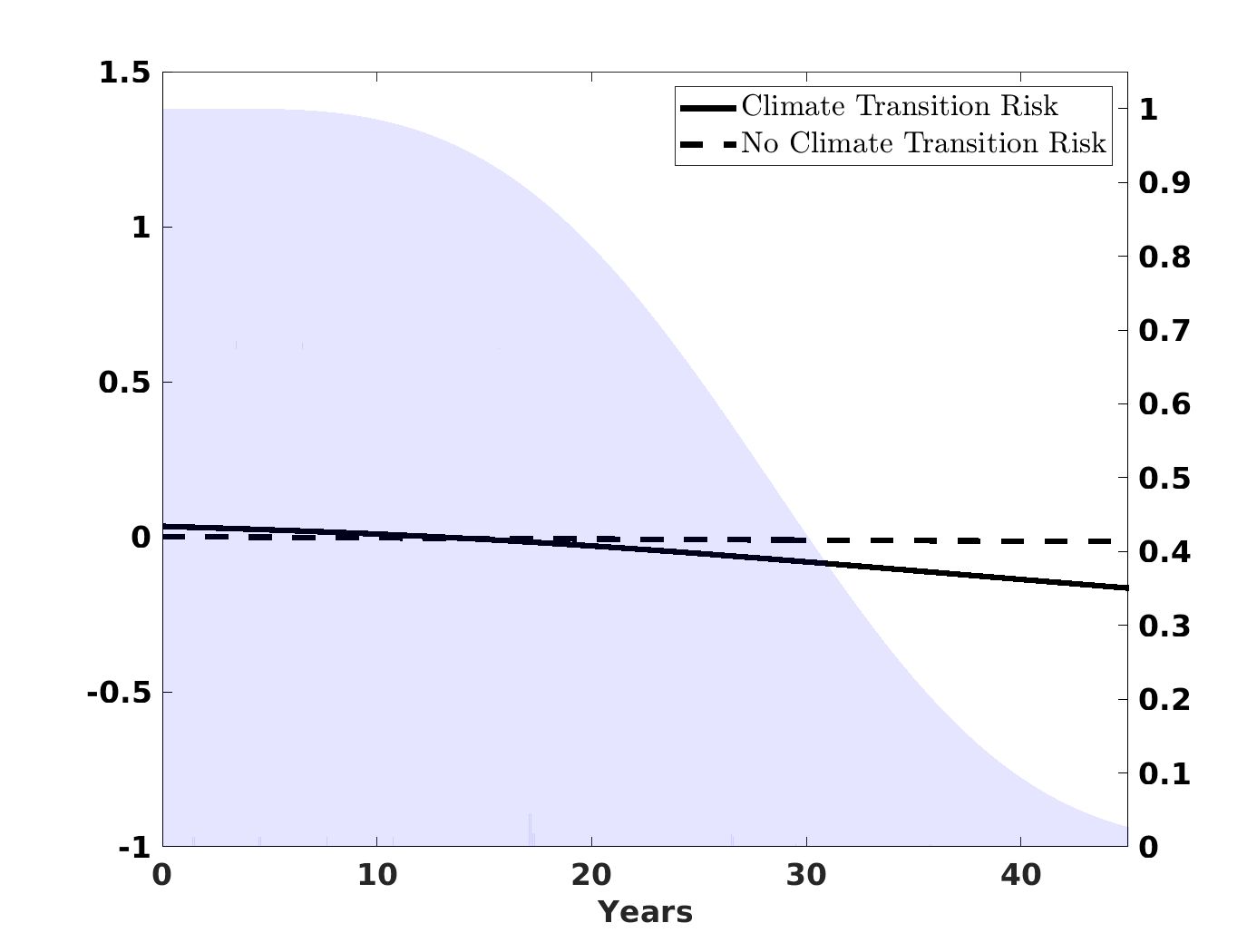}
           \caption[]{{\small Coal Spot Price: $P_{1,t}$}}
        \end{subfigure}

        \begin{subfigure}[b]{0.328\textwidth}
            \centering
            \includegraphics[width=\textwidth]{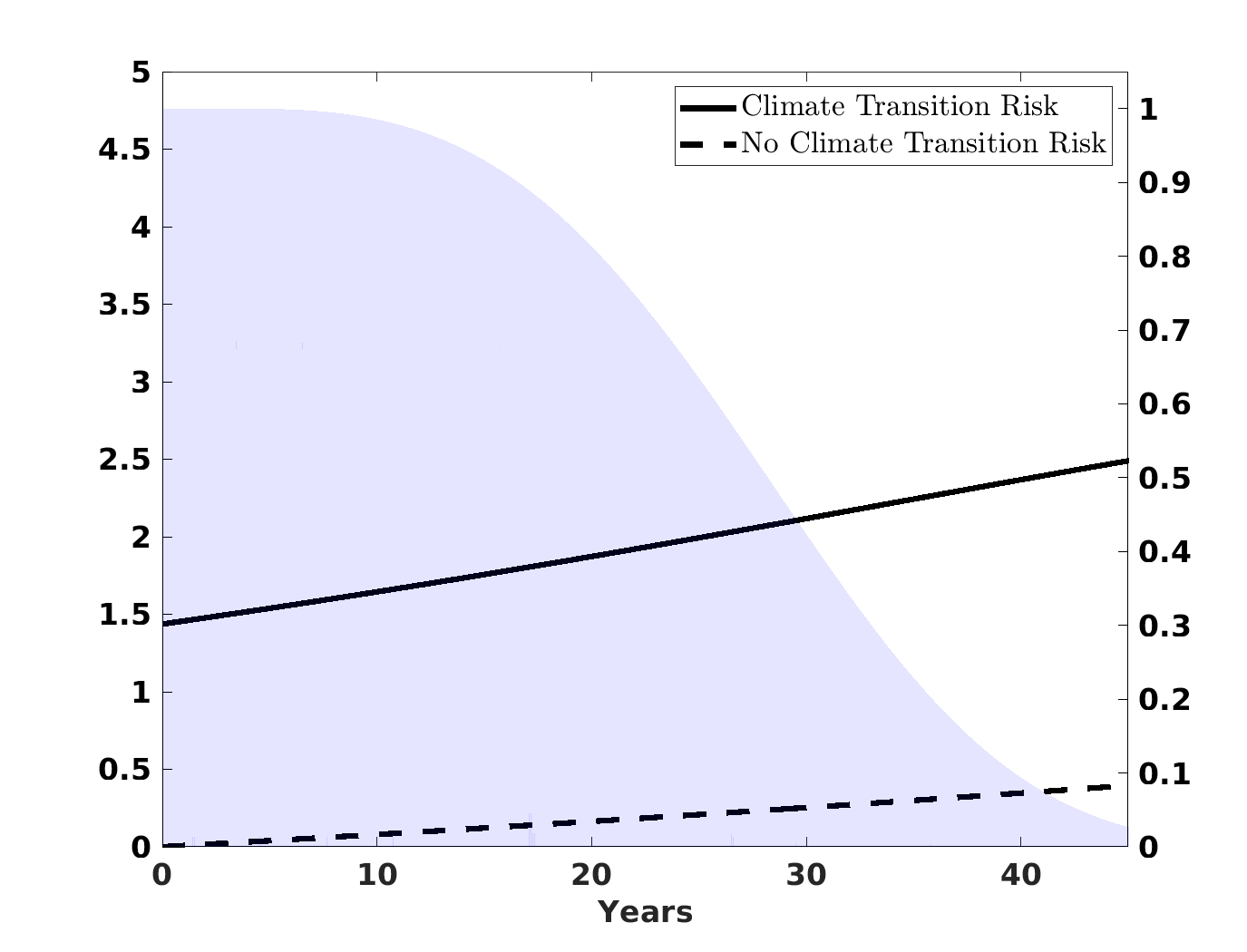}
            \caption[]{{\small Green Firm Price: $S^{(3)}_{t}$}}              
        \end{subfigure}
        \begin{subfigure}[b]{0.328\textwidth}  
            \centering 
            \includegraphics[width=\textwidth]{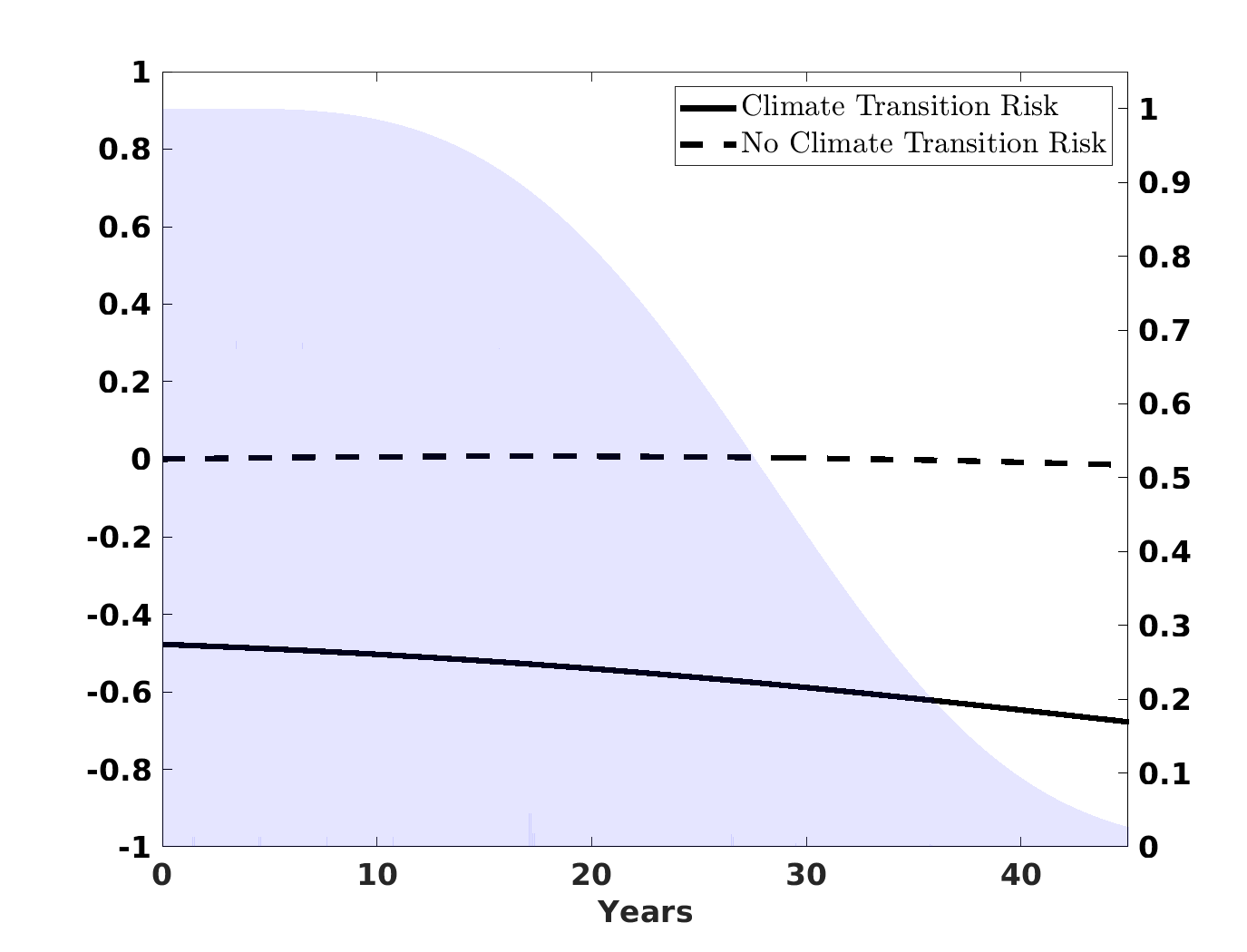}
           \caption[]{{\small Oil Firm Price: $S^{(1)}_{t}$}}
        \end{subfigure}
        \begin{subfigure}[b]{0.328\textwidth}  
            \centering 
            \includegraphics[width=\textwidth]{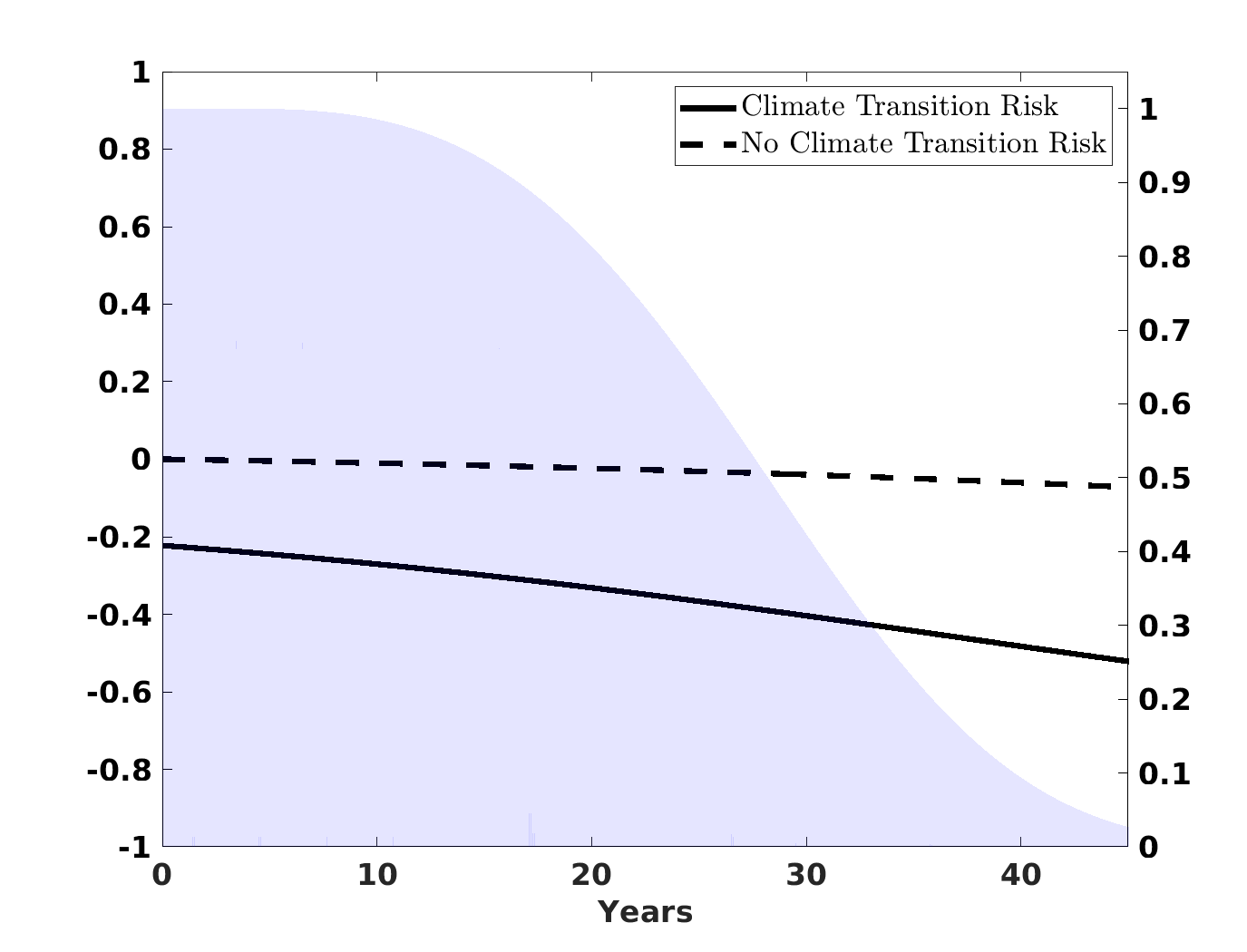}
           \caption[]{{\small Coal Firm Price: $S^{(2)}_{t}$}}
        \end{subfigure}  
        
        \vspace{-0.25cm}
\end{center}

\begin{footnotesize}
Figure \ref{fig:multi2_model_sims} shows the simulated outcomes for the two-step ``technology shock'' model based on the numerical solutions. Panels (a) through (c) show the temperature anomaly, oil production, and coal production. Panels (d) through (f) show the green investment choice, oil spot price, and coal spot price. Panels (g) and (i) show the green firm price, oil firm price, and coal firm price. Solid lines represent results for the Climate Transition Risk scenario where $\lambda_t = \lambda(T_t)$ and dashed lines represent results for the No Climate Transition Risk scenario where $\lambda_t = 0$. The blue shaded region shows the cumulative probability of no transition shock occurring.
\end{footnotesize} 

\end{figure}

% \subsubsection{Single Fossil Fuel Type}

% \begin{landscape}

\begin{figure}[!pht]

%\vspace{-1.0cm}
% \vspace{-0.5cm}

\caption{Macroeconomic and Asset Pricing Outcomes - Coal Only ``Technology'' Shock} \label{fig:multi2_model_sims}
\begin{center}
% {\scriptsize \textbf{Panel A: Hybrid Taxation/Technology Transition Scenario}}\\
        \begin{subfigure}[b]{0.328\textwidth}
            \centering
            \includegraphics[width=\textwidth]{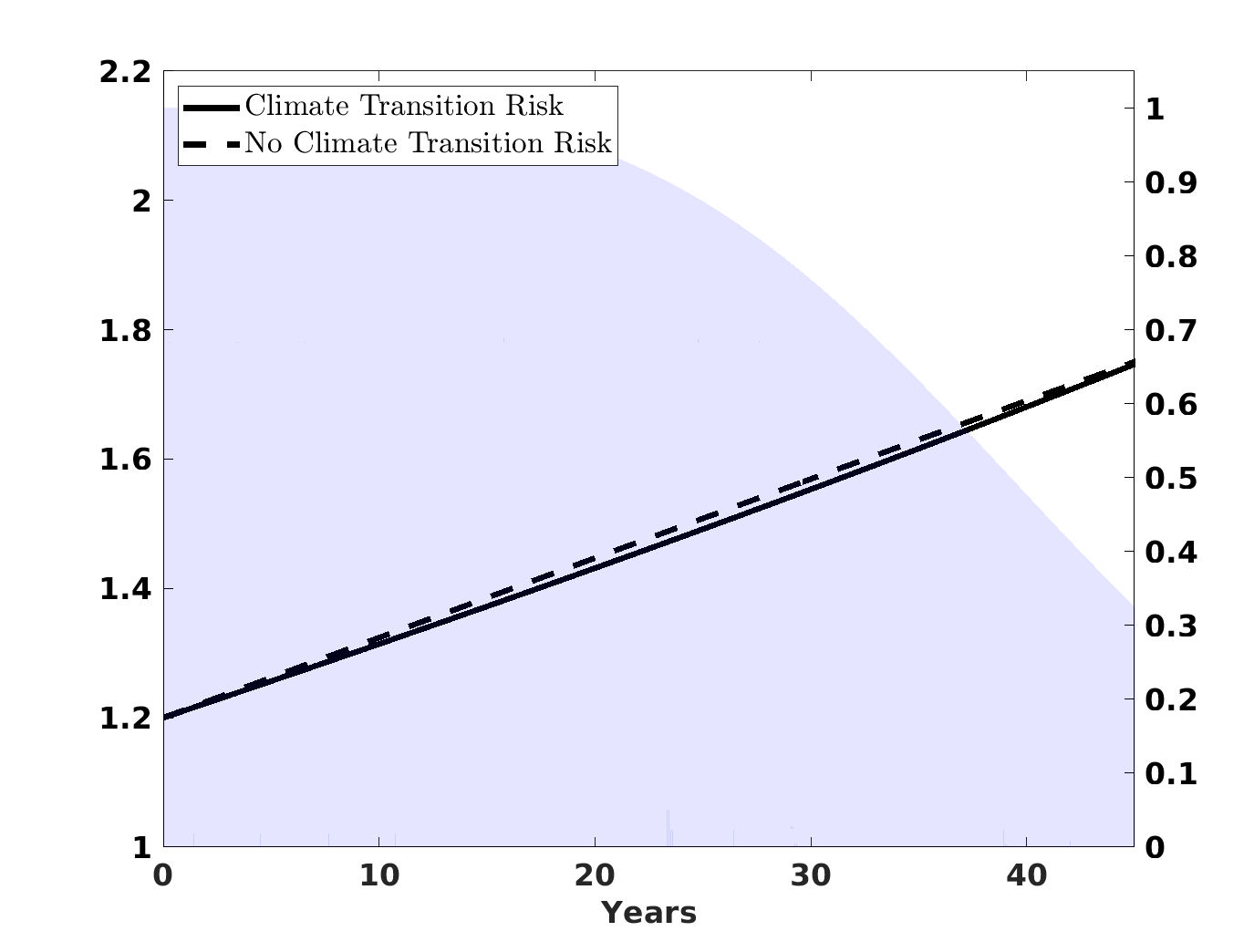}
            \caption[]{{\small Temperature: $Y_t$}}              
        \end{subfigure}
        \begin{subfigure}[b]{0.328\textwidth}  
           %  \centering 
           %  \includegraphics[width=\textwidth]{figures_submit/oil_justcoal_roff__submit.png}
           % \caption[]{{\small Oil Production: $E_{1,t}$}}
        \end{subfigure}
        \begin{subfigure}[b]{0.328\textwidth}  
            \centering 
            \includegraphics[width=\textwidth]{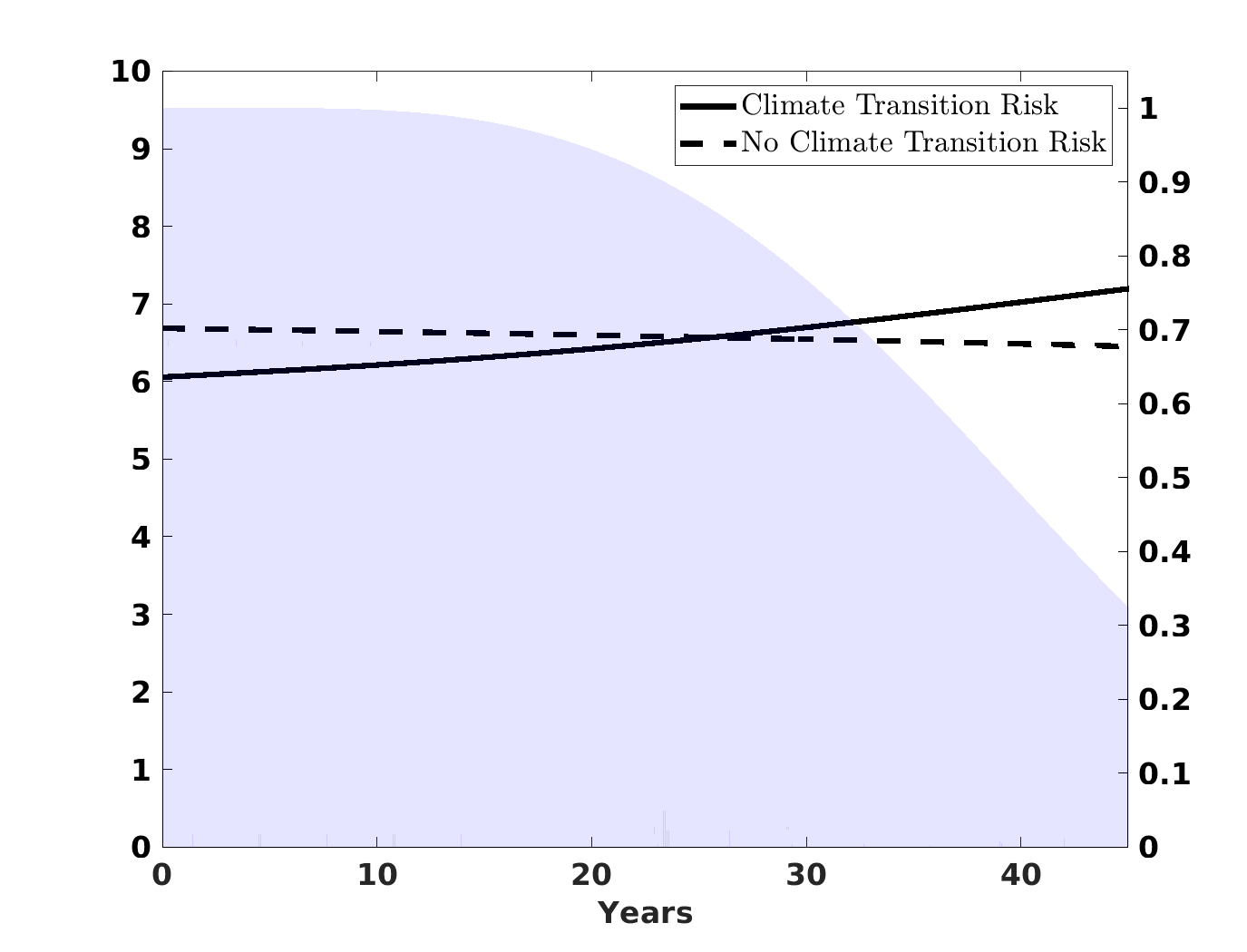}
           \caption[]{{\small Coal Production: $E_{2,t}$}}
        \end{subfigure}     
        
        \begin{subfigure}[b]{0.328\textwidth}
            \centering
            \includegraphics[width=\textwidth]{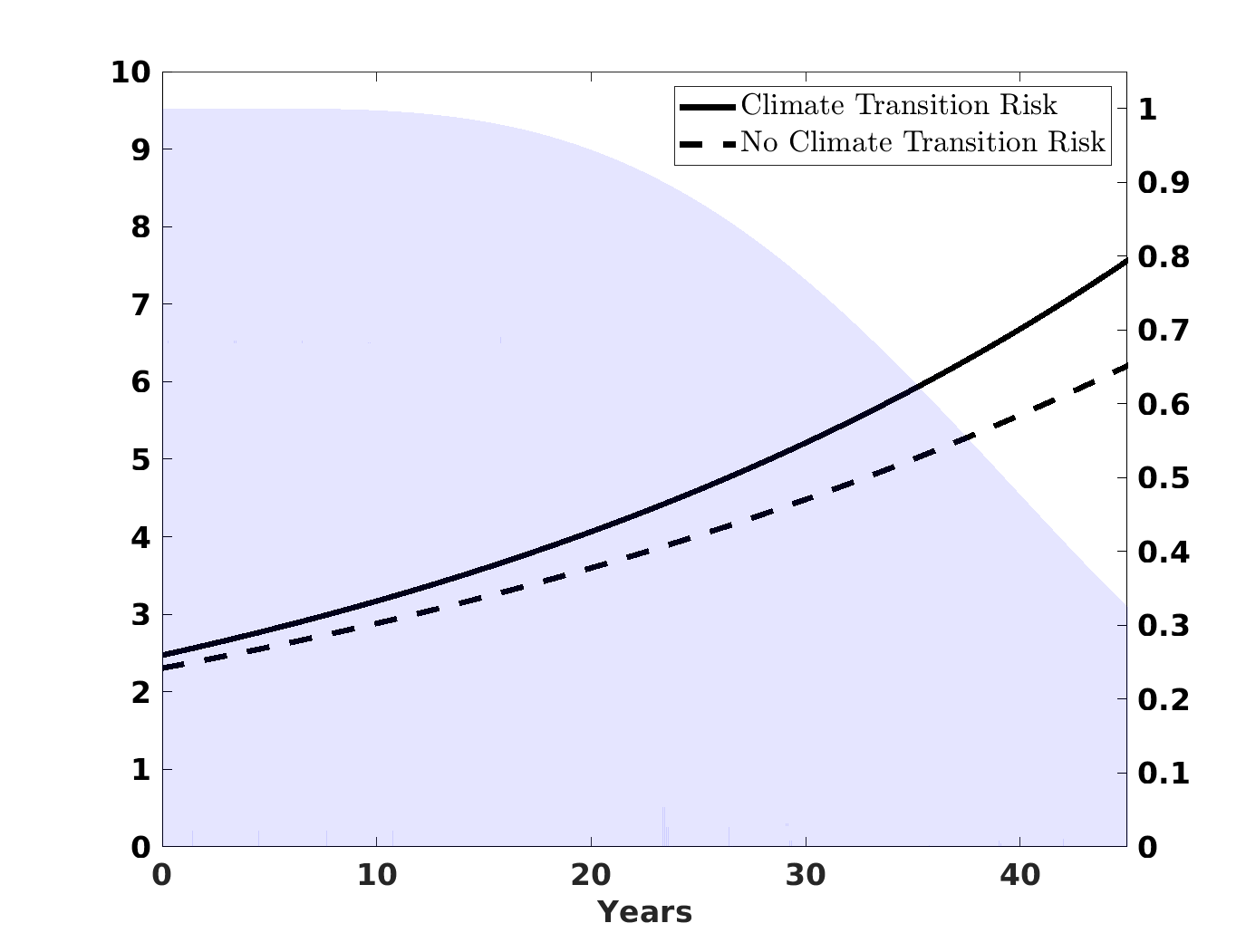}
            \caption[]{{\small Green Investment: $I_{G,t}$}}              
        \end{subfigure}
        \begin{subfigure}[b]{0.328\textwidth}  
           %  \centering 
           %  \includegraphics[width=\textwidth]{figures_submit/p1_justcoal_roff__submit.png}
           % \caption[]{{\small Oil Spot Price: $P_{1,t}$}}
        \end{subfigure}
        \begin{subfigure}[b]{0.328\textwidth}  
            \centering 
            \includegraphics[width=\textwidth]{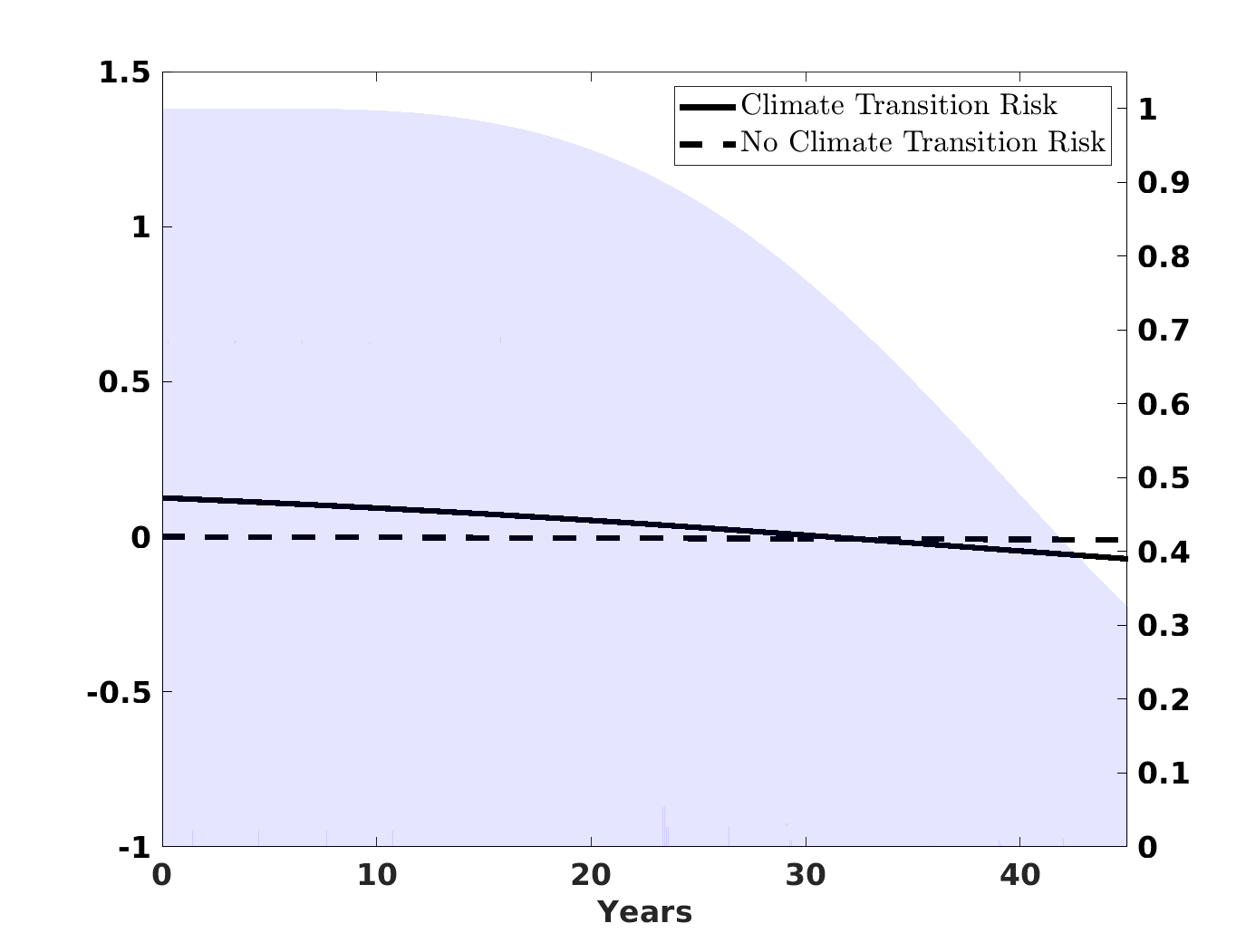}
           \caption[]{{\small Coal Spot Price: $P_{1,t}$}}
        \end{subfigure}

        \begin{subfigure}[b]{0.328\textwidth}
            \centering
            \includegraphics[width=\textwidth]{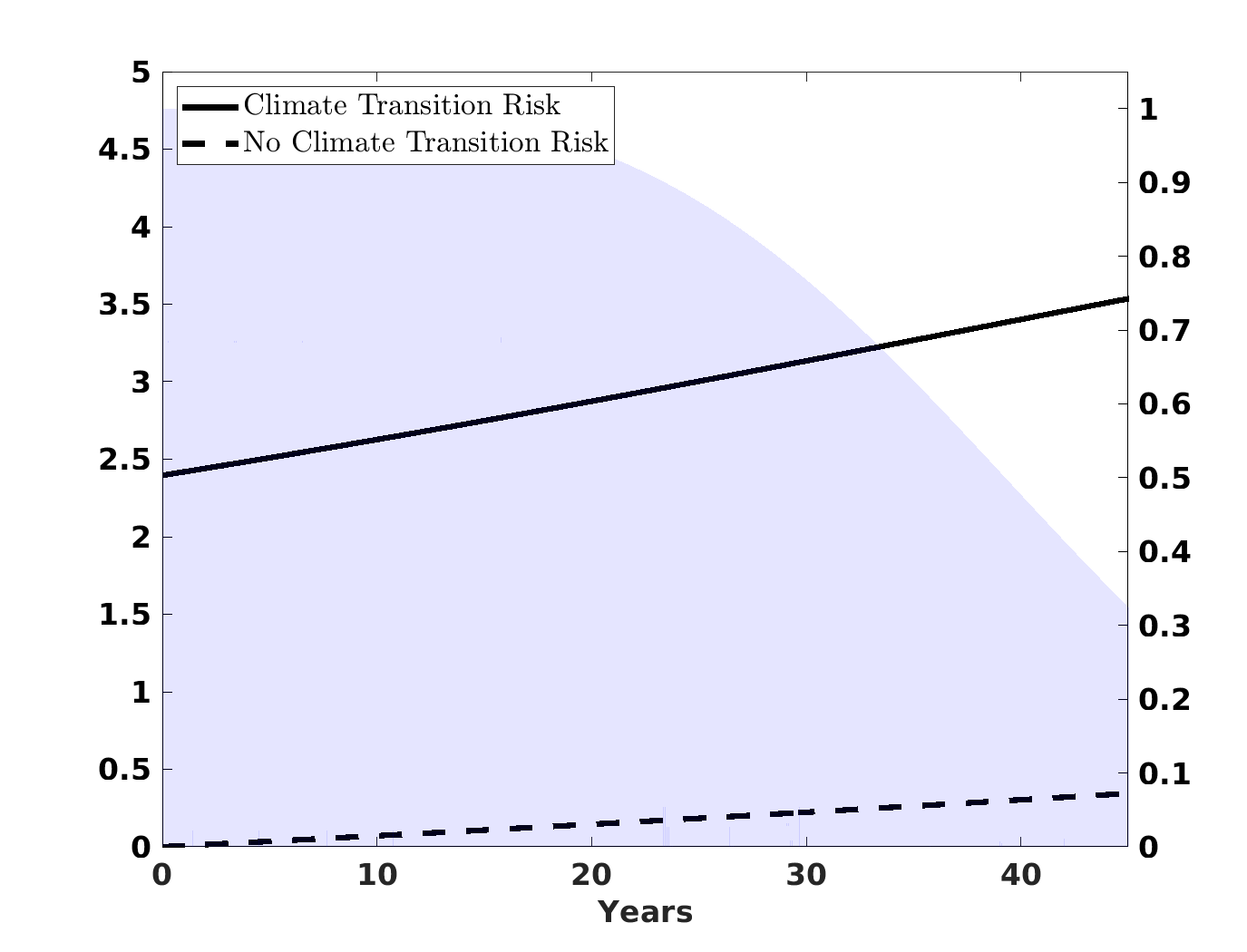}
            \caption[]{{\small Green Firm Price: $S^{(3)}_{t}$}}              
        \end{subfigure}
        \begin{subfigure}[b]{0.328\textwidth}  
           %  \centering 
           %  \includegraphics[width=\textwidth]{figures_submit/s1_justcoal_roff__submit.png}
           % \caption[]{{\small Oil Firm Price: $S^{(1)}_{t}$}}
        \end{subfigure}
        \begin{subfigure}[b]{0.328\textwidth}  
            \centering 
            \includegraphics[width=\textwidth]{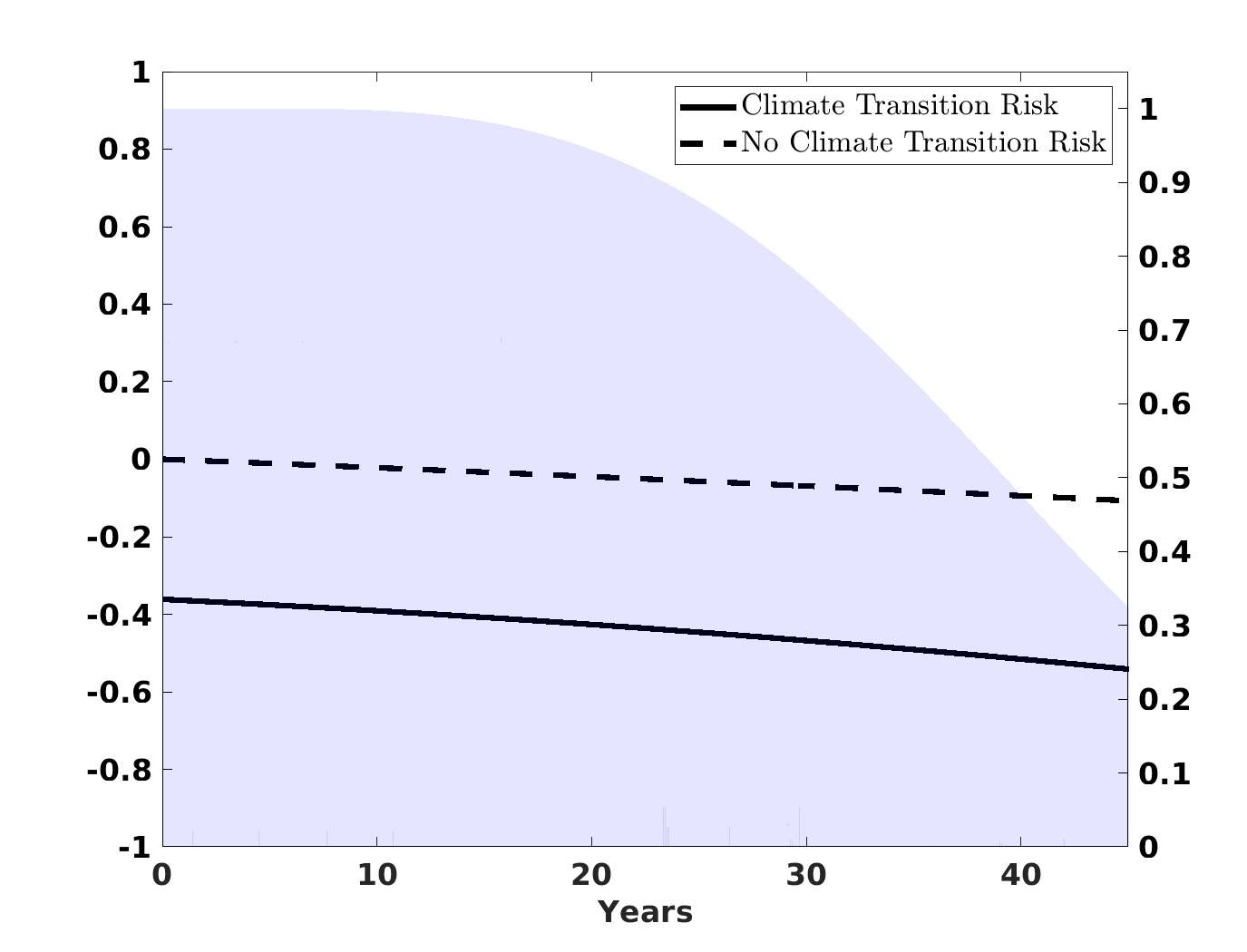}
           \caption[]{{\small Coal Firm Price: $S^{(2)}_{t}$}}
        \end{subfigure}  
        
        \vspace{-0.25cm}
\end{center}

\begin{footnotesize}
Figure \ref{fig:multi2_model_sims} shows the simulated outcomes for the ``coal only technology shock''  model based on the numerical solutions. Panels (a) through (c) show the temperature anomaly, oil production, and coal production. Panels (d) through (f) show the green investment choice, oil spot price, and coal spot price. Panels (g) and (i) show the green firm price, oil firm price, and coal firm price. Solid lines represent results for the Climate Transition Risk scenario where $\lambda_t = \lambda(T_t)$ and dashed lines represent results for the No Climate Transition Risk scenario where $\lambda_t = 0$. The blue shaded region shows the cumulative probability of no transition shock occurring.
\end{footnotesize} 

\end{figure}

% \begin{landscape}

\begin{figure}[!pht]

%\vspace{-1.0cm}
% \vspace{-0.5cm}

\caption{Macroeconomic and Asset Pricing Outcomes - Oil Only ``Technology'' Shock} \label{fig:multi2_model_sims}
\begin{center}
% {\scriptsize \textbf{Panel A: Hybrid Taxation/Technology Transition Scenario}}\\
        \begin{subfigure}[b]{0.328\textwidth}
            \centering
            \includegraphics[width=\textwidth]{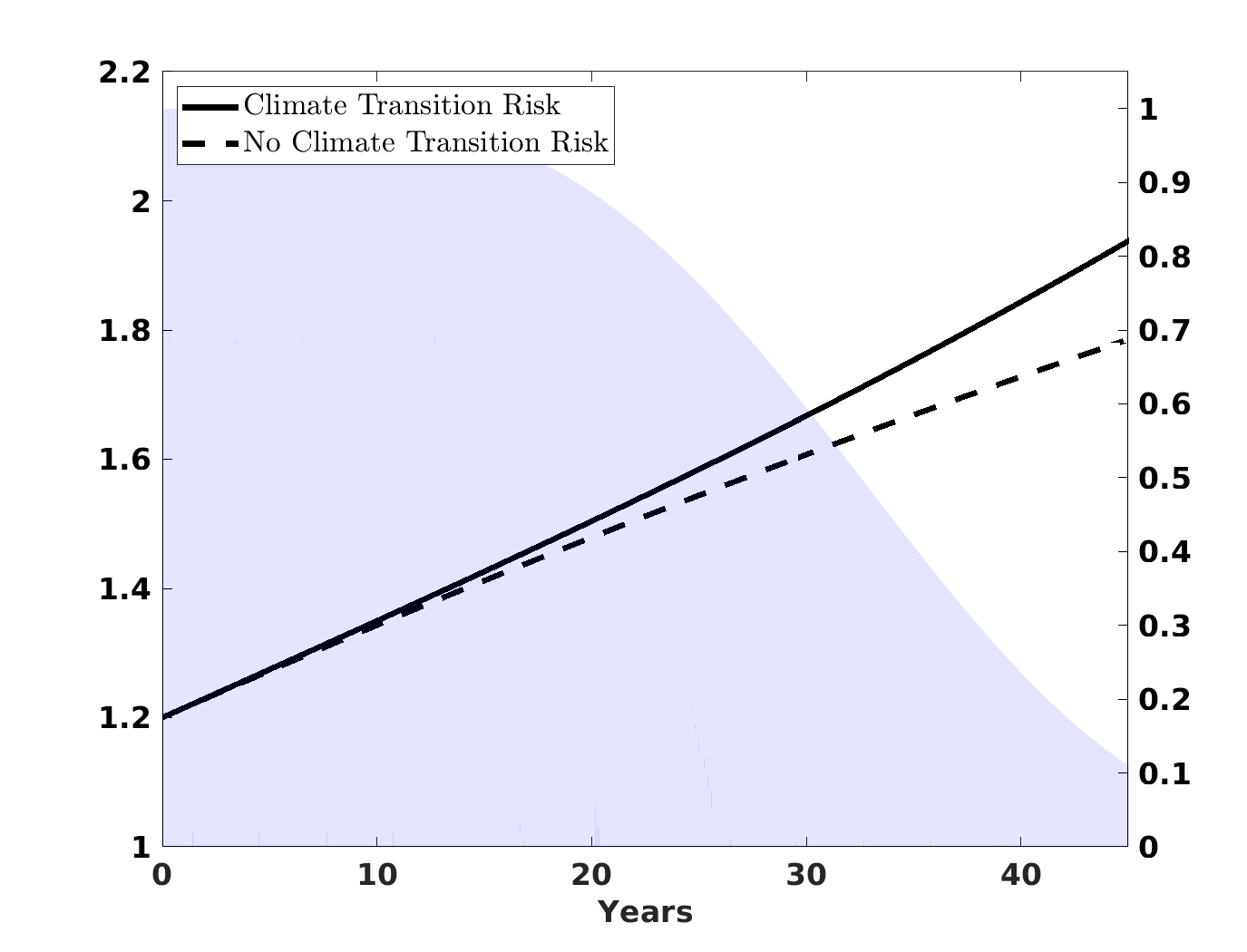}
            \caption[]{{\small Temperature: $Y_t$}}              
        \end{subfigure}
        \begin{subfigure}[b]{0.328\textwidth}  
            \centering 
            \includegraphics[width=\textwidth]{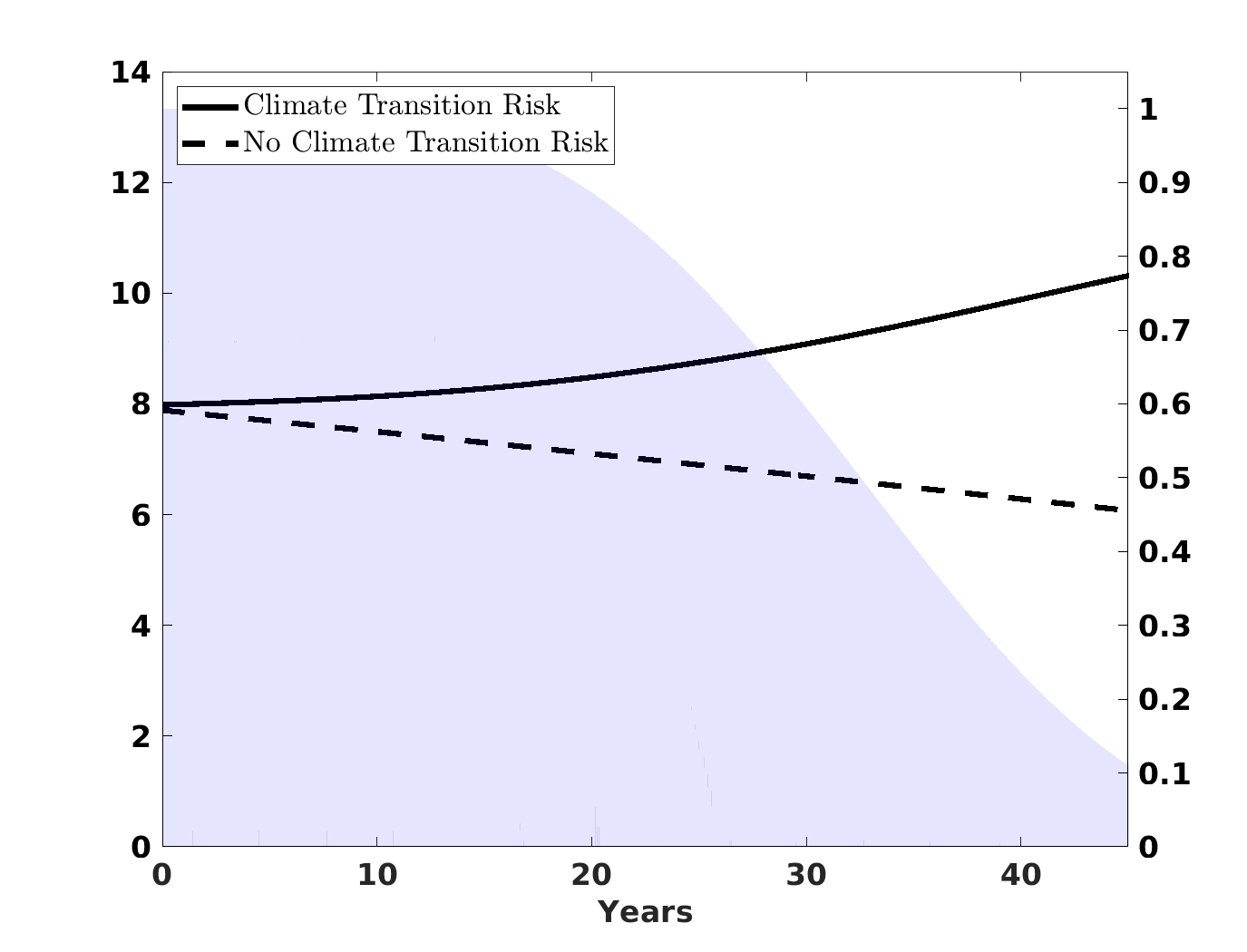}
           \caption[]{{\small Oil Production: $E_{1,t}$}}
        \end{subfigure}
        \begin{subfigure}[b]{0.328\textwidth}  
           %  \centering 
           %  \includegraphics[width=\textwidth]{figures_submit/coal_justoil_roff__submit.png}
           % \caption[]{{\small Coal Production: $E_{2,t}$}}
        \end{subfigure}     
        
        \begin{subfigure}[b]{0.328\textwidth}
            \centering
            \includegraphics[width=\textwidth]{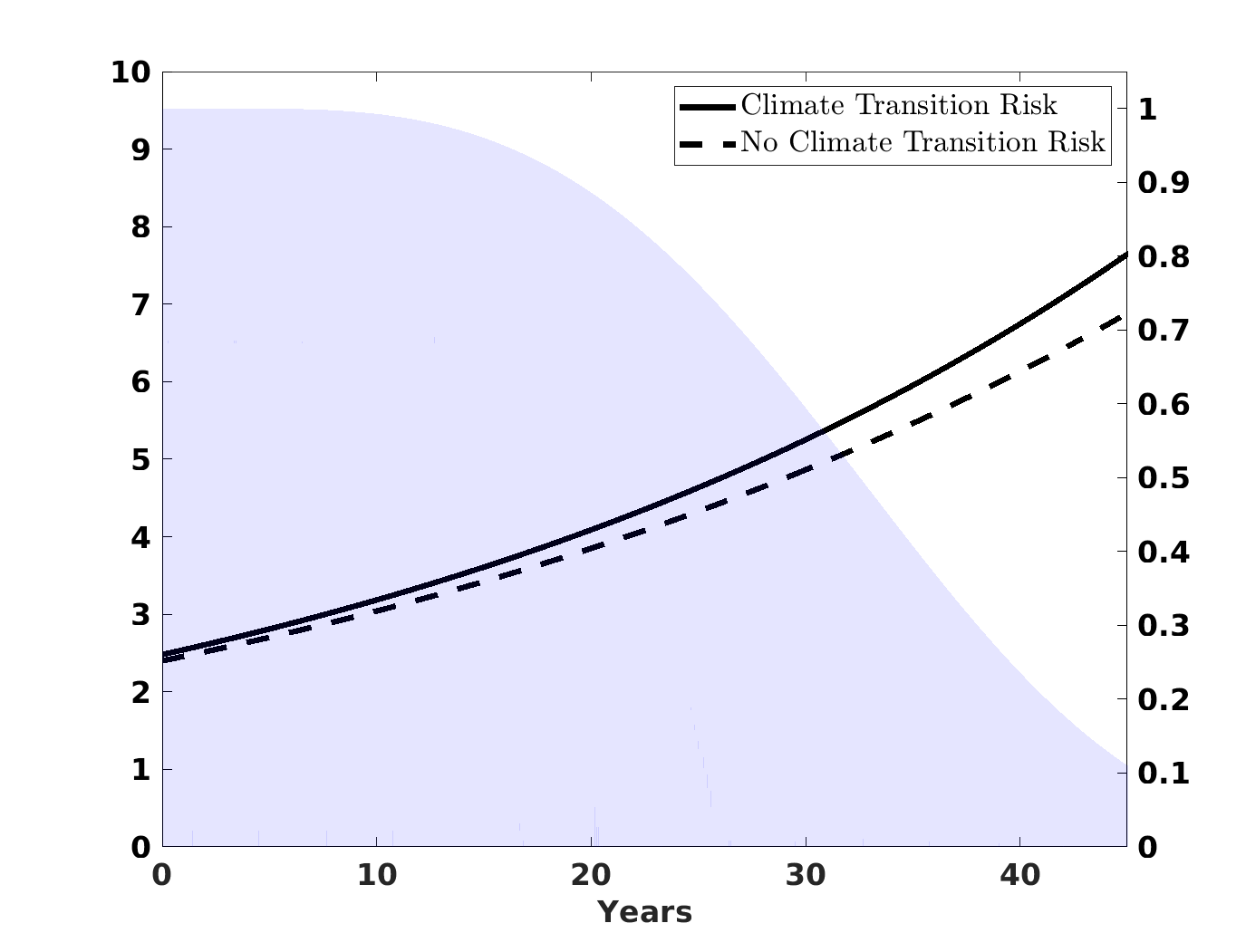}
            \caption[]{{\small Green Investment: $I_{G,t}$}}              
        \end{subfigure}
        \begin{subfigure}[b]{0.328\textwidth}  
            \centering 
            \includegraphics[width=\textwidth]{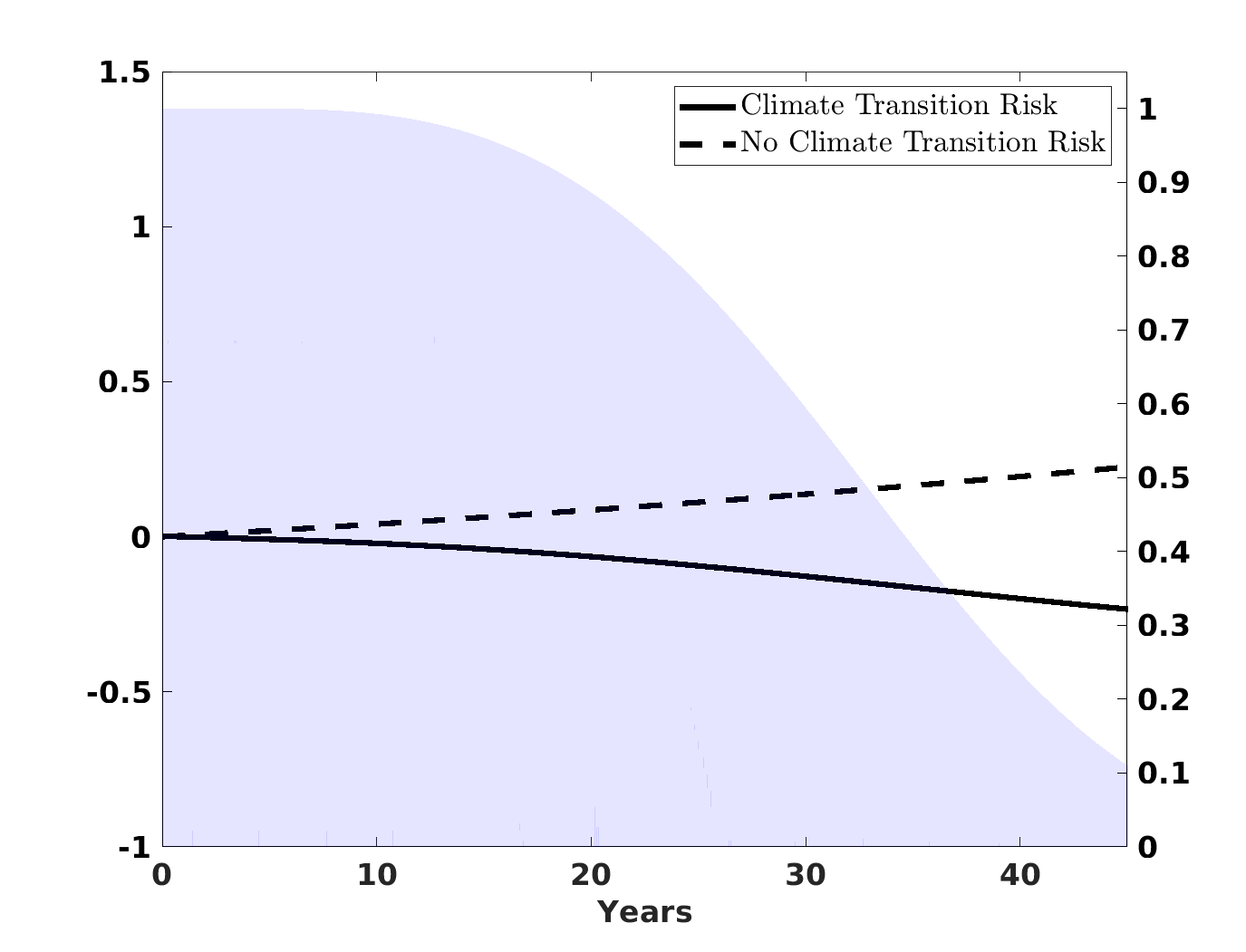}
           \caption[]{{\small Oil Spot Price: $P_{1,t}$}}
        \end{subfigure}
        \begin{subfigure}[b]{0.328\textwidth}  
           %  \centering 
           %  \includegraphics[width=\textwidth]{figures_submit/p2_justoil_roff__submit.png}
           % \caption[]{{\small Coal Spot Price: $P_{1,t}$}}
        \end{subfigure}

        \begin{subfigure}[b]{0.328\textwidth}
            \centering
            \includegraphics[width=\textwidth]{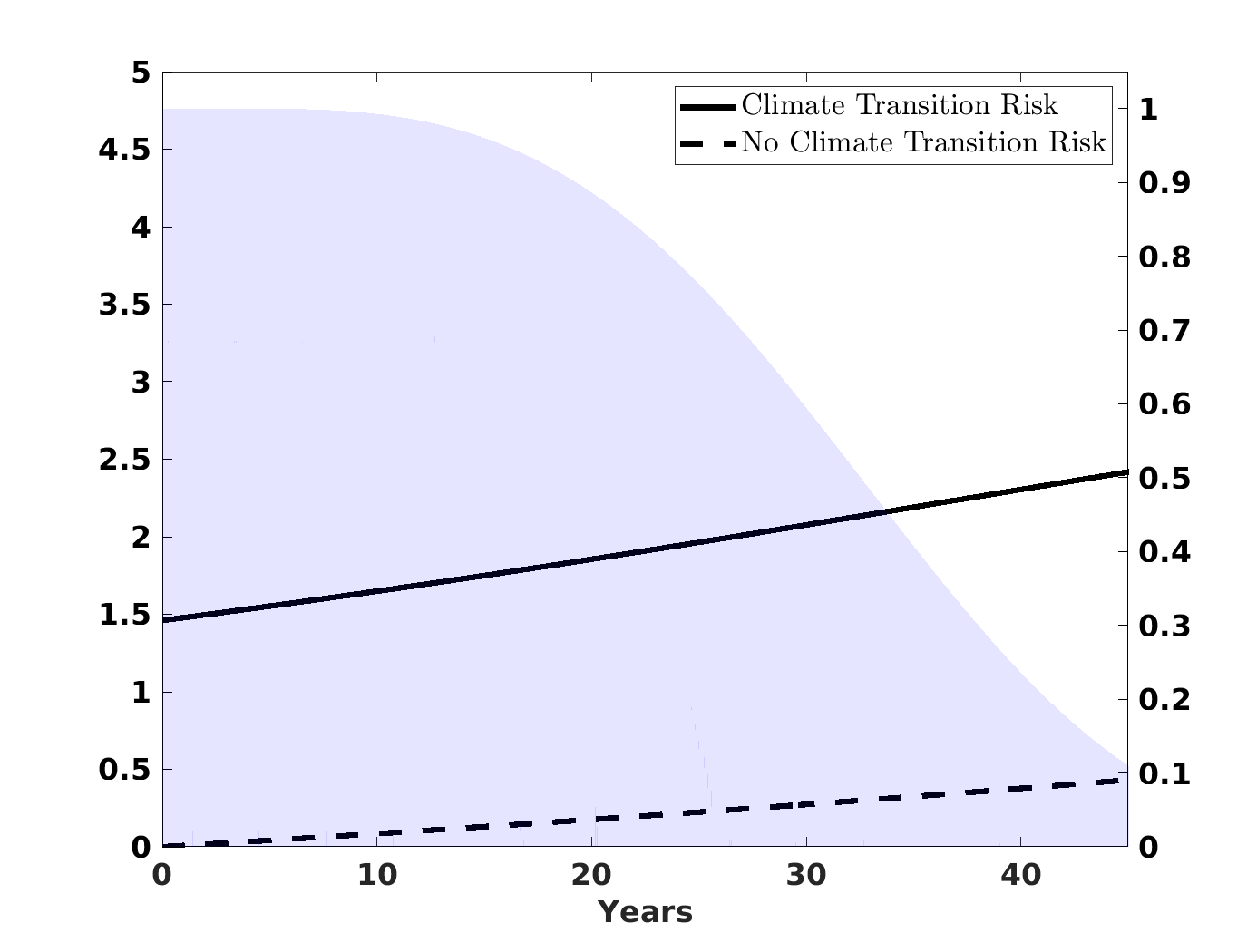}
            \caption[]{{\small Green Firm Price: $S^{(3)}_{t}$}}              
        \end{subfigure}
        \begin{subfigure}[b]{0.328\textwidth}  
            \centering 
            \includegraphics[width=\textwidth]{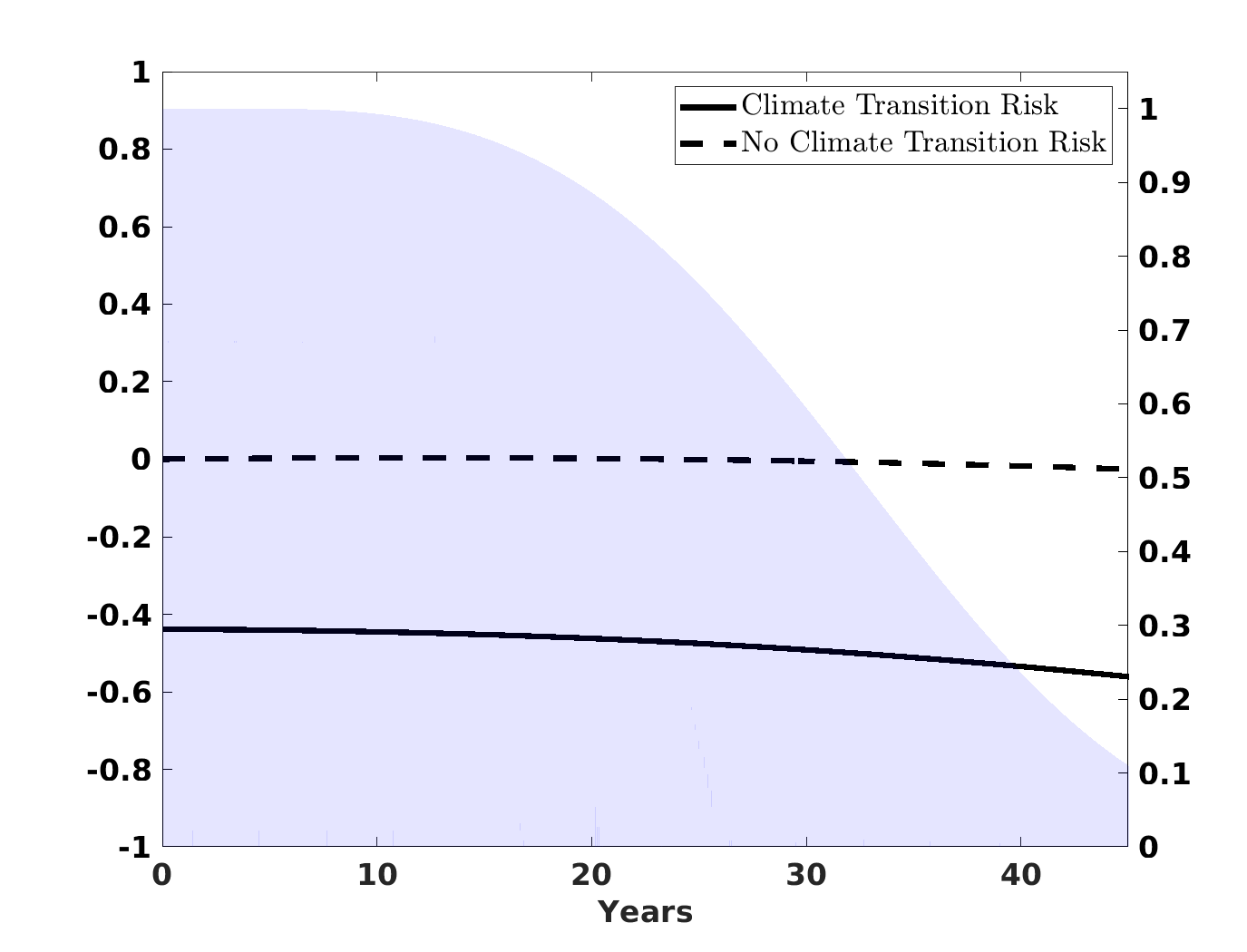}
           \caption[]{{\small Oil Firm Price: $S^{(1)}_{t}$}}
        \end{subfigure}
        \begin{subfigure}[b]{0.328\textwidth}  
            % \centering 
           %  \includegraphics[width=\textwidth]{figures_submit/s2_justoil_roff__submit.png}
           % \caption[]{{\small Coal Firm Price: $S^{(2)}_{t}$}}
        \end{subfigure}  
        
        \vspace{-0.25cm}
\end{center}

\begin{footnotesize}
Figure \ref{fig:multi2_model_sims} shows the simulated outcomes for the ``oil only technology shock''  model based on the numerical solutions. Panels (a) through (c) show the temperature anomaly, oil production, and coal production. Panels (d) through (f) show the green investment choice, oil spot price, and coal spot price. Panels (g) and (i) show the green firm price, oil firm price, and coal firm price. Solid lines represent results for the Climate Transition Risk scenario where $\lambda_t = \lambda(T_t)$ and dashed lines represent results for the No Climate Transition Risk scenario where $\lambda_t = 0$. The blue shaded region shows the cumulative probability of no transition shock occurring.
\end{footnotesize} 

\end{figure}

% \subsubsection{Coal then Oil Transition Shocks}

% \begin{landscape}

\begin{figure}[!pht]

%\vspace{-1.0cm}
% \vspace{-0.5cm}

\caption{Macroeconomic and Asset Pricing Outcomes - Coal then Oil ``Taxation'' Shocks} \label{fig:multi2_model_sims}
\begin{center}
% {\scriptsize \textbf{Panel A: Hybrid Taxation/Technology Transition Scenario}}\\
        \begin{subfigure}[b]{0.328\textwidth}
            \centering
            \includegraphics[width=\textwidth]{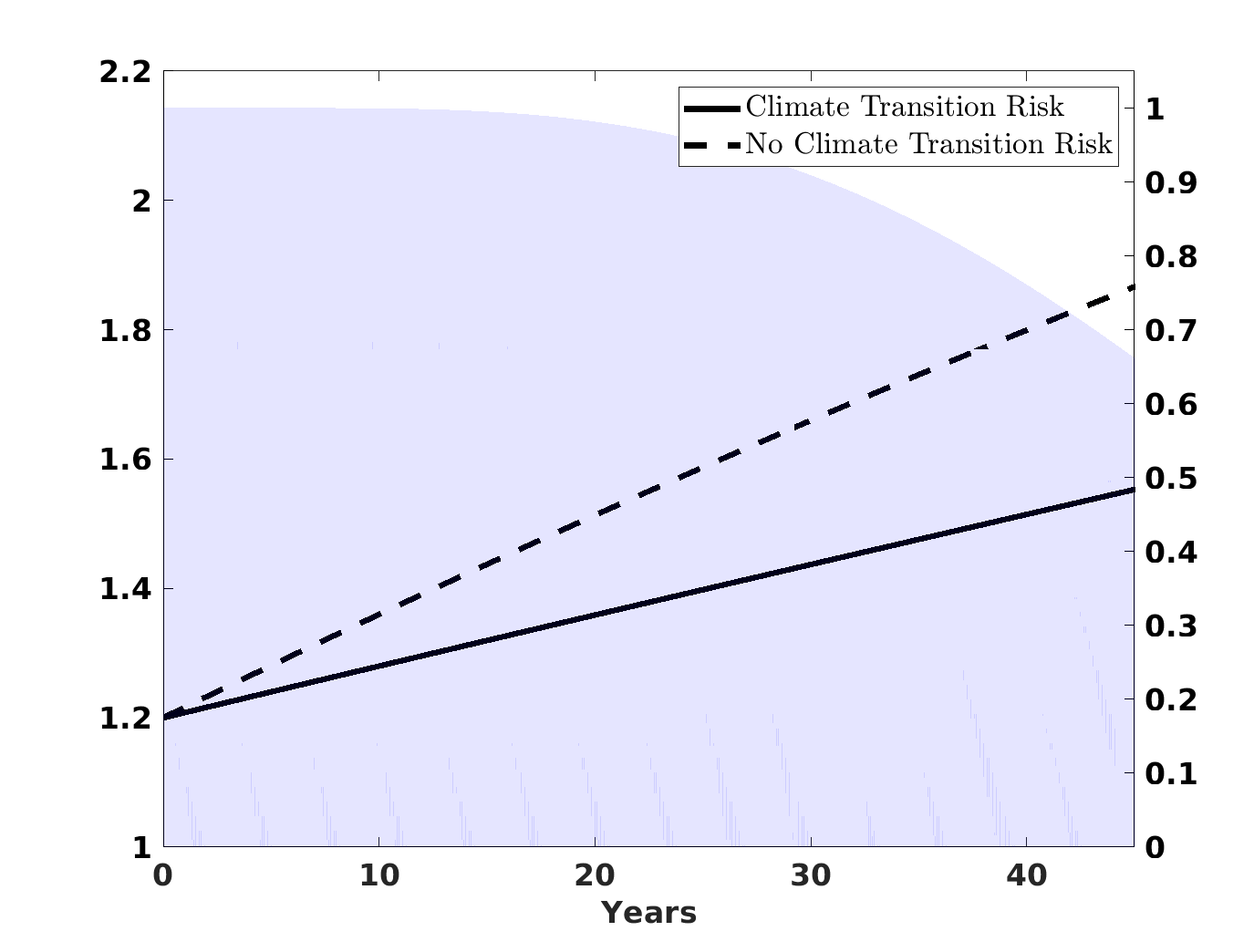}
            \caption[]{{\small Temperature: $Y_t$}}              
        \end{subfigure}
        \begin{subfigure}[b]{0.328\textwidth}  
            \centering 
            \includegraphics[width=\textwidth]{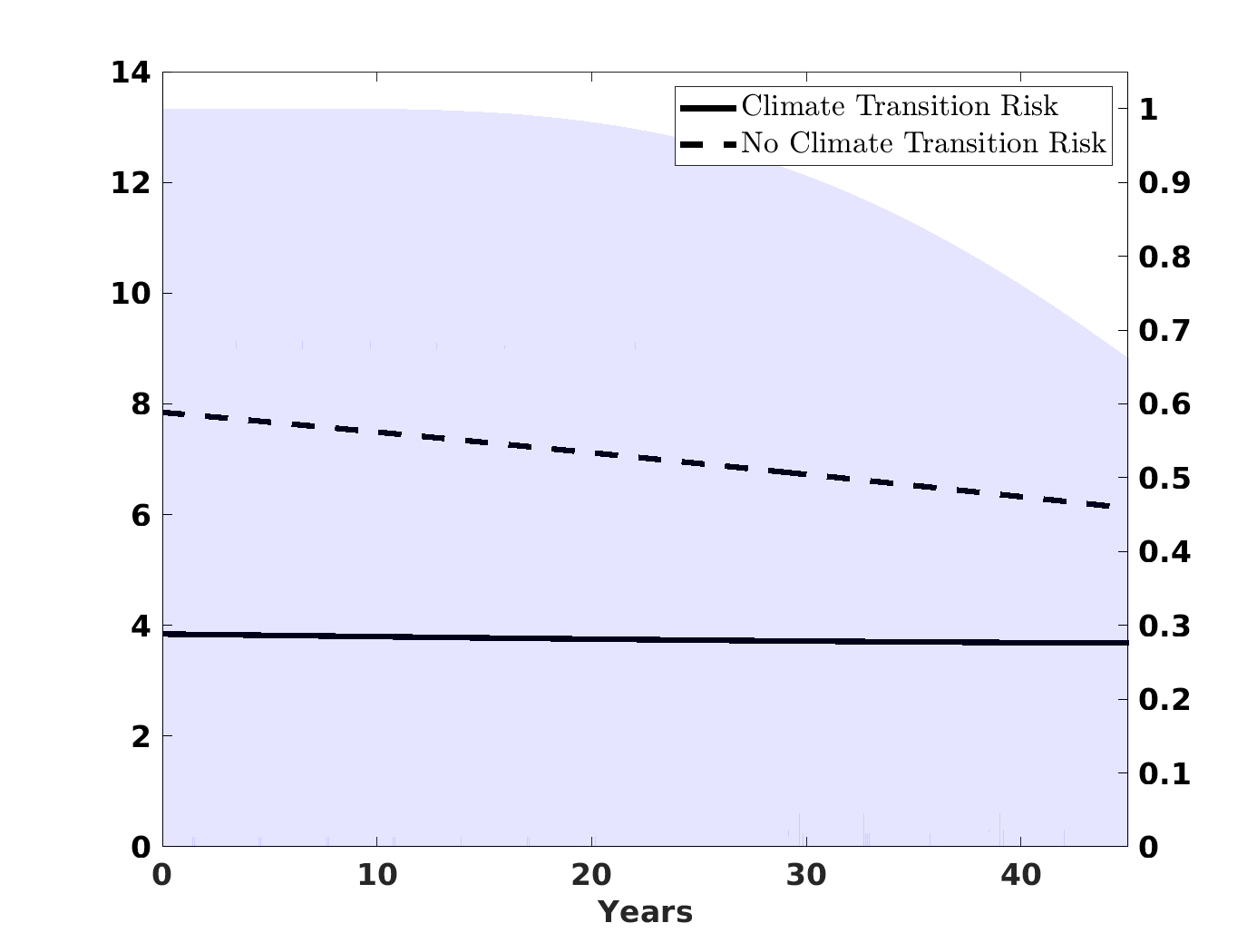}
           \caption[]{{\small Oil Production: $E_{1,t}$}}
        \end{subfigure}
        \begin{subfigure}[b]{0.328\textwidth}  
            \centering 
            \includegraphics[width=\textwidth]{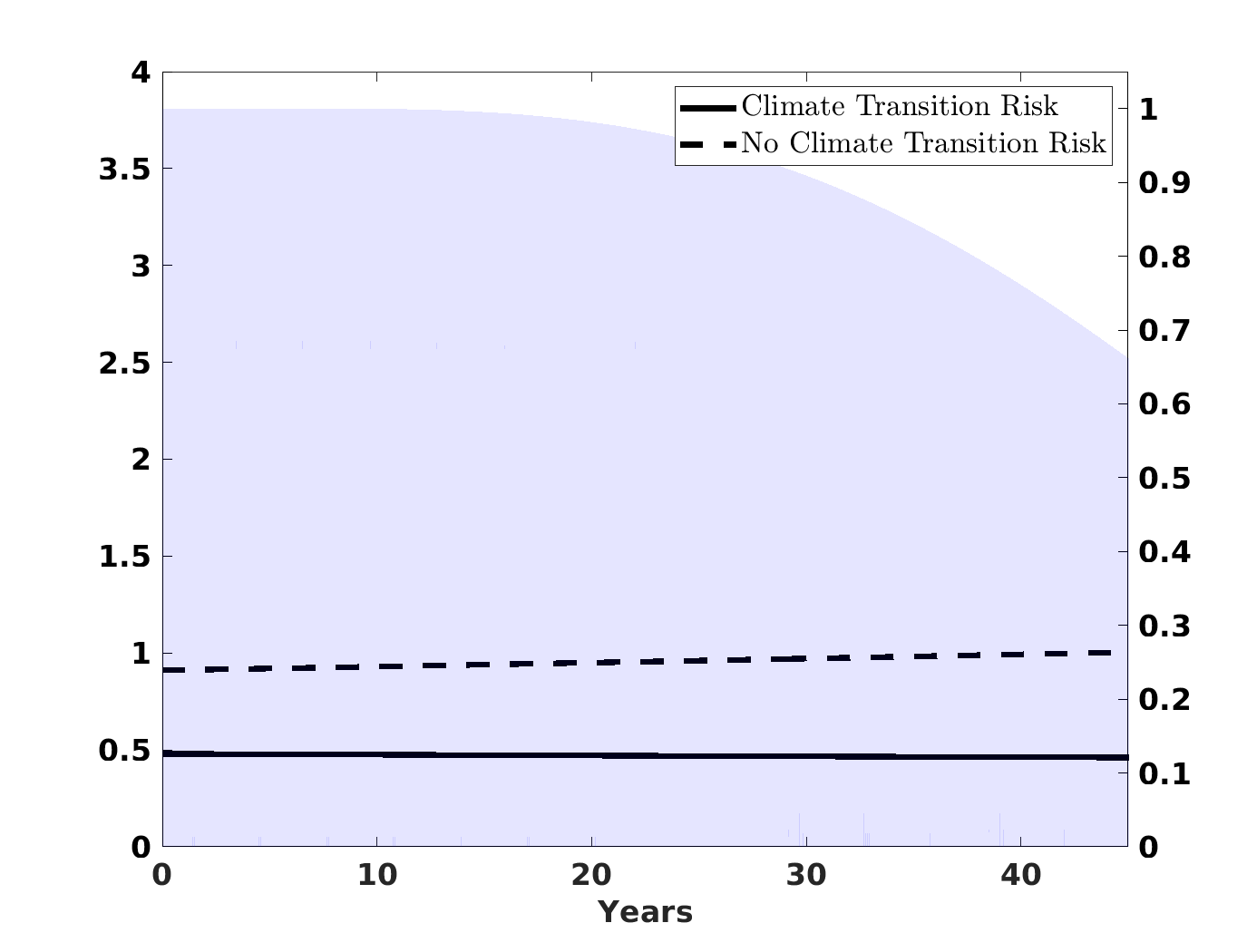}
           \caption[]{{\small Coal Production: $E_{2,t}$}}
        \end{subfigure}     
        
        \begin{subfigure}[b]{0.328\textwidth}
            \centering
            \includegraphics[width=\textwidth]{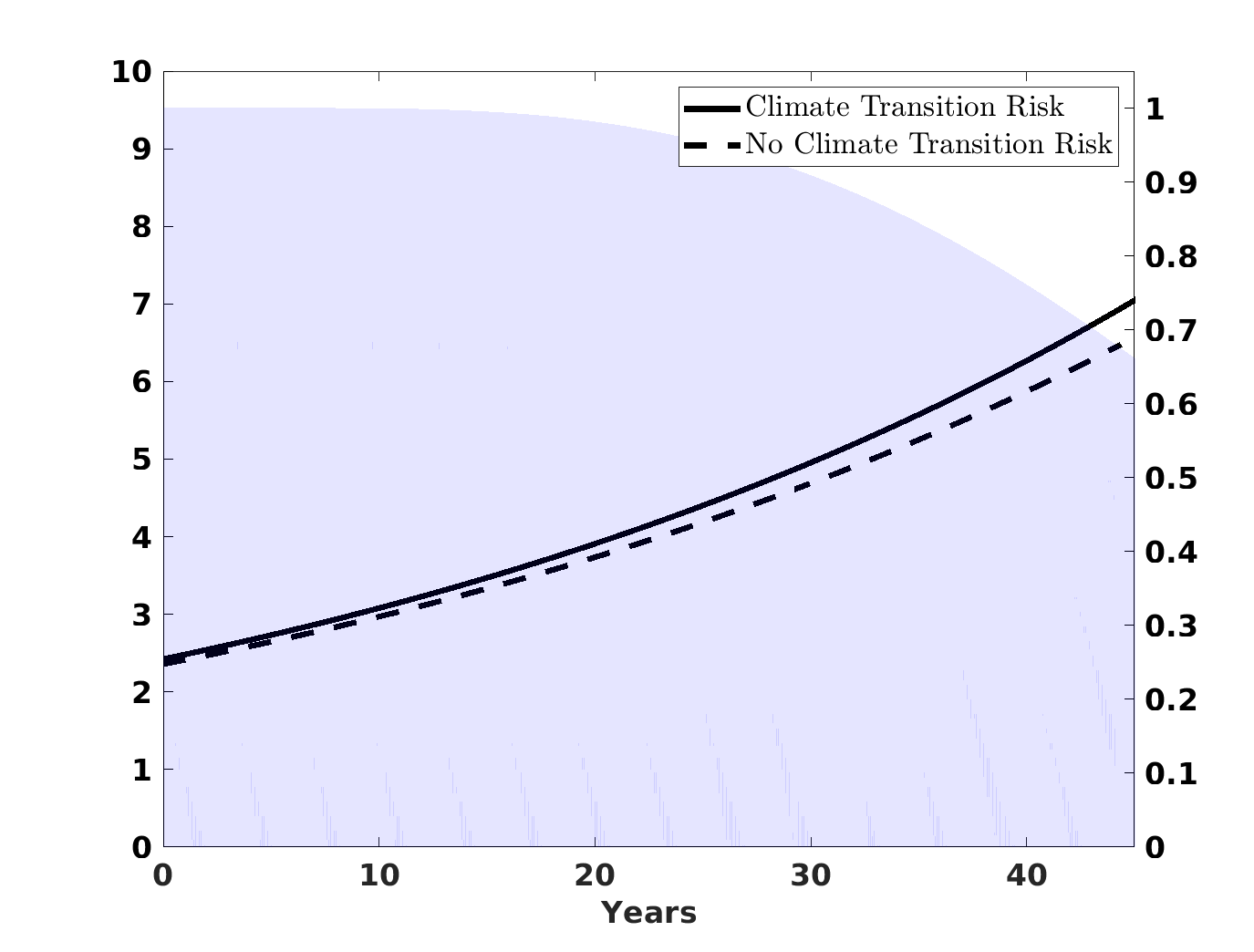}
            \caption[]{{\small Green Investment: $I_{G,t}$}}              
        \end{subfigure}
        \begin{subfigure}[b]{0.328\textwidth}  
            \centering 
            \includegraphics[width=\textwidth]{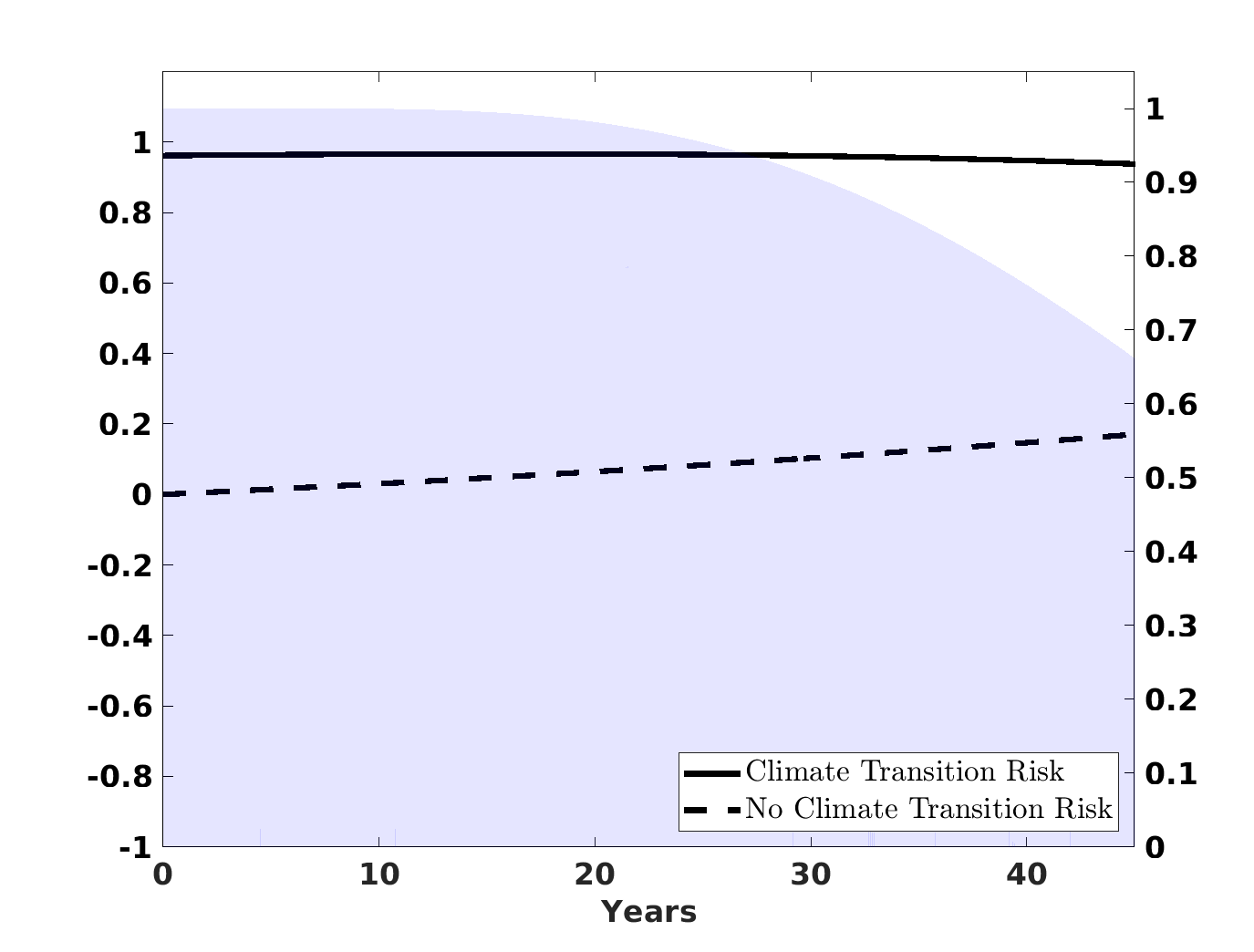}
           \caption[]{{\small Oil Spot Price: $P_{1,t}$}}
        \end{subfigure}
        \begin{subfigure}[b]{0.328\textwidth}  
            \centering 
            \includegraphics[width=\textwidth]{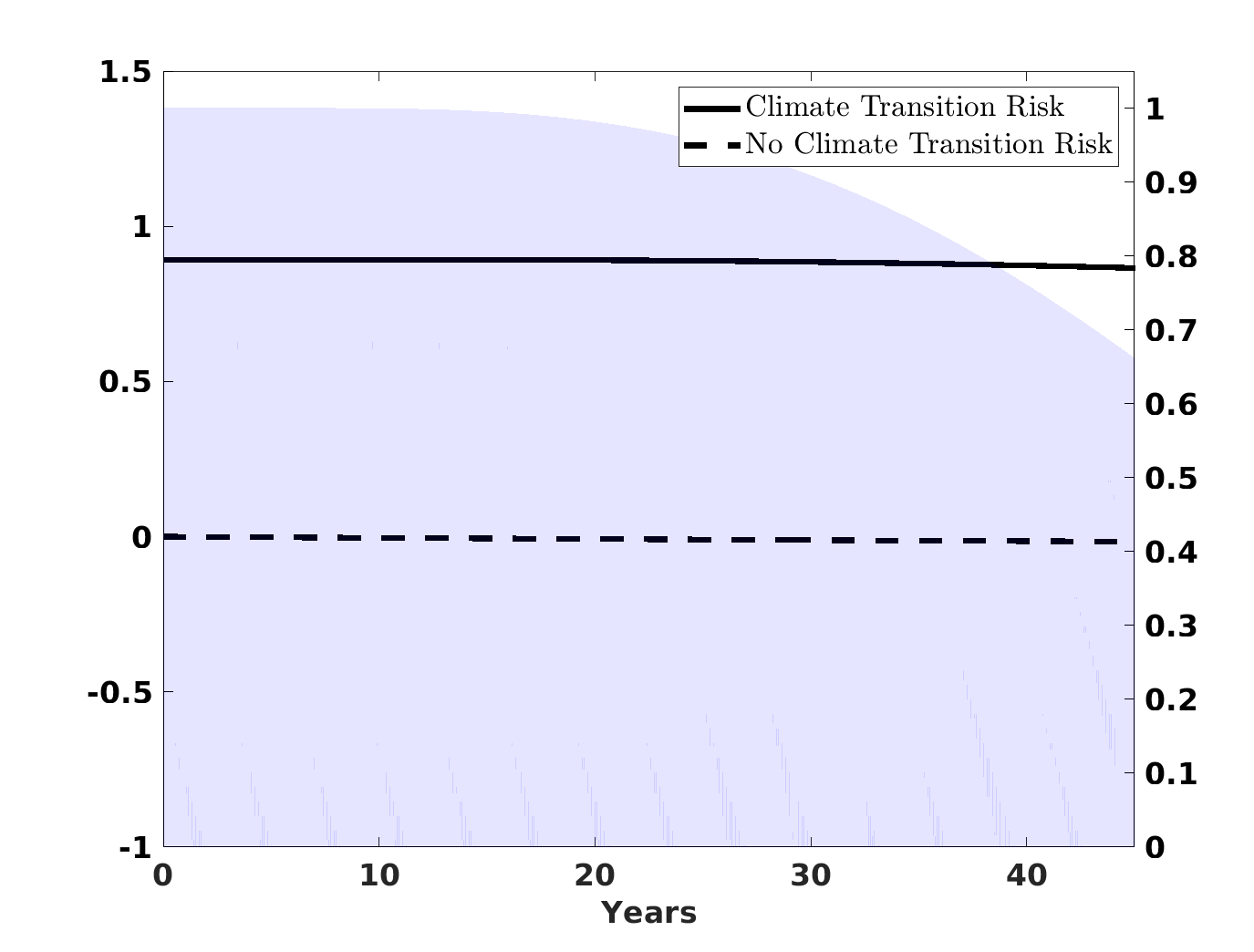}
           \caption[]{{\small Coal Spot Price: $P_{1,t}$}}
        \end{subfigure}

        \begin{subfigure}[b]{0.328\textwidth}
            \centering
            \includegraphics[width=\textwidth]{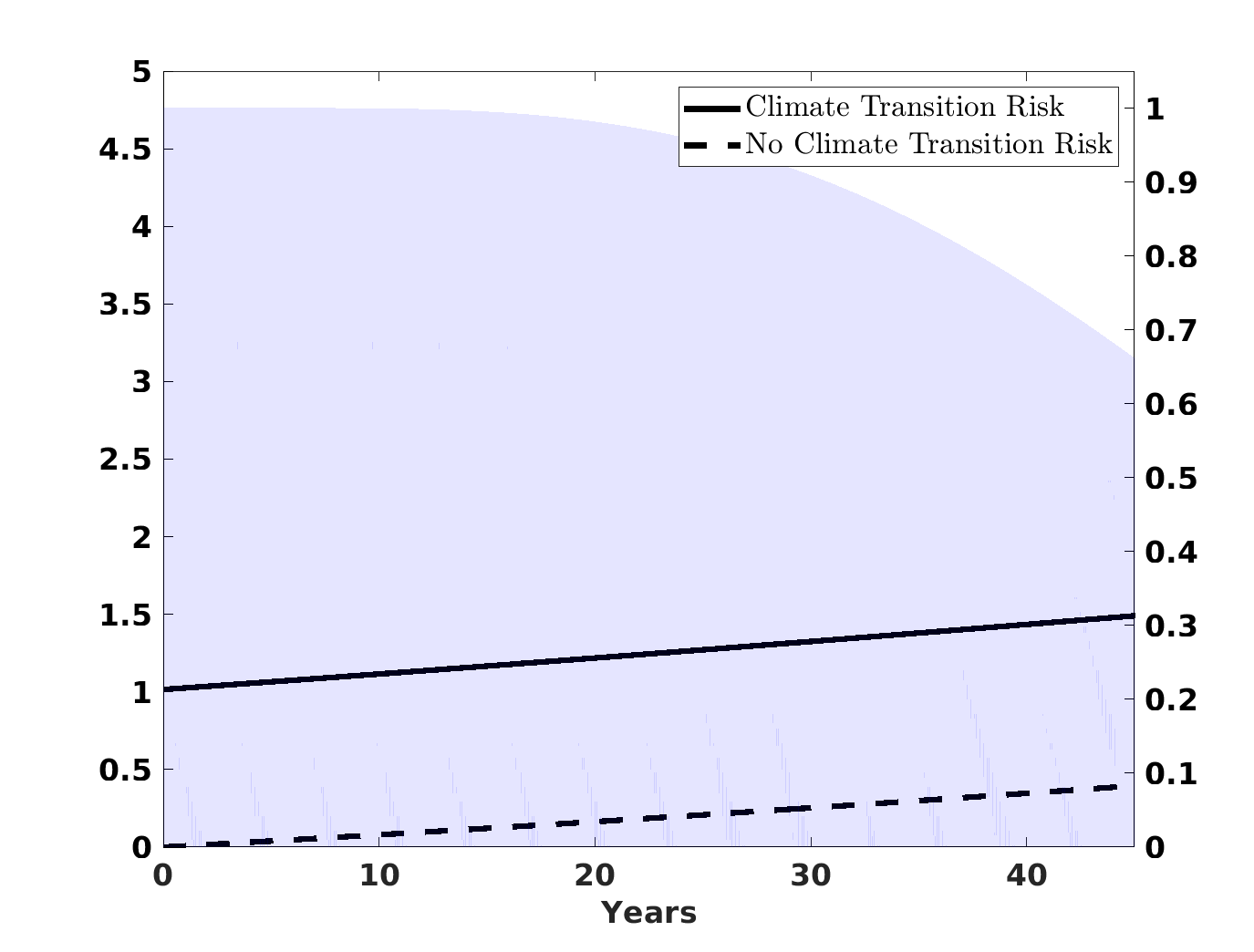}
            \caption[]{{\small Green Firm Price: $S^{(3)}_{t}$}}              
        \end{subfigure}
        \begin{subfigure}[b]{0.328\textwidth}  
            \centering 
            \includegraphics[width=\textwidth]{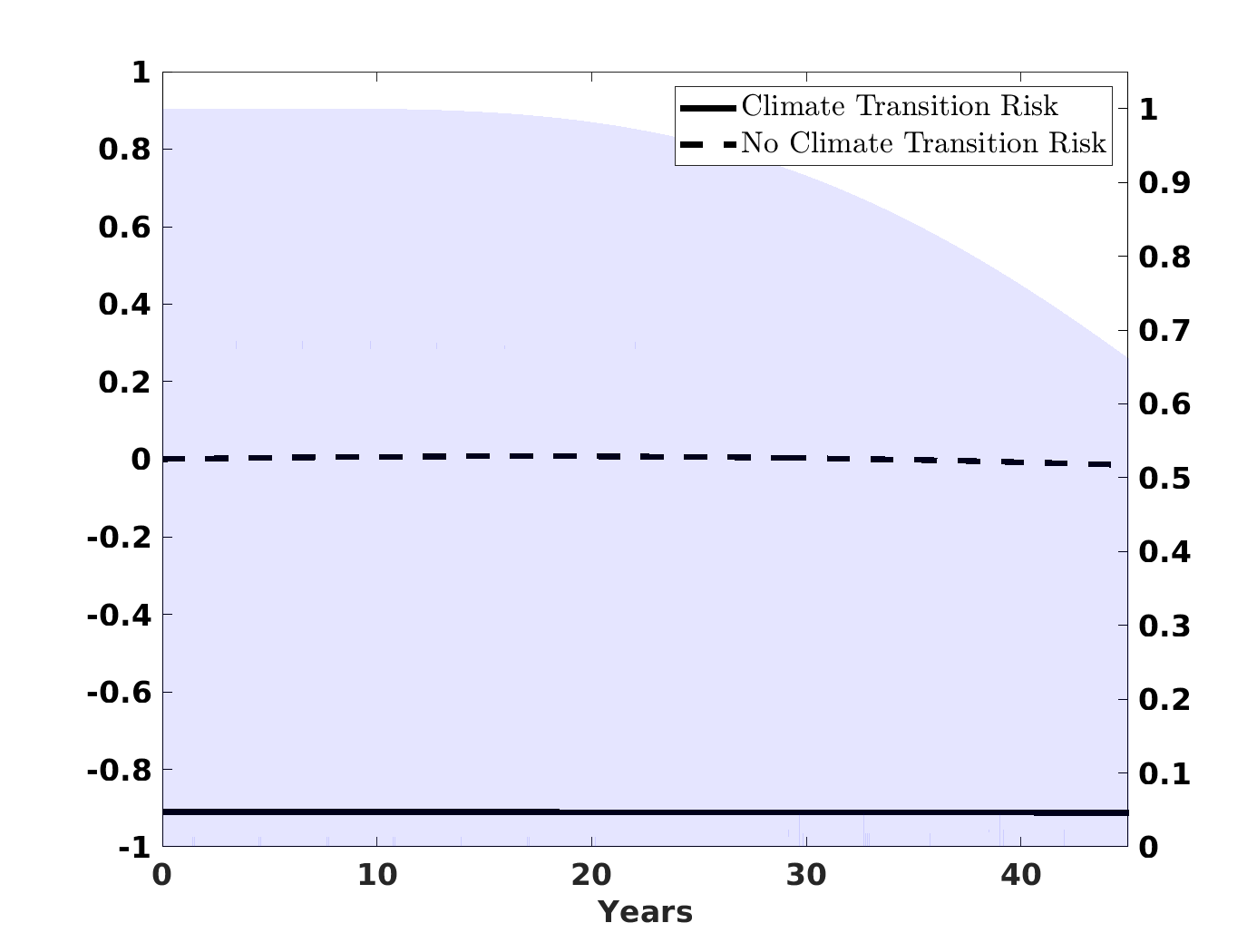}
           \caption[]{{\small Oil Firm Price: $S^{(1)}_{t}$}}
        \end{subfigure}
        \begin{subfigure}[b]{0.328\textwidth}  
            \centering 
            \includegraphics[width=\textwidth]{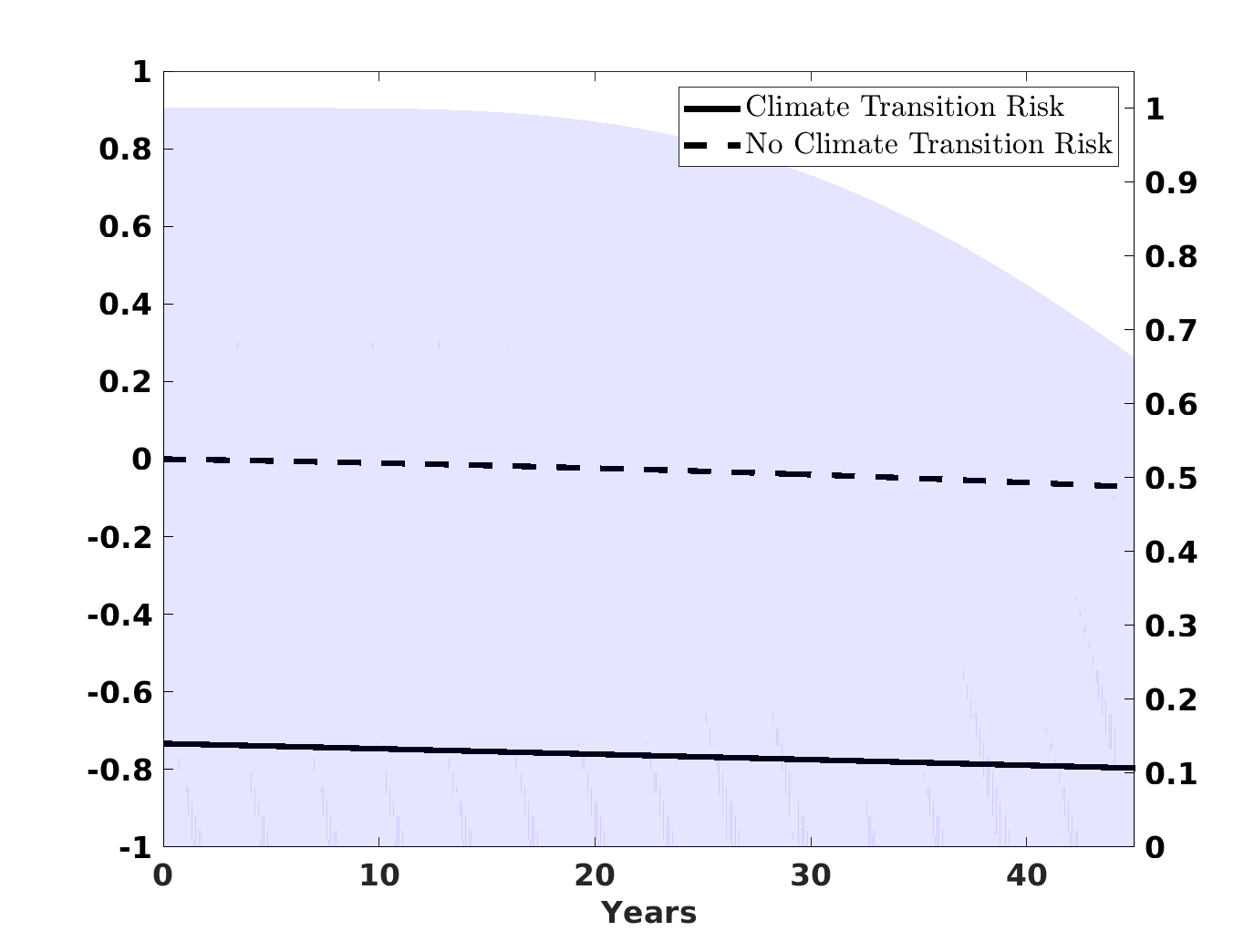}
           \caption[]{{\small Coal Firm Price: $S^{(2)}_{t}$}}
        \end{subfigure}  
        
        \vspace{-0.25cm}
\end{center}

\begin{footnotesize}
Figure \ref{fig:multi2_model_sims} shows the simulated outcomes for the ``coal then oil taxation shock''  model based on the numerical solutions. Panels (a) through (c) show the temperature anomaly, oil production, and coal production. Panels (d) through (f) show the green investment choice, oil spot price, and coal spot price. Panels (g) and (i) show the green firm price, oil firm price, and coal firm price. Solid lines represent results for the Climate Transition Risk scenario where $\lambda_t = \lambda(T_t)$ and dashed lines represent results for the No Climate Transition Risk scenario where $\lambda_t = 0$. The blue shaded region shows the cumulative probability of no transition shock occurring.
\end{footnotesize} 

\end{figure}

% \begin{landscape}

\begin{figure}[!pht]

%\vspace{-1.0cm}
% \vspace{-0.5cm}

\caption{Macroeconomic and Asset Pricing Outcomes - Coal then Oil ``Technology'' Shocks} \label{fig:multi2_model_sims}
\begin{center}
% {\scriptsize \textbf{Panel A: Hybrid Taxation/Technology Transition Scenario}}\\
        \begin{subfigure}[b]{0.328\textwidth}
            \centering
            \includegraphics[width=\textwidth]{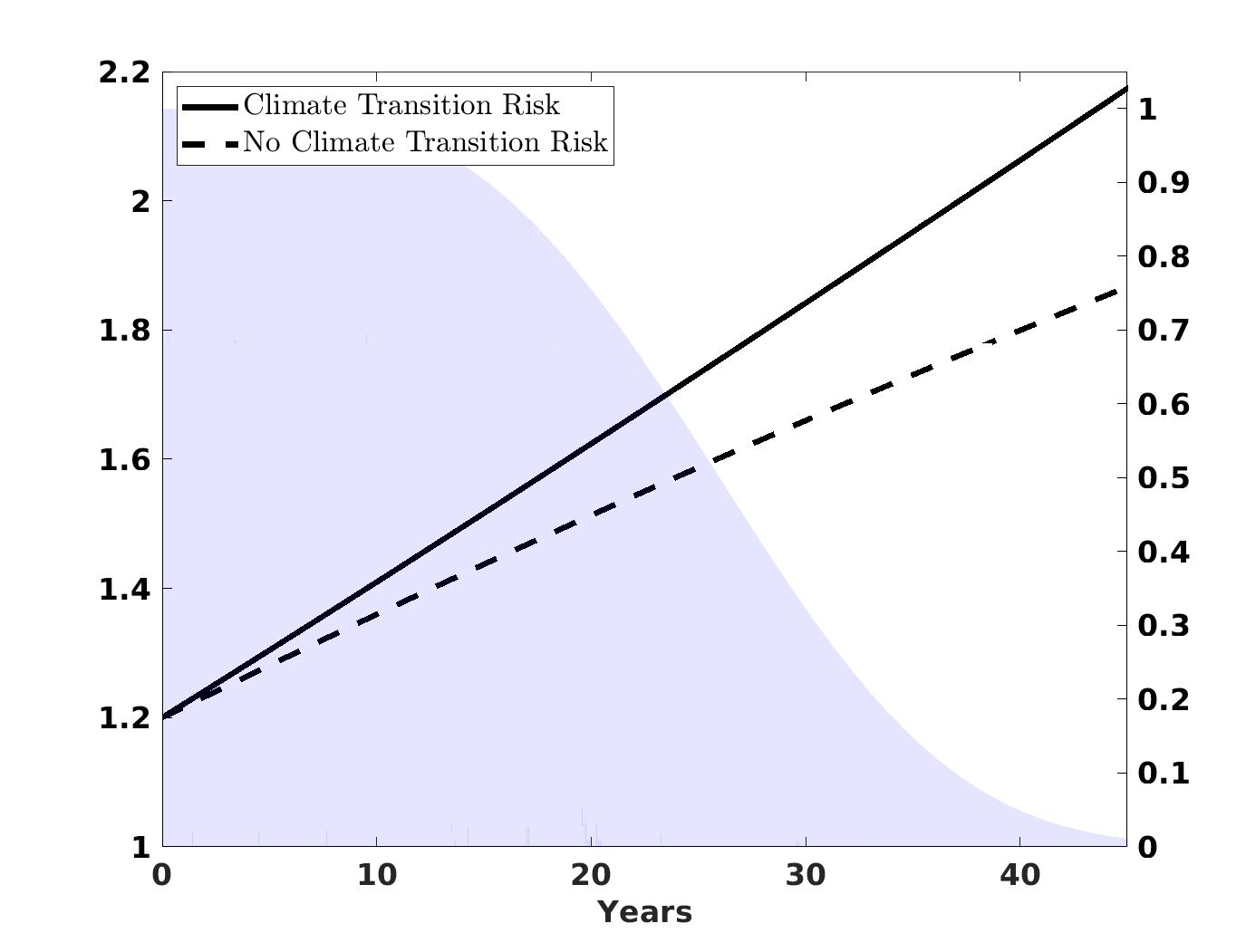}
            \caption[]{{\small Temperature: $Y_t$}}              
        \end{subfigure}
        \begin{subfigure}[b]{0.328\textwidth}  
            \centering 
            \includegraphics[width=\textwidth]{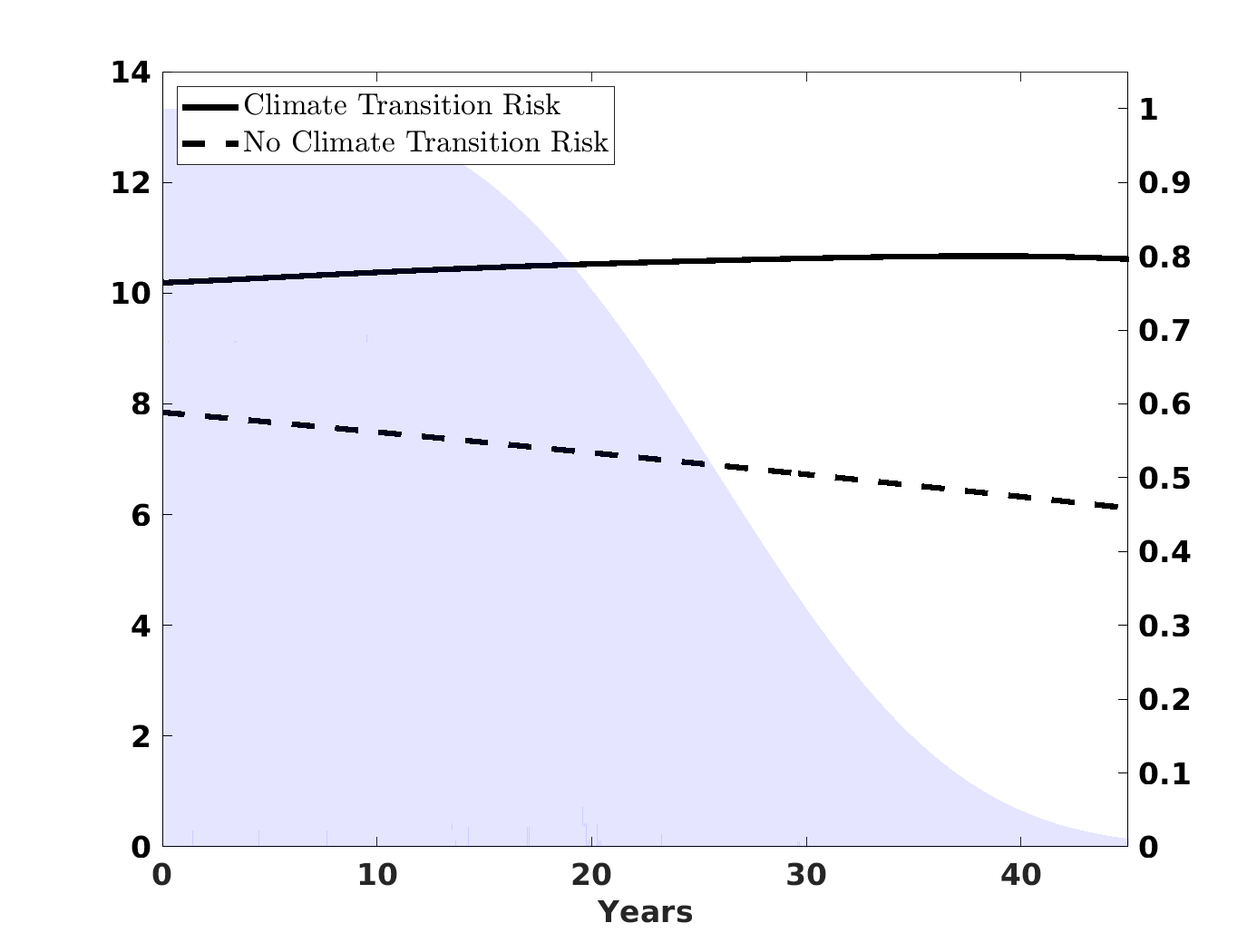}
           \caption[]{{\small Oil Production: $E_{1,t}$}}
        \end{subfigure}
        \begin{subfigure}[b]{0.328\textwidth}  
            \centering 
            \includegraphics[width=\textwidth]{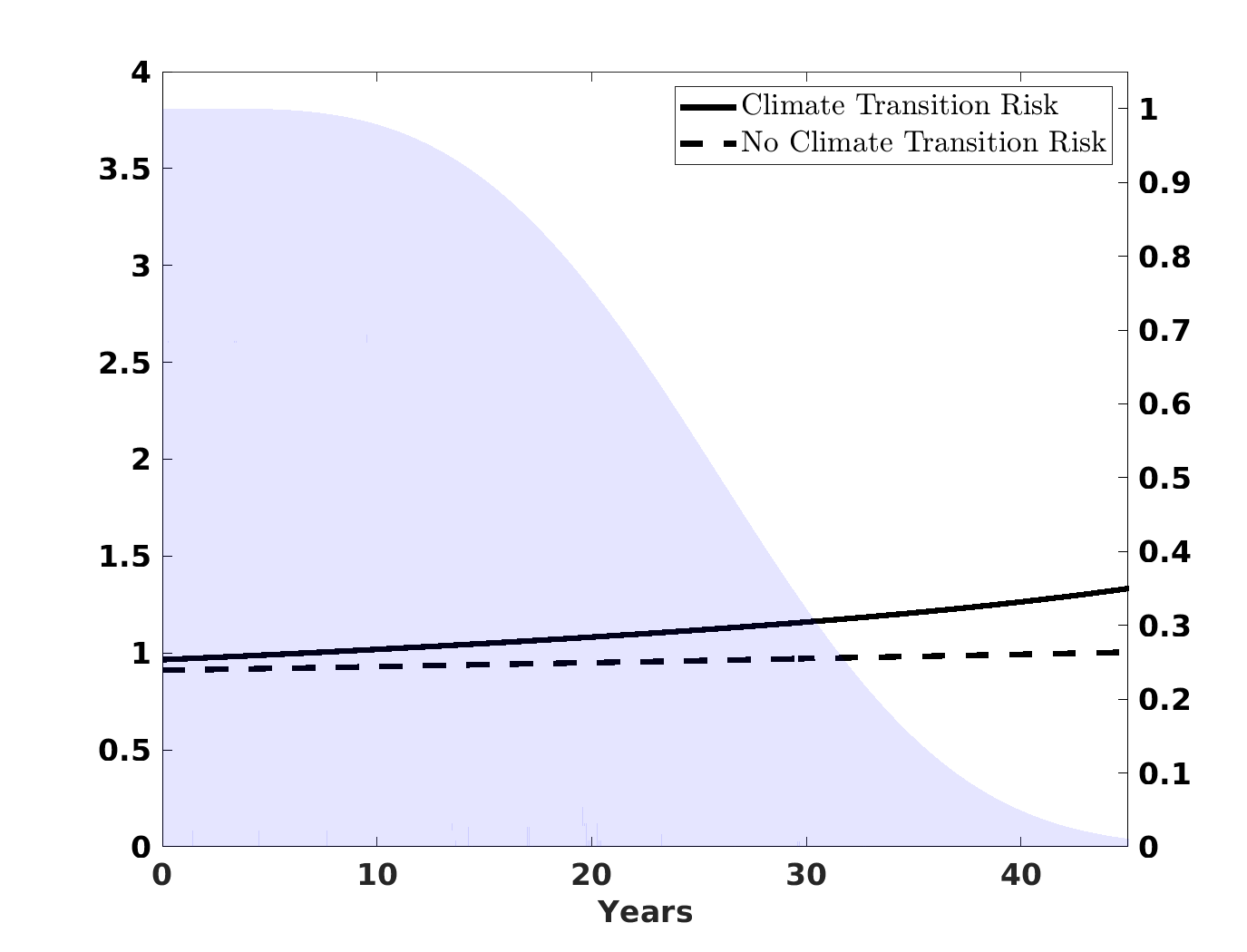}
           \caption[]{{\small Coal Production: $E_{2,t}$}}
        \end{subfigure}     
        
        \begin{subfigure}[b]{0.328\textwidth}
            \centering
            \includegraphics[width=\textwidth]{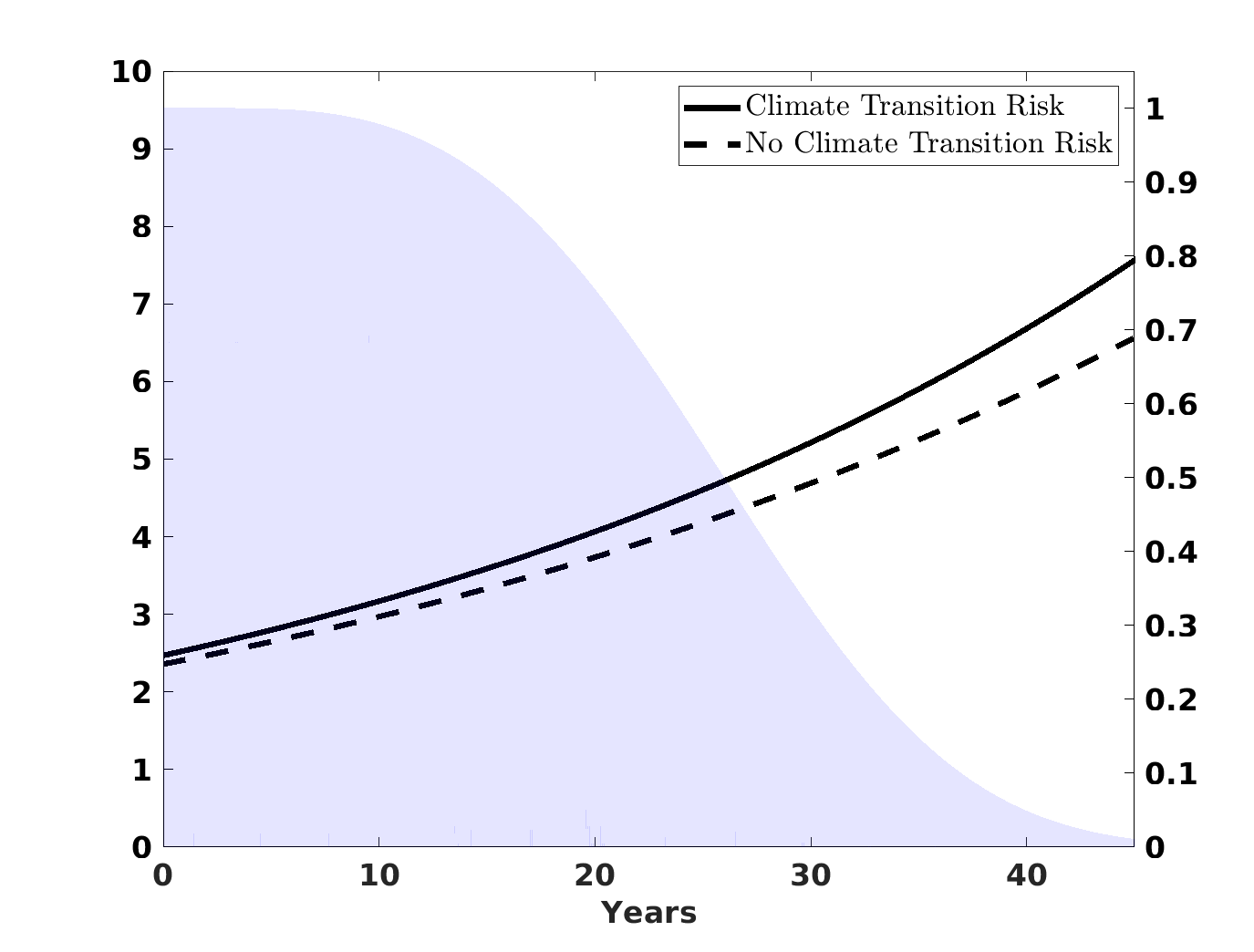}
            \caption[]{{\small Green Investment: $I_{G,t}$}}              
        \end{subfigure}
        \begin{subfigure}[b]{0.328\textwidth}  
            \centering 
            \includegraphics[width=\textwidth]{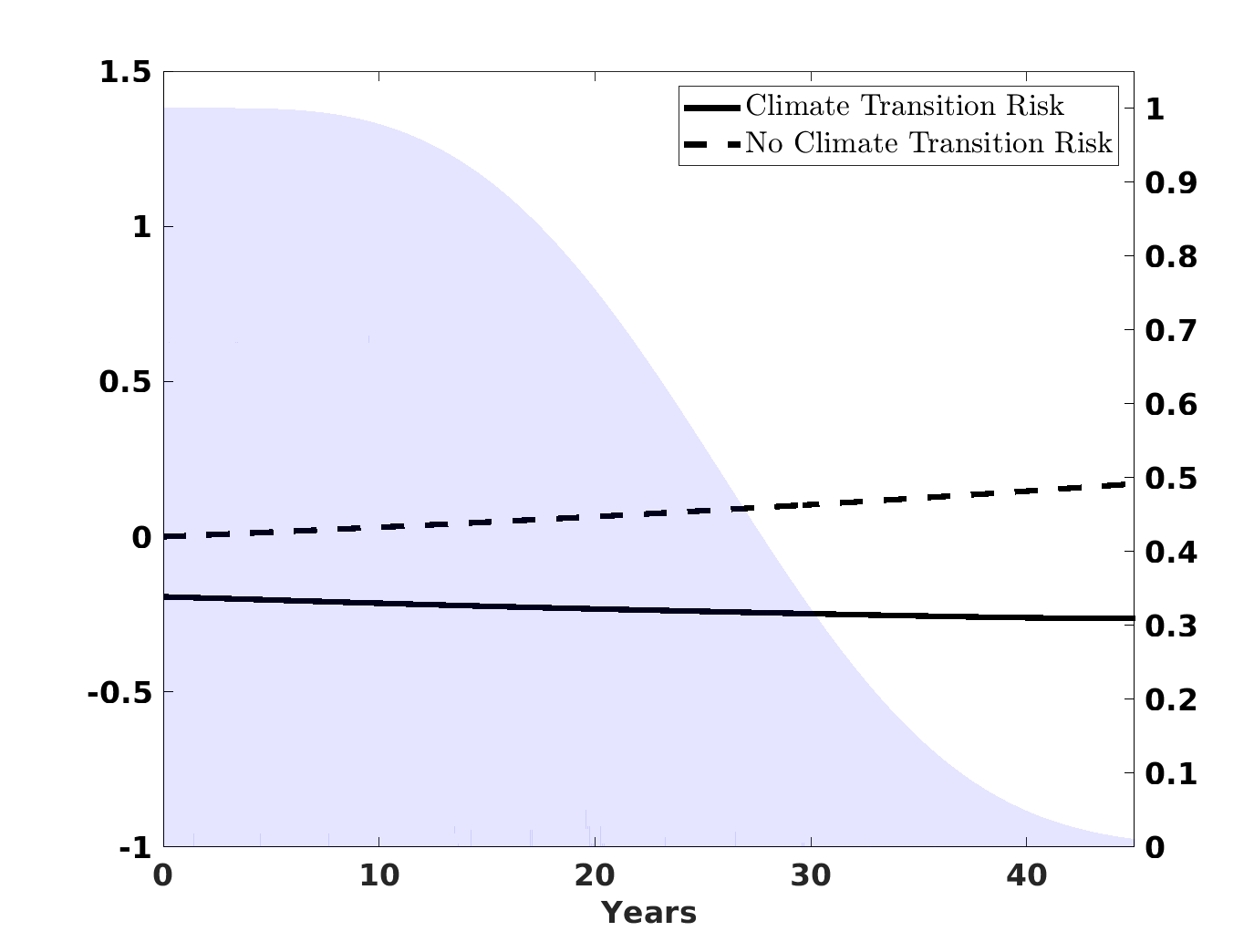}
           \caption[]{{\small Oil Spot Price: $P_{1,t}$}}
        \end{subfigure}
        \begin{subfigure}[b]{0.328\textwidth}  
            \centering 
            \includegraphics[width=\textwidth]{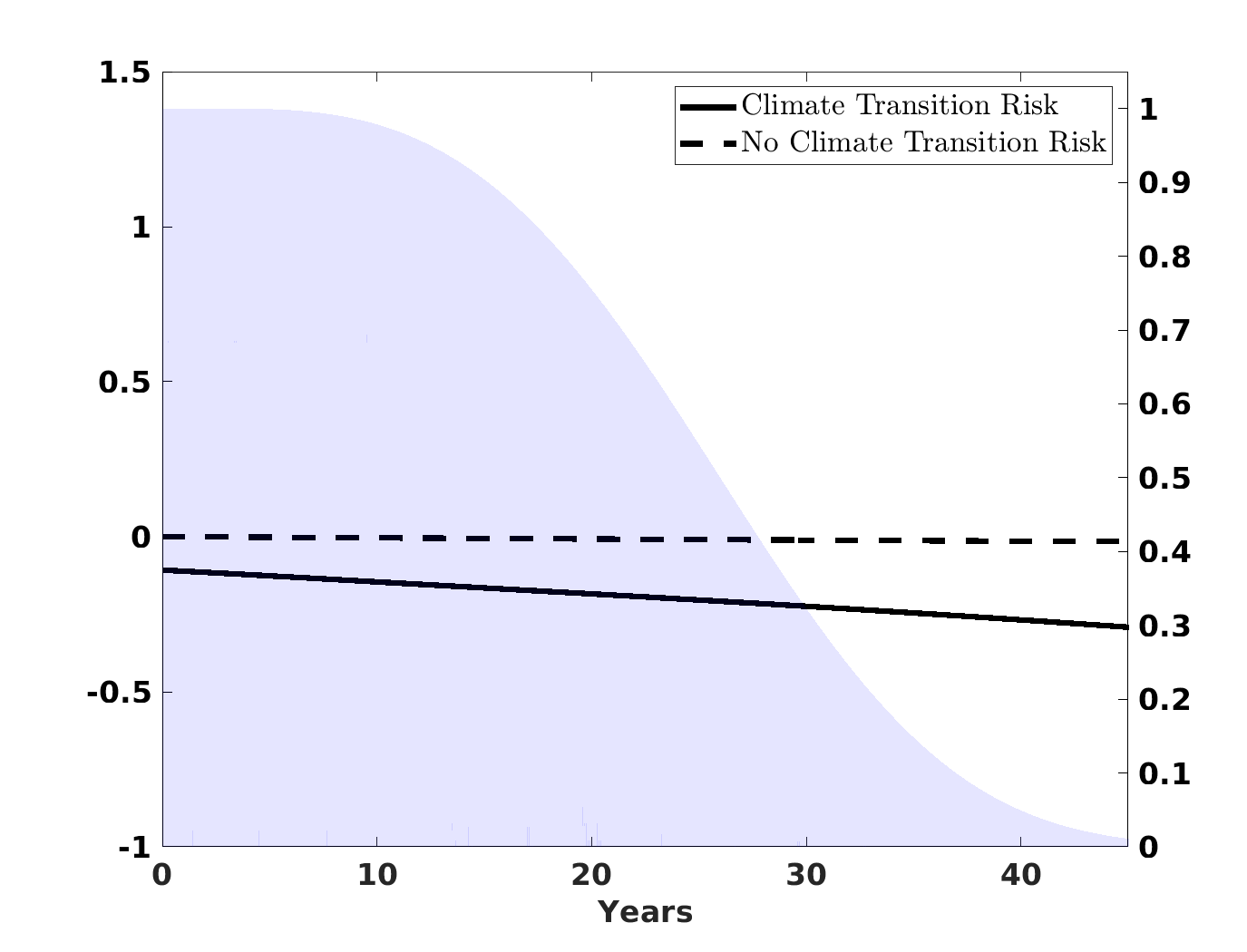}
           \caption[]{{\small Coal Spot Price: $P_{1,t}$}}
        \end{subfigure}

        \begin{subfigure}[b]{0.328\textwidth}
            \centering
            \includegraphics[width=\textwidth]{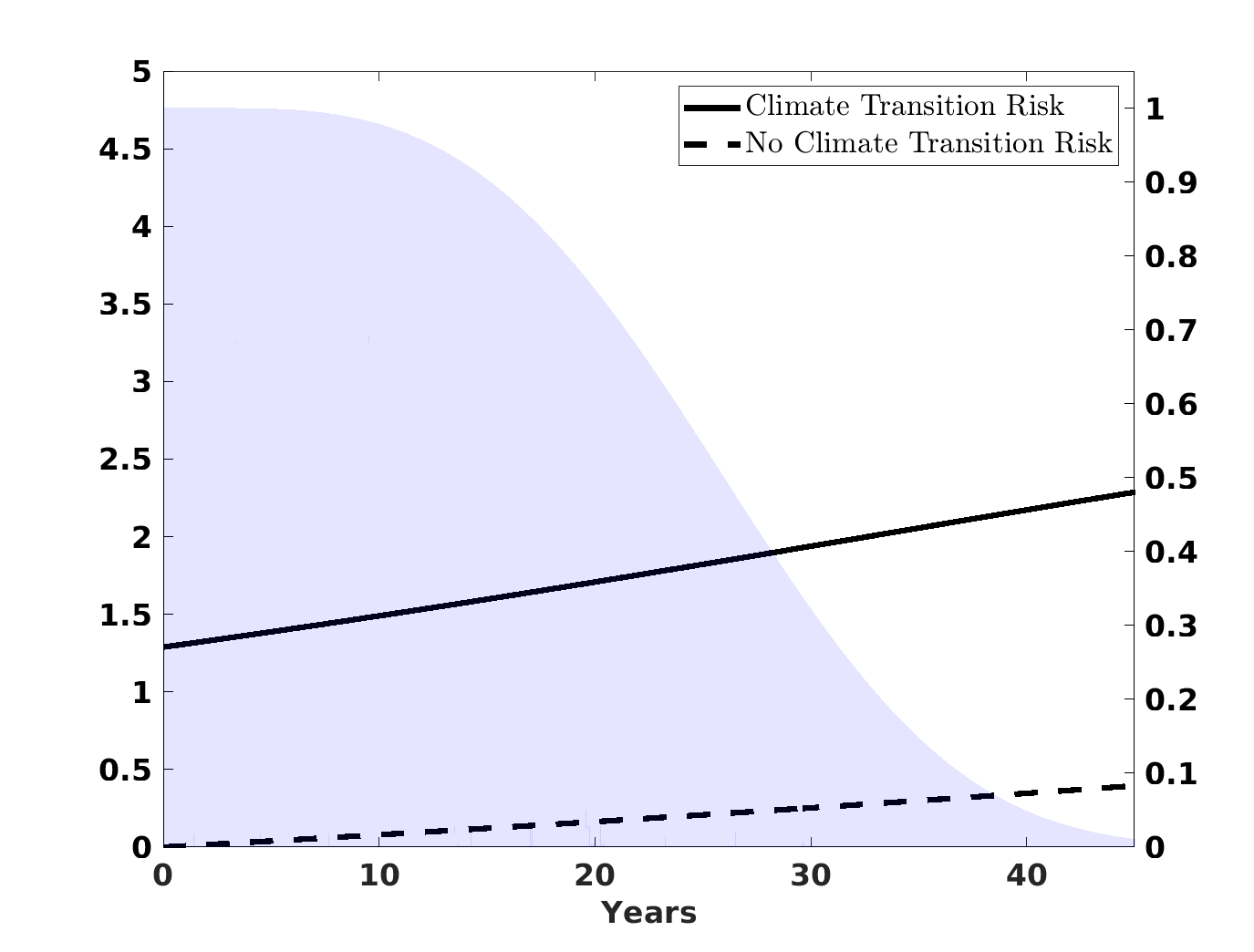}
            \caption[]{{\small Green Firm Price: $S^{(3)}_{t}$}}              
        \end{subfigure}
        \begin{subfigure}[b]{0.328\textwidth}  
            \centering 
            \includegraphics[width=\textwidth]{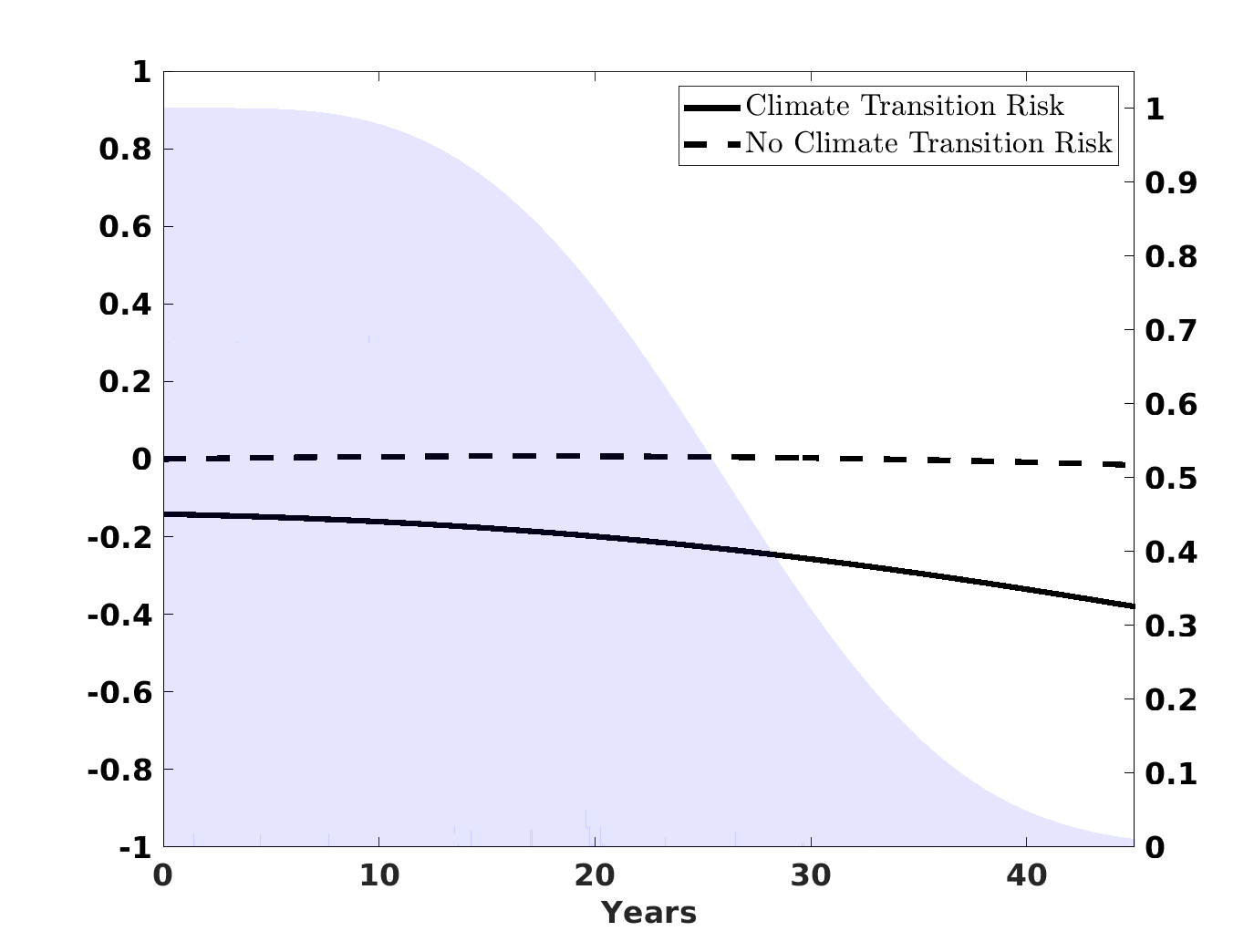}
           \caption[]{{\small Oil Firm Price: $S^{(1)}_{t}$}}
        \end{subfigure}
        \begin{subfigure}[b]{0.328\textwidth}  
            \centering 
            \includegraphics[width=\textwidth]{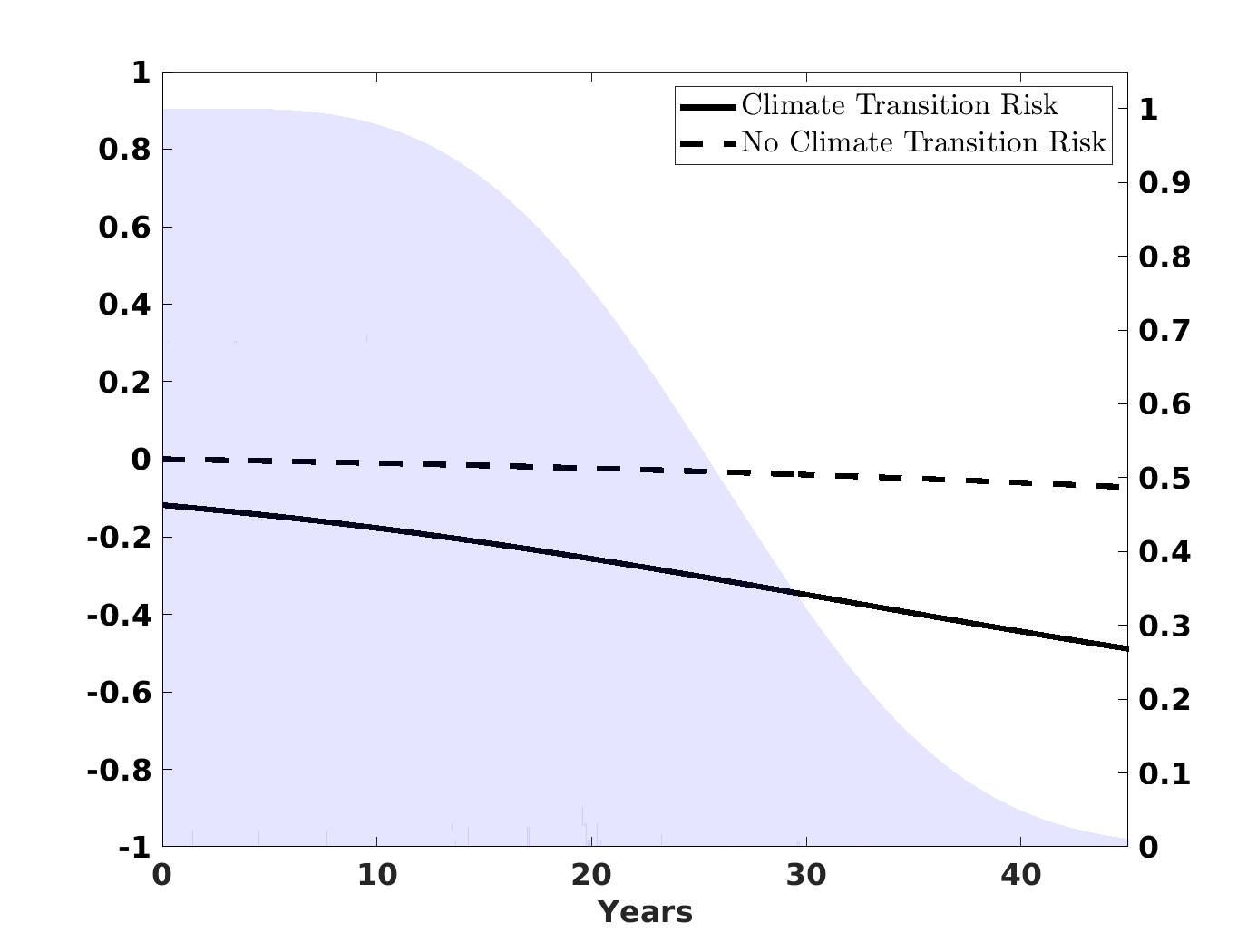}
           \caption[]{{\small Coal Firm Price: $S^{(2)}_{t}$}}
        \end{subfigure}  
        
        \vspace{-0.25cm}
\end{center}

\begin{footnotesize}
Figure \ref{fig:multi2_model_sims} shows the simulated outcomes for the ``coal then oil technology shock''  model based on the numerical solutions. Panels (a) through (c) show the temperature anomaly, oil production, and coal production. Panels (d) through (f) show the green investment choice, oil spot price, and coal spot price. Panels (g) and (i) show the green firm price, oil firm price, and coal firm price. Solid lines represent results for the Climate Transition Risk scenario where $\lambda_t = \lambda(T_t)$ and dashed lines represent results for the No Climate Transition Risk scenario where $\lambda_t = 0$. The blue shaded region shows the cumulative probability of no transition shock occurring.
\end{footnotesize} 

\end{figure}

\newpage
\clearpage

\section{Empirical Analysis}

\subsection{Event Study Analysis}

I now provide a detailed explanation of the event study analysis. The estimation framework builds on the framework of \cite{koijen2016financial} to identify how events that shift climate-linked transition risk expectations impact cumulative abnormal returns of stock returns for sector-based portfolios of US equities. At the event level, the estimation exploits cross-sectional variation in climate-linked transition risk exposure across different sectors. The aggregate responses then exploit time-series variation in the response of US equities to climate-linked transition risk events across time. The formal estimation procedure is set-up as follows:

\begin{enumerate}
    \item Using daily returns for the 49 sector portfolios provided on Ken French's website, I estimate abnormal returns as unexplained differences from a market model regression for each sector $i$ using the daily returns between 10 and 260 trading days (approximately one year) leading up to the event date:
\begin{eqnarray*}
R_{i,t} = \alpha_i +\beta_{i,Mkt} (R_{Mkt,t}-R_{f,t}) +\epsilon_{i,t}
\end{eqnarray*}

\item I then aggregate the residuals for each sector $\epsilon_{i,t}$, starting from 10 days before the event to 10 days after the event in order to get the cumulative abnormal returns:
\begin{eqnarray*}
CAR_{i,t-10 \rightarrow t+10} = \sum_{t=-10}^{10} \epsilon_{i,t} = \sum_{t=-10}^{10} (R_{i,t} - \alpha_i +\beta_{i,Mkt} (R_{Mkt,t}-R_{f,t}))
\end{eqnarray*}

I use 10 days on each side of the event window to capture any anticipation or leakage of information before the event. This is somewhat larger than the  5 trading days before and after each event window used by \cite{koijen2016financial}. However, numerous events in my list, such as the IPCC meetings and AR releases, are anticipated and carried out over various days and so the expanded window allows for the shock impact to build up for delayed responses and cumulative effects. \cite{mackinlay1997event} and others have similarly used event window sizes of this length or larger. Results are qualitatively and quantitatively similar if I use 15 days on each side of the event.

\item Next, I compute the measure of climate-linked transition risk exposure $\beta_{i,ClimateTransition}$ by estimating sectors' exposures to oil price innovations $\Delta P_{oil,t}$ from a time-series regression for each sector over the full time series:
\begin{eqnarray*}
R_{i,t} = \alpha_i +\beta_{i,Mkt} (R_{Mkt,t}-R_{f,t}) + \beta_{i,ClimateTransition} \Delta P_{oil,t} + \varepsilon_{i,t}
\end{eqnarray*}

This proxy is based on the theoretical model prediction that changes in the likelihood of a future climate-linked transition  risk realization influences production and prices for the fossil fuel sector. Because oil prices directly reflect fossil fuel production, are available at a daily frequency, and are in direct units comparison to returns, I choose oil price innovations as my proxy. 

\end{enumerate}

These steps provide the necessary inputs needed to derive the quantitative implications. To provide intuition, I first highlight the estimation results for event-level outcomes related to the announcement, publication, and eventual dismantling of the Clean Power Plan, an Obama administration policy limiting carbon emissions for power plants in the US that was linked with US participation in the 2015 Paris Climate Agreement. To estimate the effects on asset prices around these events, I run the following cross-sectional regression
\begin{eqnarray*}
CAR_{i,t-10 \rightarrow t+10}^{event} = \delta_{0}^{event} + \delta_{1}^{event} \frac{\beta_{i,OilPrice}}{\sigma(\beta_{i,OilPrice})} + e_{i}^{event}
\end{eqnarray*}
where the superscript $event$ denotes that observations and estimates are conditional on a given event date from the climate-linked transition risk event list. The climate-linked transition risk betas $\beta_{i,ClimateTransition}$ are normalized by the cross-sectional standard deviation of the beta estimates $\sigma(\beta_{i,ClimateTransition})$ so that the coefficient $\delta_1$ measures the percent change in cumulative abnormal returns around a climate-related transition risk event for a one-standard deviation increase in the climate-linked transition risk exposure.

\begin{table}[!ht]
\caption{Event Study Analysis of Events Related to the Clean Power Plan} \label{fig:MultiEventStudy}
\begin{center}
%\vspace{0.25cm}
\begin{tabular}{l c c c}
\hline
\hline
Date & Description & $\delta_1$ Estimate & t-stat \\
\hline
%CPP Announce & & & \\
08/03/2015 & \multicolumn{1}{p{8cm}}{President Obama announces the final version of the Clean Power Plan.} & -0.32 & -0.48 \\
10/23/2015 & \multicolumn{1}{p{8cm}}{The Carbon Emissions rules for the Clean Power Plan are published.} & -3.24 & -2.25 \\
12/11/2015$^*$ & \multicolumn{1}{p{8cm}}{196 UNFCCC participating members agree to the Paris Accords at the conclusion of COP21.} & 0.12 & 0.10 \\
02/09/2016 & \multicolumn{1}{p{8cm}}{The US Supreme Court ordered a hold on EPA enforcement of the Clean Power Plan until lower court rules on lawsuit against the plan.} & 5.62 & 3.18 \\
11/08/2016 & \multicolumn{1}{p{8cm}}{Donald Trump is elected as the 45th president of the United States of America.} & 0.30 & 0.23 \\
06/01/2017 & \multicolumn{1}{p{8cm}}{President Trump announces the US will withdrawal from the Paris Agreement; numerous US states announce intentions to uphold the Paris Agreement under US Climate Alliance. } & -2.26 & -6.47 \\
03/28/2017 & \multicolumn{1}{p{8cm}}{President Trump signs an executive order to have the head of the EPA, Scott Pruitt, review the Clean Power Plan.} & 1.97 & 4.59 \\
\hline
\hline
\end{tabular}
\begin{footnotesize}
*12/11/2015 was not a trading day, so I use the first trading day after this event (12/14/2015). 
\end{footnotesize}  
\end{center} 

\vspace{0.25cm}

\begin{footnotesize}
Table \ref{fig:MultiEventStudy} shows the relationship between returns and climate-linked transition events. The estimates shows the response of cumulative abnormal returns across sectors following specific climate-linked transition risk events based on the normalized sector exposure to climate policy risk. The events shown center actions related to the Clean Power Plan, the Paris Climate Accords, and the election of Donald Trump as the US President in 2016. The t-stats for the estimated coefficients use heteroskedasticity-robust standard errors. 
\end{footnotesize}  

\end{table}

Table \ref{fig:MultiEventStudy} shows the estimation results for each event. Dates that coincided with increases in the likelihood of a future climate transition that will restrict the use of fossil fuels include when President Obama announced the Clean Power Plan (08/03/2015), when the carbon emissions rules for the Clean Power Plan were published (10/23/2015), and when the Paris Agreement was adopted at the end of the IPCC's COP21 meeting (12/11/2015). Of these, the most significant is when the final rules are published for the Clean Power Plan, with a statistically significant $\delta_1$ estimate of $-3.24$. The dates that coincided with decreases in the likelihood of future climate transition that will restrict the use of fossil fuels include when the US Supreme Court put a hold on the EPA enforcing the Clean Power Plan (02/09/2019), when Donald Trump was elected as president of the United States of America (11/08/2016), and when President Trump signed an executive order to review the Clean Power Plan (03/28/2017). All of these estimates are positive, with the US Supreme Court ruling and Trump's executive order leading to statistically significant $\delta_1$ estimates of $5.62$ and $1.97$, respectively. 

The final event was on 06/01/2017, when President Trump announced that the US would withdraw from the Paris Accord. However, shortly thereafter on that same day, numerous states announced their intention to comply with the Clean Power Plan and Paris Agreement objectives under the US Climate Alliance. Ex ante, the asset pricing response is therefore ambiguous given that these two events suggest different expectations for future climate transition. The estimate result is statistically significant and negative, with an estimated $\delta_1$ value of $-2.26$, revealing that markets interpreted the coordinated response by states to continue following the Clean Power Plan and Paris Agreement regulations as a strong signal for an increased likelihood of future climate policy action and stranded assets risk. These events help further validate the model results and highlight the ability of asset prices to help identify expected future outcomes related to climate transition risk.

%%%%%%%%%%%%%%%%%%%%%%%%%%%%%%%%%%%%%%%%%%%%%%%%%%%%

Finally, the estimates in the main text focusing on the aggregate quantitative relationship between climate transition risk exposure and cumulative abnormal returns around shocks to the likelihood of a future climate transition are estimated using the following panel regression:
\begin{eqnarray*}
CAR_{i,t-10 \rightarrow t+10} = \delta_{0} + \delta_{1} \frac{\beta_{i,ClimateTransition}}{\sigma(\beta_{i,ClimateTransition})} + e_{i}
\end{eqnarray*} 

My aggregate response analysis focuses on the three separate panels noted in the main text: all events in the full time sample of 1974-2019, only events in the more recent time subsample starting in 2009, and only events in the earlier subsample before 2009. Based on the model, I expect increased asset pricing implications of climate transition events in the recent subsample where temperature increases and climate change concerns are higher. The coefficients $\delta_{0}$ and $\delta_{1}$ are the average estimated values across all climate transition events in the sample\footnote{In order to aggregate all transition-related event outcomes in a consistent manner, I use $-CAR_{t-10 \rightarrow t+10}$ for events I identify as decreasing the likelihood of a future climate transition.}. All estimates use heteroskedastic-robust standard errors to calculate t-statistics. Results for the aggregated panel estimates are given in Figure \ref{fig:PanelEventStudy}. Figures \ref{fig:PanelEventStudyb} and \ref{fig:PanelEventStudyc} further refine our results via a scatter plot of the average climate transition risk exposure-cumulative abnormal return relationship across industries for the recent, transition-relevant sample (Figure \ref{fig:PanelEventStudyb}) and the dynamic asset pricing response of ``dirty'' firms around climate-transition-related events. Specifically, the figure shows $CAR_{\beta,t-10 \rightarrow t+j}$ for $j = \{-10,...,10\}$ (Figure \ref{fig:PanelEventStudyc}). Note that $CAR_{\beta,t-10 \rightarrow t+j}$ is the average across all climate-transition-related events of the cumulative abnormal returns for a type of climate-transition-risk factor-mimicking portfolio. The portfolio is constructed as a weighted average of sector portfolio returns, where the weights are calculated by re-scaling the $\beta_{i,ClimateTransition}$ estimates found previously to sum to one. Thus $CAR_{\beta}$ is a measure of expected future climate transition consequences based on aggregate information captured by variation in equity prices and oil prices around climate transition related events.

This empirical evidence confirms the following theoretical model predictions: an increase in the likelihood of a future climate-linked transition leads to lower cumulative abnormal returns for firms with higher transition risk exposure; the magnitude of these effects increases over time, validating the state-dependent response predicted also predicted by the model; and overall that asset prices are a valuable source of information for identifying climate change risks, including those related to a climate-linked transition risk.

\subsection{Vector Autoregression Analysis}

I now detail the estimation setting for the  structural vector autoregression (VAR) for the augmented model of the global oil market. These results focus on providing essential evidence on the dynamic fossil fuel sector responses to climate-linked transition risk shocks linked to my theoretical model and expectations about the type of transition risk mechanism anticipated by the aggregate economy. The augmented global oil market VAR I use in my analysis builds on the proposed framework of \cite{kilian2009not}. The model is given by:
\begin{eqnarray*}
y_{t} & = & \nu + \sum_{j=1}^N A_j y_{t-j} + u_{t}
\end{eqnarray*}

where the vector of endogenous state variable vector $y_t$ is defined by
\begin{eqnarray*}
y_{t} & = & [ClimateTransition_{t}, \Delta prod_t, rea_t, \Delta p_{t}^{oil} ] '
\end{eqnarray*}

$ClimateTransition_{t}$ is the time-series of $CAR^{event}_{\beta,t-10 \rightarrow t+10}$ values for all climate-transition-related events, and zero otherwise, aggregated up to the monthly frequency\footnote{I discuss results for alternative $ClimateTransition_t$ definitions in Section \ref{Robustness} and the Online Appendix.}, $\Delta prod_t$ is the percent change in global oil production available from the EIA, $\Delta rea_t$ is a measure of real economic activity given by innovations in the log OECD industrial production index, and $\Delta p_{t}^{oil}$ is log differences in the real West Texas Intermediate (WTI) monthly closing price for crude oil\footnote{I include month dummies to control for seasonality, which has little impact on the results.}. The choice of variables used in my main analysis follows \cite{baumeister2019structural}, though the results are qualitatively the same and quantitatively similar if I use Kilian's specification with his updated measure of real economic activity based on an index of nominal shipping freight rates and the refiner's acquisition price for the oil price.

I use a Cholesky decomposition of the estimated variance-covariance matrix for identification of the structural shocks. The Cholesky decomposition identification strategy imposes a recursive interpretation of the impact of the shocks. The general representation and interpretation for this framework is given by:
\begin{equation*}
u_t = B
\begin{bmatrix}
   \epsilon_{\text{climate transition}}, &
   \epsilon_{\text{oil supply}}, &
   \epsilon_{\text{aggregate demand}}, &
   \epsilon_{\text{oil-specific demand}} 
\end{bmatrix}'
\end{equation*}

where B is the lower triangular matrix derived from the Cholesky decomposition of the estimated variance-covariance matrix $\hat{\Sigma}$, i.e., $B B' = E_t[u_t'u_t] = \hat{\Sigma}$. I outline the specific interpretation and identification of each shock in what follows.

From the VAR estimates and the recursive identification structure, I derive impulse response functions (IRFs), or the cumulative responses to a given structural variable shock, which are the results I use to examine the validity of the transition risk mechanism. Note that if the $ClimateTransition_{t}$ shocks are irrelevant, then we should see no impact on production or pricing outcomes. From these estimates, we are able to identify two key outcomes: climate-linked transition risk has a statistically significant and economically meaningful impact on both prices and production for the recent subsample, validating the importance and state-dependent effects of the expected transition risk predicted by the model; and the positive response of oil production to the shock to climate-related transition risk highlights that firms are likely expecting climate-linked transition risk (at least partially) in the form of a technology shock. These results provide novel inference about the market expectations of climate-linked transition risk for the oil sector, and provides valuable insight for policymakers, investors, and households in order to respond to climate transition risk.

\subsection{Robustness of Estimation Identification} \label{Robustness}

I discuss a number of robustness tests that explore specific assumptions for the empirical estimation and provide additional evidence in support of the outcomes and identification.

\subsubsection{\textit{Index Decomposition and Alternative Index Measures.}} 

\begin{figure}[!phtb]
%    \centering
\caption{Impulse Response Functions for Transition Likelihood Shock}\label{fig:var_climpol_global}
    
\begin{subfigure}{0.48\linewidth}
\includegraphics[width=\linewidth]{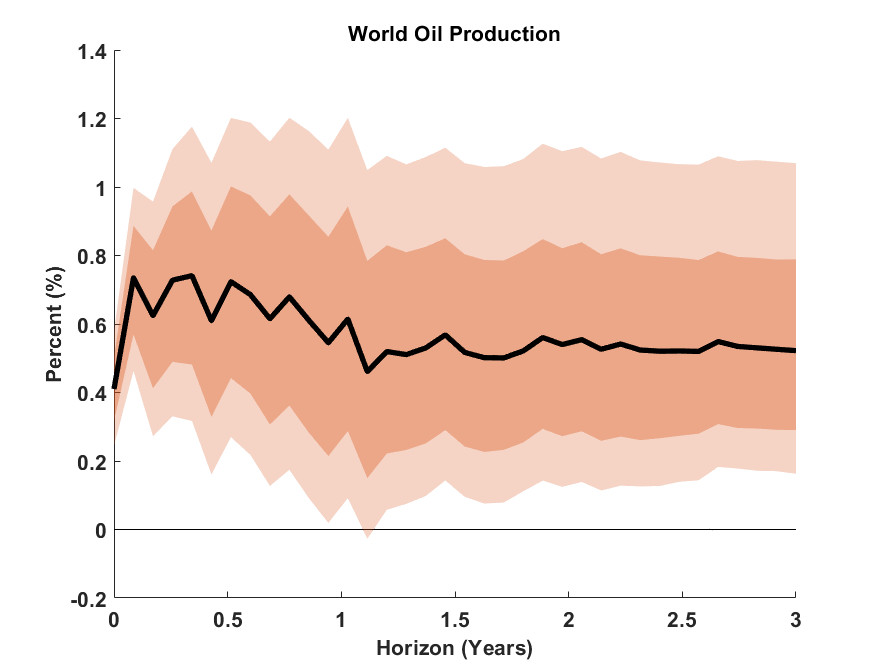}
\caption{Oil Production IRF, 2009-2019}
\end{subfigure}
    \hfill
\begin{subfigure}{0.48\linewidth}
\includegraphics[width=\linewidth]{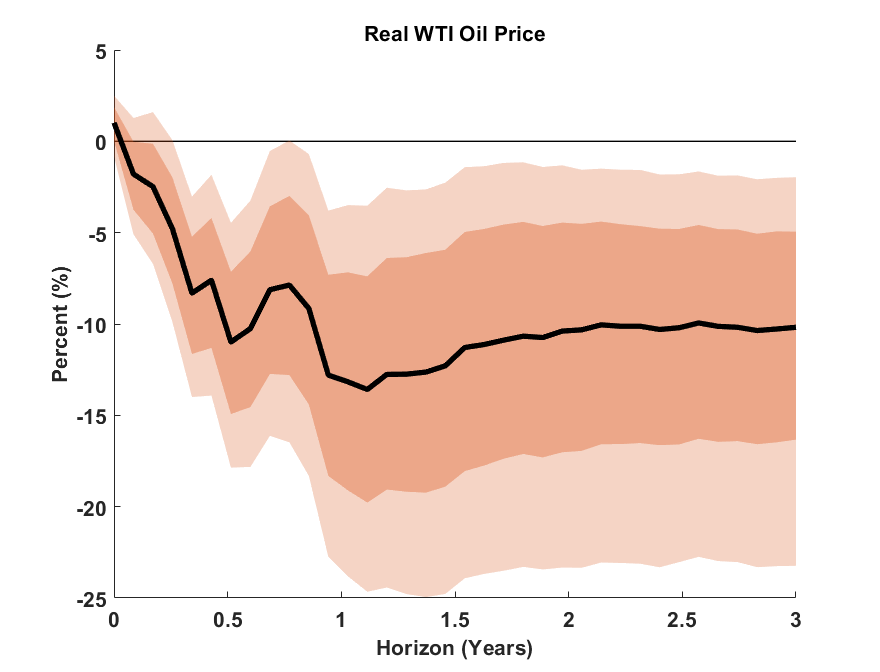}
\caption{Oil Price IRF, 2009-2019}
\end{subfigure}

\medskip

\begin{subfigure}{0.48\linewidth}
\includegraphics[width=\linewidth]{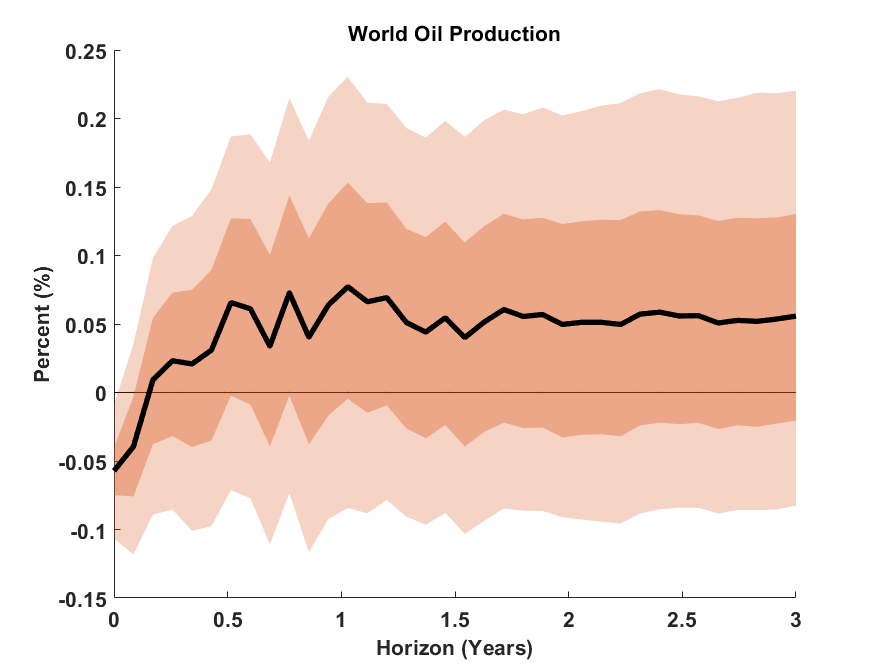}
\caption{Oil Production IRF, 2009-2019}
\end{subfigure}
    \hfill
\begin{subfigure}{0.48\linewidth}
\includegraphics[width=\linewidth]{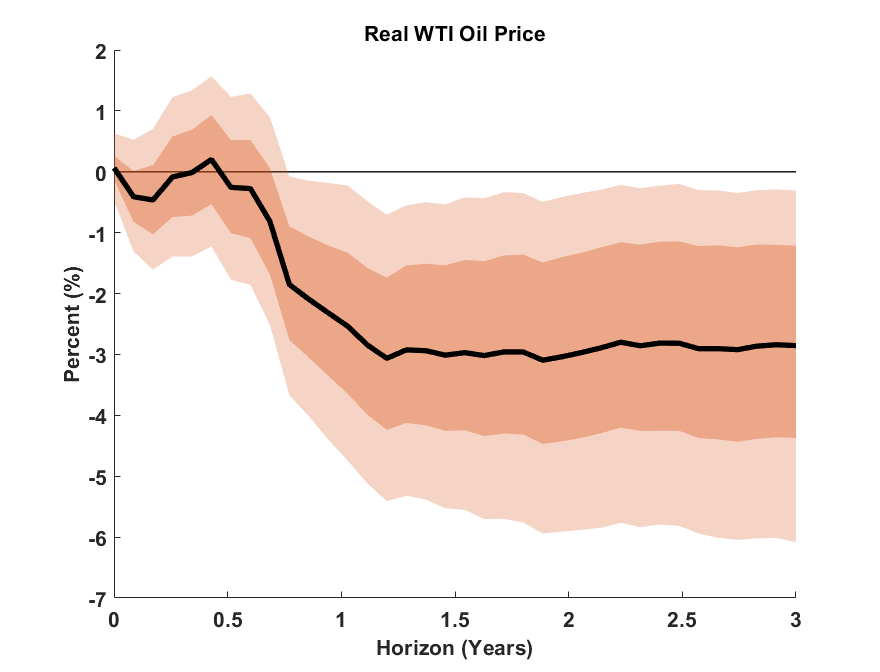}
\caption{Oil Price IRF, 2009-2019}
\end{subfigure}

\vspace{0.25cm}

\begin{footnotesize}

Figure \ref{fig:var_climpol_global} shows the estimated impulse response functions for global oil production and the WTI spot price of oil for a shock to alternative $ClimateTransition_t$ measures. The black line is the estimated IRF, the dark red shaded region represents the one-standard deviation error band, and the light red shaded region represents the two-standard deviation error band. The top panels are for $ClimateTransition_t$ constructed using only global policy events, and the bottom panels are $ClimateTransition_t$ constructed using only the sectors for the top quintile of $\beta_{i,ClimateTransition}$ estimates, excluding gold (i.e., coal, oil, mines, steel, machines, fabricated products, ships, and construction). The estimates are for the transition-risk-relevant time subsample (2009-2019). Bootstrapped error bands are calculated using 10,000 simulated samples. The VAR is estimated using 12 lags. See text for the full VAR specification used and definition of variables.

\end{footnotesize}
\end{figure}

To explore the $ClimateTransition_t$ index in greater depth, I decompose the index by reconstructing the measure using different subsets of event categories: US Policy events, global policy events, disaster events, and technological change events. I re-estimate the augmented global oil market VAR using different combinations of these categories, and find that the largest and most significant effects are estimated from the $ClimateTransition_t$ variable constructed using only global policy events. Figure \ref{fig:var_climpol_global} (top panels) shows the IRFs for this setting. The oil price and production responses are statistically significant and larger in magnitude when compared to the baseline specification. These results highlight the importance of anticipated global actions for the model-implied mechanism, but also highlight that the baseline results, estimated using a relatively inclusive index to avoid ``cherry-picking'' outliers, are likely a lower bound for the effects of climate transition risk.

%\ref{var_climpol_decomp}

\begin{figure}[!phtb]
%    \centering
\caption{Impulse Response Functions for Transition Likelihood Shock}\label{fig:var_climpol_alt}
    
\begin{subfigure}{0.48\linewidth}
\includegraphics[width=\linewidth]{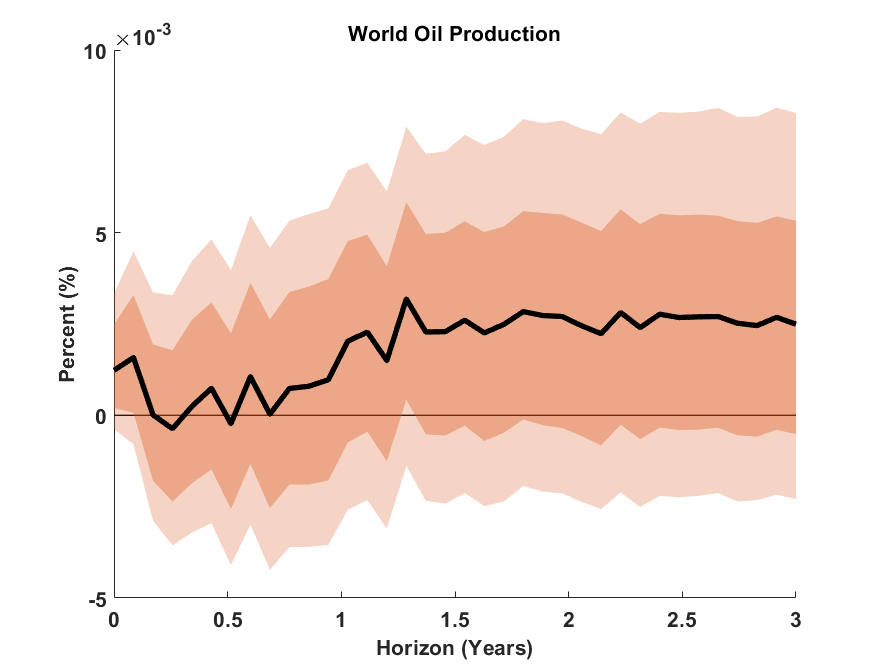}
\caption{Oil Production IRF, 2009-2019}
\end{subfigure}
    \hfill
\begin{subfigure}{0.48\linewidth}
\includegraphics[width=\linewidth]{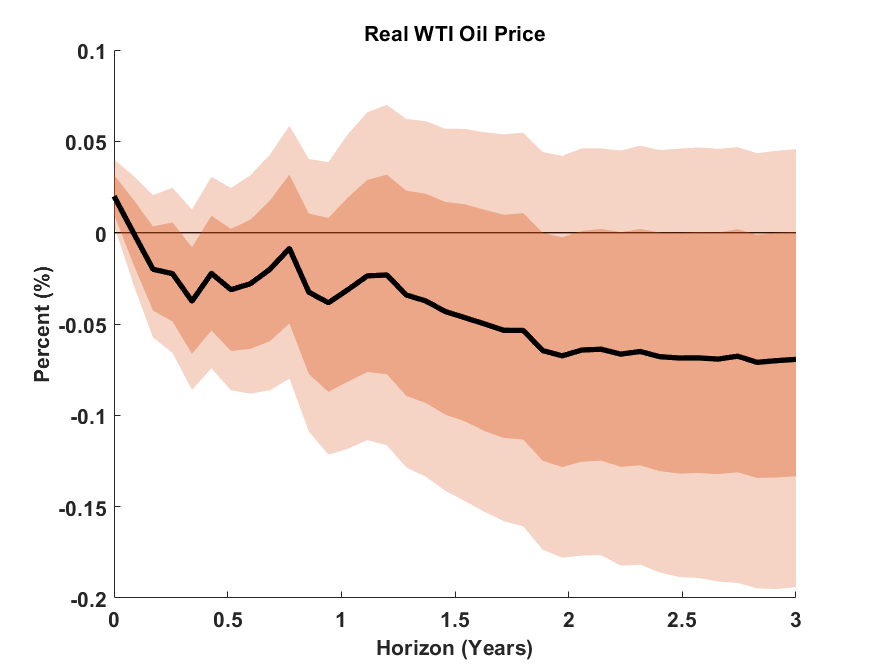}
\caption{Oil Price IRF, 2009-2019}
\end{subfigure}

\medskip

\begin{subfigure}{0.48\linewidth}
\includegraphics[width=\linewidth]{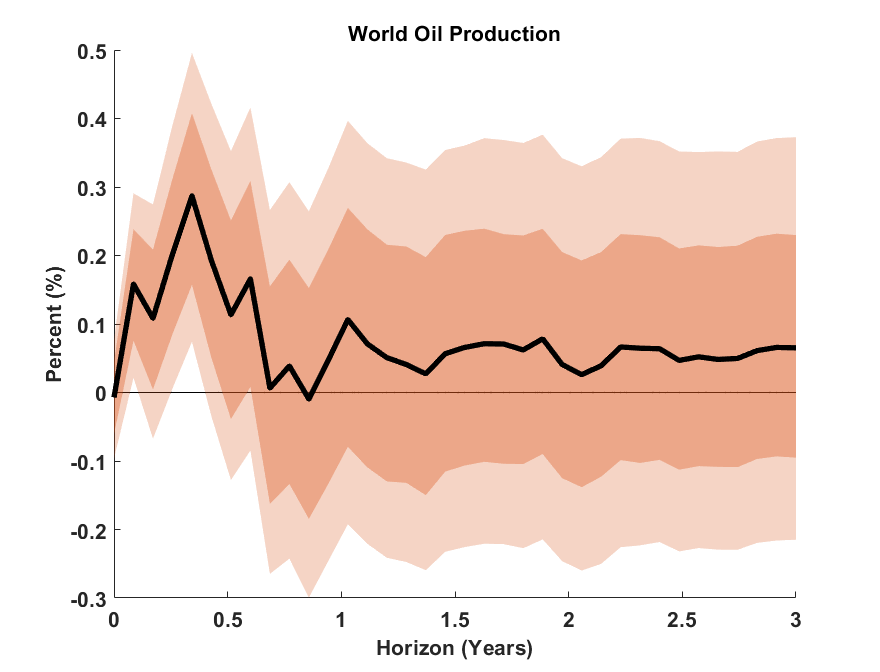}
\caption{Oil Production IRF, 2009-2019}
\end{subfigure}
    \hfill
\begin{subfigure}{0.48\linewidth}
\includegraphics[width=\linewidth]{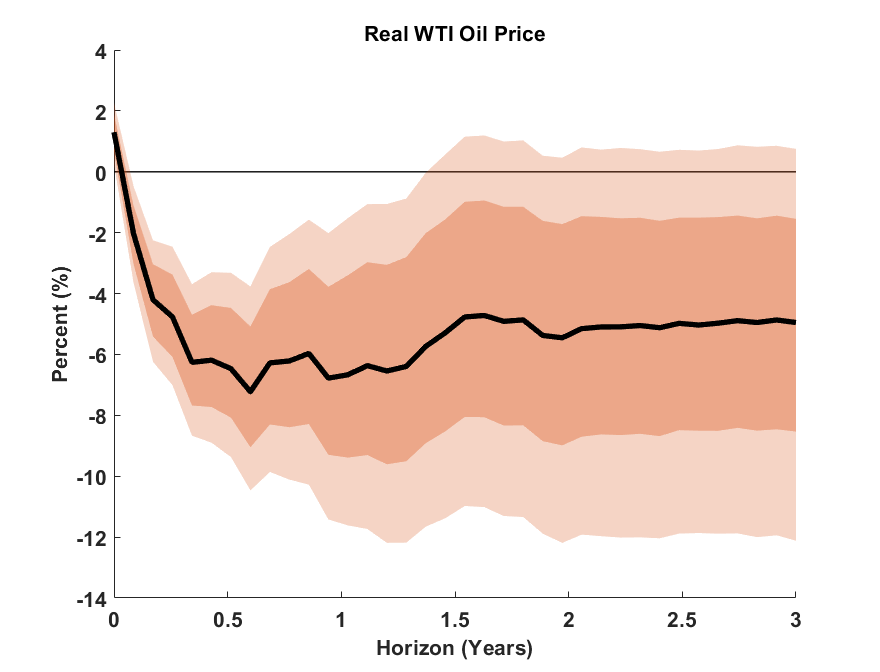}
\caption{Oil Price IRF, 2009-2019}
\end{subfigure}

\vspace{0.25cm}

\begin{footnotesize}

Figure \ref{fig:var_climpol_alt} shows the estimated impulse response functions for global oil production and the WTI spot price of oil for a shock to alternative $ClimateTransition_t$ measures. The black line is the estimated IRF, the dark red shaded region represents the one-standard deviation error band, and the light red shaded region represents the two-standard deviation error band. The top panels are for $ClimateTransition_t$ constructed using only the dummy event index, and the bottom panels are for $ClimateTransition_t$ constructed using only the high-transition-risk-exposure weighted portfolio returns. The estimates are for the transition-risk-relevant time subsample (2009-2019). Bootstrapped error bands are calculated using 10,000 simulated samples. The VAR is estimated using 12 lags. The high-transition-risk-exposure weighted portfolio abnormal returns are constructed from a market model regression using the entire time-series of data. See text for the full VAR specification used and definition of variables.
\end{footnotesize}
\end{figure}

I also consider three additional $ClimateTransition_t$ index construction methods. The first uses only the dummy variable version of the index without interacting with the return measure. The second definition of $ClimateTransition_t$ uses cumulative abnormal returns of the factor mimicking portfolio directly without the dummy event index interaction. Figure \ref{fig:var_climpol_alt} gives these results. In each case, the results are qualitatively similar, but are attenuated and less statistically significant, if at all, when compared to the baseline. The third uses only sectors in the top quintile of $\beta_{i,ClimateTransition}$ estimates and excludes gold (i.e., coal, oil, mines, steel, machines, fabricated products, ships, and construction) to focus on return outcomes explicitly linked to transition risk. The point estimates magnitudes are doubled, though slightly less statistically significant for production beyond 1 year, when compared to the baseline. Estimates for this alternative are shown in the bottom panels of Figure \ref{fig:var_climpol_global}. These results provide additional empirical support for the mechanism in my analysis, while emphasizing the value of information contained in asset prices and the narrative component of the index for identifying the transition risk effect.

%\ref{var_alt_climpol}

Next, I re-estimate the VAR using the measure of innovation in climate change news from \cite{ardia2023climate}\footnote{I thank the authors for making this data freely available at \url{https://sentometrics-research.com}.}, which builds on the index built by \cite{engle2020hedging}, as an $ClimateTransition_t$ measure. Figure \ref{fig:var_climpol_opec} (top panels) shows the results using this alternative index for the more recent, transition-focused subsample (2009-2019). The results are again qualitatively similar to the baseline specification, though the effects are statistical insignificant and point estimates are close to null. These results highlight the importance of identifying the distinct climate transition risk mechanism, rather than a broad climate change news channel.

%\ref{var_ardia}

\subsubsection{\textit{Alternative Fossil Fuels and Regions.}} 

\begin{figure}[!phtb]
%    \centering
\caption{Impulse Response Functions for Transition Likelihood Shock}\label{fig:var_climpol_altff}
    
\begin{subfigure}{0.48\linewidth}
\includegraphics[width=\linewidth]{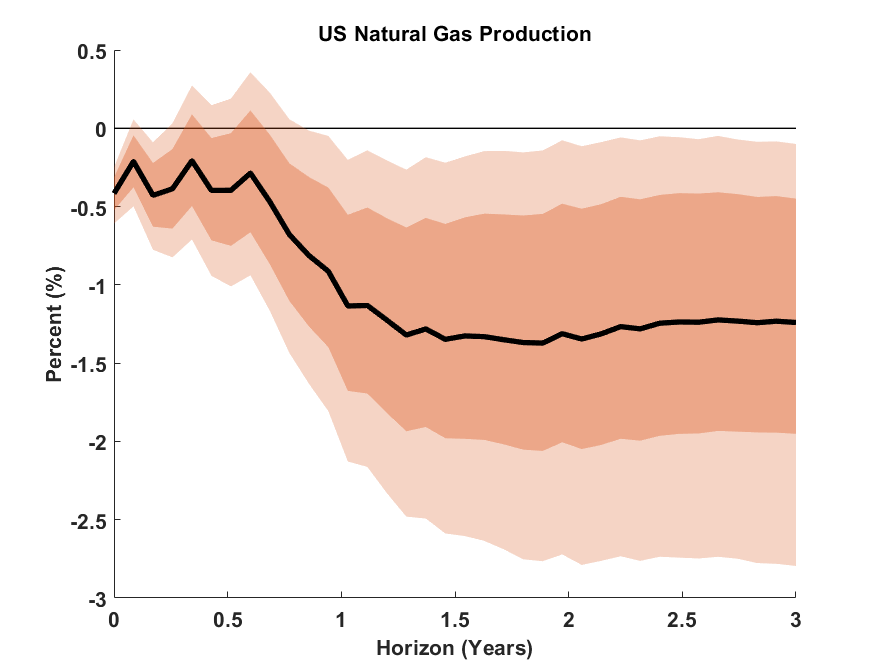}
\caption{NG Production IRF, 2009-2019}
\end{subfigure}
    \hfill
\begin{subfigure}{0.48\linewidth}
\includegraphics[width=\linewidth]{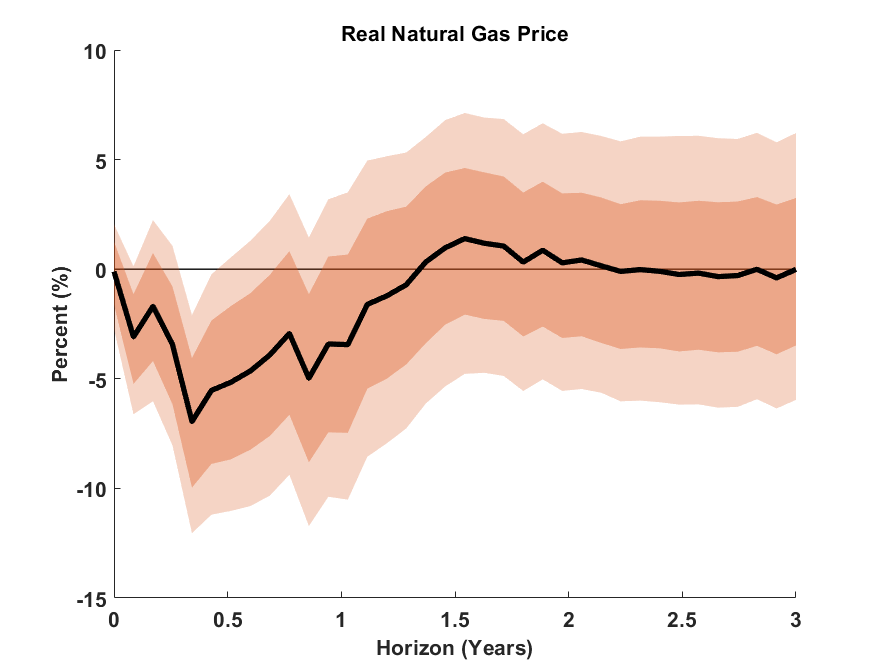}
\caption{NG Price IRF, 2009-2019}
\end{subfigure}

\medskip

\begin{subfigure}{0.48\linewidth}
\includegraphics[width=\linewidth]{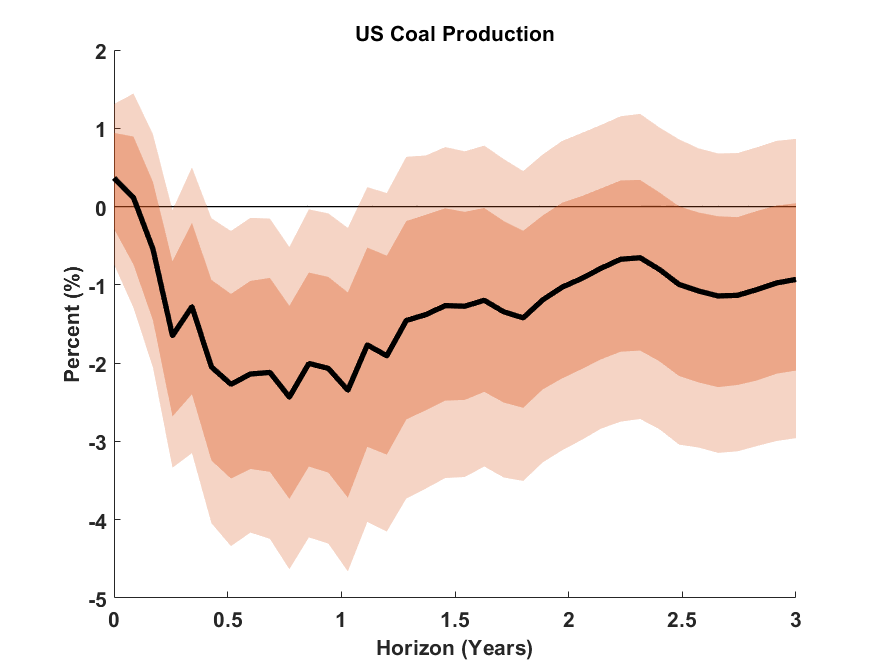}
\caption{Coal Production IRF, 2009-2019}
\end{subfigure}
    \hfill
\begin{subfigure}{0.48\linewidth}
\includegraphics[width=\linewidth]{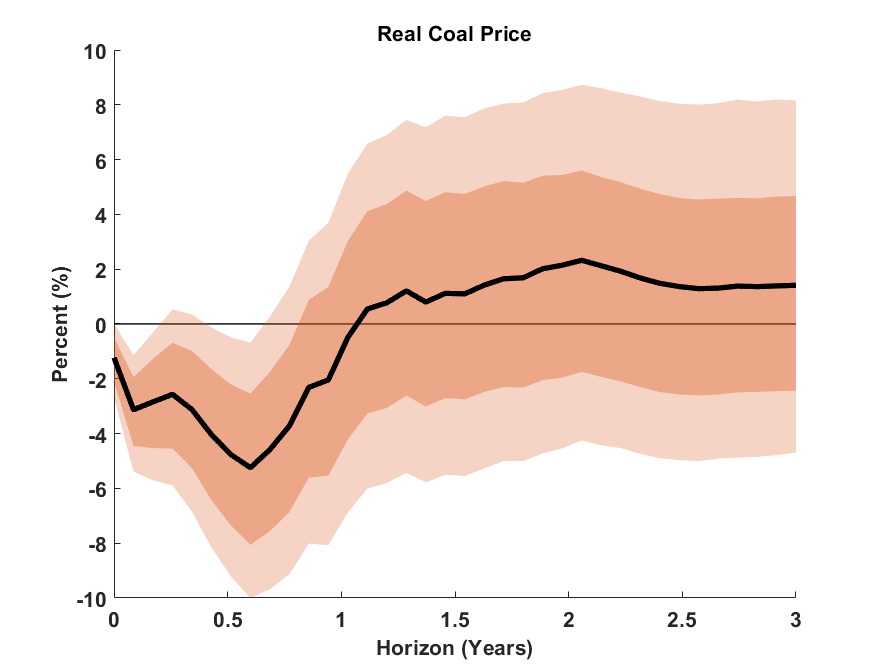}
\caption{Coal Price IRF, 2009-2019}
\end{subfigure}

\vspace{0.25cm}

\begin{footnotesize}

Figure \ref{fig:var_climpol_altff} shows the estimated impulse response functions for alternative fossil fuel production and spot prices for a shock to $ClimateTransition_t$. The black line is the estimated IRF, the dark red shaded region represents the one-standard deviation error band, and the light red shaded region represents the two-standard deviation error band. The top panels are for natural gas (NG) production and spot price variables, and the bottom panels are for coal production and spot price variables. The estimates are for the transition0-risk-relevant time subsample (2009-2019). Bootstrapped error bands are calculated using 10,000 simulated samples. The VAR is estimated using 12 lags. See text for the full VAR specification used and definition of variables.
\end{footnotesize}
\end{figure}

A useful comparison for understanding the empirical results is whether the predicted implications for the oil sector apply to alternative fossil fuels such as coal or natural gas. To address this, I repeat the VAR estimation using instead coal and natural gas for the fossil fuel variables.\footnote{The EIA only provides coal and natural gas production data at the monthly frequency for the US, and therefore these estimates only consider US production values.} The results are shown in Figure \ref{fig:var_climpol_altff}. For coal, the price response is negative and the production response is negative, and only production shows even marginal statistical significance for a brief time. This is consistent with the fact that extensive policy is already restricting coal production, matching the prediction of the transition scenario where a non-transition carbon tax is in place before the transition occurs. The price responses for natural gas are almost entirely statistically insignificant, and the production responses are marginally statistically significant at various horizons. However, the point estimates are negative for both, in line with uncertainty about how a future transition will impact the often perceived cleaner alternative of natural gas, as we well potentially being influenced by alternative forces in this sector like the fracking boom. The results highlight the importance of different climate risk mechanisms shown in the model and show that the baseline results are not driven mechanically by a common time trend.

%\ref{var_alt_fuels}

\begin{figure}[!phtb]
%    \centering
\caption{Impulse Response Functions for Transition Likelihood Shock}\label{fig:var_climpol_opec}

\begin{subfigure}{0.48\linewidth}
\includegraphics[width=\linewidth]{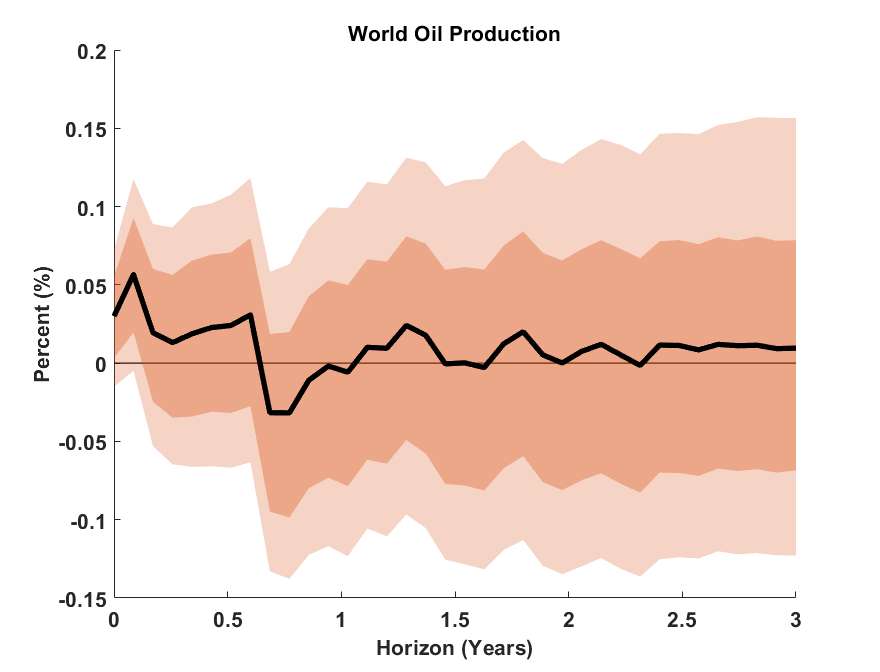}
\caption{Oil Production IRF, 2009-2019}
\end{subfigure}
    \hfill
\begin{subfigure}{0.48\linewidth}
\includegraphics[width=\linewidth]{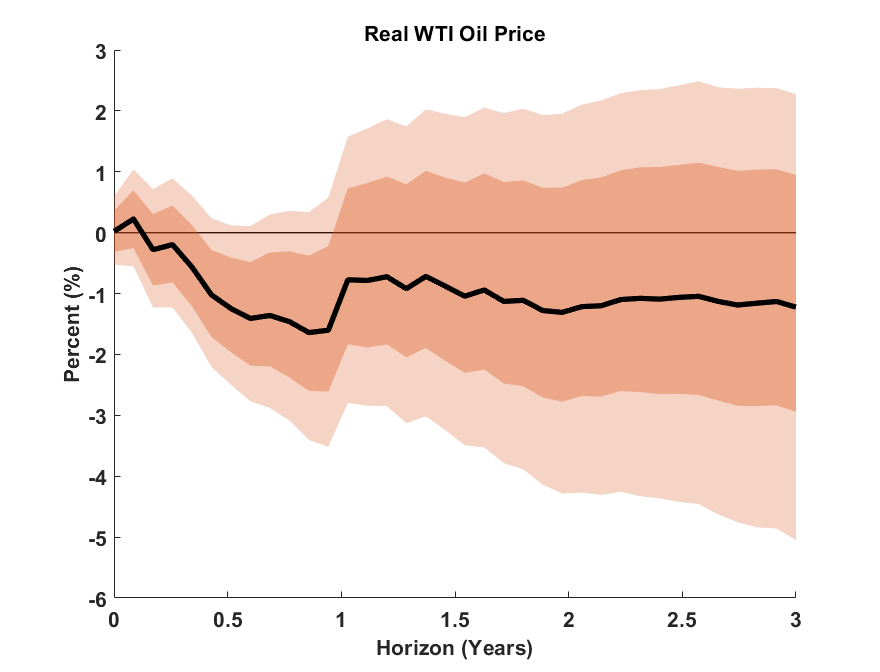}
\caption{Oil Price IRF, 2009-2019}
\end{subfigure}

\medskip
    
\begin{subfigure}{0.48\linewidth}
\includegraphics[width=\linewidth]{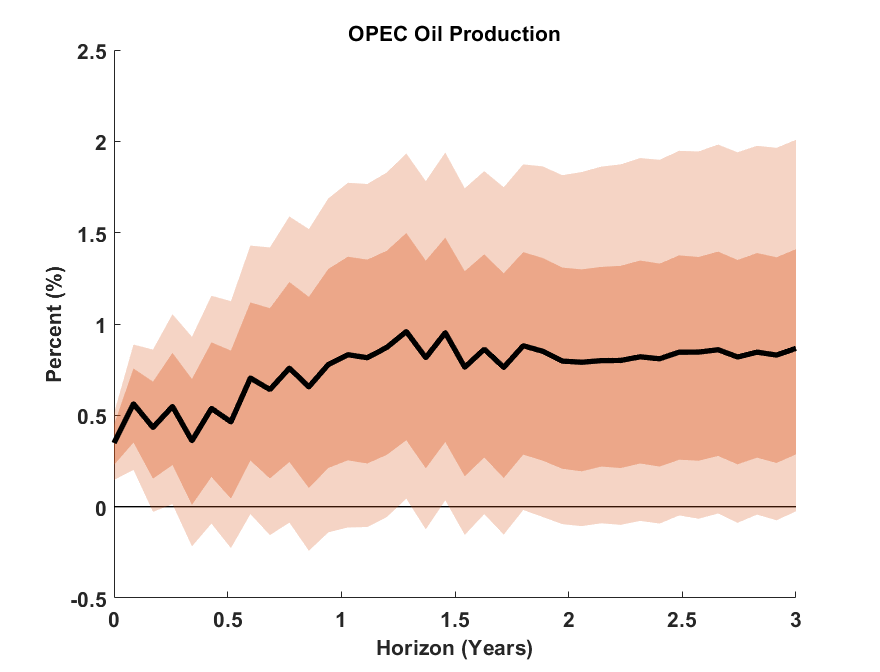}
\caption{Oil Production IRF, 2009-2019}
\end{subfigure}
    \hfill
\begin{subfigure}{0.48\linewidth}
\includegraphics[width=\linewidth]{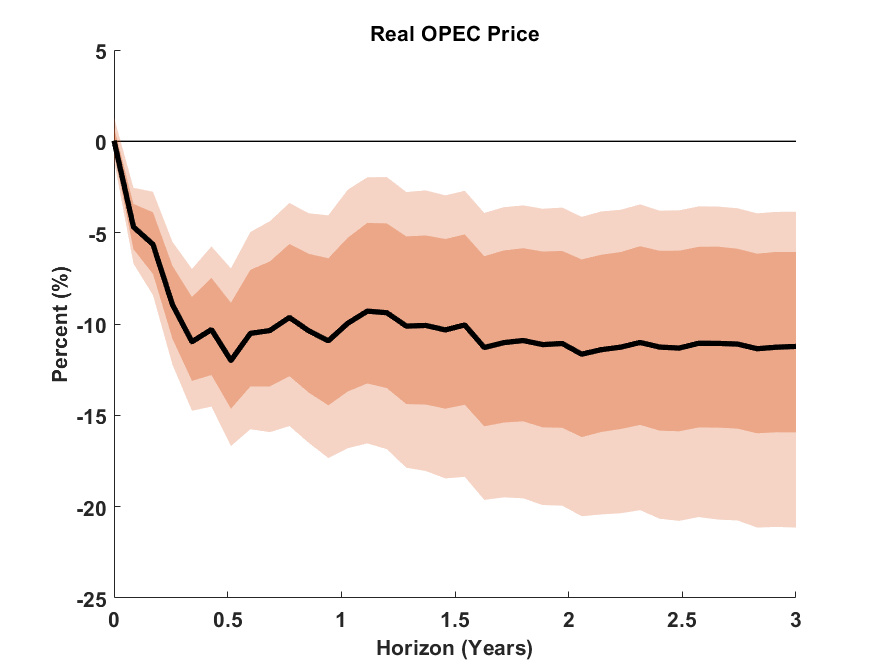}
\caption{Oil Price IRF, 2009-2019}
\end{subfigure}

\vspace{0.25cm}

\begin{footnotesize}

Figure \ref{fig:var_climpol_opec} shows the estimated impulse response functions for oil production and the spot price of oil for a shock to an alternative $ClimateTransition_t$ measure or the baseline $ClimateTransition_t$ but an alternative region for oil production and the spot price of oil. The black line is the estimated IRF, the dark red shaded region represents the one-standard deviation error band, and the light red shaded region represents the two-standard deviation error band. The top panels are for $ClimateTransition_t$ constructed using the \cite{ardia2023climate} Climate Change News Index, and the bottom panels are for OPEC oil production and the OPEC-basket spot price of oil for a shock to the baseline $ClimateTransition_t$ measure. The estimates are for the transition-risk-relevant time subsample (2009-2019). Bootstrapped error bands are calculated using 10,000 simulated samples. The VAR is estimated using 12 lags. See text for the full VAR specification used and definition of variables.
\end{footnotesize}
\end{figure}

Another feature I test empirically is variation in oil production competition and transition concerns for different regions. To do this, I re-estimate the VAR using OPEC production and prices. Figure \ref{fig:var_climpol_opec} (bottom panels) provides these estimates. The price and production responses are both consistent with the baseline results, and though point estimates are slightly larger for each, the production response is not statistically significant for almost all of the response horizon. While the insignificance of the estimates are roughly consistent with the model prediction that reduced competition diminishes the ``run'' effect, the outcomes also suggest the potential influence of transition risk, given the qualitative outcomes and size the of the point estimates, for OPEC during this time period when market competition increased and oil supply was high with the US becoming a significant oil producer and OPEC countries largely unable to agree on production cuts.

%\ref{var_alt_regions}
    
\newpage
\clearpage    

\subsection{Climate Transition Index Details}

%\vspace{-0.5cm}

\begin{table}[!ht]
\caption{Climate-Linked Transition Event List} \label{table:EventList1}
\begin{center}
%\vspace{0.25cm}
\vspace{-0.65cm}
\begin{tabular}{l c c c c}
\hline
\hline
Date & Event & Shock Sign & Category & Source \\
\hline
{\scriptsize 03/09/1982} & \multicolumn{1}{p{5.5cm}}{{\scriptsize World's largest wind farm in California's Altamont Pass begins operations}} & {\scriptsize  +} & {\scriptsize tech. innov. } & {\scriptsize ProCon.org } \\ 
%(Solar One)
{\scriptsize 04/12/1982} & \multicolumn{1}{p{5.5cm}}{{\scriptsize First large scale solar plant (Solar One)}} & {\scriptsize  +} & {\scriptsize tech. innov.} & {\scriptsize ProCon.org } \\ 
{\scriptsize 10/01/1982} & \multicolumn{1}{p{5.5cm}}{{\scriptsize First complete nuclear reactor decontamination/decommissioning in US}} & {\scriptsize  +} & {\scriptsize tech. innov.} & {\scriptsize ProCon.org } \\ 
{\scriptsize 01/07/1983} & \multicolumn{1}{p{5.5cm}}{{\scriptsize  Nuclear Waste Policy Act}} & {\scriptsize  +} & {\scriptsize US policy } & {\scriptsize Wikipedia } \\
{\scriptsize 11/06/1984} & \multicolumn{1}{p{5.5cm}}{{\scriptsize Ronald Reagan elected as POTUS}} & {\scriptsize  -} & {\scriptsize US policy } & {\scriptsize Wikipedia } \\ 
{\scriptsize 03/22/1985} & \multicolumn{1}{p{5.5cm}}{{\scriptsize Vienna Convention for the protection of the ozone layer}} & {\scriptsize  +} & {\scriptsize global policy } & {\scriptsize Wikipedia } \\ 
{\scriptsize 04/28/1986} & \multicolumn{1}{p{5.5cm}}{{\scriptsize Largest ever nuclear accident at Chernobyl}} & {\scriptsize -} & {\scriptsize disaster} & {\scriptsize ProCon.org } \\ 
{\scriptsize 09/16/1987} & \multicolumn{1}{p{5.5cm}}{{\scriptsize Montreal Protocol}} & {\scriptsize  +} & {\scriptsize global policy } & {\scriptsize Wikipedia } \\ 
{\scriptsize 10/19/1987} & \multicolumn{1}{p{5.5cm}}{{\scriptsize Brundtland Report}} & {\scriptsize  +} & {\scriptsize global policy } & {\scriptsize Wikipedia } \\ 
{\scriptsize 06/24/1988} & \multicolumn{1}{p{5.5cm}}{{\scriptsize James Hansen testifies to U.S. Senate that man-made global warming has begun}} & {\scriptsize  +} & {\scriptsize US policy } & {\scriptsize Wikipedia } \\
{\scriptsize 11/08/1988} & \multicolumn{1}{p{5.5cm}}{{\scriptsize George H. W. Bush elected as POTUS}} & {\scriptsize -} & {\scriptsize US policy } & {\scriptsize Wikipedia } \\ 
{\scriptsize 11/11/1988} & \multicolumn{1}{p{5.5cm}}{{\scriptsize IPCC established}} & {\scriptsize  +} & {\scriptsize global policy } & {\scriptsize Wikipedia, IPCC } \\
{\scriptsize 03/27/1989} & \multicolumn{1}{p{5.5cm}}{{\scriptsize Exxon Valdez disaster in Alaska becomes the largest oil spill in US waters}} & {\scriptsize  +} & {\scriptsize disaster } & {\scriptsize ProCon.org } \\ 
{\scriptsize 01/23/1990} & \multicolumn{1}{p{5.5cm}}{{\scriptsize Congress passes act to stimulate development of hydrogen power}} & {\scriptsize  +} & {\scriptsize US policy } & {\scriptsize ProCon.org } \\ 
{\scriptsize 08/30/1990} & \multicolumn{1}{p{5.5cm}}{{\scriptsize IPCC First AR published}} & {\scriptsize  +} & {\scriptsize global policy } & {\scriptsize Wikipedia, IPCC } \\
{\scriptsize 11/07/1990} & \multicolumn{1}{p{5.5cm}}{{\scriptsize IPCC and Second World Climate Conference call for global treaty}} & {\scriptsize  +} & {\scriptsize global policy } & {\scriptsize IPCC } \\
{\scriptsize 12/11/1990} & \multicolumn{1}{p{5.5cm}}{{\scriptsize UN General Assembly negotiations on a framework convention begin}} & {\scriptsize  +} & {\scriptsize global policy } & {\scriptsize IPCC } \\
{\scriptsize 09/03/1991} & \multicolumn{1}{p{5.5cm}}{{\scriptsize Publication of The First Global Revolution by the Club of Rome}} & {\scriptsize  +} & {\scriptsize global policy } & {\scriptsize Wikipedia } \\ 
{\scriptsize 05/11/1992} & \multicolumn{1}{p{5.5cm}}{{\scriptsize UNFCCC Convention adopted}} & {\scriptsize  +} & {\scriptsize global policy } & {\scriptsize IPCC } \\
{\scriptsize 06/15/1992} & \multicolumn{1}{p{5.5cm}}{{\scriptsize UNFCCC formed/Rio Earth Summit}} & {\scriptsize  +} & {\scriptsize global policy } & {\scriptsize IPCC } \\
{\scriptsize 06/15/1992} & \multicolumn{1}{p{5.5cm}}{{\scriptsize IPCC Supplementary Report published}} & {\scriptsize  +} & {\scriptsize global policy } & {\scriptsize Wikipedia, IPCC } \\ 
{\scriptsize 10/26/1992} & \multicolumn{1}{p{5.5cm}}{{\scriptsize Energy Policy Act}} & {\scriptsize  +} & {\scriptsize US policy } & {\scriptsize Wikipedia } \\
{\scriptsize 11/03/1992} & \multicolumn{1}{p{5.5cm}}{{\scriptsize Bill Clinton elected as POTUS}} & {\scriptsize  +} & {\scriptsize US policy } & {\scriptsize Wikipedia } \\ 
{\scriptsize 03/21/1994} & \multicolumn{1}{p{5.5cm}}{{\scriptsize UNFCCC enters into force}} & {\scriptsize  +} & {\scriptsize global policy } & {\scriptsize IPC } \\
{\scriptsize 04/07/1995} & \multicolumn{1}{p{5.5cm}}{{\scriptsize COP1 - Berlin, Germany}} & {\scriptsize  +} & {\scriptsize global policy } & {\scriptsize Wikipedia, IPCC } \\
{\scriptsize 12/15/1995} & \multicolumn{1}{p{5.5cm}}{{\scriptsize IPCC Second AR published}} & {\scriptsize  +} & {\scriptsize global policy } & {\scriptsize Wikipedia, IPCC } \\
{\scriptsize 06/05/1996} & \multicolumn{1}{p{5.5cm}}{{\scriptsize Solar Two plant demonstrates low cost method of storing solar energy}} & {\scriptsize  +} & {\scriptsize tech. innov.} & {\scriptsize ProCon.org } \\ 
{\scriptsize 06/26/1996} & \multicolumn{1}{p{5.5cm}}{{\scriptsize EU adopts max $2^{\circ}$C anomaly GMT target}} & {\scriptsize  +} & {\scriptsize global policy } & {\scriptsize Wikipedia } \\ 
\hline
\hline
\end{tabular}
\end{center}
\end{table}

\setcounter{table}{0}

\vspace{-1.5cm}

\begin{table}[!ht]
\caption{Climate-Linked Transition Event List (Continued)} \label{table:EventList2}
\begin{center}
%\vspace{0.25cm}
\vspace{-0.65cm}
\begin{tabular}{l c c c c}
\hline
\hline
Date & Event & Shock Sign & Category & Source \\
\hline
{\scriptsize 07/19/1996} & \multicolumn{1}{p{5.5cm}}{{\scriptsize COP2 - Geneva, Switzerland}} & {\scriptsize  +} & {\scriptsize global policy } & {\scriptsize Wikipedia, IPCC } \\
{\scriptsize 10/09/1996} & \multicolumn{1}{p{5.5cm}}{{\scriptsize Hydrogen Future Act}} & {\scriptsize  +} & {\scriptsize US policy } & {\scriptsize Wikipedia } \\ 
{\scriptsize 11/05/1996} & \multicolumn{1}{p{5.5cm}}{{\scriptsize Bill Clinton elected as POTUS}} & {\scriptsize  +} & {\scriptsize US policy } & {\scriptsize Wikipedia } \\ 
{\scriptsize 12/05/1996} & \multicolumn{1}{p{5.5cm}}{{\scriptsize GM EV1 electric car available to public}} & {\scriptsize  +} & {\scriptsize tech. innov.} & {\scriptsize ProCon.org } \\
{\scriptsize 06/25/1997} & \multicolumn{1}{p{5.5cm}}{{\scriptsize US Senate passes Byrd–Hagel Resolution}} & {\scriptsize -} & {\scriptsize US policy } & {\scriptsize Wikipedia } \\ 
{\scriptsize 12/11/1997} & \multicolumn{1}{p{5.5cm}}{{\scriptsize COP3/Kyoto Protocol agreed}} & {\scriptsize  +} & {\scriptsize global policy } & {\scriptsize Wikipedia, IPCC } \\
{\scriptsize 11/16/1998} & \multicolumn{1}{p{5.5cm}}{{\scriptsize COP4 - Buenos Aires, Argentina}} & {\scriptsize  +} & {\scriptsize global policy } & {\scriptsize Wikipedia, IPCC } \\
{\scriptsize 11/05/1999} & \multicolumn{1}{p{5.5cm}}{{\scriptsize COP5 - Bonn, Germany}} & {\scriptsize  +} & {\scriptsize global policy } & {\scriptsize Wikipedia, IPCC } \\
{\scriptsize 11/07/2000} & \multicolumn{1}{p{5.5cm}}{{\scriptsize George W. Bush elected as POTUS}} & {\scriptsize  -} & {\scriptsize US policy } & {\scriptsize Wikipedia } \\ 
{\scriptsize 11/27/2000} & \multicolumn{1}{p{5.5cm}}{{\scriptsize COP6 - The Hague, Netherlands}} & {\scriptsize  +} & {\scriptsize global policy } & {\scriptsize Wikipedia, IPCC } \\
{\scriptsize 03/28/2001} & \multicolumn{1}{p{5.5cm}}{{\scriptsize Bush withdraws from Kyoto negotiations}} & {\scriptsize  -} & {\scriptsize US policy } & {\scriptsize Wikipedia } \\ 
{\scriptsize 04/06/2001} & \multicolumn{1}{p{5.5cm}}{{\scriptsize IPCC Third AR published}} & {\scriptsize  +} & {\scriptsize global policy } & {\scriptsize Wikipedia, IPCC } \\ 
{\scriptsize 07/27/2001} & \multicolumn{1}{p{5.5cm}}{{\scriptsize COP6 - Bonn, Germany}} & {\scriptsize  +} & {\scriptsize global policy } & {\scriptsize Wikipedia, IPCC } \\
{\scriptsize 11/12/2001} & \multicolumn{1}{p{5.5cm}}{{\scriptsize COP7 - Marrakech, Morocco}} & {\scriptsize  +} & {\scriptsize global policy } & {\scriptsize Wikipedia, IPCC } \\
{\scriptsize 11/01/2002} & \multicolumn{1}{p{5.5cm}}{{\scriptsize COP8 - New Delhi, India}} & {\scriptsize  +} & {\scriptsize global policy } & {\scriptsize Wikipedia, IPCC } \\
{\scriptsize 02/06/2003} & \multicolumn{1}{p{5.5cm}}{{\scriptsize Bush unveils Hydrogen Fuel Initiative}} & {\scriptsize  +} & {\scriptsize US policy } & {\scriptsize ProCon.org } \\ 
{\scriptsize 02/27/2003} & \multicolumn{1}{p{5.5cm}}{{\scriptsize Plans announced to build world's first zero emissions coal power plant}} & {\scriptsize  +} & {\scriptsize tech. innov.} & {\scriptsize ProCon.org } \\ 
{\scriptsize 12/12/2003} & \multicolumn{1}{p{5.5cm}}{{\scriptsize COP9 - Milan, Italy}} & {\scriptsize  +} & {\scriptsize global policy } & {\scriptsize Wikipedia, IPCC } \\
{\scriptsize 11/02/2004} & \multicolumn{1}{p{5.5cm}}{{\scriptsize George W. Bush elected as POTUS}} & {\scriptsize  -} & {\scriptsize US policy } & {\scriptsize Wikipedia } \\ 
{\scriptsize 12/17/2004} & \multicolumn{1}{p{5.5cm}}{{\scriptsize COP10 - Buenos Aires, Argentina}} & {\scriptsize  +} & {\scriptsize global policy } & {\scriptsize Wikipedia, IPCC } \\
{\scriptsize 01/03/2005} & \multicolumn{1}{p{5.5cm}}{{\scriptsize EU emissions trading scheme launched}} & {\scriptsize  +} & {\scriptsize global policy } & {\scriptsize Wikipedia, IPCC } \\
{\scriptsize 02/16/2005} & \multicolumn{1}{p{5.5cm}}{{\scriptsize Kyoto Protocol enters into force}} & {\scriptsize  +} & {\scriptsize global policy } & {\scriptsize Wikipedia, IPCC } \\
{\scriptsize 07/08/2005} & \multicolumn{1}{p{5.5cm}}{{\scriptsize 31st G8 summit discusses climate change}} & {\scriptsize  +} & {\scriptsize global policy } & {\scriptsize Wikipedia } \\
{\scriptsize 08/08/2005} & \multicolumn{1}{p{5.5cm}}{{\scriptsize  Energy Policy Act}} & {\scriptsize  +} & {\scriptsize US policy } & {\scriptsize Wikipedia } \\
{\scriptsize 11/09/2005} & \multicolumn{1}{p{5.5cm}}{{\scriptsize US House prevents drilling for oil in the Arctic National Refuge}} & {\scriptsize  +} & {\scriptsize US policy } & {\scriptsize ProCon.org } \\ 
{\scriptsize 12/09/2005} & \multicolumn{1}{p{5.5cm}}{{\scriptsize COP11 - Montreal, Canada}} & {\scriptsize  +} & {\scriptsize global policy } & {\scriptsize Wikipedia, IPCC } \\
{\scriptsize 01/03/2006} & \multicolumn{1}{p{5.5cm}}{{\scriptsize Clean Development Mechanism opens}} & {\scriptsize  +} & {\scriptsize global policy } & {\scriptsize IPCC} \\
{\scriptsize 10/30/2006} & \multicolumn{1}{p{5.5cm}}{{\scriptsize Stern Review is published}} & {\scriptsize  +} & {\scriptsize global policy } & {\scriptsize Wikipedia} \\
{\scriptsize 11/17/2006} & \multicolumn{1}{p{5.5cm}}{{\scriptsize COP12 - Nairobi, Kenya}} & {\scriptsize  +} & {\scriptsize global policy } & {\scriptsize Wikipedia, IPCC } \\
{\scriptsize 02/02/2007} & \multicolumn{1}{p{5.5cm}}{{\scriptsize IPCC Fourth AR published}} & {\scriptsize  +} & {\scriptsize global policy } & {\scriptsize Wikipedia, IPCC } \\
{\scriptsize 02/16/2007} & \multicolumn{1}{p{5.5cm}}{{\scriptsize Washington Declaration}} & {\scriptsize  +} & {\scriptsize global policy } & {\scriptsize Wikipedia } \\ 
{\scriptsize 06/07/2007} & \multicolumn{1}{p{5.5cm}}{{\scriptsize 33rd G8 summit}} & {\scriptsize  +} & {\scriptsize global policy } & {\scriptsize Wikipedia} \\ 
{\scriptsize 07/31/2007} & \multicolumn{1}{p{5.5cm}}{{\scriptsize UN General Assembly plenary debate}} & {\scriptsize  +} & {\scriptsize global policy } & {\scriptsize Wikipedia } \\ 
{\scriptsize 08/09/2007} & \multicolumn{1}{p{5.5cm}}{{\scriptsize  America COMPETES Act}} & {\scriptsize  +} & {\scriptsize US policy } & {\tiny Wikipedia } \\
{\scriptsize 08/31/2007} & \multicolumn{1}{p{5.5cm}}{{\scriptsize Vienna Climate Change Talks}} & {\scriptsize  +} & {\scriptsize global policy } & {\scriptsize Wikipedia } \\ 
{\scriptsize 09/24/2007} & \multicolumn{1}{p{5.5cm}}{{\scriptsize  United Nations high-level-event}} & {\scriptsize  +} & {\scriptsize global policy } & {\scriptsize Wikipedia } \\ 
{\scriptsize 09/28/2007} & \multicolumn{1}{p{5.5cm}}{{\scriptsize Washington Conference}} & {\scriptsize  +} & {\scriptsize global policy } & {\scriptsize Wikipedia } \\ 
{\scriptsize 11/19/2007} & \multicolumn{1}{p{5.5cm}}{{\scriptsize IPCC Report concludes climate change is happening, mostly human caused}} & {\scriptsize  +} & {\scriptsize global policy } & {\scriptsize ProCon.org} \\ 
\hline
\hline
\end{tabular}
\end{center}
\end{table}

\setcounter{table}{0}

\vspace{-1.5cm}

\begin{table}[!ht]
\caption{Climate-Linked Transition Event List (Continued)} \label{table:EventList3}
\begin{center}
%\vspace{0.25cm}
\vspace{-0.65cm}
\begin{tabular}{l c c c c}
\hline
\hline
Date & Event & Shock Sign & Category & Source \\
\hline
{\scriptsize 12/17/2007} & \multicolumn{1}{p{5.5cm}}{{\scriptsize COP13 - Bali, Indonesia}} & {\scriptsize  +} & {\scriptsize global policy } & {\scriptsize Wikipedia, IPCC } \\
{\scriptsize 12/19/2007} & \multicolumn{1}{p{5.5cm}}{{\scriptsize  Energy Independence and Security Act}} & {\scriptsize  +} & {\scriptsize US policy } & {\scriptsize Wikipedia } \\
{\scriptsize 01/02/2008} & \multicolumn{1}{p{5.5cm}}{{\scriptsize Joint Implementation Mechanism starts}} & {\scriptsize  +} & {\scriptsize global policy } & {\scriptsize IPCC } \\
{\scriptsize 01/30/2008} & \multicolumn{1}{p{5.5cm}}{{\scriptsize First commercial cellulosic ethanol plant goes into production in Wyoming}} & {\scriptsize  +} & {\scriptsize tech. innov. } & {\scriptsize ProCon.org } \\ 
{\scriptsize 05/22/2008} & \multicolumn{1}{p{5.5cm}}{{\scriptsize Strategic Petroleum Reserve Fill Suspension and Consumer Protection Act}} & {\scriptsize  +} & {\scriptsize US policy } & {\scriptsize Wikipedia } \\
{\scriptsize 06/18/2008} & \multicolumn{1}{p{5.5cm}}{{\scriptsize Food, Conservation, and Energy Act}} & {\scriptsize  +} & {\scriptsize US policy } & {\scriptsize Wikipedia } \\
{\scriptsize 09/30/2008} & \multicolumn{1}{p{5.5cm}}{{\scriptsize Australian Garnaut review published}} & {\scriptsize  +} & {\scriptsize global policy } & {\scriptsize Wikipedia } \\
{\scriptsize 10/03/2008} & \multicolumn{1}{p{5.5cm}}{{\scriptsize Public Law 110-343 (Energy Improvement and Extension Act)}} & {\scriptsize  +} & {\scriptsize US policy } & {\scriptsize Wikipedia } \\
{\scriptsize 10/07/2008} & \multicolumn{1}{p{5.5cm}}{{\scriptsize National Biofuel Action Plan unveiled}} & {\scriptsize  +} & {\scriptsize US policy } & {\scriptsize ProCon.org } \\ 
{\scriptsize 11/04/2008} & \multicolumn{1}{p{5.5cm}}{{\scriptsize Barack Obama elected as POTUS}} & {\scriptsize  +} & {\scriptsize US policy } & {\scriptsize Wikipedia } \\ 
{\scriptsize 11/26/2008} & \multicolumn{1}{p{5.5cm}}{{\scriptsize UK Parliament passes Climate Change Act}} & {\scriptsize  +} & {\scriptsize global policy } & {\scriptsize Wikipedia } \\
{\scriptsize 12/15/2008} & \multicolumn{1}{p{5.5cm}}{{\scriptsize COP14 - Poznań, Poland}} & {\scriptsize  +} & {\scriptsize global policy } & {\scriptsize Wikipedia, IPCC } \\
{\scriptsize 12/22/2008} & \multicolumn{1}{p{5.5cm}}{{\scriptsize Worst coal ash spill in US history in Kingston, Tennessee}} & {\scriptsize  +} & {\scriptsize disaster } & {\scriptsize ProCon.org } \\ 
{\scriptsize 02/17/2009} & \multicolumn{1}{p{5.5cm}}{{\scriptsize  American Recovery and Reinvestment Act}} & {\scriptsize  +} & {\scriptsize US policy } & {\scriptsize Wikipedia, ProCon.org } \\
{\scriptsize 04/22/2009} & \multicolumn{1}{p{5.5cm}}{{\scriptsize First framework for wind energy development on US outer continental shelf announced}} & {\scriptsize  +} & {\scriptsize US policy } & {\scriptsize ProCon.org } \\ 
{\scriptsize 05/05/2009} & \multicolumn{1}{p{5.5cm}}{{\scriptsize Obama issues presidential directive to USDA to expand access to biofuels}} & {\scriptsize  +} & {\scriptsize US policy } & {\scriptsize ProCon.org } \\ 
{\scriptsize 05/27/2009} & \multicolumn{1}{p{5.5cm}}{{\scriptsize US announces millions in Recovery Act funding for solar and geothermal energy}} & {\scriptsize  +} & {\scriptsize US policy } & {\scriptsize ProCon.org } \\ 
{\scriptsize 06/26/2009} & \multicolumn{1}{p{5.5cm}}{{\scriptsize US House of Representatives passes the American Clean Energy and Security Act}} & {\scriptsize  +} & {\scriptsize US policy } & {\scriptsize Wikipedia } \\
{\scriptsize 09/22/2009} & \multicolumn{1}{p{5.5cm}}{{\scriptsize United Nations Secretary General's Summit on Climate Change.}} & {\scriptsize  +} & {\scriptsize global policy } & {\scriptsize Wikipedia } \\ 
{\scriptsize 10/27/2009} & \multicolumn{1}{p{5.5cm}}{{\scriptsize US invests billions to modernize grid}} & {\scriptsize  +} & {\scriptsize US policy } & {\scriptsize ProCon.org } \\ 
{\scriptsize 12/18/2009} & \multicolumn{1}{p{5.5cm}}{{\scriptsize COP15 - Copenhagen, Denmark}} & {\scriptsize  +} & {\scriptsize global policy } & {\scriptsize Wikipedia, IPCC } \\
{\scriptsize 04/20/2010} & \multicolumn{1}{p{5.5cm}}{{\scriptsize BP oil rig explodes and causes largest oil spill in US history}} & {\scriptsize  +} & {\scriptsize disaster } & {\scriptsize ProCon.org } \\ 
{\scriptsize 12/10/2010} & \multicolumn{1}{p{5.5cm}}{{\scriptsize COP16 - Cancún, Mexico}} & {\scriptsize  +} & {\scriptsize global policy } & {\scriptsize Wikipedia, IPCC } \\
{\scriptsize 03/11/2011} & \multicolumn{1}{p{5.5cm}}{{\scriptsize Earthquake off coast of Japan damages six powerplants at Fukushima Dai-ichi}} & {\scriptsize  -} & {\scriptsize disaster } & {\scriptsize ProCon.org } \\
{\scriptsize 09/01/2011} & \multicolumn{1}{p{5.5cm}}{{\scriptsize Solyndra declares bankruptcy after receiving federal loan guarantees}} & {\scriptsize  -} & {\scriptsize tech. innov. } & {\scriptsize ProCon.org } \\
{\scriptsize 12/09/2011} & \multicolumn{1}{p{5.5cm}}{{\scriptsize COP17 - Durban, South Africa}} & {\scriptsize  +} & {\scriptsize global policy } & {\scriptsize Wikipedia, IPCC } \\
{\scriptsize 02/09/2012} & \multicolumn{1}{p{5.5cm}}{{\scriptsize US NRC approves first new nuclear power plants since 1978}} & {\scriptsize  +} & {\scriptsize US policy } & {\scriptsize ProCon.org } \\
\hline
\hline
\end{tabular}
\end{center}
\end{table}

\setcounter{table}{0}

\vspace{-0.75cm}

\begin{table}[!ht]
\caption{Climate-Linked Transition Event List (Continued)} \label{table:EventList4}
\begin{center}
%\vspace{0.25cm}
\vspace{-0.65cm}
\begin{tabular}{l c c c c}
\hline
\hline
Date & Event & Shock Sign & Category & Source \\
\hline
{\scriptsize 03/27/2012} & \multicolumn{1}{p{5.5cm}}{{\scriptsize EPA announces first CAA standard for carbon pollution from new power plants}} & {\scriptsize  +} & {\scriptsize US policy } & {\scriptsize ProCon.org } \\
{\scriptsize 04/17/2012} & \multicolumn{1}{p{5.5cm}}{{\scriptsize EPA issues first ever clean air rules for natural gas produced by fracking}} & {\scriptsize  +} & {\scriptsize US policy } & {\scriptsize ProCon.org } \\
{\scriptsize 11/06/2012} & \multicolumn{1}{p{5.5cm}}{{\scriptsize Barack Obama elected as POTUS}} & {\scriptsize  +} & {\scriptsize US policy } & {\scriptsize Wikipedia } \\ 
{\scriptsize 12/07/2012} & \multicolumn{1}{p{5.5cm}}{{\scriptsize COP18 - Doha, Qatar}} & {\scriptsize  +} & {\scriptsize global policy } & {\scriptsize Wikipedia, IPCC } \\
{\scriptsize 06/25/2013} & \multicolumn{1}{p{5.5cm}}{{\scriptsize Obama releases Climate Action Plan}} & {\scriptsize  +} & {\scriptsize US policy } & {\scriptsize ProCon.org } \\
{\scriptsize 09/05/2013} & \multicolumn{1}{p{5.5cm}}{{\scriptsize Pacific Islands issue Majuro Declaration}} & {\scriptsize  +} & {\scriptsize global policy } & {\scriptsize Wikipedia } \\
{\scriptsize 09/20/2013} & \multicolumn{1}{p{5.5cm}}{{\scriptsize EPA issues new proposal to cut GHG emissions from new power plants}} & {\scriptsize  +} & {\scriptsize US policy } & {\scriptsize ProCon.org } \\
{\scriptsize 09/27/2013} & \multicolumn{1}{p{5.5cm}}{{\scriptsize IPCC Fifth AR published (Pt. 1)}} & {\scriptsize  +} & {\scriptsize global policy } & {\scriptsize Wikipedia, IPCC } \\
{\scriptsize 11/25/2013} & \multicolumn{1}{p{5.5cm}}{{\scriptsize COP19 - Warsaw, Poland}} & {\scriptsize  +} & {\scriptsize global policy } & {\scriptsize Wikipedia, IPCC } \\
{\scriptsize 02/13/2014} & \multicolumn{1}{p{5.5cm}}{{\scriptsize Ivanpah, the world's largest solar power generation plant, goes online}} & {\scriptsize  +} & {\scriptsize tech. innov. } & {\scriptsize ProCon.org } \\
{\scriptsize 03/31/2014} & \multicolumn{1}{p{5.5cm}}{{\scriptsize IPCC Fifth AR published (Pt. 2)}} & {\scriptsize  +} & {\scriptsize global policy } & {\scriptsize IPCC } \\
{\scriptsize 05/09/2014} & \multicolumn{1}{p{5.5cm}}{{\scriptsize Obama announces solar power commitments and executive actions}} & {\scriptsize  +} & {\scriptsize US policy } & {\scriptsize ProCon.org } \\
{\scriptsize 06/02/2014} & \multicolumn{1}{p{5.5cm}}{{\scriptsize EPA proposes first ever rules to reduce emissions from existing power plants}} & {\scriptsize  +} & {\scriptsize US policy } & {\scriptsize ProCon.org } \\
{\scriptsize 09/22/2014} & \multicolumn{1}{p{5.5cm}}{{\scriptsize Rockefellers and over 800 global investors announce fossil fuel divestment}} & {\scriptsize  +} & {\scriptsize global policy } & {\scriptsize ProCon.org } \\
{\scriptsize 09/23/2014} & \multicolumn{1}{p{5.5cm}}{{\scriptsize UN Secretary-General's Climate Summit}} & {\scriptsize  +} & {\scriptsize global policy } & {\scriptsize IPCC } \\
{\scriptsize 12/12/2014} & \multicolumn{1}{p{5.5cm}}{{\scriptsize COP20 - Lima, Peru}} & {\scriptsize  +} & {\scriptsize global policy } & {\scriptsize Wikipedia, IPCC } \\
{\scriptsize 08/03/2015} & \multicolumn{1}{p{5.5cm}}{{\scriptsize Obama Announces CPP}} & {\scriptsize  +} & {\scriptsize US policy } & {\scriptsize Wikipedia, ProCon.org } \\
{\scriptsize 10/23/2015} & \multicolumn{1}{p{5.5cm}}{{\scriptsize  CPP carbon emissions rules published}} & {\scriptsize  +} & {\scriptsize US policy } & {\scriptsize Wikipedia } \\
{\scriptsize 12/14/2015} & \multicolumn{1}{p{5.5cm}}{{\scriptsize  COP21/Paris Agreement adopted}} & {\scriptsize  +} & {\scriptsize global policy } & {\scriptsize Wikipedia, IPCC } \\
{\scriptsize 02/09/2016} & \multicolumn{1}{p{5.5cm}}{{\scriptsize  SCOTUS places hold on CPP}} & {\scriptsize  -} & {\scriptsize US policy } & {\scriptsize Wikipedia } \\
{\scriptsize 11/08/2016} & \multicolumn{1}{p{5.5cm}}{{\scriptsize Trump elected as POTUS}} & {\scriptsize  -} & {\scriptsize US policy } & {\scriptsize Wikipedia } \\
{\scriptsize 11/18/2016} & \multicolumn{1}{p{5.5cm}}{{\scriptsize COP22 - Marrakech, Morocco}} & {\scriptsize  +} & {\scriptsize global policy } & {\scriptsize Wikipedia, IPCC } \\
{\scriptsize 03/28/2017} & \multicolumn{1}{p{5.5cm}}{{\scriptsize Trump orders CPP review}} & {\scriptsize  -} & {\scriptsize US policy } & {\scriptsize Wikipedia, ProCon.org } \\
{\scriptsize 04/24/2017} & \multicolumn{1}{p{5.5cm}}{{\scriptsize  March for Science, People's Climate March protest Trump climate change}} & {\scriptsize  +} & {\scriptsize US policy } & {\scriptsize Wikipedia } \\
{\scriptsize 06/01/2017} & \multicolumn{1}{p{5.5cm}}{{\scriptsize Trump withdraws from Paris Accord/US Climate Alliance announced}} & {\scriptsize  +} & {\scriptsize US policy } & {\scriptsize Wikipedia } \\
{\scriptsize 07/31/2017} & \multicolumn{1}{p{5.5cm}}{{\scriptsize Two nuclear power reactors in South Carolina abandoned before completion}} & {\scriptsize  -} & {\scriptsize tech. innov. } & {\scriptsize ProCon.org } \\
{\scriptsize 11/17/2017} & \multicolumn{1}{p{5.5cm}}{{\scriptsize COP23 - Bonn, Germany}} & {\scriptsize  +} & {\scriptsize global policy } & {\scriptsize Wikipedia, IPCC } \\
{\scriptsize 12/12/2017} & \multicolumn{1}{p{5.5cm}}{{\scriptsize One Planet Summit}} & {\scriptsize  +} & {\scriptsize global policy } & {\scriptsize IPCC } \\
{\scriptsize 12/22/2017} & \multicolumn{1}{p{5.5cm}}{{\scriptsize Tax bill opens Arctic National Wildlife Refuge for oil drilling}} & {\scriptsize  -} & {\scriptsize US policy } & {\scriptsize ProCon.org } \\
\hline
\hline
\end{tabular} 
\end{center} 

\end{table}

\setcounter{table}{0}

\vspace{-0.75cm}

\begin{table}[!ht]
\caption{Climate-Linked Transition Event List (Continued)} \label{table:EventList5}
\begin{center}
%\vspace{0.25cm}
\vspace{-0.65cm}
\begin{tabular}{l c c c c}
\hline
\hline
Date & Event & Shock Sign & Category & Source \\
\hline
{\scriptsize 05/09/2018} & \multicolumn{1}{p{5.5cm}}{{\scriptsize Solar power to be required on all new California homes by 2020}} & {\scriptsize  +} & {\scriptsize US policy } & {\scriptsize ProCon.org } \\
{\scriptsize 10/08/2018} & \multicolumn{1}{p{5.5cm}}{{\scriptsize IPCC $1.5^{\circ} C$ Special Report published.}} & {\scriptsize  +} & {\scriptsize global policy } & {\scriptsize Wikipedia, IPCC } \\ 
{\scriptsize 12/14/2018} & \multicolumn{1}{p{5.5cm}}{{\scriptsize COP24 - Katowice, Poland}} & {\scriptsize  +} & {\scriptsize global policy } & {\scriptsize Wikipedia, IPCC } \\
{\scriptsize 03/22/2019} & \multicolumn{1}{p{5.5cm}}{{\scriptsize New Mexico commits to 100\% renewable energy for electricity by 2050}} & {\scriptsize  +} & {\scriptsize US policy } & {\scriptsize ProCon.org } \\
{\scriptsize 04/29/2019} & \multicolumn{1}{p{5.5cm}}{{\scriptsize Ocasio-Cortez and Markey introduce resolution calling for a Green New Deal}} & {\scriptsize  +} & {\scriptsize US policy } & {\scriptsize Wikipedia } \\
{\scriptsize 06/27/2019} & \multicolumn{1}{p{5.5cm}}{{\scriptsize SB50 - Bonn, Germany}} & {\scriptsize  +} & {\scriptsize global policy } & {\scriptsize Wikipedia, IPCC } \\
{\scriptsize 09/20/2019} & \multicolumn{1}{p{5.5cm}}{{\scriptsize Three Mile Island to close, site of worst commercial nuclear accident in US}} & {\scriptsize  -} & {\scriptsize tech. innov. } & {\scriptsize ProCon.org } \\
{\scriptsize 09/23/2019} & \multicolumn{1}{p{5.5cm}}{{\scriptsize UNSG's Climate Action Summit}} & {\scriptsize  +} & {\scriptsize global policy } & {\scriptsize Wikipedia, IPCC } \\
{\scriptsize 12/02/2019} & \multicolumn{1}{p{5.5cm}}{{\scriptsize EC issues a European Green Deal}} & {\scriptsize  +} & {\scriptsize global policy } & {\scriptsize Wikipedia } \\
{\scriptsize 12/13/2019} & \multicolumn{1}{p{5.5cm}}{{\scriptsize COP25 - Madrid, Spain}} & {\scriptsize  +} & {\scriptsize global policy } & {\scriptsize Wikipedia, IPCC } \\
\hline
\hline
\end{tabular} 
\end{center}

\end{table}

\begin{table}[!ht]
\caption{Sources Used to Construct Climate-Linked Transition Event List} \label{table:EventListSources}
\begin{center}
%\vspace{0.25cm}
\vspace{-0.65cm}
\begin{tabular}{l c c}
\hline
\hline
Source & Title & URL  \\
\hline
{\scriptsize ProCon.org} & \multicolumn{1}{p{6.0cm}}{{\scriptsize Historical Timeline: History of Alternative Energy and Fossil fuels}} & \multicolumn{1}{p{6.0cm}}{\scriptsize \url{https://alternativeenergy.procon.org/historical-timeline/} } \\
{\scriptsize IPCC} & \multicolumn{1}{p{6.0cm}}{{\scriptsize Archive of Publications and Data Reports}} & \multicolumn{1}{p{6.0cm}}{\scriptsize \url{https://archive.ipcc.ch/publications_and_data/publications_and_data_reports.shtml} } \\
{\scriptsize IPCC} & \multicolumn{1}{p{6.0cm}}{{\scriptsize UNFCCC -- 25 Years of Effort and Achievement: Key Milestones in the Evolution of International Policy Timelines}} & \multicolumn{1}{p{6.0cm}}{\scriptsize \url{https://unfccc.int/timeline/} } \\
{\scriptsize Wikipedia} & \multicolumn{1}{p{6.0cm}}{{\scriptsize United Nations Climate Change conference}} & \multicolumn{1}{p{6.0cm}}{\scriptsize  \url{https://en.wikipedia.org/wiki/United_Nations_Climate_Change_conference} } \\
{\scriptsize Wikipedia} & \multicolumn{1}{p{6.0cm}}{{\scriptsize Timeline of international climate politics}} & \multicolumn{1}{p{6.0cm}}{\scriptsize \url{https://en.wikipedia.org/wiki/Timeline_of_international_climate_politics} } \\
{\scriptsize Wikipedia} & \multicolumn{1}{p{6.0cm}}{{\scriptsize List of United State energy acts}} & \multicolumn{1}{p{6.0cm}}{\scriptsize \url{https://en.wikipedia.org/wiki/List_of_United_States_energy_acts} } \\
{\scriptsize Wikipedia} & \multicolumn{1}{p{6.0cm}}{{\scriptsize Clean Power Plan}} & \multicolumn{1}{p{6.0cm}}{\scriptsize \url{https://en.wikipedia.org/wiki/Clean_Power_Plan} } \\
{\scriptsize Wikipedia} & \multicolumn{1}{p{6.0cm}}{{\scriptsize Post–Kyoto Protocol negotiations on greenhouse gas emissions}} & \multicolumn{1}{p{6.0cm}}{\scriptsize \url{https://en.wikipedia.org/wiki/Post-Kyoto_Protocol_negotiations_on_greenhouse_gas_emissions} } \\
{\scriptsize Wikipedia} & \multicolumn{1}{p{6.0cm}}{{\scriptsize List of presidents of the United States}} & \multicolumn{1}{p{6.0cm}}{\scriptsize \url{https://en.wikipedia.org/wiki/List_of_presidents_of_the_United_States} } \\
\hline
\hline
\end{tabular} 
\end{center} 

\vspace{0.25cm}

\end{table}

\newpage
\clearpage

\subsection{Vector Autoregression Estimates} \label{VARApp}

\subsubsection[\texorpdfstring{$ClimateTransition_t$}{TEXT} Decomposition]{\texorpdfstring{$ClimateTransition_t$}{TEXT} Decomposition}
\label{decomp}

\newpage
\clearpage

\subsubsection[Alternative \texorpdfstring{$ClimateTransition_t$}{TEXT}]{Alternative \texorpdfstring{$ClimateTransition_t$}{TEXT}}
\label{alt_climtrans}

\newpage
\clearpage

\subsubsection{Text-Based \texorpdfstring{$ClimateTransition_t$}{TEXT} and Alternative Regions}
%\label{rea}

\newpage
\clearpage

\subsubsection{Alternative Fossil Fuels}
\label{alt_ff}

\end{appendices}

\end{document}